\def\ncco{Nd$_{\rm 2-x}$Ce$_{\rm x}$CuO$_{\rm 4-\delta}$}
\def\nca{Nd$_{\rm 1.85}$Ce$_{\rm 0.15}$CuO$_{\rm 4-\delta}$}
\def\ncb{Nd$_{\rm 1.92}$Ce$_{\rm 0.08}$CuO$_{\rm 4}$}
\def\nco{Nd$_{\rm 2}$CuO$_{\rm 4}$}
\def\lsco{La$_{\rm 2-x}$Sr$_{\rm x}$CuO$_{\rm 4}$}
\def\lscoc{La$_{\rm 1.85}$Sr$_{\rm 0.15}$CuO$_{\rm 4}$}
\def\lco{La$_{\rm 2}$CuO$_{\rm 4}$}
\def\bisco{Bi$_{\rm 2}$Sr$_{\rm 2}$CaCu$_{\rm 2}$O$_{\rm 8}$}
\documentclass[amstex,float,a4paper,twoside,12pt,cite,wrapfig]{bookp}
\usepackage{url}
\usepackage{graphicx}
\usepackage{fancyheadings}
\usepackage[numbers,sort&compress,comma]{natbib}
\usepackage{subfigure}
\usepackage[perpage,stable,symbol*]{footmisc}
\usepackage[dutch,USenglish]{babel}

\renewcommand{\chaptermark}[1]%
   {\markboth{{Chapter \thechapter.\ ~~ #1}}{}}
\renewcommand{\sectionmark}[1]%
   {\markright{{\thesection.\ #1}}}
\renewcommand\bibsection{\chapter*{\refname}
\addcontentsline{toc}{chapter}{\refname}} 
\newcommand{\helv}{%
   \fontfamily{phv}\fontseries{b}\fontsize{12}{15}\selectfont}
\newcommand{\helf}{%
   \fontfamily{phv}\fontseries{n}\fontsize{12}{15}\selectfont}
\begin{document}

\thispagestyle{empty} \vspace*{-1.2cm}
\begin{center}
 \vspace*{3.3cm}
{\bf \Huge New light on EuO thin films}\\
\vspace{0.5cm} {\bf \Large Preparation, transport, magnetism and
spectroscopy of a ferromagnetic
semiconductor}\\
\vspace*{15.7cm}

{\Large {\bf P. G. Steeneken\\}}
\end{center}
\newpage
\begin{center}
\end{center}
\thispagestyle{empty}
\newpage

\thispagestyle{empty} \vspace*{-1.2cm}
\begin{center}
{\large Rijksuniversiteit Groningen}\\ \vspace*{2.5cm}
{\bf \Huge New light on EuO thin films}\\
\vspace{0.5cm} {\bf \Large Preparation, transport, magnetism and spectroscopy of a ferromagnetic
semiconductor}\\
\vspace*{2.5cm} {\large Proefschrift}\\
\vspace*{1cm} ter verkrijging van het doctoraat in de\\ Wiskunde
en Natuurwetenschappen\\ aan de Rijksuniversiteit Groningen\\ op
gezag van de\\ Rector Magnificus, dr. F. Zwarts, \\ in het
openbaar te verdedigen op\\ vrijdag 11 oktober 2002\\ om 16.00 uur\\

\vspace*{2.5cm} door\\
\vspace*{1cm}

{\large {\bf Peter Gerard Steeneken\\} \vspace*{1cm} geboren
op 3 mei 1974\\ te Groningen}\\
\end{center}

\newpage
\thispagestyle{empty}
\begin{tabbing}
Promotores: \hspace{3cm} \= Prof. dr. G. A. Sawatzky \\
\> Prof. dr. ir. L. H. Tjeng \\
\\
Beoordelingscommissie: \> Prof. dr. M. C. Aronson\\
\> Prof. dr. R. Griessen\\ \> Prof. dr. D. van der Marel
\end{tabbing}

\vspace*{18cm} \noindent ISBN: 90-367-1695-0

\newpage
\thispagestyle{empty} \vspace*{20cm} \hspace{10cm} {\Large Voor
mijn ouders}
\newpage
\thispagestyle{empty} \noindent The colored surface on the front
cover represents a 3D graph of the logarithm of the resistivity
(from bottom to top) of a Eu-rich EuO film on ZrO$_2$ (001) at
temperatures from 10-300 K (from front to back) and at magnetic
fields of 0-5 T (from left to right). It is the same data set as
in figures \ref{Tmagres}(a) and \ref{Mmagres}(a). The back side
cover depicts the magnetization of EuO as calculated from a mean
field model in the same temperature-field range, with the
magnetization increasing from top to bottom. The degree of
alignment of the spins above the surface is also proportional to
the magnetization. Moreover their transparency disappears in the
range where Eu-rich EuO becomes metallic. The color of the
surface indicates the simultaneous red-shift. The state of the
electrons is projected by several light sources. The similarity
between the surfaces on front and back suggests a universal
relation between resistivity and magnetization below the Curie
temperature (see also figure \ref{Mmagres}(a)).
\\
\vspace*{7cm}
\begin{figure*}[h!]
\includegraphics[clip=true,width=0.85\textwidth]{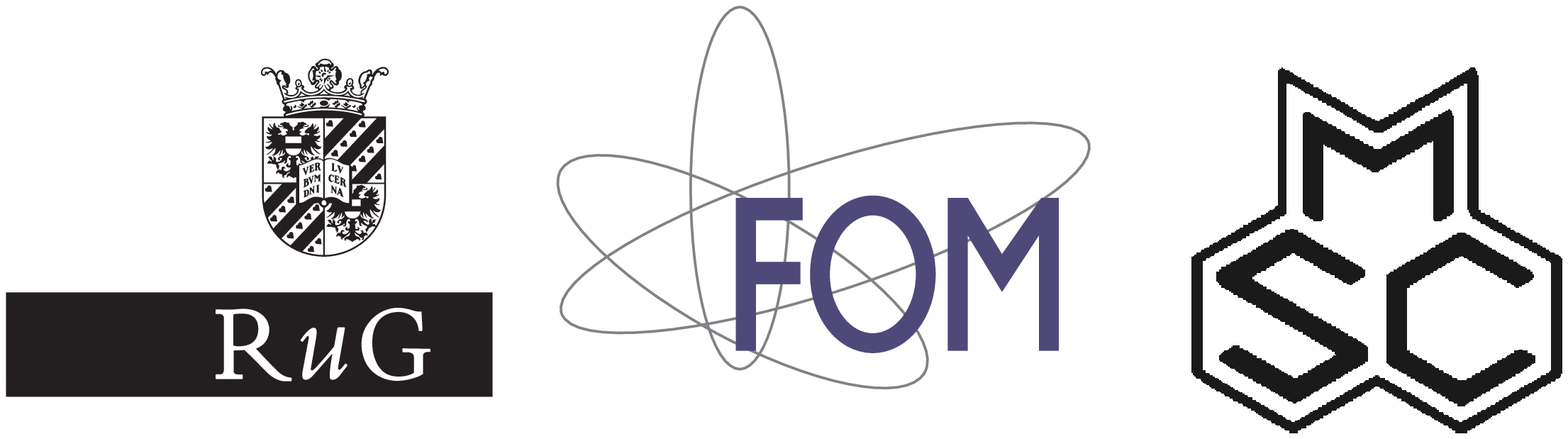}
\end{figure*}

\noindent The work described in this thesis was performed in the
Surface Science Group (part of the Materials Science Centre) at
the Solid State Physics Laboratory of the University of
Groningen, the Netherlands. The project was supported by the
Dutch Foundation for Fundamental Research on Matter (FOM) with
financial support from the Dutch Organization for the Advancement
of Pure Research
(NWO).\\
\\
\noindent Electronic edition (2002), Arxiv.org reprint in (2012).

\newpage

\pagenumbering{arabic} \tableofcontents \markboth{}{}
\def\ncco{Nd$_{\rm 2-x}$Ce$_{\rm x}$CuO$_{\rm 4-\delta}$}

\newpage


\chapter{Introduction}

\label{Introduction}


{\sl The extreme diversity in the properties of solids originates
for a large part from the multiformity of the many-electron
state. Most characteristics of condensed matter, like optical,
magnetic and transport properties, find their source mainly in
this electronic structure. For a deeper understanding of
condensed matter, knowledge of the electronic state of the system
is therefore essential.}

In this thesis, we will investigate europium monoxide (EuO), a
fascinating compound in which the dependence of macroscopic
properties on electronic structure is especially manifest, as it
experiences a phase transition at which a change in electronic
structure is accompanied by tremendous changes in magnetic,
optical and transport properties. Besides these spectacular
effects, EuO can also be considered as a model compound in
several respects, because it is a clear realization of the
existence of the Heisenberg ferromagnet \cite{Heisenberg28,Vanvleck38}, and is also considered as a model
compound for ionic ferromagnetic semiconductors. Additionally, it
is one of the simplest and clearest representatives of a much
studied class of ferromagnets in which the transition to the
ferromagnetic state at the Curie temperature is accompanied by a
change from a semiconducting to metallic conductivity.

As an introduction to the results of this thesis research, we
will review previous work on EuO, while focussing on subjects
addressed in this thesis. For additional information we refer the
reader to the many review articles that have been written on the
properties of EuO
\cite{Methfessel68,Kasuya68,Haas70,Wachter72,Kasuya72,Wachter79,Nolting79,Mauger86,Tsuda91,Nagaev01}.

\section{Introduction to the properties of EuO}
\label{introeuo}
\subsection{Structure and growth}

\begin{figure}[!htb]
\centerline{\includegraphics[width=10.5cm]{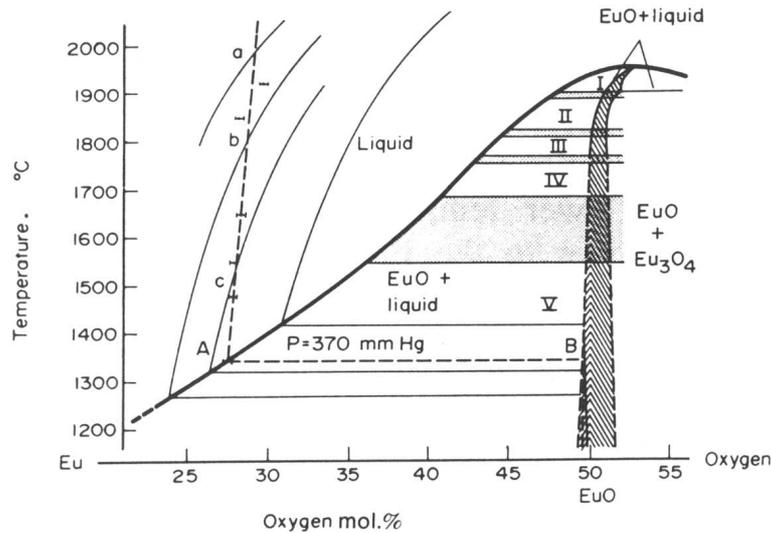}}
\caption{\label{phasediag} Phase diagram of Eu-O from
\cite{Shafer72}.}
\end{figure}

EuO has a rocksalt structure, with a room temperature lattice
constant of 5.144 \AA, which reduces to 5.127 \AA~below 10 K
\cite{Levy69}. The density of Eu atoms in EuO is 44\% higher than
in Eu metal, and is only 3\% lower than the concentration of Gd
atoms in ferromagnetic Gd metal. Europium monoxide has an ionic
Eu$^{2+}$O$^{2-}$ character, such that the electronic
configuration of europium is [Xe]$4f^75d^06s^0$ and that of
oxygen is $1s^22s^22p^6$. It was shown that homogeneous large
single crystals can be grown \cite{Guerci66,Shafer72,Fischer} at
temperatures of around 1800$^\circ$C. The phase diagram as
obtained by Shafer {\it et al.} \cite{Shafer72} is shown in
figure \ref{phasediag}. The stoichiometry of the resulting EuO
crystals depends significantly on the growth temperature, as is
indicated by the regions I to V in the phase diagram. Regions I
and II correspond to oxygen rich samples, region III to
stoichiometric EuO and regions IV and V correspond to Eu-rich
samples. If the starting composition contains excess oxygen, part
of the europium atoms will become Eu$^{3+}$ ions, and Eu
vacancies, Eu$_3$O$_4$ (see \cite{Holmes66,Holmes68,Batlogg75})
and Eu$_2$O$_3$ will form.

In many studies thin films of EuO have been grown
\cite{Ahn67,Ahn68,Ahn70,Lee70,Paparoditis71,Massenet74,Lubecka87,Guanghua96,Konno98,Iwata00,Iwata00b}.
It was shown that EuO films can be obtained by reactive
evaporation of Eu metal in a low oxygen pressure while heating
the substrate to $\sim 400^\circ$C. When such polycrystalline
films are grown on a glass substrate the grain size is $\sim$ 30
nm \cite{Massenet74}. Thin films of EuO can also be obtained by
partial oxidation of Eu metal films of less than 100 nm
thickness. Such oxidized films contain a large excess of Eu atoms
and have Eu metal inclusions, which can increase their Curie
temperature significantly \cite{Massenet74}. To prevent
uncontrolled reactions of the very reactive Eu atoms with oxygen
or water to Eu$_2$O$_3$ or Eu(OH)$_3$, it is important to prepare
and preserve the samples under (ultra) high vacuum conditions.
Alternatively the films can be covered by a protective layer,
although this can also affect their properties. The transport
properties are especially sensitive to oxidation or heating above
350$^\circ$C as can be seen from figure \ref{euoepi9}. The
ferromagnetism seems more robust, films of 20 nm which have been
exposed to air for one week still show magnetic hysteresis.
Although EuO thin films have often been grown, the growth
mechanism is not well understood. We will address this issue in
chapter \ref{euoepi}.

\subsection{Magnetism}

\subsubsection{Theories of Ferromagnetism}
Our present understanding of ferromagnetism is still strongly
influenced by the first conceptual models. The earliest attempt
at a quantitative analysis of ferromagnetism was made by Weiss
and is known as the mean (or molecular) field theory. Weiss
proposed that the local magnetic moments in a ferromagnet are
coupled by an effective macroscopic field $H_{\rm eff}$ along
$z$. In first order perturbation theory, the energy of an atom
with local moment $J$ in such a field is $E=g\mu_B H_{\rm eff}
J_z / (k_B T)=x J_z$. Therefore the free energy per ion is given
by:
\begin{equation}
\label{Weiss} \frac{1}{N} \mathcal{F} = E_0 - k_B T \ln {\left(
\sum_{J_z=-J}^J e^{-x J_z} \right)}
\end{equation}
In combination with the thermodynamic relation for the
magnetization $M=-\frac{\partial \mathcal{F}}{\partial H_{\rm
eff}}$ and the assumption that the contribution of each moment to
the effective field can be replaced by its mean contribution
($H_{\rm eff} = H_{\rm ext} + \lambda M$), the magnetization
versus temperature at any external field $H_{\rm ext}$ can be
calculated self-consistently. It can be shown that the
ferromagnetic Curie temperature, i.e. the temperature above which
the spontaneous magnetization vanishes, is given by $T_c =
\lambda N g^2 \mu_0 \mu_B^2 J(J+1) / 3 k_B$, where $N$ is the
density of magnetic moments. Above the Curie temperature, the
Curie-Weiss paramagnetic susceptibility is found: $\chi =
M/H_{\rm ext} = C / (T-\theta)$, with $\theta=T_c$ and
$C=T_c/\lambda$.

\begin{figure}[!htb]
\centerline{\includegraphics[width=11.25cm]{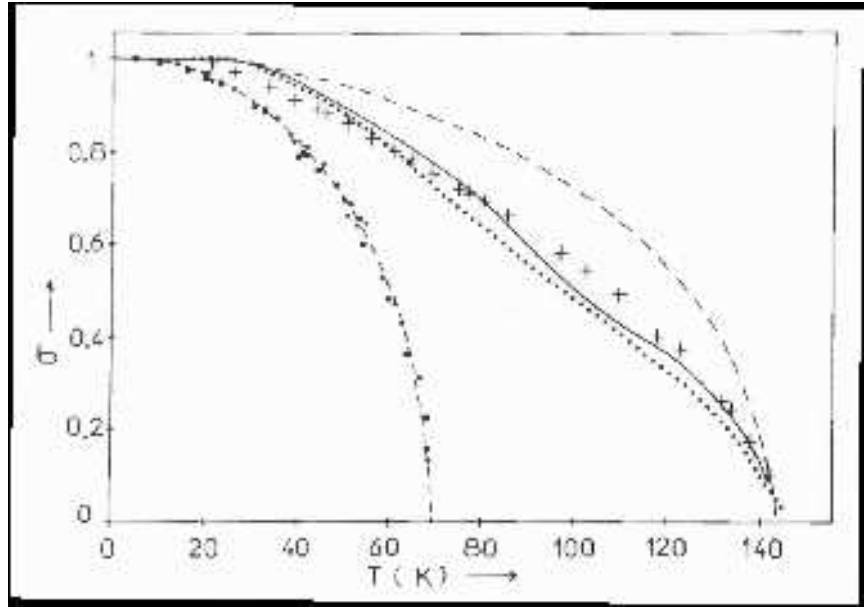}}
\caption{\label{EuOmag} Reduced magnetization $\sigma=M/M_{sat}$
versus temperature, reproduced from \cite{Mauger78}. Solid
circles and squares correspond to nearly stoichiometric single
crystals of EuO. Plus signs correspond to a 2\% Gd doped EuO
crystal. Dotted line corresponds to a Eu-rich thin EuO film grown
by oxidation of Eu metal (from \cite{Massenet74}). Dashed lines
are solutions to the mean field model for T$_c=69.5$~K and
T$_c=148$~K. The solid line is a calculation by Mauger {\it et
al.}}
\end{figure}

The mean field model is especially well-obeyed in compounds with
widely separated ions with large, spin-only moments
\cite{Fazekas99}. Moreover it works better at high coordination
numbers. For the fcc Ising model, T$_c$ predicted by the mean
field model is less than 20\% different from the exact solution
\cite{Fisher67,Ashcroft76}. As all these conditions are favorable
in stoichiometric EuO, the mean field model describes the
temperature dependence of the magnetization very well, as can be
seen in figure \ref{EuOmag}. However, even for EuO the mean field
model is unsatisfactory in several respects. It cannot account
for the difference between the ferromagnetic Curie temperature of
69.33 K \cite{Kornblit73} and the paramagnetic Curie temperature
of 79$\pm$4 K \cite{Kobler75}. Also the constant $C$, as found
from the paramagnetic susceptibility, is several percent larger
than the mean field value. Moreover, the mean field model does
not account exactly for the magnetization near the critical
temperature \cite{Wachter79}.

The Weiss theory does not explain the origin of the effective
field created by the magnetized spins. The first theoretical (and
experimental) attempts to explain ferromagnetism by Ewing
\cite{Bozorth51}, considered each atom in the ferromagnet as a
permanent magnet. The magnitude of the dipole-dipole interactions
between the atoms is however much too small to account for the
high Curie temperatures. As an example, the dipole-dipole
interaction in iron is of the order of 0.1 K, which is certainly
much too small to account for a Curie temperature of 1043 K
\cite{Fazekas99}.

Instead the mechanism which is usually responsible for
ferromagnetism, and for many other magnetic ordering effects, is
the mechanism of exchange. Exchange is a result of the
combination of the Pauli principle and Coulomb repulsion between
electrons. It can be understood by the following simple argument:
\emph{as the Pauli principle forbids spin-parallel electrons to
be in the same state, they will spatially separate, this spatial
separation will reduce their Coulomb repulsion and thus lower
their energy with respect to the spin-antiparallel
configuration.} On the other hand, the space accessible to the
electrons will often reduce in the spin-parallel configuration
and thus raise their kinetic energy. Therefore ferromagnetism
will only occur when the reduction in Coulomb energy as a result
of exchange will be larger than the kinetic energy increase.

It was shown by Heisenberg \cite{Heisenberg28} that the effective
field postulated by Weiss can be explained in terms of exchange
interactions between electrons in neighboring atoms. This led to
the much studied Heisenberg Hamiltonian:
\begin{equation}
\mathcal{H}=-J_{\rm ex} \sum_\mathbf{<i,j>}\mathbf{S_i\cdot S_j}
\end{equation}
where the brackets indicate that the sum has to be taken over
nearest neighbors and the exchange energy $J_{\rm ex}$ is
positive in the ferromagnetic Heisenberg model. From this model
it can be shown \cite{Ashcroft76} that the effective mean field
parameter $\lambda = \sum_{<i,j=0>} J_{\rm ex} / (N g^2 \mu_0
\mu_B^2)$. The Heisenberg model describes the magnetic properties
of EuO close to the Curie temperature better than mean field
theory \cite{Wachter79} and it can account for the spin-wave
spectrum of EuO \cite{Dietrich75}.

On the other hand, the Heisenberg model neglects the changes in
kinetic energy as a result of exchange. Moreover, for the
transition-metal ferromagnets the saturation magnetization is
usually significantly lower than the prediction of the Heisenberg
model of fully aligned local moments at T$=0$ K. Stoner proposed
a band model of ferromagnetism \cite{Stoner38} which resolved
these issues, and also reproduced the mean field results. Let us
describe it briefly.

For a system to become ferromagnetic, it has to develop a
spin-polarization ($n_\uparrow \neq n_\downarrow$). In a band
model the spin-dependent electron density can then be written as:
\begin{equation}
n_\uparrow = \frac{n}{2}+m \hspace{1cm}
n_\downarrow=\frac{n}{2}-m
\end{equation}
In the mean field approximation it can be shown that the exchange
interaction is given by
\begin{equation}
E_{ex}(m) \approx \frac{1}{2} \left( (n_\uparrow^2 +
n_\downarrow^2) V_{\uparrow \uparrow} + 2 n_\uparrow n_\downarrow
V_{\uparrow \downarrow} \right) =  const - m^2 U
\end{equation}
Where $U = V_{\uparrow \downarrow}-V_{\uparrow \uparrow}$ is the
difference between the repulsive Coulomb potential among
spin-antiparallel and spin-parallel electron gases\footnote{By
applying the mean field approximation to the Hubbard model, the
same expression is obtained with $U$ equal to the Hubbard $U$
\cite{Fazekas99}.}. $V_{\uparrow \uparrow}<V_{\uparrow
\downarrow}$ because the Pauli principle will maintain a larger
spatial separation between spin-parallel electrons. On the other
hand the average kinetic energy of the electrons will increase.
For small magnetization $m$ the spin-up and spin-down chemical
potentials will shift to $\mu_{\uparrow,\downarrow} = \pm
m/\rho(\varepsilon_F)$, i.e. a spin-splitting proportional to the
magnetization. $\rho(\varepsilon_F)$ is the density of states at
the Fermi level. Therefore the total energy change as a result of
the spin-polarization is given by:
\begin{equation}
\Delta E(m) = \int_0^{m/\rho(\varepsilon_F)} \varepsilon
\rho(\varepsilon_F) d\varepsilon -
\int_{-m/\rho(\varepsilon_F)}^0 \varepsilon \rho(\varepsilon_F)
d\varepsilon - U m^2 = m^2(\frac{1}{\rho(\varepsilon_F)}-U)
\end{equation}

This indicates that ferromagnetism will only occur when $U >
\frac{1}{\rho(\varepsilon_F)}$, which is called the Stoner
criterion and illustrates the competition between kinetic and
exchange interactions. The success of the Stoner model can be
understood from the fact that it accounts for fractional magnetic
moments and because it shows that ferromagnetism will more likely
occur in narrow band materials, like the transition metals, as
the bandwidth scales approximately as $W \propto
\frac{1}{\rho(\varepsilon_F)}$.

Although the Stoner model might be a good starting point for the
description of the transition metals, strong correlation effects
in the $3d$ bands have to be considered. This is a topic of
ongoing research\footnote{It seems that the conditions for
ferromagnetism in the Hubbard model are rather strict. This is
largely due to the fact that the Hubbard model in its simplest
form does not consider Hund's first rule
\cite{Held98,Fazekas99}.}. For $f$ electron systems the Stoner
model is certainly inappropriate, as $f$ electrons are so highly
localized that direct exchange interactions with electrons on
neighboring atoms are negligible. Zener \cite{Zener51,Zener51b}
proposed that localized moments might instead be aligned via
double exchange or via the exchange interactions with delocalized
conduction electrons. This RKKY exchange coupling via conduction
electrons, was shown to be of long-range, oscillatory nature
\cite{Ruderman54,Kasuya56,Yosida57} and is indeed often
responsible for the magnetic ordering phenomena in rare-earth
metals, like the ferromagnetism in Gd and Tb metal\footnote{Kondo
interactions can however lead to fascinating complications, like
the formation of a Kondo singlet phase
\cite{Fazekas99,Jensen91}.}.

\subsubsection{Ferromagnetic insulators/semiconductors}

In the case of an insulator these exchange mechanisms via free
electrons of course do not work. Additionally, Kramers
superexchange \cite{Kramers34} generally leads to
antiferromagnetic ordering. Therefore, only exchange via virtual
excitations of valence band electrons to the conduction band,
like that of Bloembergen and Rowland \cite{Bloembergen55}, can
lead to a ferromagnetic interaction. It is thus not surprising
that the existence of a ferromagnetic semiconductor or insulator
was seriously doubted in the fifties~\cite{Wachter79}. This issue
was resolved with the discovery of the ferromagnetic
semiconductors CrBr$_3$ \cite{Tsubokawa} in 1960 and EuO one year
later \cite{Matthias}. Soon afterwards the other magnetic
ordering europium chalcenogides were found, ferromagnetic EuS,
EuSe (ferro/ferri/antiferro-magnetic \cite{Griessen71}) and
antiferromagnetic EuTe.

\begin{figure}[!htb]
\vspace{0.5cm}
\centerline{\includegraphics[width=8cm]{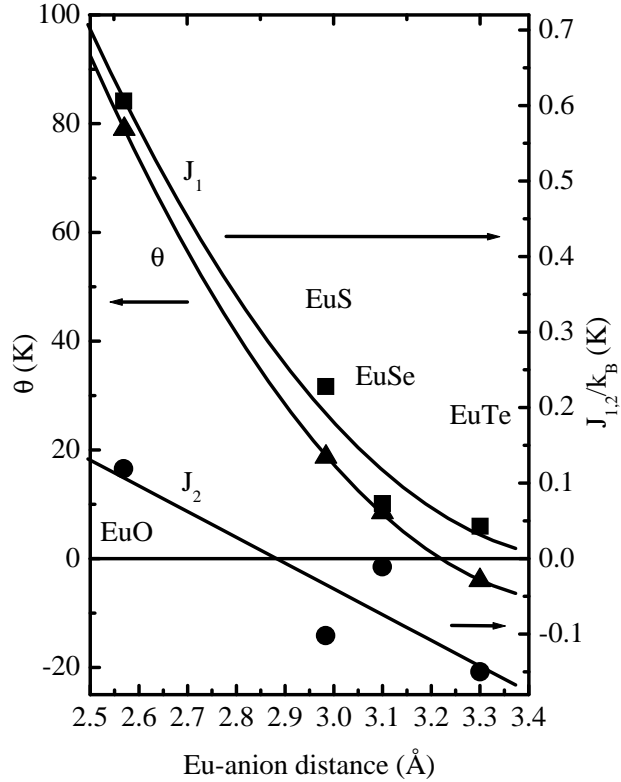}}
\caption{\label{Exch} Curie-Weiss temperature $\theta$ (triangles) and
Heisenberg neareast and next-nearest neighbor exchange parameters
$J_1$(squares) and $J_2$(circles) versus the Eu-anion distance for the
different Eu chalcenogides. Data collected by Wachter
\cite{Wachter79}.}
\end{figure}

It was shown by neutron scattering experiments, that the
Heisenberg model gives a good description of the spin-wave
spectrum of EuO \cite{Dietrich75,Passel76}. The exchange
integrals were accurately determined to be $J_1/k_B=0.606\pm
0.008$ K and $J_2/k_B=0.119\pm 0.015$ K for nearest and
next-nearest neighbors\footnote{Note however that T$_c$ predicted
from the mean field Heisenberg model is only 42 K using these
parameters.}. Also the exchange parameters of the other Eu
chalcenogides were determined and it was found that $\theta$ and
$J_1$ depend similarly on the Eu-anion distance as is shown in
figure \ref{Exch}. However, the question of the microscopic
mechanism from which these exchange interactions originate,
remains. Kasuya \cite{Kasuya70,Mauger86} proposed the following
competing exchange mechanisms in EuO:
\begin{enumerate}
\item Nearest neighbor Eu-Eu interaction
\begin{itemize}
\item A $4f$ electron is excited to the $5d$ band, experiences an
exchange interaction with the $4f$ spin on a nearest neighbor and
returns to the initial state. This generally leads to
ferromagnetic exchange.
\end{itemize}
\item Next-nearest neighbor Eu-Eu interactions
\begin{itemize}
\item The Kramers-Anderson superexchange. An $f$ electron
is transferred via oxygen to an $f$ orbital on a neighboring
atom. This exchange is antiferromagnetic and is very small,
because the transfer integrals $t_{fp}$ are small and the
electron-electron repulsion $U_f$ is large ($\sim$11 eV, see
figure \ref{XPSBIS}).
\item Superexchange via the $d-f$ interaction. Oxygen electrons
are transferred to the $d$ orbitals of neighboring Eu atoms,
there they affect the $4f$ spins via the $d-f$ exchange
interaction. The 180$^{\circ}$ Eu-chalcogenide-Eu bond angle
leads to a substantial antiferromagnetic exchange (similar to
Kramers-Anderson superexchange) and explains why $J_2$ is
negative in EuS, EuSe and EuTe.
\item A mix of both above mentioned mechanisms. Via hybridization
the $5d$ and $2p$ orbitals form bonding and antibonding molecular
orbitals. An oxygen electron is excited from the bonding to the
antibonding $5d$-$2p$ molecular orbital, which has exchange
interaction with both Eu spins, its place is taken by a Eu $4f$
electron, after which the $5d$ electron fills the $4f$
hole\footnote{This mechanism is essentially equal to the virtual
excitation of an O $2p$ electron to the conduction band, where it
contributes to the exchange, similar to the way in which free
electrons induce RKKY exchange.}. This could lead to a
ferromagnetic exchange and would explain why $J_2$ is positive in
EuO.
\end{itemize}
\end{enumerate}
Kasuya estimated the magnitude of these interactions (they were
also discussed in \cite{Kocharyan75,Khomskii97}). These estimates
should however be considered with care, as there is a large
uncertainty in the initial parameters. Recently, first principle
LDA+U calculations of the different exchange strengths by Elfimov
\cite{Elfimov03} show promising agreement with experiments. Such
calculations might finally solve the question why ferromagnetic
semiconductors can exist.

\subsubsection{Effect of doping}
EuO can be electron-doped in several ways. Divalent Eu can be
substituted by a trivalent Gd or other rare-earth ions like La,
Dy, Tb, Ho or Lu \cite{Holtzberg64,Holtzberg66,Methfessel68}, but
doping with other atoms like Ag and Cu is also possible
\cite{Lee71}. Gd doping has been most studied, as Gd also has a
half-filled $4f$ shell and was thus expected to leave the $4f$
spin-system untouched, only supplying carriers to the
solid\footnote{However, indications have been found
\cite{Bayer71,Wachter79} that Gd ions have their spin
anti-parallel to the Eu spins.}. The system could also be doped
by substituting oxygen with a single valent anion like Cl,
although this method has not been applied often \cite{Mauger86}.
Another way to electron dope EuO, is to supply an excess of Eu
atoms to the compound. As these excess Eu atoms are too large to
be incorporated as interstitials, oxygen vacancy sites will form.
In practice it turns out that it is very difficult to introduce a
large concentration of oxygen vacancies in EuO$_{1-x}$ crystals,
the vacancy concentration $x$ usually remains below 0.5\%. Higher
concentration can be reached in polycrystalline thin films that
are grown by the oxidation of Eu metal, as shown by Massenet {\it
et al.} \cite{Massenet74}. Also the concentration of rare earth
ions that can be incorporated in crystals is limited in some
cases \cite{Shafer68,Mauger86}.

\begin{figure}[!htb]
   \centering
   \begin{minipage}[b]{0.33333\textwidth}
      \centering
      \subfigure[Ferromagnetic Curie temperature of
Eu$_{1-x}$Gd$_x$O versus doping concentration $x$, from
\cite{Mauger80}.]{
         \label{Tcdop:a}
         \includegraphics[width=5cm]{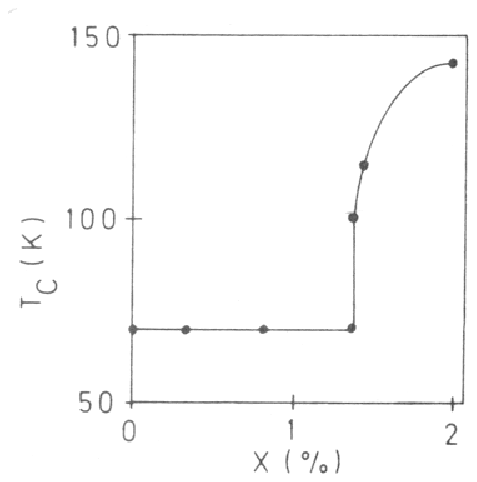}}
   \end{minipage}%
   \begin{minipage}[b]{0.33333\textwidth}
      \centering
      \subfigure[Paramagnetic Curie-Weiss
temperature $\theta$ and lattice constant of Eu$_{1-x}$Gd$_x$Se
(curve a) and Eu$_{1-x}$La$_x$Se (curve b) versus doping
concentration $x$, from \cite{Holtzberg64}.]{
         \label{Tcdop:b}
      \includegraphics[width=5cm]{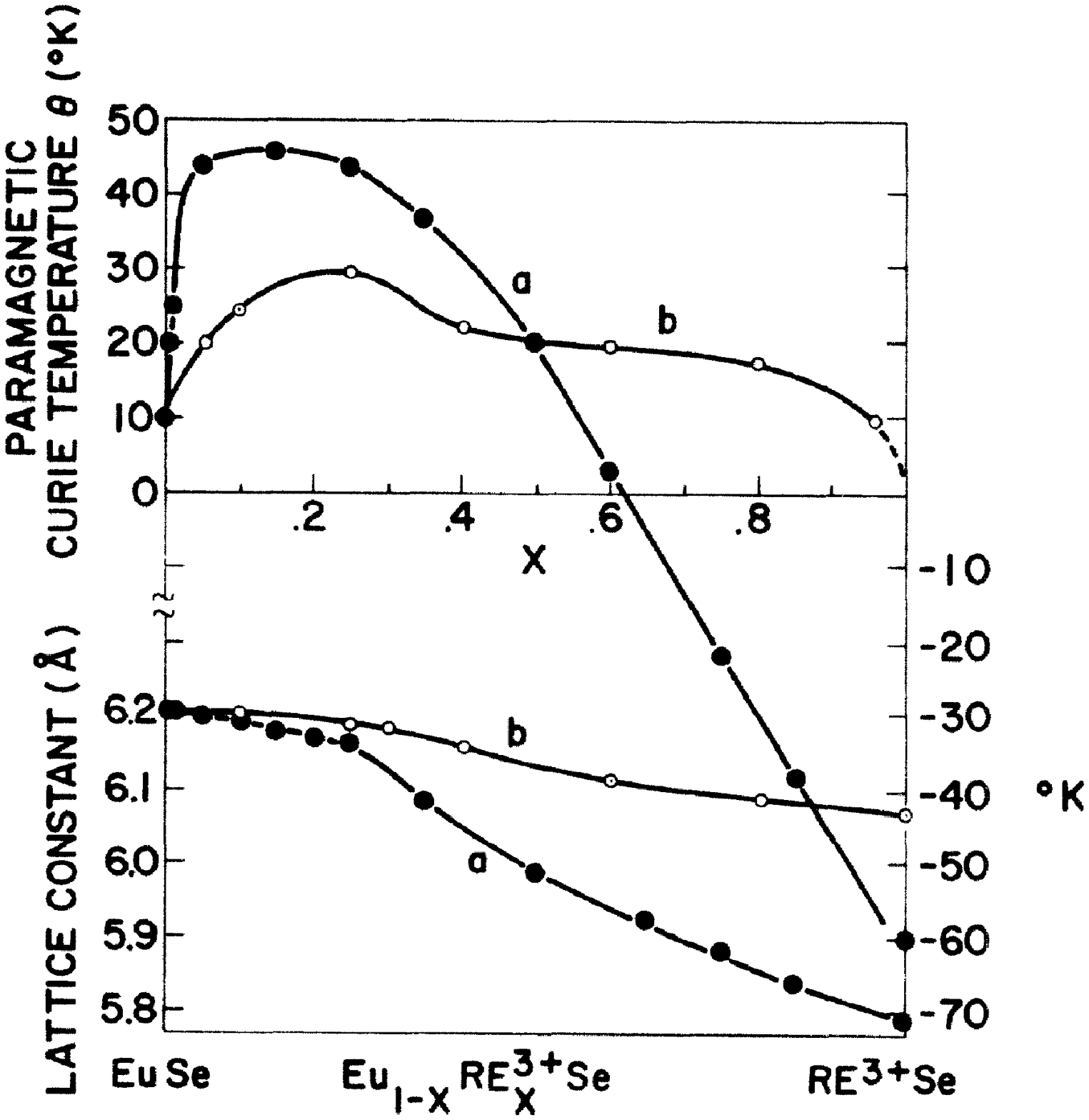}}
   \end{minipage}%
   \begin{minipage}[b]{0.33333\textwidth}
      \centering
      \subfigure[Variation of $\theta$ with the free electron concentration,
assuming an RKKY exchange mechanism \cite{Holtzberg64}.]{
         \label{Tcdop:c}
      \includegraphics[width=5cm]{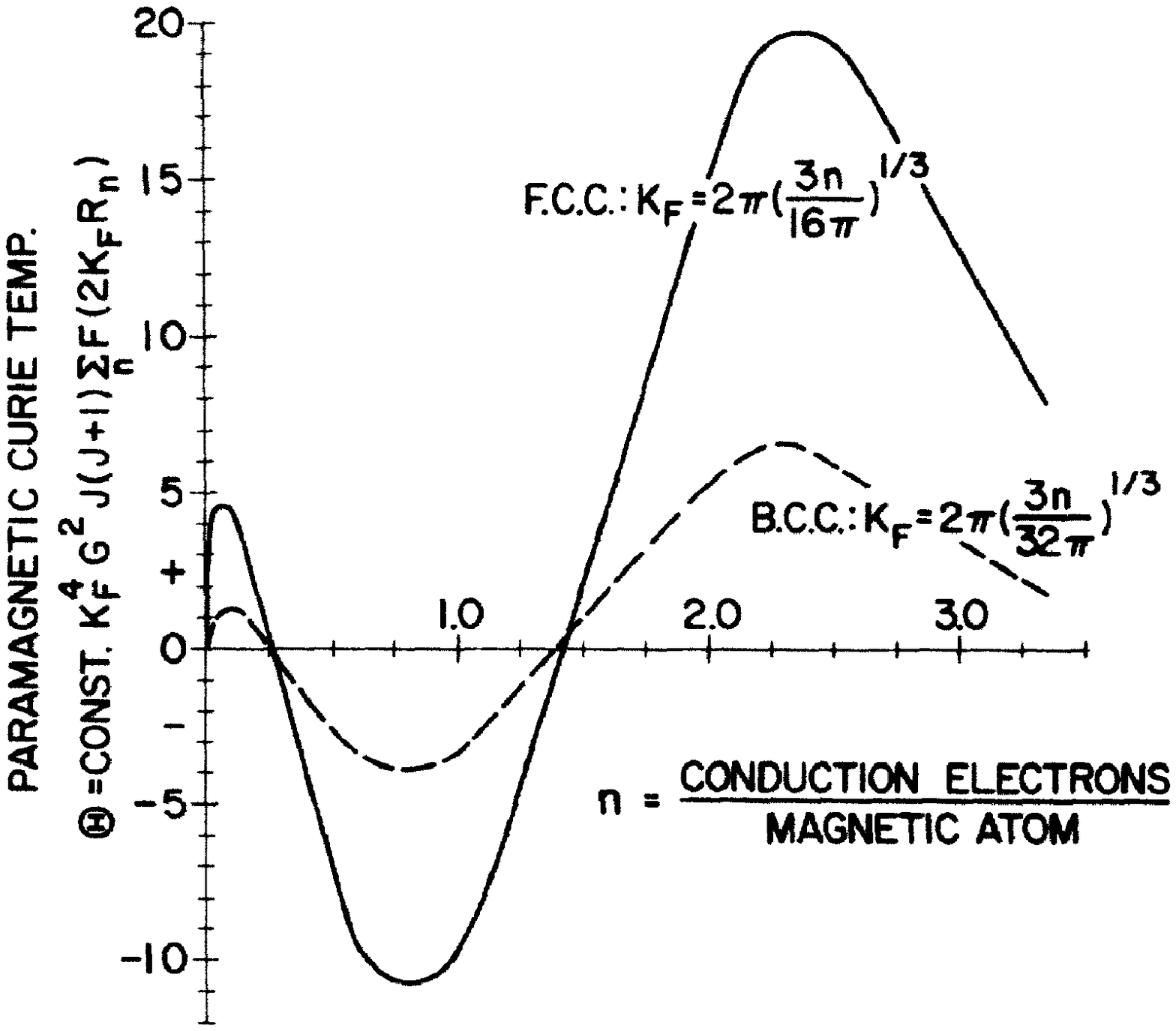}
      }
   \end{minipage}
   \caption{\label{Tcdop} Doping dependence of the Curie temperature in the Eu chalcenogides.
            Note that the
            labels in figure (c) should be, FCC: $k_F R_n = 2 \pi \left(3 \sqrt{2}
            n/8 \pi\right)^{1/3}$ and BCC: $k_F R_n = 2 \pi \left(9 \sqrt{3} n/32
            \pi\right)^{1/3}$. Here $R_n$ is the nearest neighbor distance between
             magnetic ions.}
\end{figure}

Although doping concentrations below 1\% have an enormous effect
on the transport properties of EuO, they do not seem to affect
the Curie temperature much \cite{Mauger80,Godart80,Mauger86} as
can be seen from figure \ref{Tcdop:a}. Higher doping
concentrations have a large effect on T$_c$, and can even result
in a Curie temperature as high as 148 K. On the other hand an
increase in the \emph{free} carrier concentration at room
temperature of $0.1\%$ (as determined by optics \cite{Schoenes})
is sufficient to lead to an increase in T$_c$ of $\sim$10 K. This
suggests that only the \emph{free} carriers are responsible for
increasing the ferromagnetic interactions with doping and seems
to contradict models \cite{Kasuya68,Torrance72} which ascribe the
increase of the Curie temperature to high-spin bound magnetic
polarons (BMP). It also suggests that for smaller doping
concentrations the carriers are indeed bound, as will be
discussed more extensively in chapter \ref{Magres}. The increased
exchange via free electrons is probably similar to that expected
from the RKKY interaction. The increased Curie temperature can be
qualitatively understood from the following simple argument
\cite{Khomskii00,Schoenes}. If the conduction electrons align
their spin with the $4f$ moments, they will gain an exchange
energy $\Delta_{\rm ex} \approx 0.3$ eV~\cite{Freiser}. Therefore
the total energy gain per Eu atom at a doping concentration
$x=2$\% is $x\Delta_{\rm ex} \approx k_B\times 70$ K which
corresponds reasonably well with the observed increase in T$_c$
of 79 K \cite{Mauger80}\footnote{One should be careful when
applying this argument, as there are indications that the
exchange splitting decreases with doping \cite{Schoenes}.}.
Increasing the doping level above 2\% does not seem to increase
T$_c$ anymore \cite{Shafer68}. A similar maximum T$_c$ around 2\%
electron doping was found for Eu$_{1-x}$Gd$_x$S (52 K
\cite{McGuire73,Mauger86}) and for EuSe (45 K
\cite{Holtzberg64}). At higher doping concentrations the Curie
temperature in EuS and EuSe decreases again\footnote{Similar
behavior is expected in EuO, it seems however difficult to reach
high enough doping concentrations \cite{Shafer68}.} and they
eventually become antiferromagnetic as is shown in figure
\ref{Tcdop:b}. The doping dependence of the Curie temperature
could be a result of the oscillatory concentration dependence of
the RKKY interaction between two spins separated by a distance
$R_{ij}$, $\mathcal{I}_{RKKY}(k_F, R_{ij})\propto J^2 (2 k_F
R_{ij} \cos (2 k_F R_{ij})-\sin (2 k_F R_{ij}))/(2 R_{ij})^4$
\cite{Fazekas99}. It was shown by de Gennes
\cite{deGennes58b,Fazekas99} that the Curie temperatures of the
rare earth metals can be well understood from this relation. As
doping affects the Fermi momentum $k_F$, Holtzberg {\it et~al.}
\cite{Holtzberg64} thus related the doping dependence of T$_c$ in
the Eu chalconegides to changes in the RKKY interaction with
increasing $k_F$, as is shown in figure \ref{Tcdop:c}. Although
qualitative agreement is obtained (note that Gd metal would
correspond to $n\sim 3$), effects of band-structure and the
effect of a different orientation of the doped Gd spins, might
modify these results. Moreover, the RKKY interaction is derived
for the case when the exchange splitting is small compared to the
Fermi energy, a condition which is not fulfilled at low doping
concentrations. More sophisticated calculations were done by
Mauger {\it et~al.} \cite{Mauger77,Mauger78,Mauger86} and
recently by Santos and Nolting \cite{Santos02}.

\subsubsection{Photodoping and field doping}

As the dependence of the Curie temperature on chemical doping in
the Eu chalcogenides is very strong, it might also be possible to
achieve an increase in Curie temperature by optical injection of
carriers in the conduction band. In fact, we have attempted such
experiments by focussing a high power halogen lamp on EuO thin
films, while measuring their Kerr effect with a separate laser.
The temperature of the films was kept just above the Curie
temperature (T$-$T$_c < 1$ K). However, the experiments were
unsuccessful as no remanent magnetization was achieved. Earlier
attempts to dope electrons in the conduction band with high power
pulsed lasers on EuS, have shown that an increase in
magnetization of $\Delta M/M \approx 10^{-2}$ can be induced by
light (for a review see \cite{Kovalenko86,Nagaev88}). We have
also made several preliminary attempts to introduce charge
carriers in EuO by applying a large electric field in a field
effect transistor (FET) geometry. We grew this FET by first
growing a Cr gate electrode ($\sim$ 0.25 cm$^2$) on an
Al$_2$O$_3$ substrate. This was covered by sputtering $\sim 100$
nm of Al$_2$O$_3$. On the dielectric (above the gate electrode)
we evaporated source and drain contacts of Cr. On top of this
structure we grew our EuO film, as described in chapter
\ref{euoepi}. However, no significant effect of the field on
transport or Kerr effect was observed for gate voltages up to
$\sim$ 5 V, at higher voltages, breakdown of the dielectric layer
occurred. Recently, field induced ferromagnetism has been
observed in (In,Mn)As by Ohno {\it et~al.} \cite{Ohno00}. A
difference in T$_c$ of 1 K was observed as a result of a
field-induced electron doping of $\sim 0.15\%$. Even larger
changes in the Curie temperature are expected in the Eu
chalcenogides at comparable doping concentrations. Additionally,
field induced doping would allow studies of the doping dependence
of the metal insulator transition in a much more controlled way.
Our investigations of the effect of photo-doping on the metal
insulator transition, are described in section \ref{Photocond}.
Interestingly, we found (figure \ref{photoMIT}) that the metal
insulator transition in EuO can even be induced by light.

\subsection{Optical properties}

\begin{figure}[!htb]
   \begin{tabular}{ll}
   \begin{tabular}{l}
   \begin{minipage}[b]{0.45\textwidth}
      \centering
      \subfigure[]{
         \label{epsilon:a}
         \includegraphics[width=6.75cm]{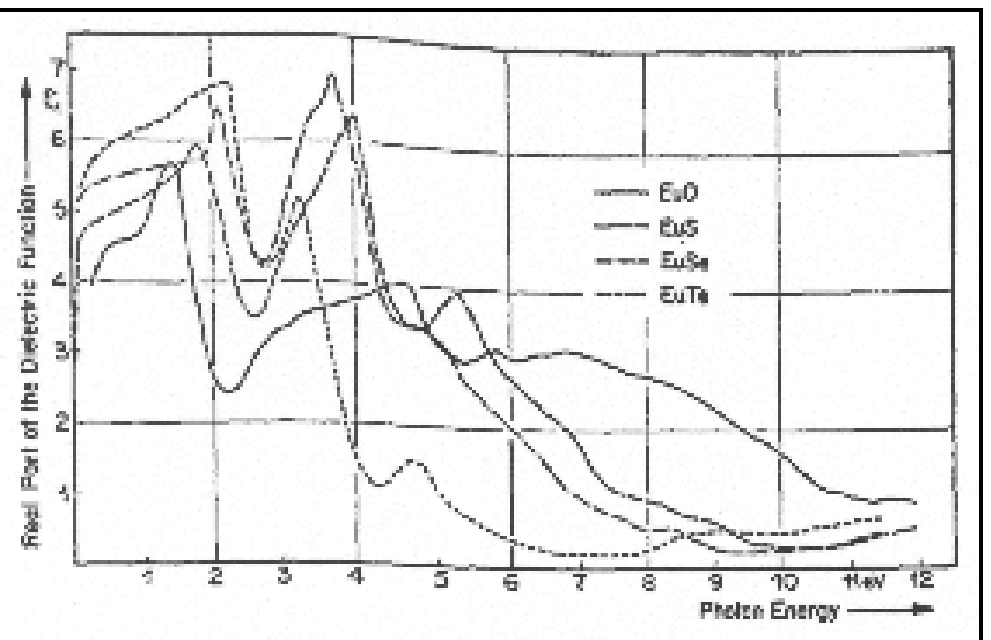}}
   \end{minipage}%
   \\
   \begin{minipage}[b]{0.45\textwidth}
      \centering
      \subfigure[]{
         \label{epsilon:c}
      \includegraphics[width=6.75cm]{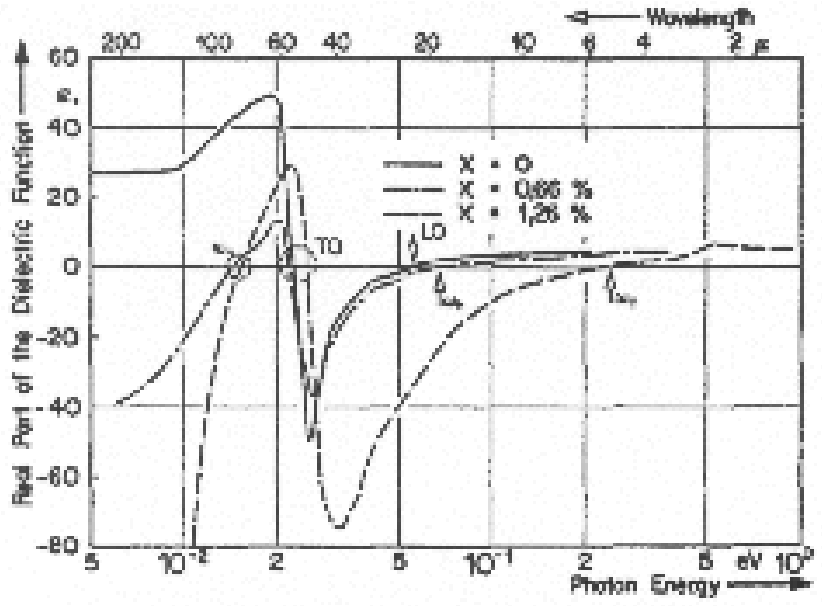}
      }
   \end{minipage}
   \end{tabular}
   &
   \begin{minipage}[c]{0.55\textwidth}
      \centering
      \subfigure[]{
         \label{epsilon:b}
      \includegraphics[width=8.25cm]{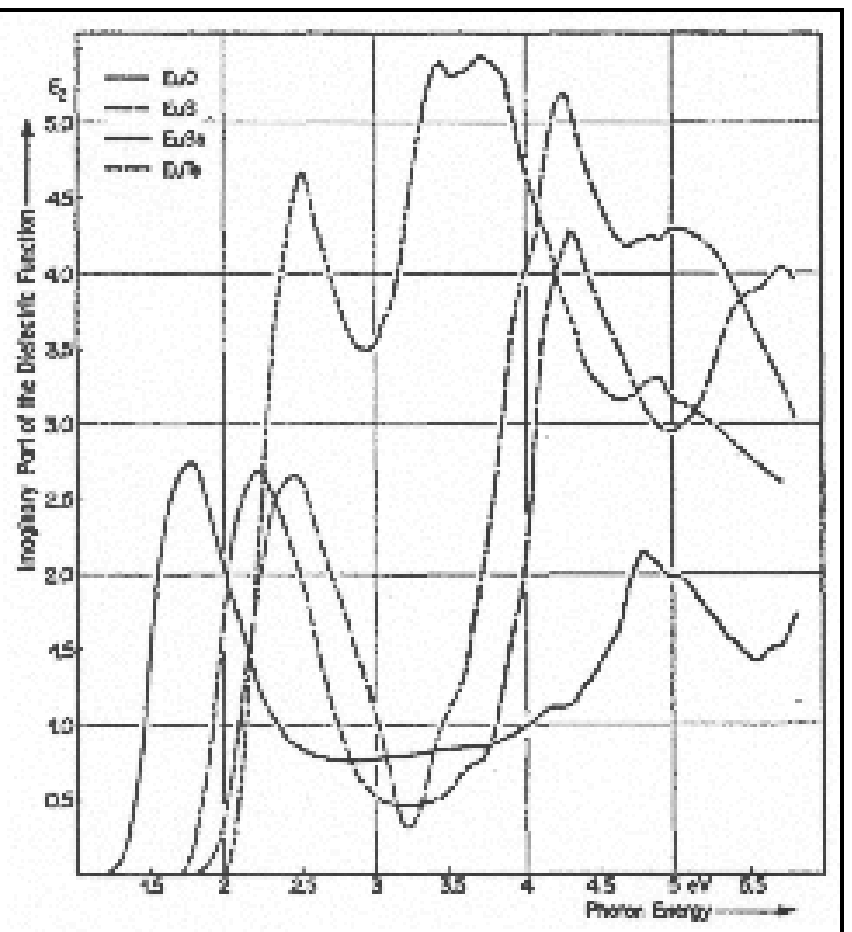}}
   \end{minipage}%
   \end{tabular}
   \caption{\label{epsilongun} Real part
of the dielectric function at 300 K of (a) EuO, EuS, EuSe and EuTe and
(b) Eu$_{1-x}$Gd$_x$O, as determined from reflectivity measurements by
a Kramers-Kronig analysis. (c) Imaginary part of the dielectric
function of the Eu chalcenogides at 300 K as determined by
ellipsometry. Reproduced from \cite{Guntherodt74}.}
\end{figure}

The optical properties of EuO have generated much interest, both
because they probe the electronic structure, and because they change
spectacularly upon lowering the temperature or applying a magnetic
field. The optical spectrum consists mainly of transitions from the
$4f$ orbitals to the $5d-6s$ conduction band across the gap of 1.15
eV. Room temperature spectra of the dielectric function of the Eu
chalcenogides are shown in figure \ref{epsilongun}.

\begin{figure}[!htb]
   \centering
   \begin{minipage}[b]{0.3888888\textwidth}
      \centering
      \subfigure[Shift of the absorption edge, from
\cite{Freiser}.]{
         \label{Optics:a}
         \includegraphics[width=5.5cm]{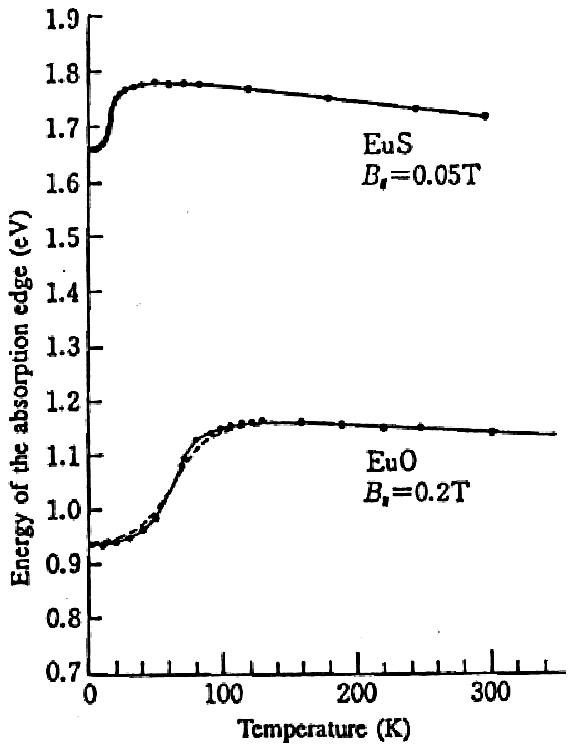}}
   \end{minipage}%
   \begin{minipage}[b]{0.55\textwidth}
      \centering
      \subfigure[Real and imaginary part of the off-diagonal component of the conductivity tensor $\sigma_{1,2xy}$, from \cite{Wang86}.]{
         \label{Optics:b}
      \includegraphics[width=8.5cm]{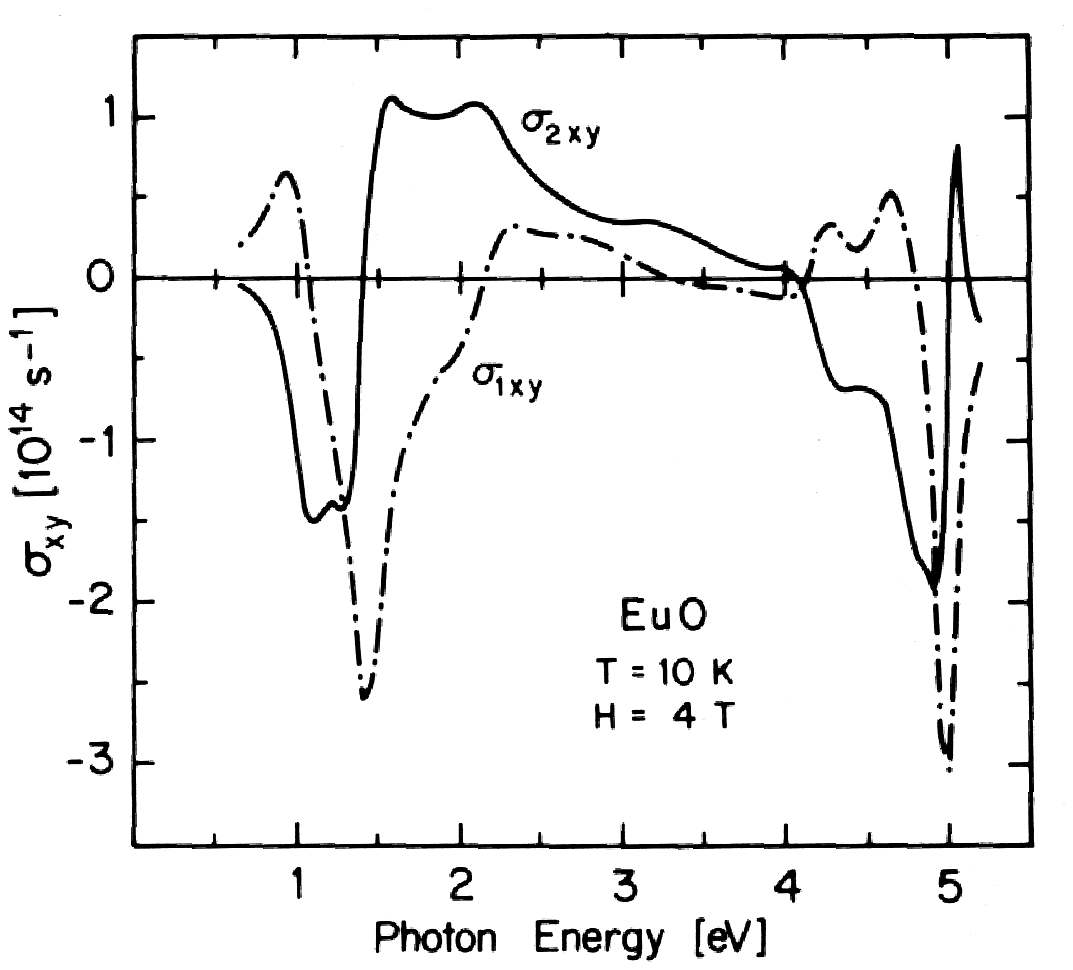}}
   \end{minipage}
   \caption{\label{Optics} Magneto-optical properties of EuO.}
\end{figure}

When the temperature is lowered, EuO becomes ferromagnetic and a
large ($\sim 0.3$~eV) shift of the optical absorption edge to
lower energies is observed \cite{Busch,Freiser,Schoenes} as is
shown in figure \ref{Optics:a}. This shift is a result of the
direct exchange interaction between the polarized Eu $4f$ spins
and the conduction electrons, which pushes the spin-down band to
higher energies and the spin-up band to lower energies, thus
reducing the gap. We have observed this effect also as a
splitting of the O $K$ edge x-ray absorption spectrum (XAS), as
will be discussed in chapter \ref{Tempxas}.
When optically exciting an electron from the $4f^7$ shell to the
$5d$ conduction band, the remaining $4f^6$ electrons are left in
one of the $J_{4f}=0-6$ final states. This effect is mainly
responsible for the large magneto-optical effects (see figure
\ref{Optics:b}) observed in EuO, as we will discuss in chapter
\ref{optprop}. These magneto-optical properties in EuO belong to
the largest known, with a Faraday rotation of 7.5$\times 10^5$
deg/cm \cite{Tu72} and a bulk Kerr rotation of 7.1$^\circ$
\cite{Wang86}, which can be strongly enhanced\footnote{The
observation of a 90$^\circ$ Kerr rotation in CeSb has also been
suggested to be a result of enhancements by (irreproducible)
surface effects \cite{Uspenskii00,Drioli99,Drioli99b}.} to
70$^\circ$ by growing films of EuO on Ag metal
\cite{Suits71,Uspenskii00}. The presence of $4f$ multiplets
strongly complicates the clarification of the electronic
structure of the $5d$ band by optics. Moreover, the electron-hole
interaction seems to result in excitonic effects which affect the
energies of the final states even more. As these effects are much
weaker for the O $1s$ shell than for the Eu $4f$ shell, the O $K$
edge XAS is more easily interpreted in terms of a one electron
picture and seems to give a clearer picture of the unoccupied
density of states, as will be discussed in chapters \ref{Spinxas}
and \ref{Tempxas}.

\subsection{Electronic structure} 
\begin{figure}[!htb]
\centerline{\includegraphics[width=8cm,angle=270]{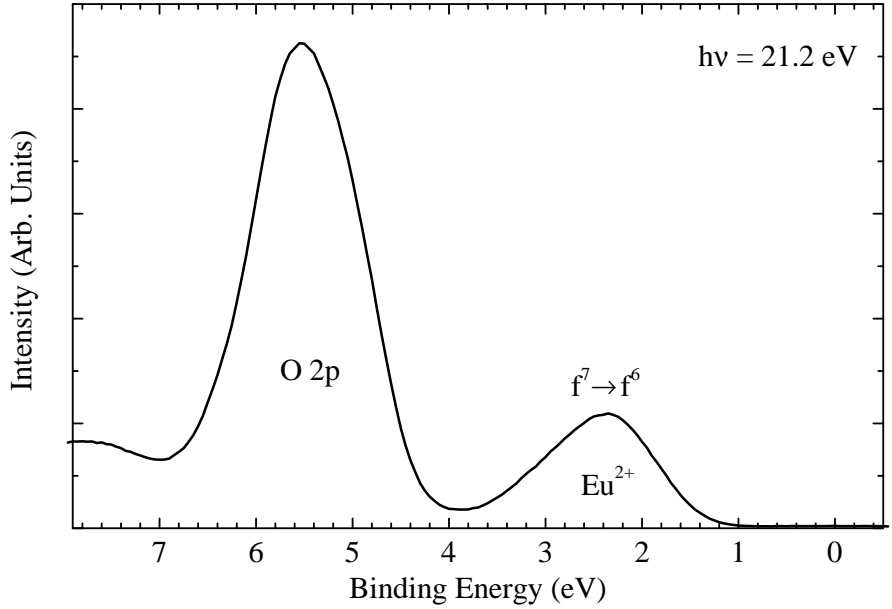}}
\caption{\label{UPS} Satellite corrected He-I UPS spectrum of a
polycrystalline stoichiometric EuO film on Ta metal. Details of
the film growth will be discussed in chapter \ref{euoepi}.}
\end{figure}

The energy levels of EuO which are
important for the understanding of the low energy features, i.e.
those which are near the chemical potential, are the O $2p$
orbitals, the Eu $4f$ orbitals, and the Eu $5d-6s$ conduction
band. As the O$^{2-}$ $2p$ and Eu$^{2+}$ $4f$ levels are
(partially) occupied, they can be well studied by electron
removal spectroscopy. Indeed these levels are clearly resolved by
photoemission spectroscopy (UPS) of EuO as is shown in figures
\ref{UPS} and \ref{euoepi2}, and as was also observed in early
works \cite{Eastman69,Busch69,Eastman71,Cotti72}.
\begin{figure}[!htb]
\centerline{\includegraphics[width=11.25cm]{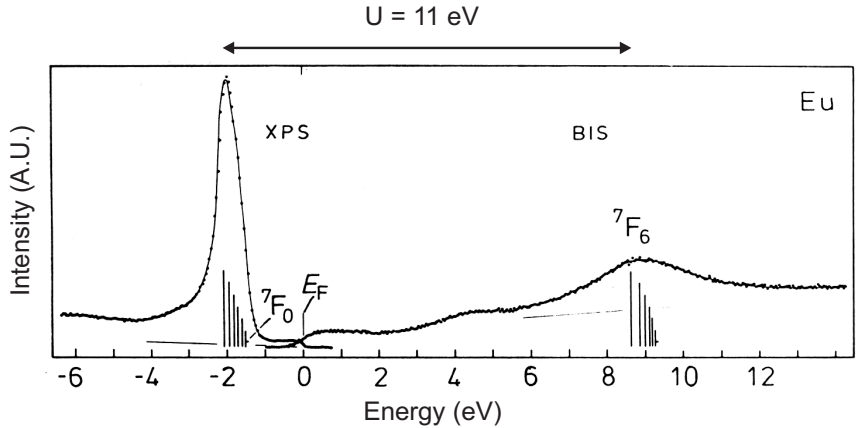}}
\caption{\label{XPSBIS} XPS and Bremsstrahlung Isochromat Spectroscopy
(BIS) of Eu metal from \cite{Lang81}.}
\end{figure}
From Hund's first rule we know that the ground state
configuration of the Eu $4f^7$ electrons will be high-spin
$S=7/2$. To remove an electron from the $4f$ shell of a Eu$^{2+}$
ion and add it to another ion will cost an energy U$^{\rm eff}$ =
E$_{4f^8}$+E$_{4f^6}-2$ E$_{4f^7}$. The Coulomb interaction
between electrons can be separated in a Coulomb repulsion term
(which is the Slater parameter $F^0$) and a Hund's rule exchange
term $J_H$. In a given configuration the repulsion between each
pair of electrons gives an energy contribution $F^0$, whereas for
each pair of spin-parallel electrons the energy is reduced by
$J_H$ as a result of exchange. Therefore, the energy of a certain
configuration (neglecting the angular multiplet splittings) with
$N_\uparrow$ ($N_\downarrow$) spin-up (spin-down) electrons is
\cite{Marel85,Marel88}:
\begin{equation}
E(N_\uparrow,N_\downarrow) = F^0
\sum_{n=1}^{N_\uparrow+N_\downarrow-1}n - J_H
\left(\sum_{n=1}^{N_\uparrow-1} n + \sum_{n=1}^{N_\downarrow-1}
n\right)
\end{equation}

\begin{figure}[!htb]
\vspace{1.4cm}
\centerline{\includegraphics[width=8cm,angle=270]{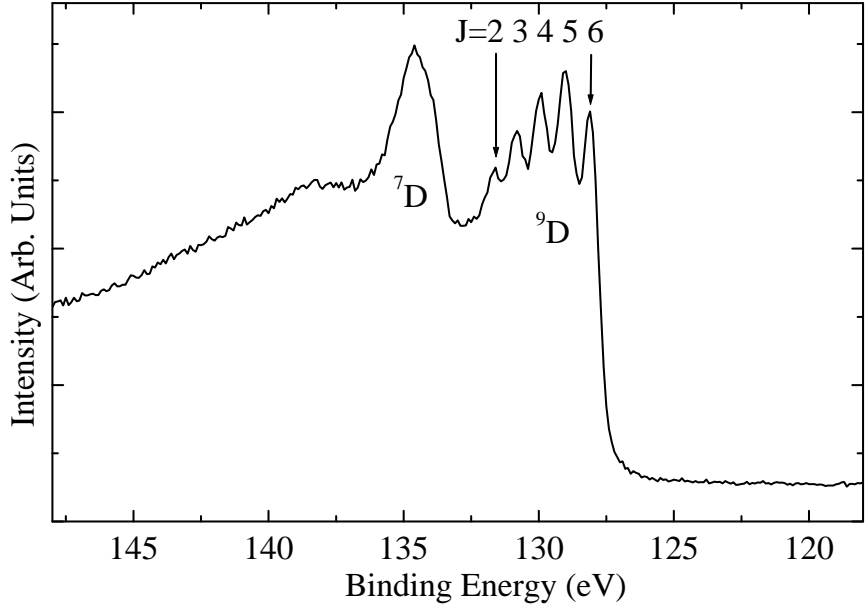}}
\caption{\label{Eu4dpes} $4d$ XPS from a polycrystalline Eu metal
film. The exchange split $^9D$ and $^7D$ levels are clearly resolved
with $2 \Delta_{{\rm ex},4d}\approx 5.5$ eV. Moreover the different
$J$ final states are clearly resolved with a splitting of 0.9 eV.}
\end{figure}
Thus, for $^8S$ Eu$^{2+}$, U$_{4f}^{\rm eff}$ = $F^0 + 6 J_H$. As $F^0
\approx 6.5$ eV and $J_H \approx 0.9$ eV, $U_{4f}^{\rm eff}\approx$
11.7 eV, which corresponds well \cite{Marel85} to the value of U$^{\rm
eff} \approx$ 11 eV found by Lang {\it et~al.} \cite{Lang81} in Eu
metal by XPS and BIS measurements as is shown in figure \ref{XPSBIS}.
Moreover, this analysis indicates that it costs $6 J_H\approx$ 5.4 eV
to flip one $4f$ spin in the ground state, showing that the high-spin
configuration is very stable. Indeed, spin-resolved photoemission
experiments show that the spin-polarization of the $4f$ levels in EuO
is nearly 100\%, as is shown in figure \ref{fig4} and was also studied
by Sattler and Siegmann \cite{Sattler72,Sattler75}. Because the $4f$
electrons are very localized, their hybridization with other orbitals
is small. Thus, the transport properties are mainly determined by the
$5d-6s$ conduction band, which is hybridized with the O $2p$ band.
Although the $4f-5d$ transfer integrals are quite small, the exchange
interactions between these orbitals is still significant and leads to
an exchange splitting of the conduction band.

The importance of $d-f$ exchange and spin-orbit coupling can be
observed in the $4d$ photoemission spectrum in figure
\ref{Eu4dpes}, although the $4f-d$ exchange and spin-orbit
coupling parameters $\Delta_{\rm ex}$ and $\xi$ of the
delocalized $5d$ orbitals are an order of magnitude smaller than
those of the more localized $4d$ orbital. Anyhow, as the $5d-4f$
exchange $\Delta_{\rm ex} \approx 0.3$ eV \cite{Busch}, the
configuration of the $4f$ spins can have a substantial influence
on the transport properties and spin-polarization of the
conduction band. This interplay between $4f$ spins and the
splitting and spin-polarization of the conduction band has
theoretically extensively been studied
\cite{Rys67,Alexander76,Haas68,Mauger86,Nolting79,Nagaev01,Schiller01,Schiller01b,Schiller01c,Schiller01d}
and will experimentally be addressed in chapters \ref{Spinxas}
and \ref{Tempxas}.

\subsection{Transport properties}
\begin{figure}[!p]
    \vspace{0cm}
   \begin{tabular}{ll}
   \begin{minipage}[b]{0.5\textwidth}
      \centering
      \subfigure[$\rho(T)$ of EuO$_{1\pm x}$ crystals, from \cite{Oliver72}.]{
         \label{trans:a}
         \includegraphics[width=4cm]{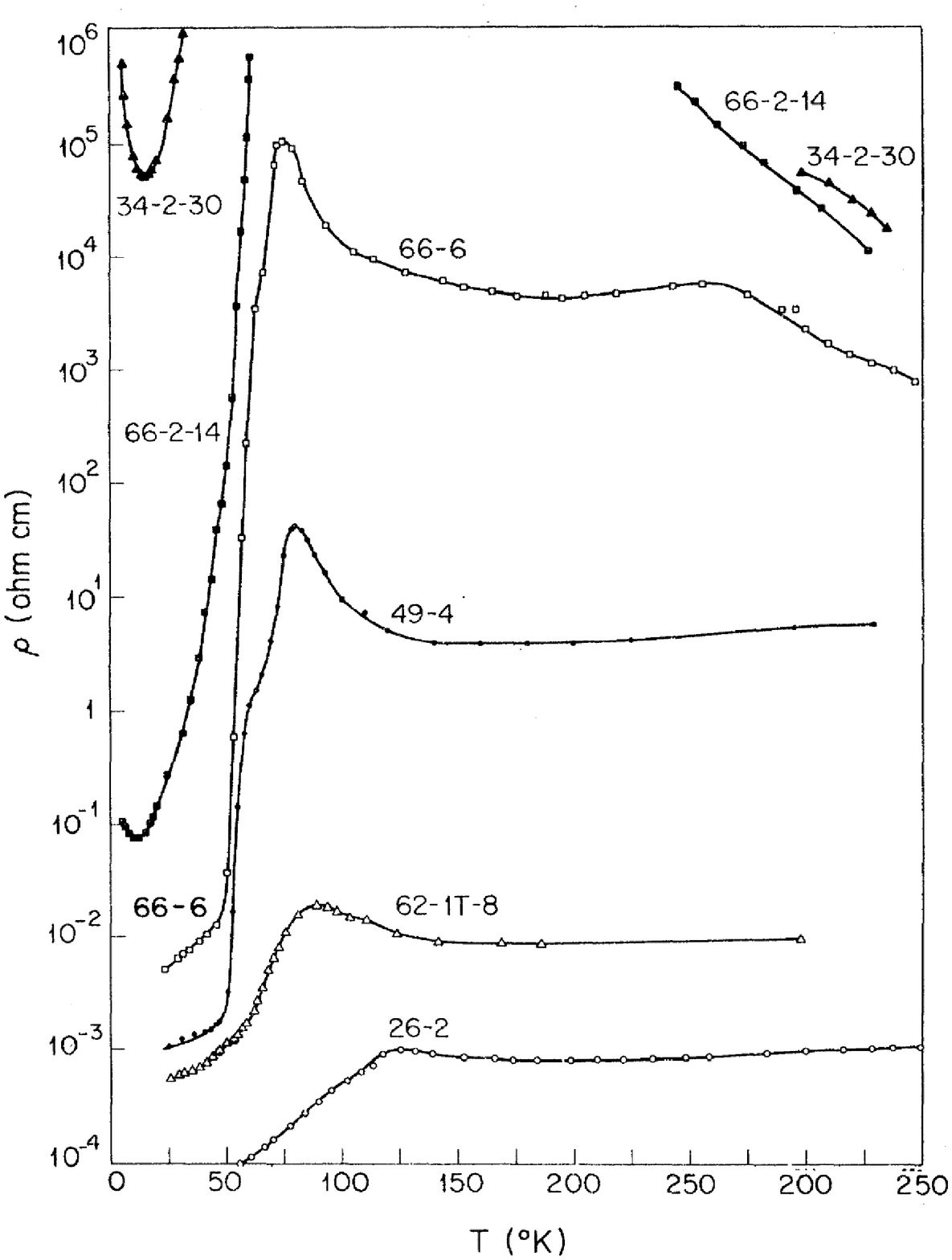}}
   \end{minipage}%
   &
   \begin{minipage}[b]{0.5\textwidth}
      \centering
      \subfigure[$\rho(T)$ of EuO$_{1\pm x}$ crystals, from \cite{Oliver72}.]{
         \label{trans:b}
      \includegraphics[width=4cm]{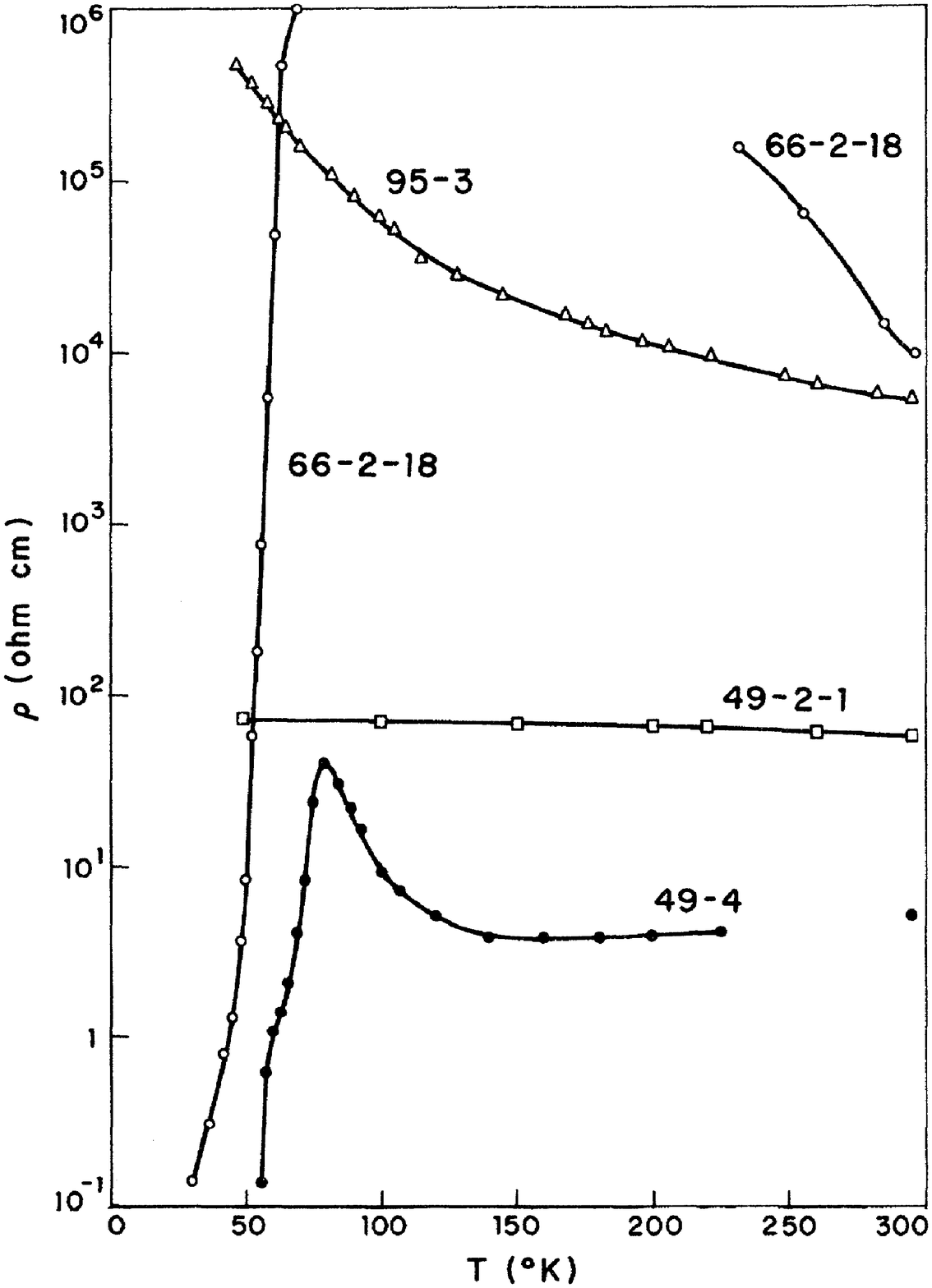}}
   \end{minipage}%
   \\
   \begin{minipage}[b]{0.5\textwidth}
      \centering
      \subfigure[$\rho(T)$ of EuO$_{1-x}$ films, from \cite{Massenet74}.]{
         \label{trans:c}
      \includegraphics[width=5cm]{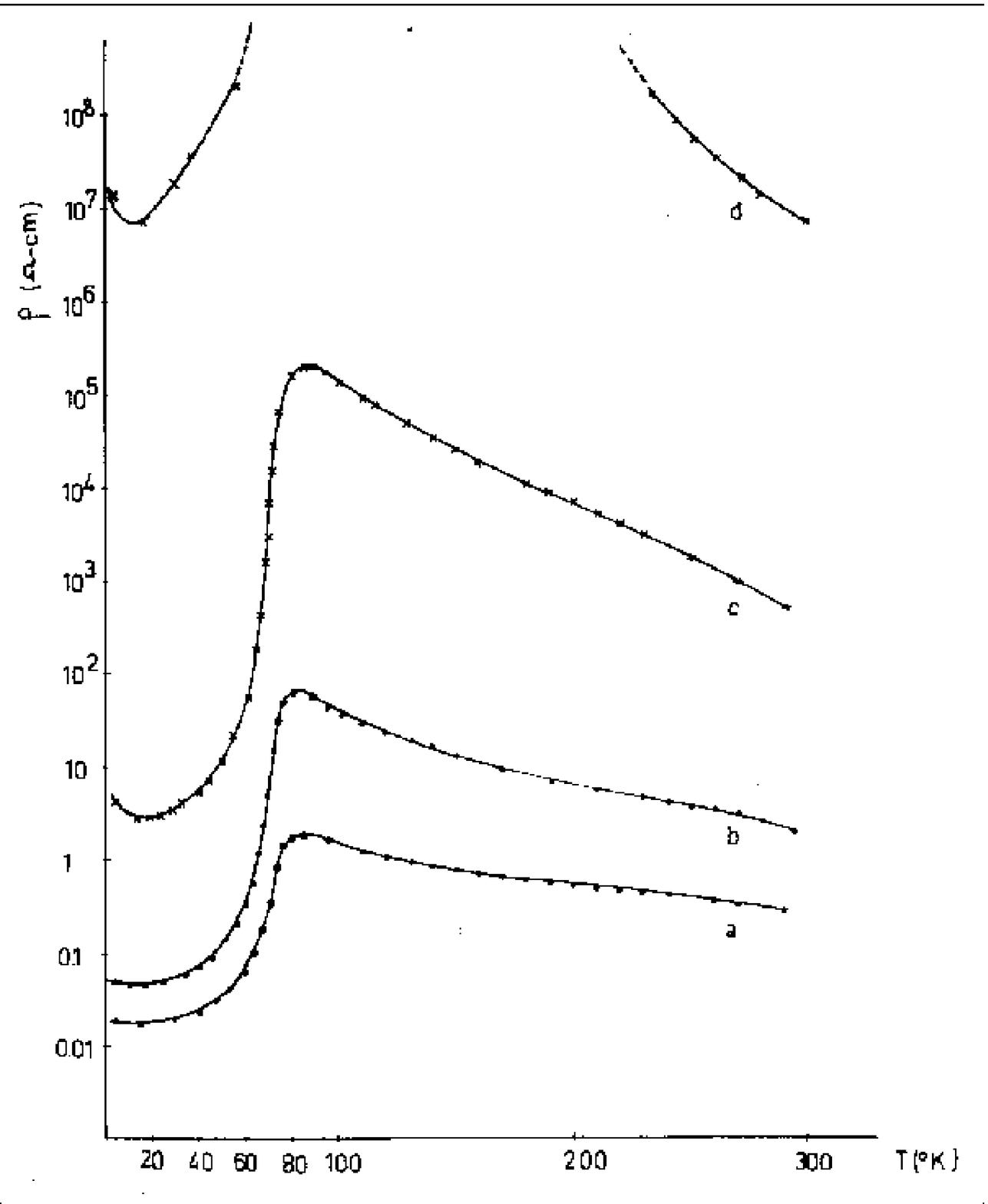}
     }
   \end{minipage}
   &
   \begin{minipage}[b]{0.5\textwidth}
      \vspace{0pt}
      \centering
      \subfigure[Magnetoresistance of EuO crystal, from \cite{Shapira}.]{
         \label{trans:e}
      \includegraphics[width=5cm]{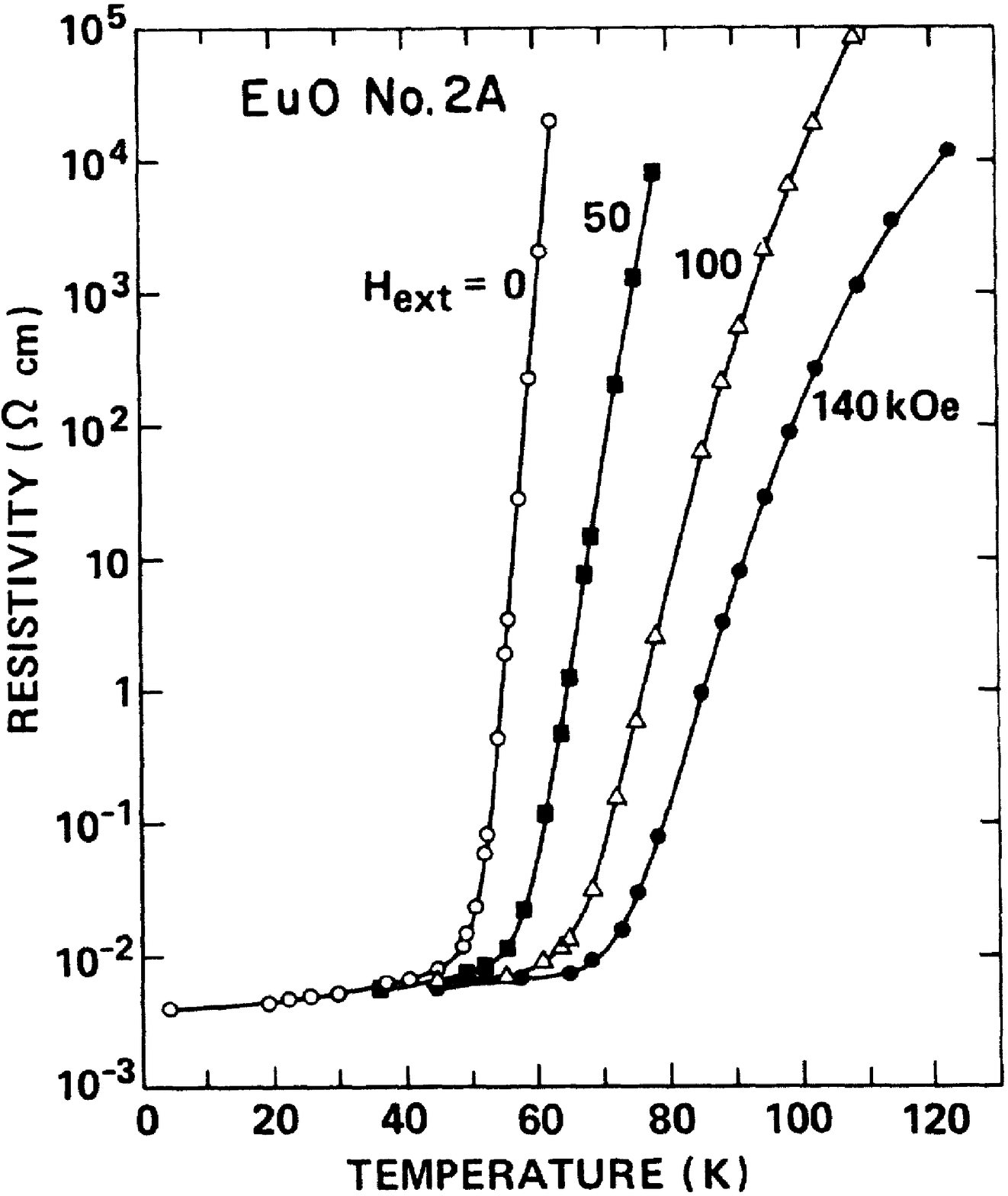}}
   \end{minipage}%
   \\
   \begin{minipage}[b]{0.5\textwidth}
      \centering
      \subfigure[$\sigma(T)$ of EuO$_{1\pm x}$ crystals, from \cite{Shafer72}.]{
         \label{trans:d}
      \includegraphics[width=5cm]{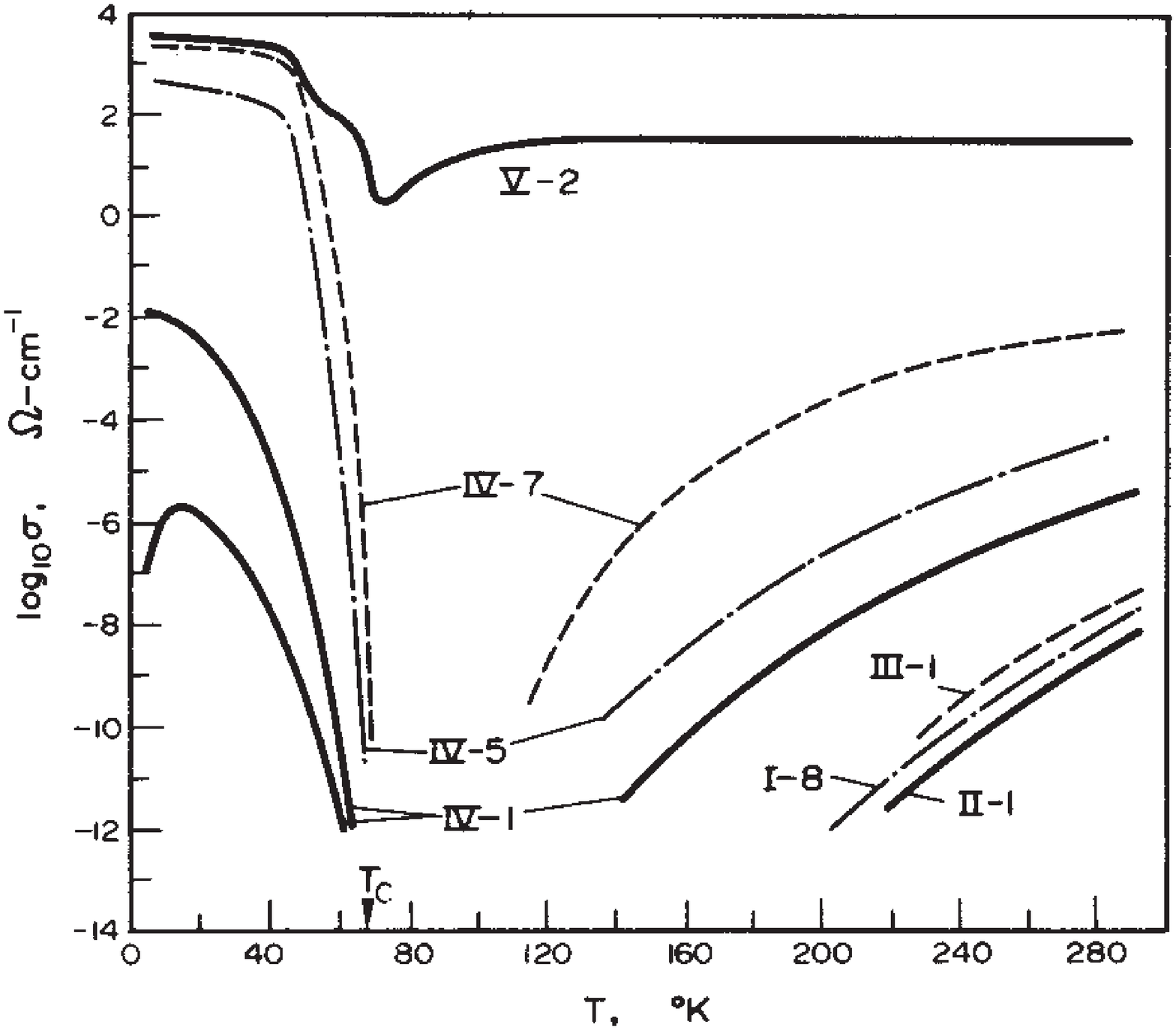}}
   \end{minipage}%
   &
   \begin{minipage}[b]{0.5\textwidth}
      \vspace{0pt}
      \centering
      \subfigure[$\rho(M)$ in La$_{0.7}$Ca$_{0.3}$MnO$_3$, from \cite{Hundley95}.]{
         \label{trans:f}
      \includegraphics[width=5cm]{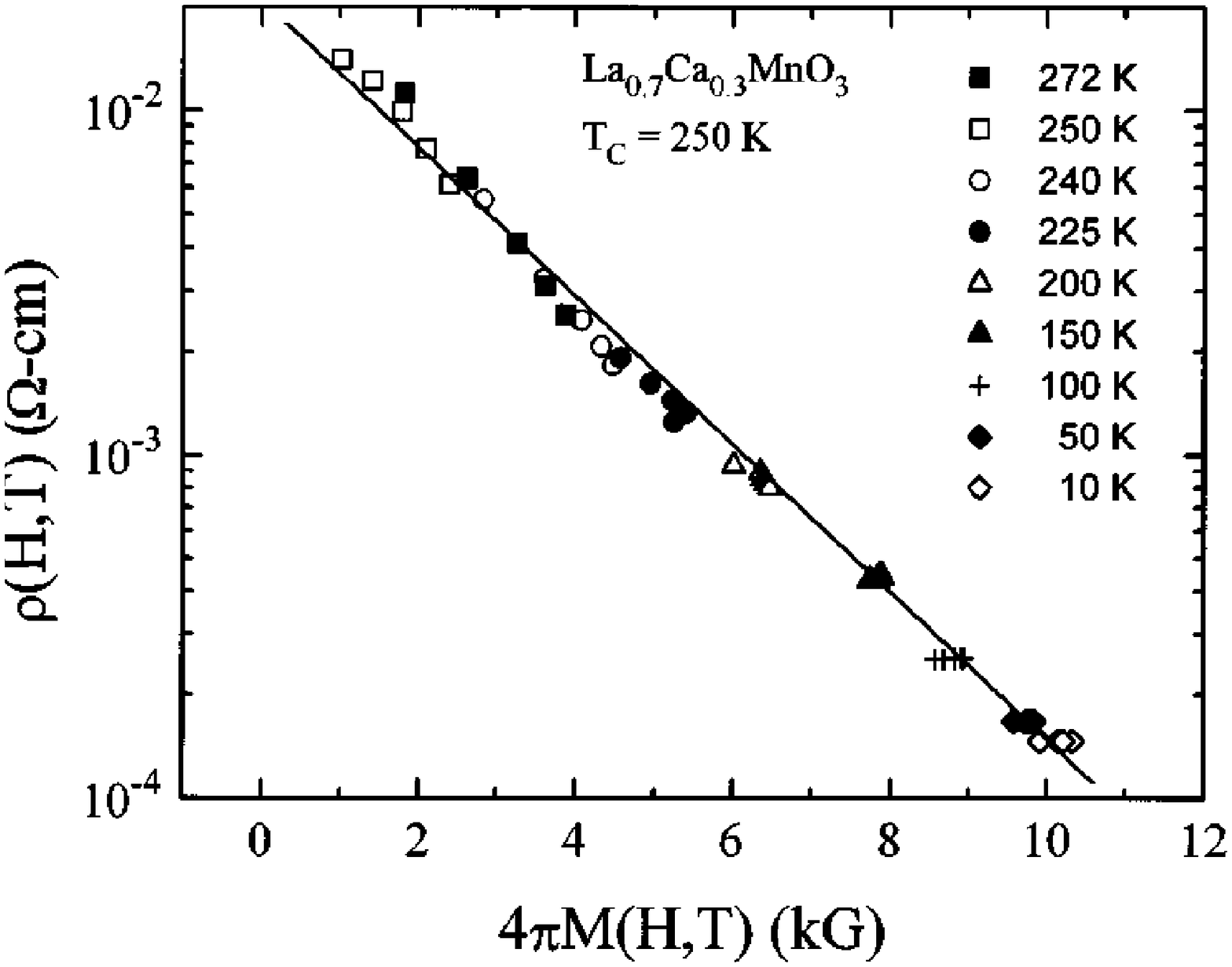}}
   \end{minipage}
   \end{tabular}
   \caption{\label{trans} Transport properties of EuO from different
   sources.}
\end{figure}

In figures \ref{trans}(a-c,e), we show a selection of resistivity
curves for different EuO$_{1\pm x}$ samples (many other samples
have been studied in literature, see e.g.
\cite{Shapira,Penney72,Llinares75,Shamokhvalov74}). For
stoichiometric or oxygen rich crystals the temperature dependent
conductivity shows a semiconductor-like behavior (curves 95-3 and
49-2-1 in figure \ref{trans:b} and curves I-8, II-1 and III-1 in
figure \ref{trans:d}). Eu-rich EuO on the other hand shows an
enormous reduction of its resistivity in the ferromagnetic state,
which can exceed 12 orders of magnitude (figure \ref{trans:d}).
However, besides these similarities there are a lot of
differences between the resistivity curves of crystals grown
under different conditions, like the magnitude and sharpness of
the resistivity change near T$_c$, the activated or non-activated
high-temperature resistivity, the resistivity cusp just above
T$_c$, the resistivity hump around 50 K and the low temperature
(T $<$ 20 K) resistivity upturn. For EuO films (see figure
\ref{trans:c}) the resistivity is again markedly different from
that of crystals, as the film resistivity decreases exponentially
with temperature and unlike crystals, neither seems to show a
temperature independent nor an activated high-temperature
resistivity. Moreover, the pronounced resistivity cusp near T$_c$
and the resistivity hump near 50 K which are observed in crystals
are not observed in films, even not in epitaxial films, as will
be discussed in chapters \ref{euoepi} and \ref{Magres}.

A second spectacular effect in Eu-rich EuO occurs upon the
application of a magnetic field: the resistivity changes by more
than six orders of magnitude when the sample temperature is close
to T$_c$, as is shown in figures \ref{trans:e} and \ref{Tmagres}.
In fact, there exists a considerable class of ferromagnetic
materials with a large negative magnetoresistance around their
Curie temperature, among which are the manganites
\cite{vanSanten50}, (Hg,Cd)Cr$_2$Se$_4$ \cite{Lehmann67}, EuB$_6$
\cite{Guy80} and the diluted ferromagnetic semiconductors like
Ga$_{1-x}$Mn$_x$As \cite{Ohno98}. A strong clue as to the origin
of these magnetoresistive effects comes from the observation that
the resistivity in some of these compounds only depends on the
magnetization and does not explicitly depend on temperature and
magnetic field (see figure \ref{trans:f} and figure
\ref{Mmagres}), because this strongly suggests that the metal
insulator transition is solely driven by a change in
magnetization.

Although the origin of this metal insulator transition is still
not unambiguously determined, several models have been proposed,
all of which relate the transition to the delocalization of doped
carriers below the Curie temperature. We will extensively review
and discuss these models in chapter \ref{Magres}. From an
experimental point of view however, much could be learned about
the mechanism of the metal insulator by investigating the
spectroscopic and microscopic structure of the vacancy sites in
Eu-rich EuO. The only experimental signatures of their presence
seem to be the peaks around 0.6 eV in the optical absorption
which are difficult to interpret (see section \ref{bindvac}). We
have tried to observe the occupied vacancy electron levels by
ultraviolet photoemission (UPS) on EuO thin films, however these
measurements were not successful, both because the density of
vacancy electrons is very small $\sim 0.3$\% and because of the
presence of He-satellites in the binding energy range where the
vacancy electrons were expected\footnote{However, at high doping
levels we detected a low intensity near the Fermi level with a
shape similar to that of Eu-metal (figure \ref{euoepi1}).}. Using
a monochromatized light source in combination with an
angle-resolved high-resolution analyzer would increase the
chances of observing the vacancy electron states. On the other
hand, the large advances in microscopic techniques in the last
decades might alow the microscopic observation and spectrocopy of
individual oxygen vacancy states in EuO. Moreover it would be
interesting to investigate the presence of larger Eu metal
clusters, as these might also strongly affect the transport
properties.

\subsection{Effect of pressure}
In rare earth compounds containing Ce, Sm, Eu, Tm or Yb, the
trivalent and divalent (tetravalent for Ce) configurations are
separated by energies of less than 2 eV \cite{Marel85,Marel88}.
In these compounds a change of valence can sometimes be
accomplished by pressure (or temperature) \cite{Jayaraman79}. A
well known example of such a transition is the $\gamma$ to
$\alpha$ phase transition in cerium. Jayaraman attributed a
sudden decrease in the EuO lattice parameter around a pressure of
300 kbar to such a valence changing Eu$^{2+}\rightarrow$
Eu$^{3+}$ transition, because the high pressure phase had a
silvery luster, indicative of a metallic state. At an even higher
pressure of 400 kbar he found a transition from NaCl to CsCl
structure. Later measurements by Zimmer {\it et~al.} did not find
the discontinuity at 300 kbar. Instead a much more continuous
change in lattice constant was observed which started around 130
kbar. Moreover, with increasing pressure, a large reduction of
the optical gap was found \cite{Wachter69,Zimmer84}, which
approached zero above 140 kbar. Although this change was
interpreted as a Eu$^{2+}\rightarrow$ Eu$^{3+}$ valence change
\cite{Zimmer84}, M\"ossbauer experiments \cite{Abd-Elmeguid90}
showed that the induced change resulted in a valence of only
$\sim$ Eu$^{2.07+}$ at high pressure. This induced concentration
of free carriers of $\sim$ 7\% is however more than sufficient to
result in metallic behavior.
\begin{figure}[!htb]
   \vspace{0.5cm}
   \centering
   \begin{minipage}[b]{0.5\textwidth}
      \centering
      \subfigure[Temperature dependent resistance
in EuO at different pressures, reproduced from \cite{DiMarzio87}.]{
         \label{pressure:a}
         \includegraphics[width=7.5cm]{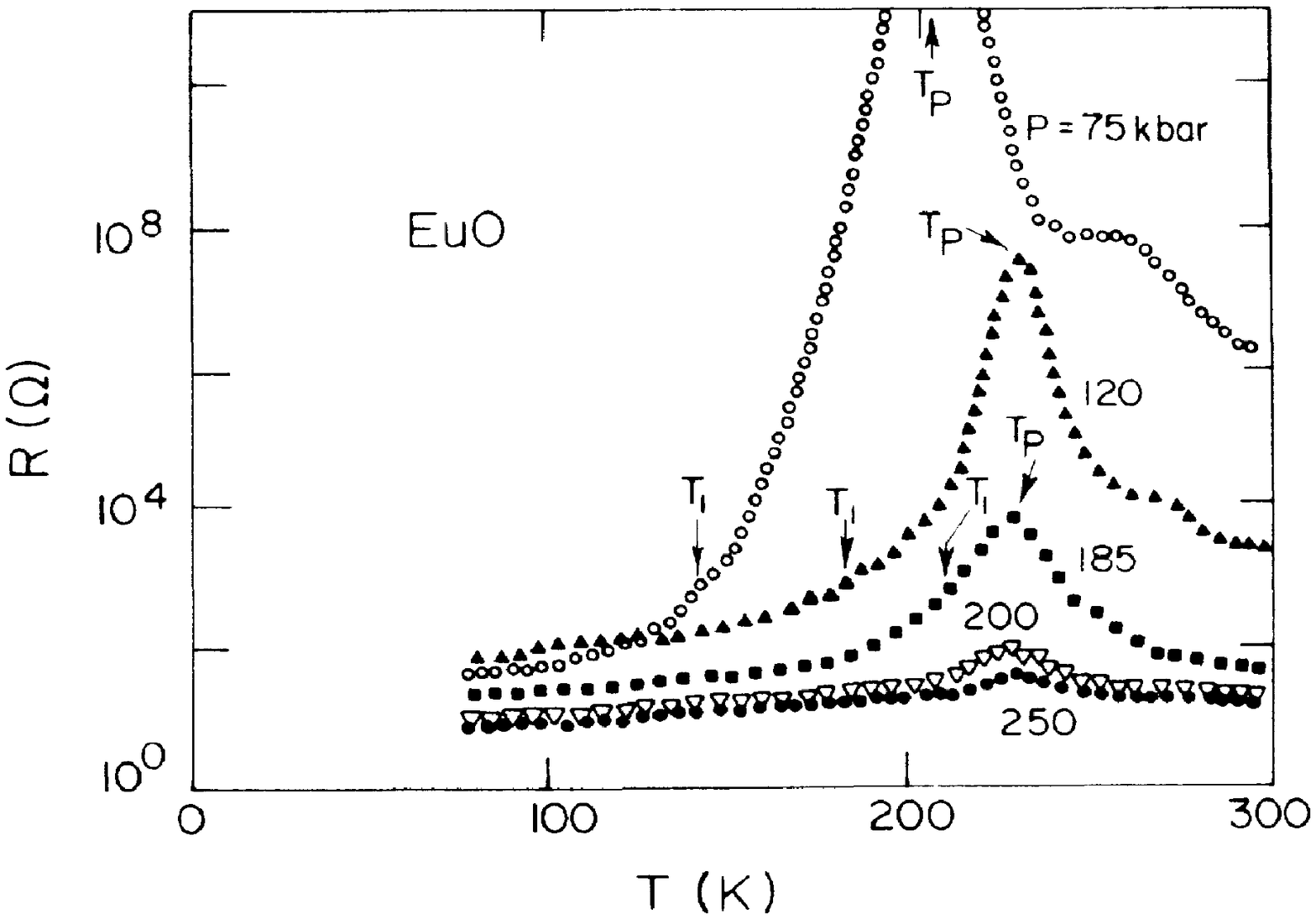}}
   \end{minipage}%
   \begin{minipage}[b]{0.5\textwidth}
      \centering
      \subfigure[Variation of Curie temperature with pressure, from
\cite{Abd-Elmeguid90}.]{
         \label{pressure:b}
      \includegraphics[width=7.5cm]{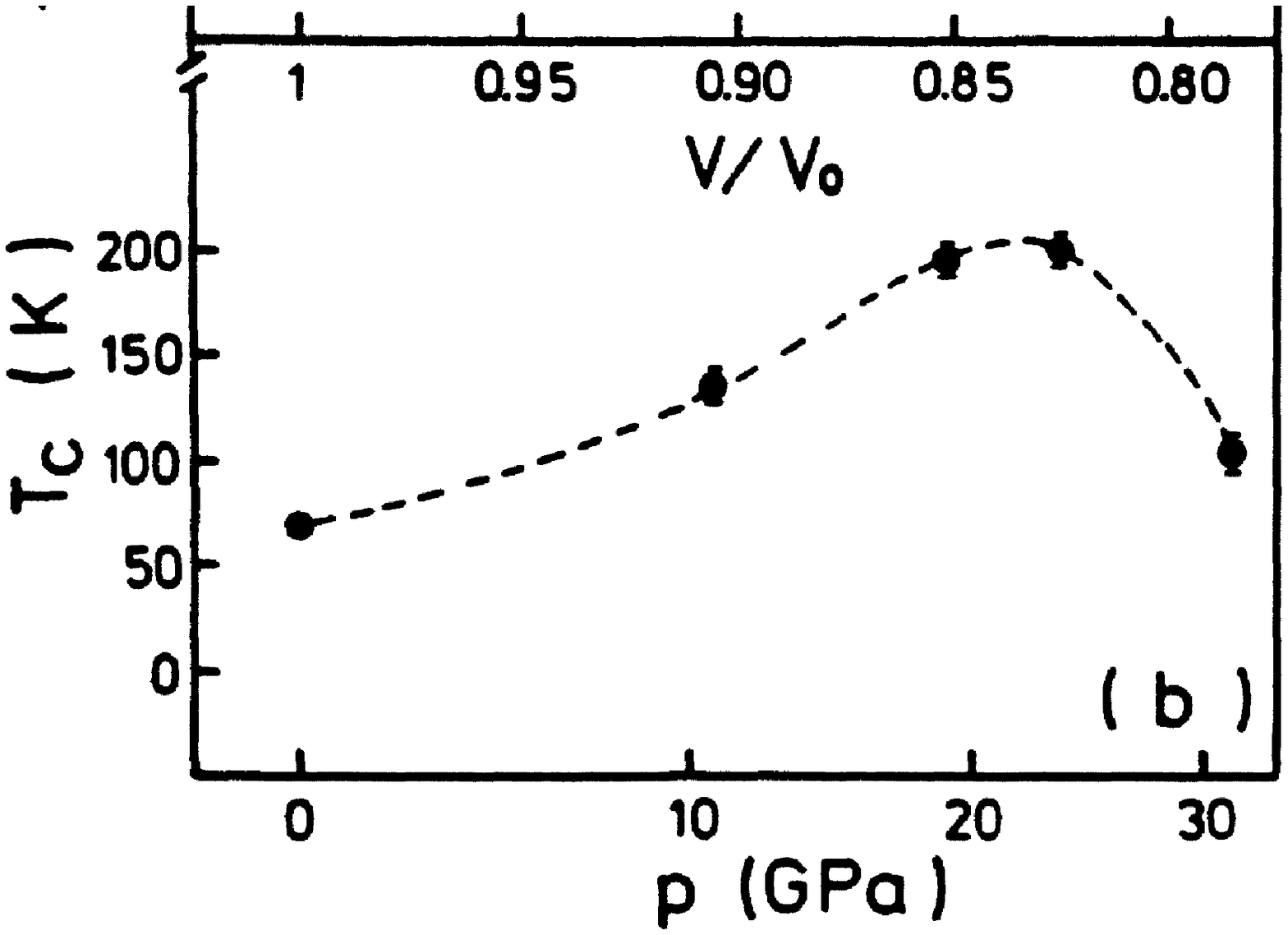}}
   \end{minipage}
   \caption{\label{pressure} EuO under pressure.}
\end{figure}
The related change in resistivity is at least as spectacular as
that induced by a magnetic field, a pressure induced change in
resistivity of more than 10 orders of magnitude is observed
around 200 K (see figure \ref{pressure:a} \cite{DiMarzio87}).
Pressure is expected to increase the exchange interactions, which
explains the increase of the Curie temperature up to $\sim 200$ K
at 200 kbar, as shown in figure
\ref{pressure:b}~\cite{Abd-Elmeguid90}. Increases in T$_c$,
accompanied by a decreasing optical gap have also been obtained
by chemical pressure, which was induced by substituting both Eu
and O for trivalent cations and anions:
Eu$_{1-x}$(Gd,Nd)$_x$O$_{1-x}$N$_x$ \cite{Etourneau80}. We note
here that it is an interesting perspective to try to induce an
effective pressure in EuO, by growing epitaxial thin films of EuO
substrates with a different lattice constant, like SrO ($a=5.16$
\AA) and CaO ($a=4.81$ \AA). Indeed, indications for modified
Curie temperatures in epitaxial films on different substrates
were found in EuO \cite{Iwata00} and EuS \cite{Stachow99}.

\section{Motivation and scope}

Considering these captivating features of EuO and similar Eu
chalcenogides EuS, EuSe and EuTe, it is not surprising that 18
years after the discovery\footnote{The existence of EuO was first
reported by Brauer \cite{Brauer53} and Eick {\it et al.}
\cite{Eick56}.} of ferromagnetism in EuO \cite{Matthias} Wachter
\cite{Wachter79} wrote in 1979: "It may very well be stated that
there hardly exists a group of compounds so thoroughly
investigated in every respect as the Eu chalcenogides." Anno
2002, it might thus appear very difficult to improve the
experimental and theoretical understanding of these compounds.
However, spectroscopic techniques have developed enormously
during the last 25 years as a result of improvements in
synchrotron radiation sources, electron spectroscopy and ultra
high vacuum technology. Yet, not many studies of EuO using these
new methods have been done, mostly because the discovery of
high-temperature superconductivity has shifted the attention of
many solid state physicists towards the cuprates and the
structurally similar manganite systems. Therefore, most of what
is known about the electronic structure of EuO is still based on
old optical studies and band-structure calculations.

\begin{itemize}
\item In this thesis, we will use new spectroscopies to
improve our knowledge of the electronic structure of EuO. In
particular, we will use x-ray absorption (XAS) at the O $K$ edge
to investigate the temperature dependent unoccupied density of
states in EuO (chapter \ref{Tempxas}). To study the effect of
ferromagnetism on the spin-polarization of the conduction band we
develop a new spectroscopic technique: spin-resolved x-ray
absorption spectroscopy\footnote{Since our experiment on EuO,
research projects have started to study the spin-polarization of
the unoccupied states in CrO$_2$ \cite{Huang02}, Fe$_3$O$_4$,
La$_{1-x}$Mn$_x$O$_3$ and Co doped TiO$_2$, using this new
technique.} (chapter \ref{Spinxas}). Moreover, we try to improve
the understanding of the relation between the spin and orbital
polarization of the conduction band in a ferromagnet, by
measuring the x-ray magnetic circular dichroism of EuO at the O
$K$ edge (chapter \ref{oxmcd}).
\end{itemize}

From a technological perspective, the exponentially increasing
data densities on magnetic recording media require large
enhancements of the sensitivity of magnetic reading heads, which
have led to a strong current interest in materials showing
colossal magnetoresistance effects. Additionally, developments in
magneto-optical recording have increased the demand for
ferromagnets which show large magneto-optical effects as media
for magneto-optical recording. Moreover, as decreasing circuit
sizes push silicon technology towards its fundamental limits,
scientists are seeking ways to exploit the properties of
electrons in a more efficient way. This has led to large research
efforts in the field of spintronics, which tries to make use of
not just the charge but also the spin of the electrons. This
might provide new logic, storage and sensor applications.
Moreover, spintronics devices might show completely new
phenomena, which are interesting from a more fundamental point of
view. However, to make use of the electron spin, one needs ways
to control and align it. In this respect, ferromagnetic
semiconductors which have a highly spin-polarized conduction band
are very interesting, as they combine the ability to
spin-polarize electrons, with a good spin-injection efficiency
\cite{Fiederling99,Ohno99} into semiconductor
devices\footnote{Contrary to ferromagnetic metals, which usually
have lower spin-polarization and a low injection efficiency to
semiconductors \cite{Filip00}.}. EuO is very interesting, because
it possesses these three technologically important features: CMR,
large magneto-optical effects and a highly spin-polarized
conduction band. While the rather low Curie temperature may make
EuO unsuitable for common household applications, it is an
extremely good model compound to develop new device concepts and
to study the basic physics that determines the performance of
devices. To manufacture such devices, the ability to grow
high-quality epitaxial EuO thin films with a controlled
stoichiometry is however essential.

\begin{itemize}
\item We make use of the superior control of growth conditions granted
by molecular beam epitaxy under ultra high vacuum (UHV)
conditions \cite{Herman96} to grow such high quality EuO films.
This new preparation route allows for the growth of EuO at lower
substrate temperatures than the 1800$^\circ$C required for bulk
single crystals. In particular, the 300-400$^\circ$C substrate
temperatures needed, are very compatible with the manufacturing
conditions for electronic devices. The effect of growth
conditions on the structure and stoichiometry of the films will
be studied in order to be able to control the stoichiometry and
electron doping of the films without requiring an extremely
accurate control of the ratio of beam fluxes (doping changes of
$\sim 0.1$\% have very large effects on the transport
properties). Moreover this investigation can provide insight in
the growth kinetics that is responsible for the growth of EuO
films (see chapter \ref{euoepi}). The technological interest is
not our only motivation to grow EuO films. EuO in thin film form
is also essential for spin-polarized electron spectroscopies like
spin-resolved photoemission and x-ray absorption, as the small
sample volume guarantees a minimal influence of the magnetic
moment of the sample on the electron-trajectories. Moreover, it
might be that EuO thin films might show intrinsically different
behavior than EuO in bulk form (see chapter \ref{Magres}).
Additionally, in situ growth of thin films with controlled doping
offers large flexibility to study doping dependent spectroscopic
and macroscopic properties without exposing the samples to
atmospheric conditions, and without requiring specialized high
temperature crystal growth facilities and expertise.

\begin{figure}[!htb]
\centerline{\includegraphics[width=8cm]{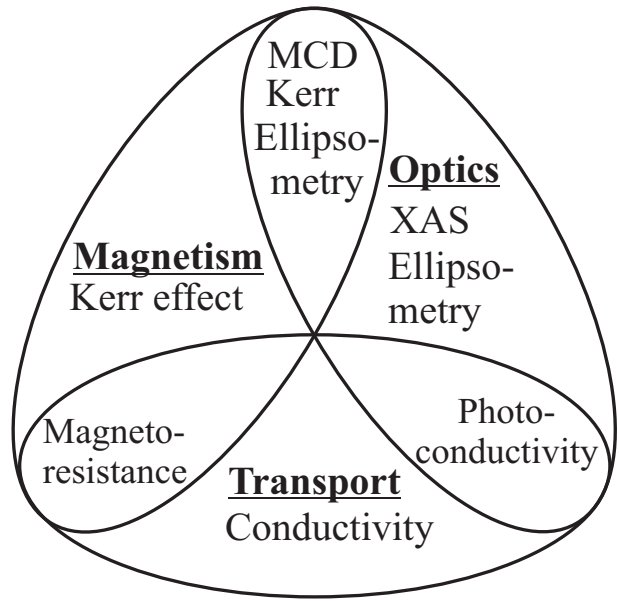}}
\caption{\label{Exps} Summary of experimental methods used in this
thesis to study magnetic, optical, transport and interrelated
properties of EuO.}
\end{figure}

\item Despite the fact that the magnetic, optical and transport
properties of EuO have been investigated extensively for more
than forty years, we will study these properties and the
interrelated magneto-optical, magnetotransport and phototransport
effects in EuO films, using various techniques as
summarized\footnote{Although ellipsometric measurements of EuO
are mentioned in the figure, the results are not presented in
this thesis, however we will briefly discuss them here: the
optical spectrum of a $\sim 50$~nm EuO film on Cr metal has been
studied using a Woollam VASE ellipsometer between 0.8 and 4~eV.
After converting these ellipsometric results to a bulk dielectric
function (taking the dielectric function of the Cr substrate into
account), good agreement with the real and imaginary part of the
dielectric constant as measured by G{\"u}ntherodt
\cite{Guntherodt74} was found (see figure \ref{epsilongun}).
Moreover, the low temperature shift of the absorption edge was
reproduced. Additionally an exploratory study of the
magneto-optical constants was performed by measuring the
ellipsometric changes when the film was magnetized in-plane by a
horseshoe magnet, in the plane of the light (longitudinal Kerr
effect configuration). Software was developed to extract the
off-diagonal components of the dielectric constant from the
magnetization induced changes in the ellipsometric signal,
fitting the curves using the Fresnel coefficients for a
magneto-optical thin film \cite{Yang93}. Good agreement with the
data by Wang {\it et~al.} \cite{Wang86} was obtained (see figure
\ref{Optics:b}). These data are not presented in this thesis
because both film quality (film had been exposed to air for one
week) and magnetic field control were insufficiently reliable.}
in figure \ref{Exps}. The motivation for this study is twofold:
the information that is obtained from these studies characterizes
the film properties and thus tells us more about the effect of
the growth conditions on the film properties. Secondly, there
exist theoretical controversies in the models that describe these
properties of EuO. This is the case for models that describe the
optical and magneto-optical properties (discussed in chapter
\ref{optprop}), but even more so for models of transport,
magnetotransport and phototransport (discussed in chapter
\ref{Magres}). The measurement of these properties in films of
which we have determined the growth conditions and structural
characteristics, in combination with the improved knowledge of
their electronic structure from electron spectroscopy, can help
us to come to a better understanding of the microscopic origins
of these properties.
\end{itemize}

In chapter \ref{NCCO} we will discuss photoemission, resonant
photoemission and XAS experiments which deal with the nature of
the states near the chemical potential in the electron doped
cuprate superconductor \ncco . Although this subject lies outside
the main line of this thesis, it will be presented, as it has
been a substantial part of the research efforts. However, the
performed research also included several interesting
participations which are outside the scope of this thesis. These
projects include: electron energy loss studies (EELS) of cuprate
and manganate compounds which have led to a better understanding
of EELS and photoemission spectroscopy
\cite{Schulte01,Schulte02}. Workfunction measurements on the
manganites \cite{Schulte01b}. Electron spectroscopic studies of
(Bi,Pb)SrCoO compounds that clarified the electronic structure of
the triangular CoO$_2$ lattice \cite{Mizokawa01}. Spin-resolved
photoemission measurements on silver which allowed verification
of theories on the Fano effect \cite{Minar01}. Spin-resolved
circular polarized resonant photoemission experiments that
studied the singlet character of the states near the chemical
potential in several High-T$_c$
cuprates~\cite{Tjernberg02,Ghiringhelli02b}. Besides some
projects on EuO already mentioned in this chapter, there were
also several other exploratory projects that have not been
published, including XPS and temperature dependent XAS on YVO$_3$
single crystals (large changes were observed around the phase
transition at 77 K), and XPS measurements on superconducting
MgB$_2$ thin films.

\section{Experimental}
Details on the experimental methods used in this thesis work can
be found in the respective chapters. As it is outside the scope
of this thesis to give a complete review of all the experimental
techniques used, we provide several references to books and
reviews on these techniques here. For details on photoemission
and photoelectron spectroscopy, see
\cite{Hufner95,SawatskyPes,SawatskyAug}, x-ray absorption and
x-ray magnetic circular dichroism \cite{deGroot01}, molecular
beam epitaxy and ultra high vacuum technology
\cite{Herman96,Luth93}, optical properties of solids
\cite{Wooten72}, magneto-optics \cite{Reim90} and for more
details on the temperature dependent transport measurements see
\cite{Hesper00}.

The thin EuO films studied in this thesis were grown in three
different facilities. The study of their structure and the
development of the growth recipe was done in a mini-MBE set-up
(for drawings and details see \cite{Altieri99}), which included
effusion cells, a temperature controlled sample-manipulator, an
annealing stage, gas inlets and a Staib RHEED gun. RHEED movies
could be recorded during growth by a CCD camera. This chamber was
attached to XPS and UPS spectrometer chambers which were
separated by a preparation chamber with load-lock. The XPS
spectrometer from Vacuum Generators/Surface Science had a
monochromatized Al-$K_\alpha$ x-ray source ($h\nu = 1486.6$ eV).
Photoemitted electrons were collected by a hemispherical electron
energy analyzer with a multichannel detector. The collection
angle was 55$^\circ$ with respect to the surface normal of the
samples. The overall resolution of the XPS was around 0.5 eV as
determined by measuring the Fermi level of silver. The UPS
measurements were performed with a differentially pumped Omicron
helium discharge lamp and a VG Clam 2 electron analyzer with an
acceptance angle of about $8^\circ$ and a resolution of about 0.1
eV. Moreover, in this chamber LEED measurements could be
performed with an Omicron Spectaleed system.

Once the growth procedures were well developed, EuO films were
also grown in the preparation chamber of a combined
photoemission-conductivity facility which has been described in
detail by Hesper \cite{Hesper00}. Conductivity of the films could
be monitored over a temperature range of 700 K, which allowed
measurements of the film conductivity at low temperatures but
also during growth. At high temperature, the holder on which the
sample was mounted had to be thermally decoupled from the
cryostat \cite{Hesper00}. The film conductivity was measured
using 4-point or 2-point contact configurations. The conductivity
was measured at low frequencies ($\sim$ 17~Hz) by Stanford
Research Systems SR-830 digital lock-in amplifiers or by DC
measurements with a Keithley 6512 electrometer. Both current and
voltage controlled measurements were done. For the photoemission
measurements in this set-up, He-I ($h\nu=21.22$ eV) and He-II
($h\nu=40.81$ eV) radiation was produced by an Omicron/Focus
HIS-13 gas discharge lamp. Photoelectrons were detected by a 150
mm VSW hemispherical electron analyzer which, after optimization,
reached resolutions of 7-8 meV.

The measurements presented in chapters \ref{Spinxas}, \ref{oxmcd}
and \ref{NCCO} were performed at beamline ID12B at the European
Synchrotron Research Facility (ESRF) in Grenoble (this beamline
has moved and is now named ID08). A detailed description of the
synchrotron, undulators, monochromators and the spin-polarized
electron detector \cite{Ghiringhelli99} was given by Ghiringhelli
\cite{Ghiringhelli02}. EuO films were grown at the beamline using
a growth procedure similar to that used in the set-ups in
Groningen. XAS of EuO films was done at the ESRF in total
electron yield mode, whereas XAS measurements of single crystals
of EuO were done at the Dragon beamline at the Synchrotron
Radiation Research Center (SRRC) in Taiwan in fluorescence yield
mode.

\markboth{}{Experimental}
\chapter{The magneto-optical spectrum of EuO}
\label{optprop}



{\it Until recently, most of what was known about the electronic
structure of EuO was based on optical measurements and band
structure calculations. The interpretation of the optical and
magneto-optical spectra has appeared to be rather difficult and
several contradictory explanations of these spectra have been
given. Before we will discuss measurements of the electronic
structure using new spectroscopic methods in the next chapters,
we will discuss how we can understand the optical properties.}

\section{Spectrum of Eu$^{2+}$ ion}
From band structure calculations \cite{Cho70} and early
photoemission studies \cite{Eastman69,Busch69,Eastman71,Cotti72}
it was found that the highest valence band orbitals of EuO are
the Eu $4f$ orbitals. Calculations \cite{Cho70,Elfimov03} also
show that the conduction band is formed mainly by Eu $5d$.
Therefore the lowest energy interband transitions in EuO are from
the Eu $4f$ orbitals to the Eu $5d$ conduction band. Before we
treat the optical spectrum of EuO let us first consider the
energies of these transitions in a Eu$^{2+}$ atom. Hund's first
rule tells us that the lowest energy state of the Eu $4f^7$ atom
will have a maximum spin of $S=7/2$. As the exchange interactions
between $4f$ electrons are large, the ground state will thus have
a very pure $^8S$ character, with $L=0$ (and thus $l_z=\sum
m_l=0$), $S=7/2$ and $J=7/2$. The excited states have a Eu
$4f^65d^1$ configuration. As transitions from an $L=0$ initial
state to an $L=0$ final state are not allowed (parity has to
change), $\Delta L=+1$ and in Russel-Saunders coupling the final
state has $L=1$, $S=7/2$ and thus $J=|S-L|\dots
S+L=\frac{5}{2},\frac{7}{2},\frac{9}{2}$. In the presence of
spin-orbit coupling, spin and orbital angular momentum mix and
the final states $J=7/2$ and $J=5/2$ now also have some $S=5/2$
character, with the spin of the $5d$ electron directed opposite
to that of the $4f$ electrons. As a result of the less favorable
spin-configuration these 2 'spin-flip' final states will have a
higher energy. Therefore we expect 5 different final state
energies. In the excitation spectrum of free Eu$^{2+}$ ions Sugar
and Spector \cite{Sugar} found 6 strong peaks (intensity $>$
100), one of which they attributed to a $J=9/2$ final state, two
were assigned to a $J=7/2$ final state and three to the $J=5/2$
final state, corresponding to the optical selection rule $\Delta
J = 0, \pm 1$. Two of the three $J=5/2$ final states are
separated only by 132 cm$^{-1}$ and therefore probably correspond
to the same type of transition. The number of main peaks is thus
nicely consistent with the 5 transitions expected from optical
selection rules.
\section{Europium atoms in crystals}
When a europium atom is incorporated in a lattice its optical
spectrum is strongly modified as a result of the crystal field.
The effect of the crystal field on the $4f$ electrons is however
much smaller than that on the $5d$ electrons. Therefore the $4f$
electrons should be treated \cite{Fazekas99} by a weak crystal
field with

exchange splittings $>$ spin-orbit coupling $>$ crystal fields.

\noindent And the $5d$ electrons are subjected to a strong
crystal field with

crystal field $>$ exchange splitting $>$ spin-orbit coupling.

\begin{figure}[!htb]
\centerline{\includegraphics[width=11.25cm]{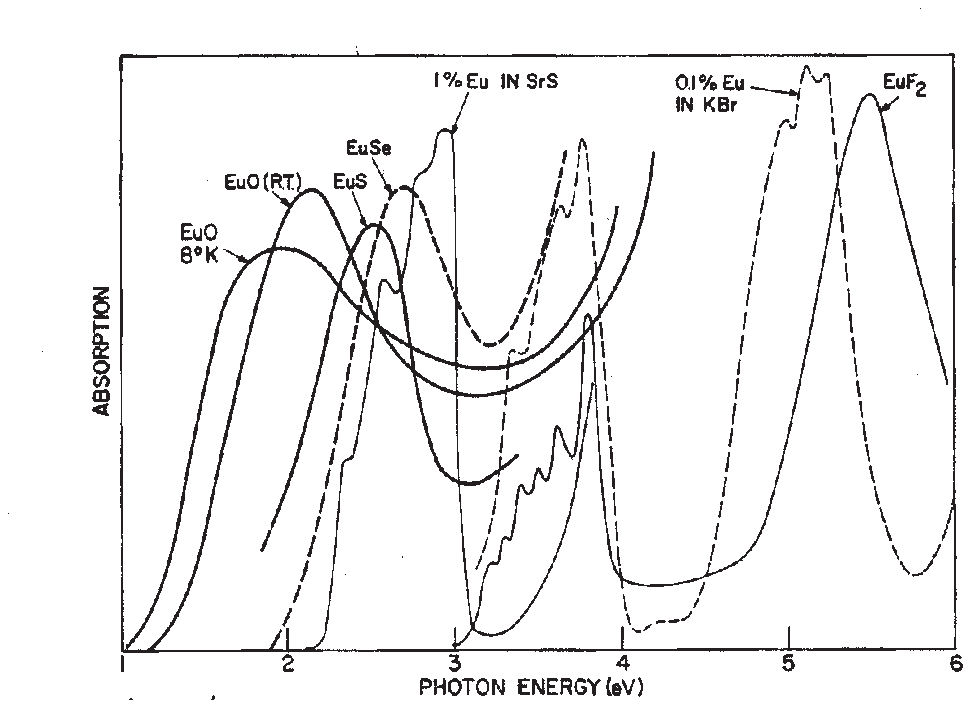}}
\caption{\label{freiser} Absorption spectra of several materials
containing Eu$^{2+}$ ions. From \cite{Freiser}.}
\end{figure}

\noindent Therefore the final state energy will be mainly
determined by the crystal field splitting of the $5d$ electron
and by the total angular momentum $J_{4f}$ of the 6 remaining
$4f$ electrons. Additionally the exchange interaction between the
$5d$ electron and the $4f$ electron can change the final state
energy. However as spin-flip transitions are only weakly allowed
via spin-orbit coupling, these spin-flip final states will only
slightly affect the absorption spectrum. From this evaluation we
therefore expect the optical spectrum of Eu ions in a lattice to
consist of 2 peaks corresponding to the $e_g$ and $t_{2g}$ final
states each of which is split into a septet corresponding to the
$J=0\dots 6$ states of the $4f^6$ configurations. Such structures
were convincingly found in the optical spectra of EuF$_2$ and of
Eu doped SrS and KBr by Freiser {\it et al.}
\cite{Freiser68,Freiser} as is shown in figure \ref{freiser}. It
should be noted that EuF$_2$ has a fluorite structure and as a
result of the crystal field the $e_g$ orbitals have the lowest
energy. The other compounds have a rocksalt structure with Eu
$t_{2g}$ as the lowest unoccupied states. The authors interpreted
the data along the same lines as we described above. In a later
work, Kasuya \cite{Kasuya72} arrived at similar conclusions, with
the multiplet structure arising as a result of different
$J_{4f}=0\dots 6$. In the europium chalcenogides the sharp
multiplet structures were generally not observed, instead a broad
peak with a width of around 0.7 eV is usually found. Possibly the
$4f$ multiplets are smeared out in the chalcenogides as a result
of $5d$ band formation. It is however also possible that the
optical spectra are blurred by surface oxidation of the reactive
materials. In fact in situ reflectivity measurements by
G{\"untherodt} \cite{Guntherodt74} on polished EuS crystals
showed very similar multiplet structures in the imaginary part of
the dielectric function indicating that the europium
chalcenogides should be interpretated similarly to EuF$_2$.

In the work by Kasuya the $4f^65d^1$ energy levels were labelled
by values $J_{\rm Kasuya}\equiv S + L_{4f}$ where $L_{4f}$ was
the orbital angular momentum of the $4f$ electrons. This
labelling was rather unfortunate as it was interpreted by
Schoenes and coworkers
\cite{Schoenes75,Schoenes85,Reim90,Wachter79,Drioli99b} that the
atomic optical selection rules for $J$ could be applied to
Kasuya's $J$, with only the final states with $J_{\rm
Kasuya}=\frac{5}{2}, \frac{7}{2}, \frac{9}{2}$ being optically
allowed. The application of the optical selection rules to
$J_{\rm Kasuya}$ is however incorrect as the orbital angular
momentum $L_{5d}$ of the $5d$ electron is neglected in the
process.

\section{Transition probabilities}

The magneto-optical spectra in EuO is mainly different from
spectra of other compounds containing Eu$^{2+}$ ions due to
interactions of the excited $5d$ electron with its surroundings.
Such interactions can result in the formation of a Eu $5d$ band
and increase the $5d$ exchange splitting in the ferromagnetic
phase. However the main features that determine the
magneto-optical spectrum in these compounds are still the $4f^6$
multiplets in the final state. As the $4f$ shell is rather
insensitive to its surroundings, these effects will be similar
for all compounds containing Eu$^{2+}$ and their spectra should
thus be explained along similar lines. Numerous studies
\cite{Weakliem72,Weakliem70,Dimmock70,Shen64a,Shen64,Shen64b,Ferre72,Ferre74,Kasuya72,Schoenes75}
have addressed the subject of the optical absorption spectrum of
Eu$^{2+}$. A very good agreement between calculated and measured
spectra of Eu$^{2+}$ impurities was found by Weakliem
\cite{Weakliem72} who diagonalized the full 196 $\times$ 196
matrix of $f^6 (^7F_J) 5d^1 (e_g)$ final states, including
spin-orbit and $f-d$ Coulomb interactions. He found that a best
fit with the experimental Eu$^{2+}$:CaF$_2$ data was obtained by
taking the $f-d$ exchange integral G$_1$=14 meV and a $5d$
spin-orbit coupling of 31 meV. The smallness of these parameters
suggests that the effects of $5d-4f$ exchange and $5d$ spin-orbit
coupling on the magneto-optical spectrum are weak. This motivated
us to calculate the magneto-optical spectrum in a simpler
approximation, neglecting these effects. Although the results are
less generally valid, we present them here as they provide a
simpler insight in the essentials that determine the
magneto-optical spectrum.

\subsection{Calculations}

The calculation is performed assuming that the angular part of
the $4f$ and $5d$ orbitals are still well approximated by
hydrogenic wave functions. In the initial state the $4f$ shell
contains seven electrons with $m_l=-l\dots l$ and $l=3$ (effects
of crystal field splitting on the $4f$ orbitals are neglected).
After optical absorption, one of these electrons is excited to
the $5d$ band and is thus in an $e_g$ or $t_{2g}$ orbital with
$l=2$. The optical dipole transition rate $R_{i\rightarrow f}$ is
given \cite{Gasiorowicz74} by:
\begin{equation}
R_{i\rightarrow f} \propto \left|<\phi_f| \mathbf{\varepsilon
\cdot r}| \phi_i>\right|^2 \label{transrate}
\end{equation}
where $\varepsilon = (\epsilon_x,\epsilon_y,\epsilon_z)$ is the
polarization vector of the light. As the radial integrals are the
same for all $4f-5d$ transitions, the transition rates only
depend on the angular components and can be expressed as
integrals of the spherical harmonics $Y_{l,m_l}$ as follows:
\begin{eqnarray}
\nonumber <\phi_f|\mathbf{\varepsilon \cdot r}|\phi_i> &\propto&
\int{d\Omega \phi_f^*(\theta,\phi)\mathbf{\varepsilon \cdot
\hat{r}}Y_{3,l_{zi}}(\theta,\phi)} \\
\mathbf{\varepsilon \cdot \hat{r}} &=&
\sqrt{\frac{4\pi}{3}}\left(\epsilon_z Y_{1,0} + \frac{-\epsilon_x
+ i \epsilon_y}{\sqrt{2}} Y_{1,1} + \frac{\epsilon_x+i
\epsilon_y}{\sqrt{2}} Y_{1,-1}\right) \label{transintegr}
\end{eqnarray}
The angular parts of the $t_{2g}$ and $e_g$ final states $\phi_f$
are \cite{Sugano70,Fazekas99}:
\begin{eqnarray}
\phi_{t_{2g}}&& \nonumber\\
xy&:&  \left(-iY_{2,2}+iY_{2,-2}\right)/\sqrt{2} \nonumber\\
xz&:&  \left(-Y_{2,1}+Y_{2,-1}\right)/\sqrt{2} \nonumber\\
yz&:& \left(iY_{2,1}+iY_{2,-1}\right)/\sqrt{2} \label{d-orbits} \\
\phi_{e_g}&& \nonumber\\
x^2-y^2&:& \left(Y_{2,2}+Y_{2,-2}\right)/\sqrt{2} \nonumber\\
3z^2-r^2&:& Y_{2,0} \nonumber
\end{eqnarray}
We choose the quantization axis ($z$-axis) parallel to the atom's
magnetization in the initial $^8S_{7/2}$ ground state, i.e. such
that $j_z=\frac{7}{2}$. Using equations (\ref{transrate}),
(\ref{transintegr}) and (\ref{d-orbits}) we have calculated the
transitions rates $R_{m_l,t_{2g}}=\sum_{t_{2g}}
R_{{\phi_{4f,m_l}} \rightarrow \phi_{t_{2g}}}$ and
$R_{m_l,e_{g}}=\sum_{e_{g}} R_{{\phi_{4f,m_l}} \rightarrow
\phi_{e_g}}$ for different polarization vectors with respect to
the quantization axis. The integrals were analytically evaluated
for linear polarized light parallel ($\varepsilon_\parallel$) and
perpendicular ($\varepsilon_\perp$) to the quantization axis, for
left and right-circularly polarized light ($\varepsilon_{\pm},
\Delta m_l=\pm 1$, with $\varepsilon \perp z$) and for an angular
average over all light polarizations ($\varepsilon_{avg}$). The
results are shown in table \ref{transrates}.

\begin{table}[!htb]
\centerline{\begin{tabular}{|c||c|c|c|c|c||c|c|c|c|c||c|}
\cline{2-11}
\multicolumn{1}{c}{} & \multicolumn{5}{|c||}{$R_{m_l,t_{2g}}\times 420$} & \multicolumn{5}{c|}{$R_{m_l,e_{g}}\times 420$} \\
\hline $m_l$ & $\varepsilon_\perp$ & $\varepsilon_\parallel$ &
$\varepsilon_{+}$ & $\varepsilon_{-}$ &$\varepsilon_{avg}$
& $\varepsilon_\perp$ & $\varepsilon_\parallel$ &  $\varepsilon_{+}$ & $\varepsilon_{-}$ &$\varepsilon_{avg}$ & $j_{z,4f}$\\
\hline
3 & 45 & 0 & 0 & 90 & 30& 45& 0 & 0 &90 & 30 & 0\\
\hline
2 & 60 & 30 & 0 & 120& 50& 0& 30& 0& 0&10 & 1\\
\hline
1 & 3 & 96 &  6&  0& 34& 39& 0& 6& 72&26 & 2\\
\hline
0 & 36 & 0 &  36&  36& 24& 0& 108&0 &0 &36 & 3\\
\hline
-1 & 3 & 96 &  0& 6& 34& 39& 0& 72& 6&26 & 4\\
\hline
-2 & 60 & 30 & 120& 0 &50 & 0& 30&0 &0 &10 & 5\\
\hline
-3 & 45 & 0 &  90& 0 & 30& 45& 0& 90& 0&30 & 6\\
\hline
\end{tabular}}
\caption{\label{transrates} Transition rates with certain
$j_{z,4f}$.}
\end{table}

\begin{table}[!htb]
\centerline{\begin{tabular}{|c||c|c|c|c|c||c|c|c|c|c|}
\cline{2-11}
\multicolumn{1}{c}{} & \multicolumn{5}{|c||}{$R_{J_{4f},t_{2g}}\times 44100$} & \multicolumn{5}{c|}{$R_{J_{4f},e_{g}}\times 44100$} \\
\hline $J_{4f}$ & $\varepsilon_\perp$ & $\varepsilon_\parallel$ &
$\varepsilon_{+}$ & $\varepsilon_{-}$ &$\varepsilon_{avg}$
& $\varepsilon_\perp$ & $\varepsilon_\parallel$ &  $\varepsilon_{+}$ & $\varepsilon_{-}$ &$\varepsilon_{avg}$ \\
\hline
0 & 675 & 0    & 0  & 1350 & 450  & 675 & 0    & 0     &1350 & 450 \\
\hline
1 & 1725 & 525 & 0  & 3450 & 1325 & 675 & 525 & 0   &1350 & 625 \\
\hline
2 & 1788 & 2541 & 126& 3450 & 2039 & 1494 & 525 & 126 & 2862& 1171 \\
\hline
3 & 2733 & 2541 & 1071& 4395& 2669& 1494& 3360 & 126 & 2862 &2116 \\
\hline
4 & 2838 & 5901 & 1071& 4605& 3859& 2859& 3360& 2646 & 3072 &3026 \\
\hline
5 & 5988 & 7476 & 7371& 4605 &6484 & 2859& 4935& 2646 &3072 &3551 \\
\hline
6 &10713 & 7476 & 16821& 4605 & 9634& 7584& 4935& 12096& 3072& 6701 \\
\hline
\end{tabular}}
\caption{\label{Jtransrates} Transition rates to states with
certain $J_{4f}$.}
\end{table}

\begin{figure}[!htb]
\centerline{\includegraphics[width=11.25cm]{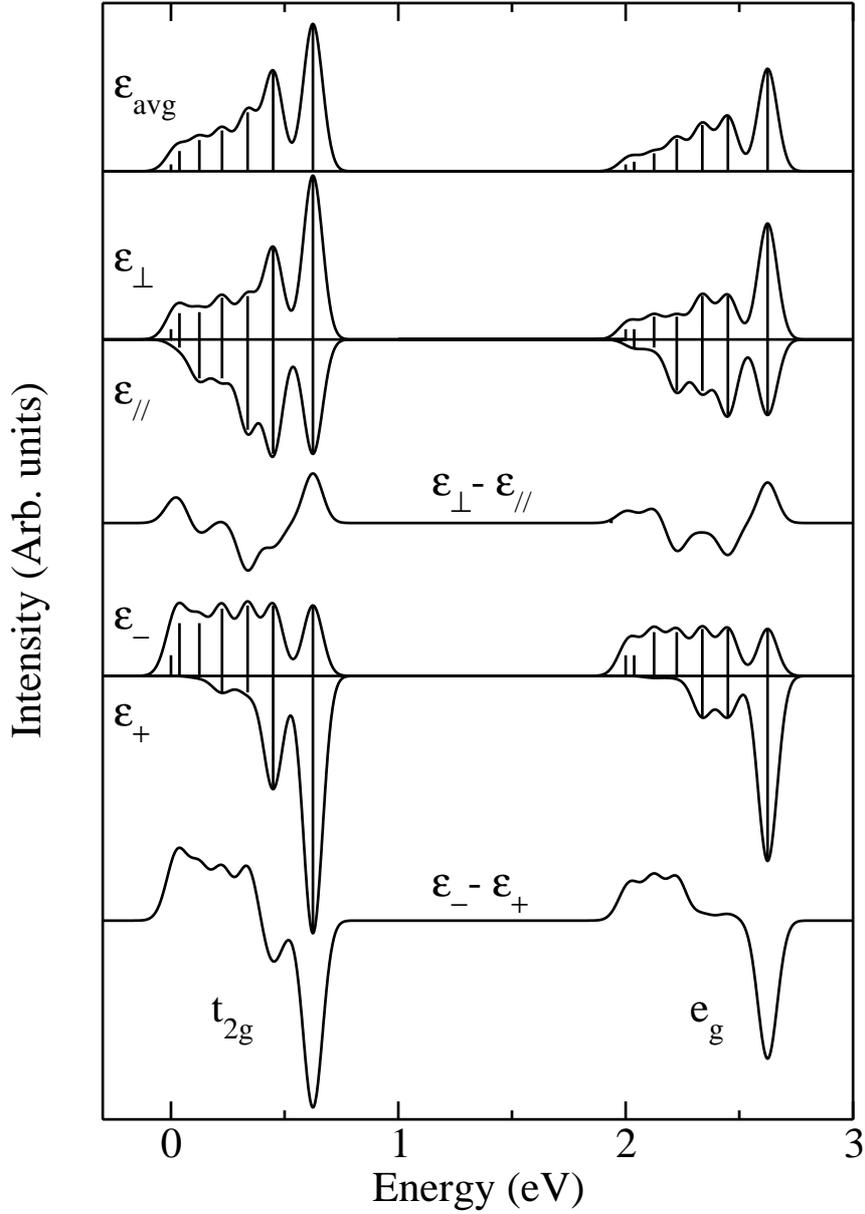}}
\caption{\label{calcoptprop} Calculated Eu$^{2+}$ $4f\rightarrow
5d$ spectra for linearly polarized light perpendicular
($\varepsilon_\perp$) and parallel ($\varepsilon_\parallel$) to
the $z$-axis, left and right-circular polarized light
($\varepsilon_\pm$) and unpolarized light ($\varepsilon_{avg}$).
Also shown are the magnetic linear ($\varepsilon_\perp -
\varepsilon_\parallel$) and circular ($\varepsilon_- -
\varepsilon_+$) dichroism spectra. The $z$-axis is taken such
that $j_{z,4f}=\frac{7}{2}$ in the initial state.}
\end{figure}
The $4f$-shell has $s_{z,4f}=7/2$ in the initial state, one
electron is removed and thus in the final state $s_{z,4f}=3$. In
the initial state $l_{z,4f}=\sum m_{l,4f}=0$, removing one
electron with orbital angular momentum $m_l$ thus leads to
$l_{z,4f} = -m_l$ in the final state, therefore
$j_{z,4f}=s_{z,4f}+l_{z,4f}=3-m_l$ as indicated in table
\ref{transrates}. As the 3 $t_{2g}$ orbitals are essentially
degenerate just like the 2 $e_g$ orbitals, the final state energy
is mainly determined by the crystal field splitting and the total
angular momentum $J_{4f}$ of the remaining $4f$ electrons. As
there are $2J_{4f}+1$ state functions $\phi_{J_{4f},j_z}$ with
$j_z=-J_{4f}\dots J_{4f}$, each of the final states with a
specific $j_z$ is a superposition
\begin{equation}
\phi_{j_z}=\frac{1}{\sqrt{7-|j_z|}} \sum_{J_{4f} \geq |j_z|}^6
\phi_{J_{4f},j_z}
\end{equation}
Therefore the transition rate to a final state with a specific
$J_{4f}$ is
\begin{equation}
R_{J_{4f}} \propto \sum_{|j_z|\leq J_{4f}}
\frac{R_{j_z}}{7-|j_z|}
\end{equation}

\subsection{Comparison with Eu$^{2+}$ impurities}

\begin{figure}[!htb]
\centerline{\includegraphics[width=11.25cm]{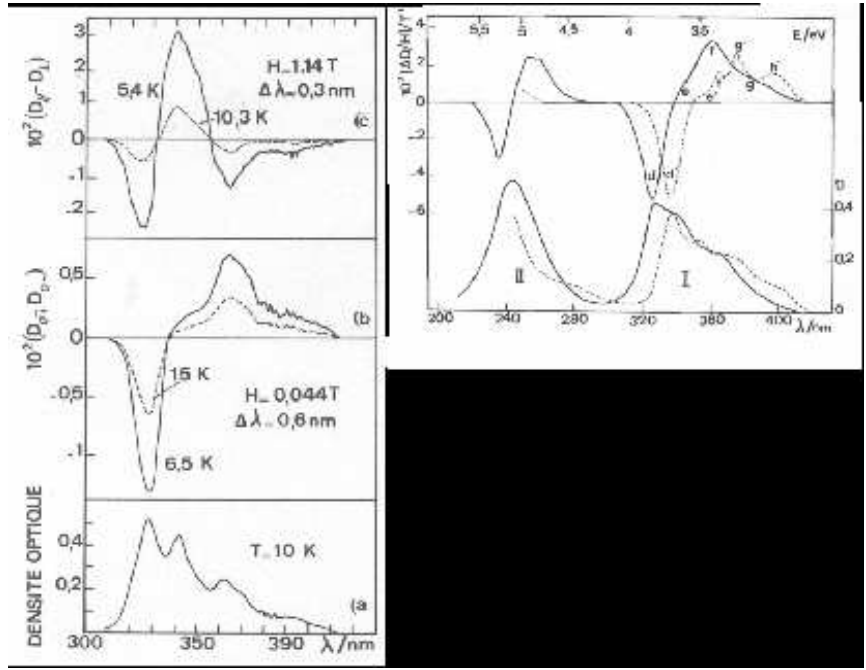}}
\caption{\label{Ferre2} Optical absorption and MCD and MLD at $5d
t_{2g}$ in Eu$^{2+}$:KCl at 10 K (left) and MCD of
Eu$^{2+}$:CaF$_2$ at room temperature (right), reproduced from
Ferr\'e {\it et al.} \cite{Ferre72,Ferre74}. Note that $D_{\sigma
\pm}$ corresponds to $\varepsilon_\mp$ and $D_{\parallel,\perp}$
to $\varepsilon_{\parallel,\perp}$.}
\end{figure}

\noindent Thus we find the transition probabilities to the
different $J$ final states as tabulated in table
\ref{Jtransrates}. In figure \ref{calcoptprop} we show the
resulting calculated spectra, making use of the experimentally
obtained energy splittings \cite{Freiser68} between the different
$J$ multiplets and a crystal field splitting 10$D_q$ of 2 eV.
Gaussian broadened spectra are also shown. The absorption
intensity distribution of the calculated multiplets is similar to
that measured in EuF$_2$ (figure \ref{freiser}). Large spectral
changes occur when the polarization of the light is changed. In
fact at the bottom of the absorption edge, the absorption of
circularly polarized light is 100\% helicity polarized, i.e.
$(\alpha_{\varepsilon_-}-\alpha_{\varepsilon_+})/(\alpha_{\varepsilon_-}+\alpha_{\varepsilon_+})=$~1.
The calculation shows how the $4f \rightarrow 5d$ transitions
from the $^8S$ ground state can account for the very large
magneto-optical effects in compounds containing Eu$^{2+}$ ions,
effects which are among the strongest known for any class of
compounds \cite{Shafer63,Shen64,Suits66,Greiner66}. The obtained
spectra are indeed similar to the more sophisticated calculations
by Weakliem \cite{Weakliem72} and account very well for the
observed magneto-optical spectra of Eu$^{2+}$:CaF$_2$ and
Eu$^{2+}$:KCl like in figure \ref{Ferre2}
\cite{Ferre74,Weakliem72}.

\begin{figure}[!htb]
\centerline{\includegraphics[width=8cm]{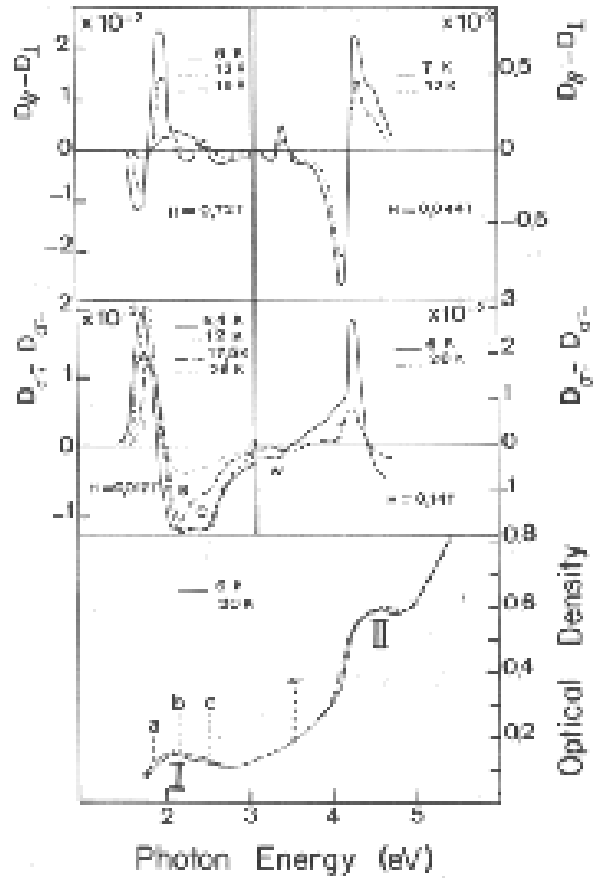}}
\caption{\label{Ferre} Optical absorption, MCD and MLD of thin
film of EuS deposited onto CaF$_2$, reproduced from Ferr\'e {\it
et al.} \cite{Ferre72,Ferre74}. Note that $D_{\sigma \pm}$
corresponds to $\varepsilon_\mp$ and $D_{\parallel,\perp}$ to
$\varepsilon_{\parallel,\perp}$.}
\end{figure}

\subsection{Comparison with europium chalcenogides}

Although the model reproduces the multiplets in the spectra of
EuF$_2$ and of compounds containing Eu$^{2+}$ impurities very
well, the absorption spectra of the Eu chalcenogides are much
broader. Possibly this is due to band formation. To model such
band formation, we have convoluted the $5d$ density of states as
obtained by LDA+U of EuO (see \cite{Elfimov03} and figure
\ref{fig5}), with the calculated multiplet structures. The
resulting spectra have absorption peak energies of $\sim$ 3-4 eV
and do not reproduce the measured absorption spectrum of EuO,
which peaks around 2 eV. This indicates that excitonic effects
pull spectral weight towards the band bottom as was suggested by
Kasuya \cite{Kasuya68}. Recently an attempt was made to model the
magneto-optical properties of EuO using an energy-band theory
\cite{Oppeneer96}, which resulted in a too broad theoretical
spectrum, possibly as a result of the same excitonic effects,
which tend to reduce the effect of band formation on the
broadening of the energy levels. This excitonic effect also
explains why the ionic approximation, in which we calculated the
magneto-optical spectrum of Eu$^{2+}$, still provides a
reasonable description of the measured magneto-optical spectrum
in the chalcenogides \cite{Ferre72,Schoenes75,Wang86,Reim90} as
can be seen by comparing figures \ref{Ferre} and
\ref{calcoptprop}.

\begin{figure}[!htb]
\centerline{\includegraphics[width=11.25cm]{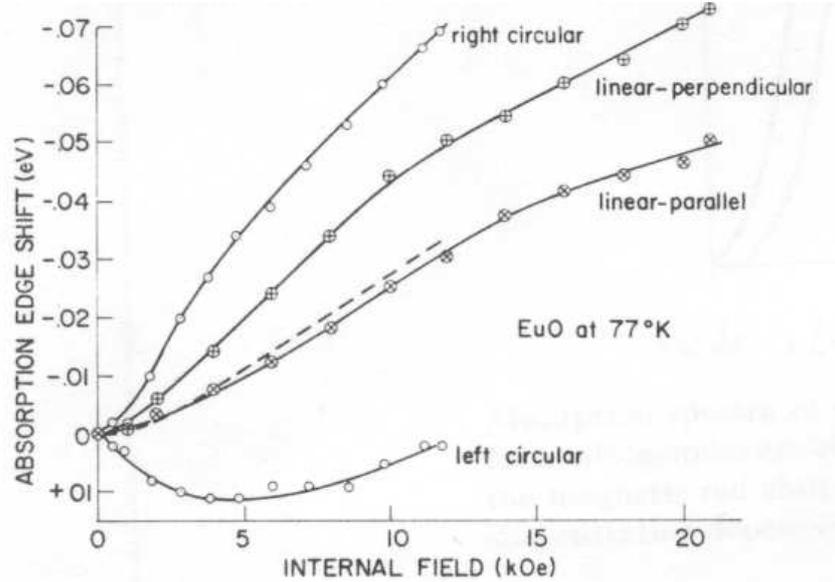}}
\caption{\label{Freisermshift} Absorption edge shift of EuO at 77
K for different magnetization directions and light polarizations
as function of the internal magnetic field, reproduced from
Freiser {\it et al.} \cite{Freiser}.}
\end{figure}

From the calculated spectra we can also understand why the shift
of the absorption edge strongly depends on the polarization of
the light when the sample is magnetized. In figure
\ref{Freisermshift} we reproduce data by Freiser {\it et al.}
taken at 77 K. Besides a shift of the absorption edge as a result
of the exchange splitting of the conduction band with increasing
magnetization, magnetic dichroism also results in a shift of the
absorption edge. The effect of dichroism on the shift can be
understood by comparing figures \ref{calcoptprop} and
\ref{Freisermshift}. When no internal magnetic field is present,
the Eu spins will be randomly oriented and the optical spectrum
will be similar to that of $\varepsilon_{avg}$. Upon application
of a field the Eu moments will align with the magnetization and
the absorption edge will shift as a result of the exchange
splitting, as will be discussed in more detail in chapter
\ref{Tempxas}. Additionally a shift of the absorption edge is
induced by dichroism. As the absorption spectrum in figure
\ref{calcoptprop} for $\varepsilon_{\perp}$ is the most similar
to that for $\varepsilon_{avg}$, the smallest dichroic
contribution is expected for a light polarization vector
$\varepsilon_{\perp}$. For $\varepsilon_\parallel$ a small shift
of the bottom edge of the conduction band to higher energies is
expected while for $\varepsilon_+$ the dichroic shift of this
edge to higher energies even exceeds the reduction due to
exchange splitting. For right-circular polarized light
($\varepsilon_-$) the shift of the absorption edge is increased
by dichroism as observed.

The calculated spectra contradict the claim of Freiser {\it et
al.} \cite{Freiser} that the spectra with the polarization vector
parallel to the magnetization ($\varepsilon_\parallel$) are free
of dichroism. This is an important observation as it means that
to measure the temperature dependence of the absorption edge as a
result of the exchange splitting, the configuration with the
magnetic field parallel to the light polarization is not optimal
as it might introduce a small reduction of the shift due to
dichroic effects (Freiser {\it et al.} use a field
H$_\parallel=2$~kOe). Instead measurements with zero total
magnetization would be preferable, as $\varepsilon_{avg}$ would
be measured at all temperatures.

\section{Conclusions}

We have analytically calculated the magneto-optical spectrum of
$4f \rightarrow 5d$ transitions on Eu$^{2+}$ ions using several
simplifying approximations. The calculated spectra agree very
well to the measured spectra of Eu$^{2+}$ impurities in crystals
and also agree well to more sophisticated calculations. Thus we
have shown that the multiplet structures and large
magneto-optical effects in these compounds are essentially a
result of the polarization dependent transition rates to
different $J_{4f}$ final states when exciting a $4f$ electron to
the $5d$ band. We argue that the magneto-optical effects in the
Eu chalcenogides have the same origin. Moreover we showed that
the polarization dependence of the shift of the absorption edge
can be well understood from the calculated dichroic effects.


%

\chapter{Growth of EuO films with controlled properties}
\label{euoepi}


{\sl We report on the growth and properties of epitaxial and
polycrystalline thin films of the ferromagnetic semiconductor
EuO. The films were grown by MBE on several substrates. Their
epitaxial structure was studied by RHEED and LEED. The
composition of the films was checked by photoemission and the
effect of growth conditions on the transport and magnetic
properties were studied by in situ conductivity and Kerr-effect
measurements. Based on our findings we propose a mechanism in
which the growth is limited by the oxygen flux. This mechanism
strongly facilitates the growth of stoichiometric EuO if the
substrate temperature is high enough to reevaporate Eu atoms
($\sim 350^\circ$C in vacuum). Moreover, we show that the doping
in Eu-rich EuO films can be tuned by the growth temperature, thus
allowing control over the magnitude of the metal-insulator
transition.}

\section{Introduction}

EuO is a ferromagnetic semiconductor with a Curie temperature of
about 69 K, which was discovered in 1961 by Matthias {\it et al.}
\cite{Matthias}. EuO has a rocksalt NaCl structure with a lattice
parameter of 5.144 \AA~at room temperature. The transition to the
ferromagnetic state is accompanied by a shift of the optical
absorption edge \cite{Busch,Freiser,Schoenes} and in Eu-rich EuO
a spectacular metal-insulator transition (MIT) and colossal
magnetoresistance (CMR) occur \cite{Oliver72,Shapira}. Moreover
EuO shows very large magneto-optical effects. The last decades
the interplay of magnetic, optical and transport properties has
attracted much interest of both theorists and experimentalists,
especially because EuO is a model compound for a whole class of
compounds with similar properties. It was shown long ago that
good single crystals of EuO can be grown at high temperatures
around 1800$^\circ$C \cite{Shafer72,Fischer}. We have obtained a
recipe for the growth of high quality epitaxial EuO films at much
lower temperatures. The ability to grow such films is especially
important for device and multilayer research, because EuO could
prove to be a very efficient spin polarized source for
semiconductor spintronics research \cite{Hao}.

In this chapter we study the epitaxial growth of EuO on the (001)
plane of 16 mol\% Y$_2$O$_3$ stabilized cubic ZrO$_2$ (YSZ) which
has a lattice constant of 5.139~\AA. This ensures an almost
perfect lattice matching of the substrate with the EuO film,
which is essential for the growth of ultrathin films without
defects due to relaxation. We also study films grown on (001) MgO
and ($1\bar{1}02$) Al$_2$O$_3$. Besides the effect of the
relative Eu and O fluxes, we examine the effect of the substrate
temperature (T$_s$) on the growth. Thus we demonstrate that the
growth of films with controlled stoichiometry is much facilitated
by using a temperature regulated distillation effect. We show
that the growth conditions have a very big influence on the
transport properties, and in particular on the metal-insulator
transition.


\section{Experimental}
\label{experimental} 
\begin{figure}[!htb]
\centerline{\includegraphics[width=11.25cm]{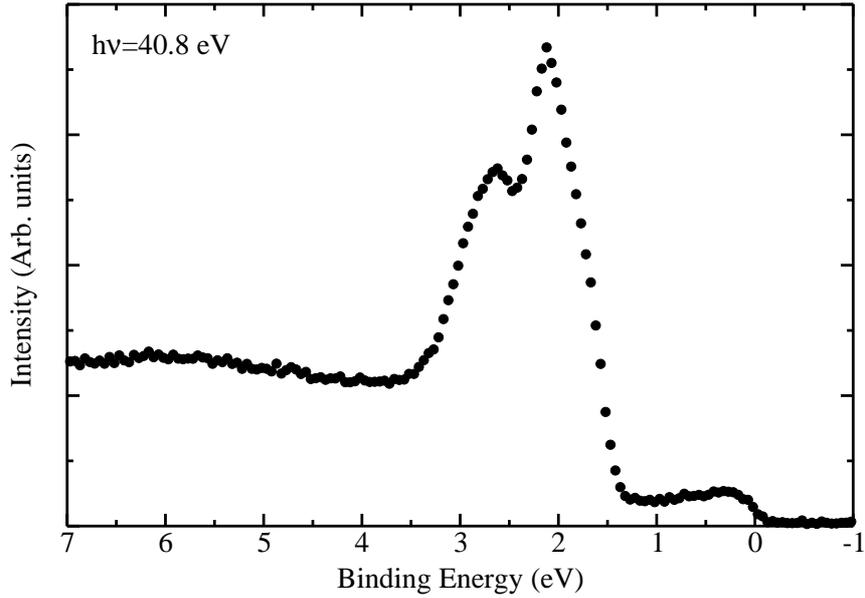}}
\caption{\label{euoepi1} Satellite corrected He-II photoemission
spectrum of a Eu metal film on polycrystalline tantalum. The
spectrum shows the Fermi level and the Eu $4f$ bulk (2.1 eV) and
surface (2.7 eV) states.}
\end{figure}

EuO films were grown by molecular beam
epitaxy (MBE) in two different UHV setups, one of which was
dedicated to low-temperature conductivity, Kerr-effect and UPS
measurements (base pressure below 5$\times$10$^{-11}$ mbar), the
other was equipped with RHEED, LEED, XPS and UPS facilities (base
pressure below 2$\times$10$^{-10}$ mbar). To grow the EuO films,
high purity Eu metal (supplied by Ames Laboratory) was evaporated
from an effusion cell which was kept at a constant temperature
between 380$^\circ$C and 500$^\circ$C depending on the type and
filling of the effusion cell. The Eu flux (J$_{Eu}$) was measured
using a quartz crystal thickness monitor at the sample position.
Simultaneously high purity oxygen gas was supplied through a leak
valve which was located far from the sample, thus allowing
accurate control of very low background pressures of oxygen. The
flux of oxygen atoms at room temperature was approximated from
the kinetic theory of gases \cite{Luth93,Herman96} to be
J$_{O}=5.37 \times 10^{20} \times P_{ox}$~cm$^{-2}$s$^{-1}$,
where $P_{ox}$ is the partial pressure of oxygen in mbar as
measured by an ionization pressure gauge which is located at
similar distance from the oxygen inlet and cryopump as the
sample. We estimated the film thicknesses by assuming a sticking
factor of 1 for oxygen and stoichiometric EuO films, the
correctness of this assumption will be discussed later. We
estimate the error in the value of the europium and oxygen fluxes
to be about 15\%. Despite thorough vacuum outgassing, Eu
evaporation caused the pressure in the chamber to increase to
3-5$\times 10^{-9}$ mbar, which was confirmed by a residual gas
mass-spectrometer to be a result of $H_2$ gas from the source
material. According to the supplier the Eu metal indeed contained
an estimated 300 ppm of H atoms. To check the effect of hydrogen
contamination, we have grown thin films of Eu metal on clean
polycrystalline Ta. As can be seen from figure \ref{euoepi1} the
UPS spectrum is similar to that in other photoemission studies of
Eu metal \cite{Olson95,Wahl99}, suggesting that the hydrogen
content of the films is small\footnote{If the hydrogen
contamination is high, semiconducting EuH$_2$ might be formed
\cite{Bischof83,Gasgnier74,Olson95,Wahl99}.}.
\

Reflection high-energy electron diffraction (RHEED) measurements
were conducted during growth at a primary beam energy of 20 keV
and an incident angle of 0.3$^\circ$-1.2$^\circ$ with respect to
the surface. The primary beam was partially blocked by a small
movable metal plate close to the RHEED screen to avoid obscuring
of the RHEED features by the high intensity primary beam. Before
growth, the substrates were annealed in an oxygen background
pressure of $\sim1\times10^{-6}$ mbar at temperatures of
650$^\circ$C for MgO and 820$^\circ$C for YSZ. The color of the
YSZ substrates turned slightly brownish after this annealing
procedure. The samples were resistively heated and their
temperature was monitored by a thermocouple close to the sample
position. XPS measurements were performed using monochromated Al
$K \alpha$ radiation at an energy resolution of the electron
analyzer of 0.5 eV.


Chromium metal contacts with a thickness of 100 nm were
evaporated on the substrates to measure the conductivity of the
EuO films. Sample resistances were determined by 2 and 4 point DC
measurements or by a low-frequency lock-in technique. The sample
temperature was monitored by a Pt film resistor at high and
moderate temperatures and by a Si-diode at low temperatures.
Kerr-effect measurements were performed using a 1.5 mW HeNe
(h$\nu=1.9$~eV) laser. The in-plane polarized light was modulated
by a chopper. Its angle of incidence was between 40$^\circ$ and
50$^\circ$. The outgoing beam was analyzed by a polarizer
(=analyzer) which was rotated with respect to the incident
polarization by an angle between 45$^\circ$ and 80$^\circ$
depending on the magnitude of the Kerr rotation. The transmitted
signal was measured by a photodiode coupled to a lock-in
amplifier. The birefringence of the cryostat windows was
eliminated by a $\frac{\lambda}{4}$-plate which was placed in the
optical path\footnote{To eliminate the birefringence of the
windows, we placed polarizer and analyzer at 90$^\circ$ to have a
minimal signal on the detector. Then a $\frac{\lambda}{4}$-plate
was inserted and rotated to minimize the intensity on the
detector to an even lower value. When the film is magnetized this
results in a Kerr-rotation but also in a Kerr-ellipticity which
are both proportional to the magnetization. In the described
setup we cannot distinguish between these two effects, therefore
we have assumed the ellipticity to be zero and calculated the
"Kerr rotation" in figures \ref{euoepi10} and \ref{euoepi11},
which is thus actually a convolution of the Kerr rotation and
ellipticity. At fixed temperature this "Kerr rotation" is
proportional to the magnetization, however when changing the
temperature, the diagonal parts of the dielectric constant also
change, which can result in a different Kerr rotation of the
magneto-optical film at the same magnetization. This might
explain the strange shape of the magnetization curve (a) in
figure \ref{euoepi10}.}. The sample was magnetized by an in situ
liquid nitrogen cooled electromagnet. From the changes with
magnetic field of the diode photocurrent, the optical
Kerr-rotation was determined.

\section{Results}
\subsection{Photoemission}
\begin{figure}[!htb]
\centerline{\includegraphics[width=8cm]{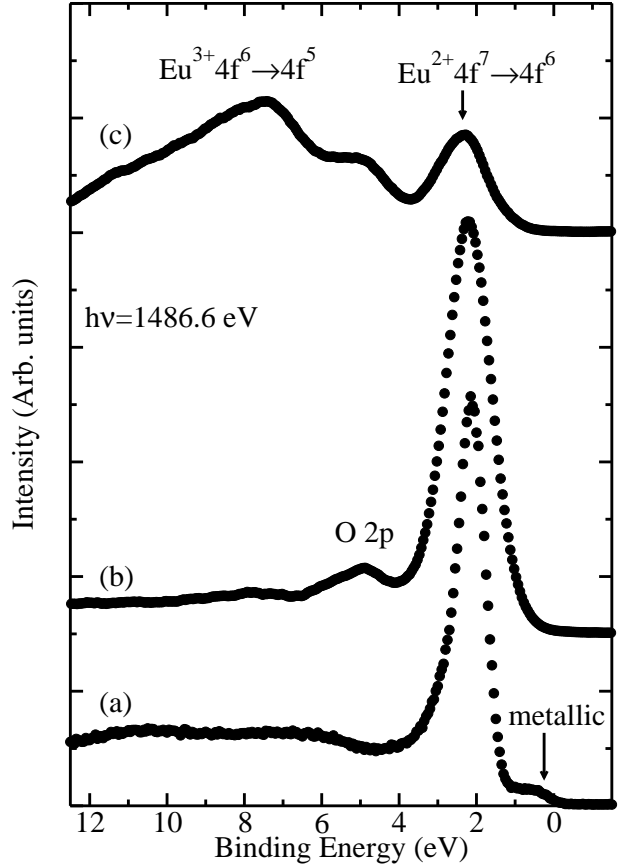}}
\caption{\label{euoepi2} Valence band XPS spectra of EuO$_{1\pm
x}$ films on polycrystalline Ta substrates. (a) Eu metal film,
grown at T$_{s}=25^\circ$C. Intensity at the Fermi level
indicates the presence of free electrons. (b) EuO film grown by
evaporation of Eu (J$_{Eu}=2.0\times 10^{13}$ cm$^{-2}$s$^{-1}$)
in a partial pressure of oxygen gas P$_{ox}=1.0\times 10^{-7}$
mbar (J$_{O}=5\times$10$^{13}$~cm$^{-2}$s$^{-1}$) with
T$_{s}=400^\circ$C. (c) Same as (b) but now grown with
J$_{O}=7\times$10$^{13}$~cm$^{-2}$s$^{-1}$. An increase in the
Eu$^{3+}$ character and decrease of Eu$^{2+}$ is clearly
observed.}
\end{figure}

The EuO films were grown on several substrates at different
temperatures and fluxes of oxygen and europium, as is indicated
in table \ref{euoepit1}. To show the effect of oxygen on the film
stoichiometry we present in figure \ref{euoepi2} the Al $K
\alpha$ valence band XPS spectra of the different oxidation
states of Eu that arise by varying the oxygen pressure.
\begin{table}[!htb]
\centerline{\begin{tabular}{rccccr}
\# & Substr. & T$_s$ ($^\circ$C) & J$_{O}$(cm$^{-2}$s$^{-1}$) & J$_{Eu}$(cm$^{-2}$s$^{-1})$\\
\hline
1 & MgO & 400 & $5.4 \times 10^{12}$ & $1.2 \times 10^{13}$ \\
2 & MgO & 300 & $5.4 \times 10^{12}$ & $1.2 \times 10^{13}$ \\
3 & YSZ & 550 & $1.1 \times 10^{13}$ & $1.2 \times 10^{13}$ \\
4 & YSZ & 300 & $1.1 \times 10^{13}$ & $1.2 \times 10^{13}$ \\
5 & YSZ & 450 & $5.4 \times 10^{12}$ & $1.2 \times 10^{13}$ \\
6 & YSZ & 450 & $1.1 \times 10^{13}$ & $1.2 \times 10^{13}$ \\
7 & YSZ & 450 & $2.2 \times 10^{13}$ & $1.2 \times 10^{13}$ \\
8 & YSZ & 350 & $7.0 \times 10^{12}$ & $1.5 \times 10^{13}$ \\
9 & YSZ & 300 & $7.0 \times 10^{12}$ & $1.5 \times 10^{13}$ \\
10 & Al$_2$O$_3$ & 350 & $ 5.4 \times 10^{12}$ & $1.2 \times 10^{13}$ \\
11 & Al$_2$O$_3$ & 300 & $ 5.4 \times 10^{12}$ & $1.2 \times 10^{13}$ \\
12 & Al$_2$O$_3$ & 340 & $ 5.4 \times 10^{12}$ & $1.3 \times 10^{13}$ \\
\end{tabular}}
\caption{\label{euoepit1} Growth conditions of discussed films,
film thicknesses are $\sim$20-30 nm.}
\end{table}
Photoemission is a very sensitive probe of the valency of Eu
because the energies of the Eu$^{2+}$ $4f^7 \rightarrow 4f^6$
electron removal states are very different from those of the
Eu$^{3+} 4f^6 \rightarrow 4f^5$ states. As can be seen from
figure \ref{euoepi2}(c), an excess of oxygen results in PES
intensity between 6 and 12 eV which is due to the presence of
Eu$^{3+}$. Actually we verified that excessive oxidation of the
films indeed results in the disappearance of the Eu$^{2+}$ peak
at 2 eV binding energy and only Eu$^{3+}$ signal and O $2p$
signal remains. Similarly the presence of Eu$^{3+}$ can be
sensitively probed by XPS of the Eu $4d$ or $3d$ shell, as the
multiplet structure with a half filled $4f^7$ shell is very
different than for a $4f^6$ configuration
\cite{Ogasawara94,Orchard78,Kowalczyk74}.

PES can also be used to monitor the presence of Eu metal, because
free electrons will appear as PES intensity at the Fermi level
like in figure \ref{euoepi2}(a). Figure \ref{euoepi2}(b) shows a
typical photoemission spectrum of stoichiometric EuO, showing
only spectral weight from Eu$^{2+}$ $4f$ and from the O $2p$
levels. Interestingly, EuO films grown under the same conditions
but at lower oxygen pressure show photoemission spectra which are
very similar to the spectrum in figure \ref{euoepi2}(b). This
suggests that the effect of oxygen pressure on the stoichiometry
is very small as long as it stays below a critical limit. For
films \#1-9 the XPS valence band spectrum was taken to verify the
absence of Eu$^{3+}$ and the absence of high concentrations of
free electrons. We only observed Eu$^{3+}$ signatures in film
\#7.

\subsection{Reevaporation of europium atoms}

\begin{figure}[!htb]
\vspace{0.5cm}
\centerline{\includegraphics[width=11.25cm]{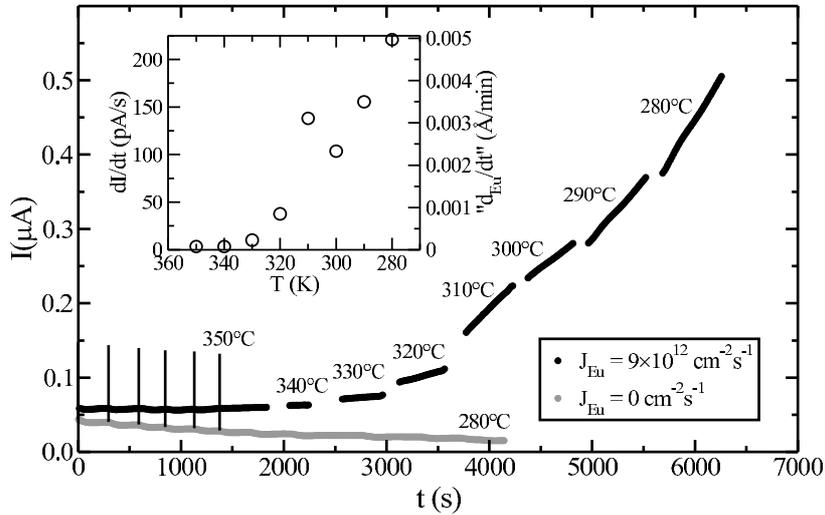}}
\caption{\label{euoepi3} Current at fixed voltage through a 10 nm
thick EuO film at different temperatures while evaporating Eu
atoms on it. The cooling curve without evaporating Eu metal is
also shown in grey. The film is cooled from 400$^\circ$C at
$t=0$~s to 280$^\circ$C in steps of 10$^\circ$C. The first 5
steps are indicated by vertical black lines, lower temperatures
are indicated in the graph. Inset: time derivative of the sample
current versus film temperature. The right axis indicates a
growth rate estimated by assuming uniform Eu films with a
resistivity of bulk Eu metal ($\rho_{350^\circ C}\approx 150 \mu
\Omega cm$).}
\end{figure}
To obtain a better picture of the growth kinetics, and in
particular the effect of substrate temperature on the
accumulation of Eu metal on a EuO surface, we measured the
current at constant voltage, through a EuO film of 10 nm while
evaporating Eu metal on it (J$_{Eu}=9\times 10^{12}$
cm$^{-2}$s$^{-1}$), but without supplying oxygen ($J_O=0$). As
shown in figure \ref{euoepi3}, the resistance of the film stays
about constant while cooling it from 400$^\circ$C to
340$^\circ$C. However at temperatures of 340$^\circ$C and below
the current through the film at constant temperature starts
rising linearly with time. We ascribe this to the accumulation of
excess Eu in the EuO surface layer region being formed. The inset
of the figure shows the time derivative of the current for
different film temperatures. This measurement indicates that at
this Eu flux, Eu metal can only accumulate below 340$^\circ$C on
the EuO surface. Moreover it indicates that the accumulation rate
strongly depends on temperature. If we assume that the extra
conduction is due to the presence of a uniform thin film of
metallic Eu ($\rho_{350C} \approx 150 \mu \Omega$cm
\cite{Curry60}) on the surface, we find a growth rate of this
film as given by the right axis of the graph in the inset. The
actual growth rate is surely bigger, as the resistivity of metal
films of less than a monolayer is strongly increased by surface
scattering and non-uniformities. However, it is probable that no
conducting Eu metal channel can be formed at these low coverages,
and that the current increase is solely due to free carriers from
the accumulated Eu metal which dope the underlaying EuO film.


\subsection{RHEED and LEED}

\begin{figure}[!htb]
\centerline{\includegraphics[width=11.25cm]{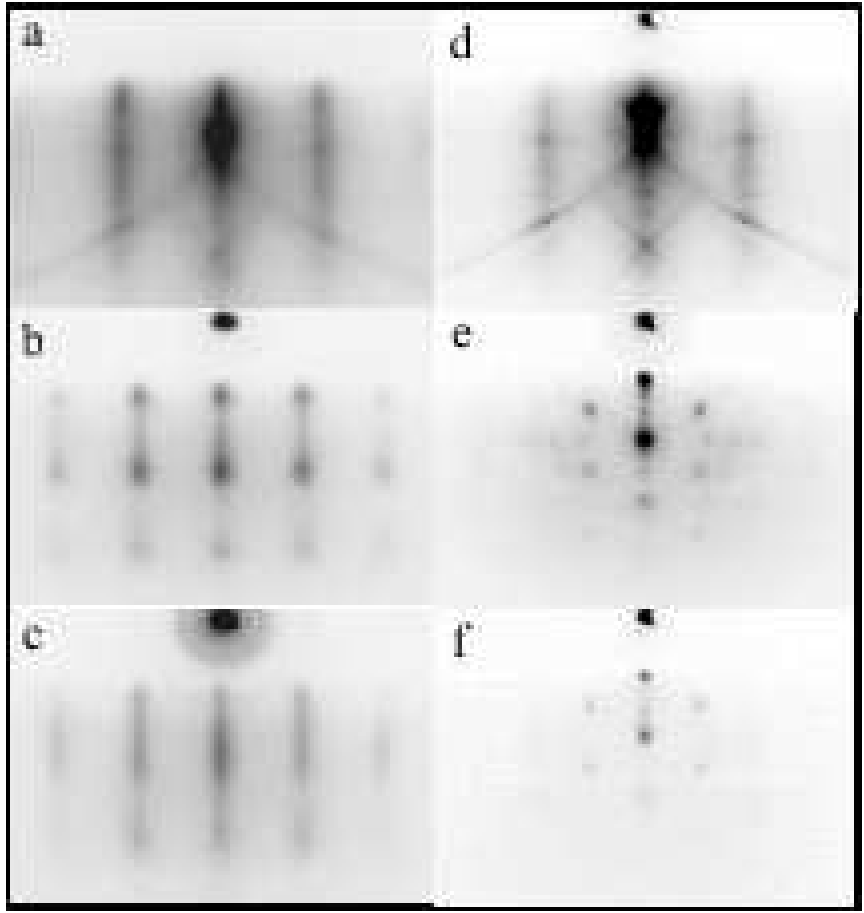}}
\caption{\label{euoepi4} RHEED patterns during EuO growth on the
(001) surface of cleaved MgO, with the incident beam along the
MgO [100] direction. Growth conditions were: J$_{Eu} = 1.2 \times
10^{13}$ cm$^{-2}$s$^{-1}$, J$_{O} = 5.37 \times 10^{12}$
cm$^{-2}$s$^{-1}$. The left column shows sample \#1 grown at
T$_{s} = 400^\circ$C, before growth (a), after 2250 s (b) and
after 4500 s (c). The right column shows the growth of sample \#2
at T$_{s} = 300^\circ$C, before growth (d), after 200 s (e) and
after 2250 s (f).}
\end{figure}

To obtain epitaxial films, we have grown EuO on the cleaved (001)
surface of MgO and the chemically polished (001) surface of YSZ.
We studied the surface structure of the EuO films during growth
with RHEED. Figure \ref{euoepi4} shows RHEED patterns during
growth on MgO substrates. The RHEED pattern during growth of
films grown at 400$^\circ$C is similar to that reported by Iwata
\emph{et al.} \cite{Iwata00}. After starting the growth of film
\#1 the MgO streaks quickly disappear and a square pattern of
spots appears in figure \ref{euoepi4}(b), with the distance
between the spots a factor of 1.24$\pm$0.05 smaller than the
distance between the MgO streaks. This is consistent with a quick
relaxation to the bulk lattice constant of EuO, as the ratio of
lattice constants $a_{EuO}/a_{MgO}$=1.22. Moreover it shows that
the [100] directions of substrate and film are aligned.
Diffraction patterns are spot-like, indicative of island growth.
After considerable time, the patterns become more streaky,
demonstrating a flattening of the surface. At lower substrate
temperatures we observed a different growth mode, as is
demonstrated in the right column of figure \ref{euoepi4} for film
\#2, which was grown at T$_{s} = 300^\circ$C. Quickly after
growth started, a hexagonal-like pattern emerged of which the
spots sharpened first and then gradually faded away. The films
grown at lower temperatures have a much rougher surface,
resulting in transmission electron diffraction (TED) patterns.
The origin of this hexagonal pattern is unclear to us, however it
seems to resemble the transmission electron diffraction pattern
expected for an incident electron beam perpendicular to the (111)
plane.

\begin{figure}[!htb]
\centerline{\includegraphics[width=11.25cm]{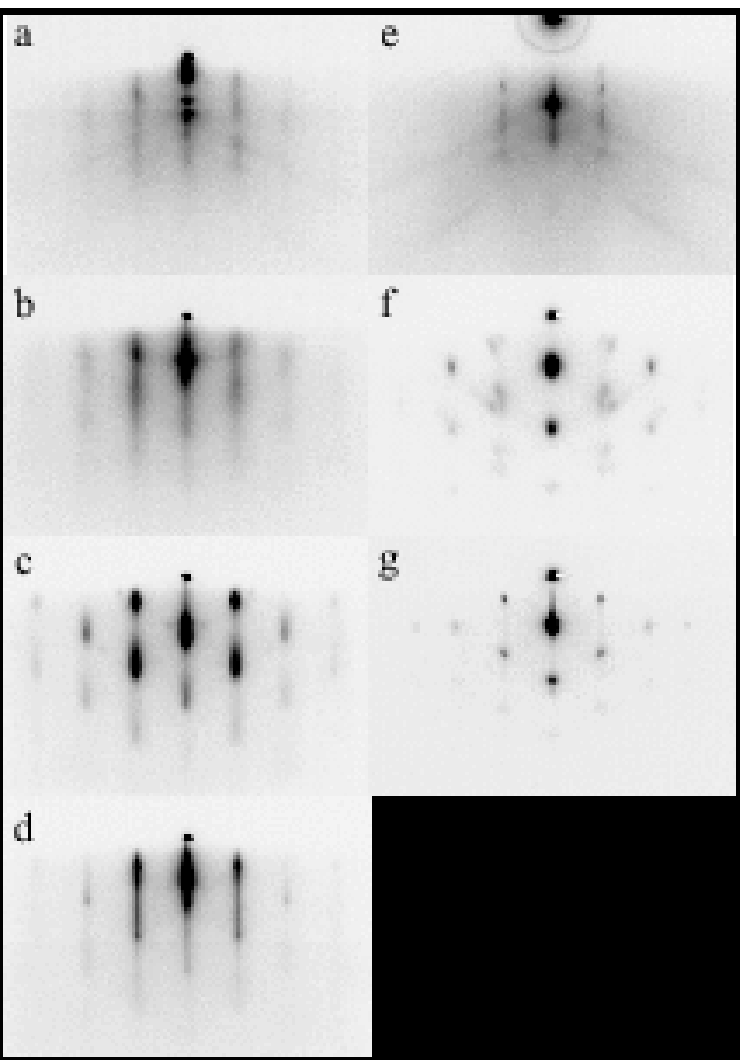}}
\caption{\label{euoepi5} RHEED patterns during EuO growth on the
(001) surface of chemically polished YSZ, with the incident beam
along the ZrO$_2$ [110] direction. Growth conditions were:
J$_{Eu} = 1.2 \times 10^{13}$ cm$^{-2}$s$^{-1}$, J$_{O} = 1.74
\times 10^{13}$ cm$^{-2}$s$^{-1}$. The left column shows sample
\#3 grown at T$_{s} = 550^\circ$C, before growth (a), after 200 s
(b), after 500 s (c), and after 4500 s (d). The right column
shows the growth of sample \#4 under the same conditions but with
T$_{s} = 300^\circ$C, before growth (e), after 500 s (f), and
after 1000 s (g).}
\end{figure}
RHEED of YSZ was performed with the primary beam along the [110]
axis. The left panel of figure \ref{euoepi5} shows RHEED during
growth at a substrate temperature of 550$^\circ$C. After the
start of growth the RHEED streaks of the substrate stayed streaky
and at the same position, indicating a homogeneous layer by layer
growth of EuO(001) on the YSZ(001) surface. After several
monolayers spots of TED features appeared, suggesting island
growth. The rectangular arranged spots seem to correspond to TED
along the [110] direction of an fcc crystal, indicating that
islands still grow epitaxially and homogeneously with the
substrate. Gradually the RHEED streaks return and sharpen up to a
pattern consistent with a flat EuO(001) surface. At lower
temperatures the RHEED was different as shown in the right panel
of figure \ref{euoepi5}. Soon after growth transmission spots
appeared due to roughening of the surface, with an unclear
arrangement of spots. After 1000~s the spots rearranged, leaving
the hexagonal-like RHEED pattern, similar to that observed on
MgO, even though the incident beam direction with respect to the
substrate was different in this case.
\begin{figure}[!htb]
\centerline{\includegraphics[width=11.25cm]{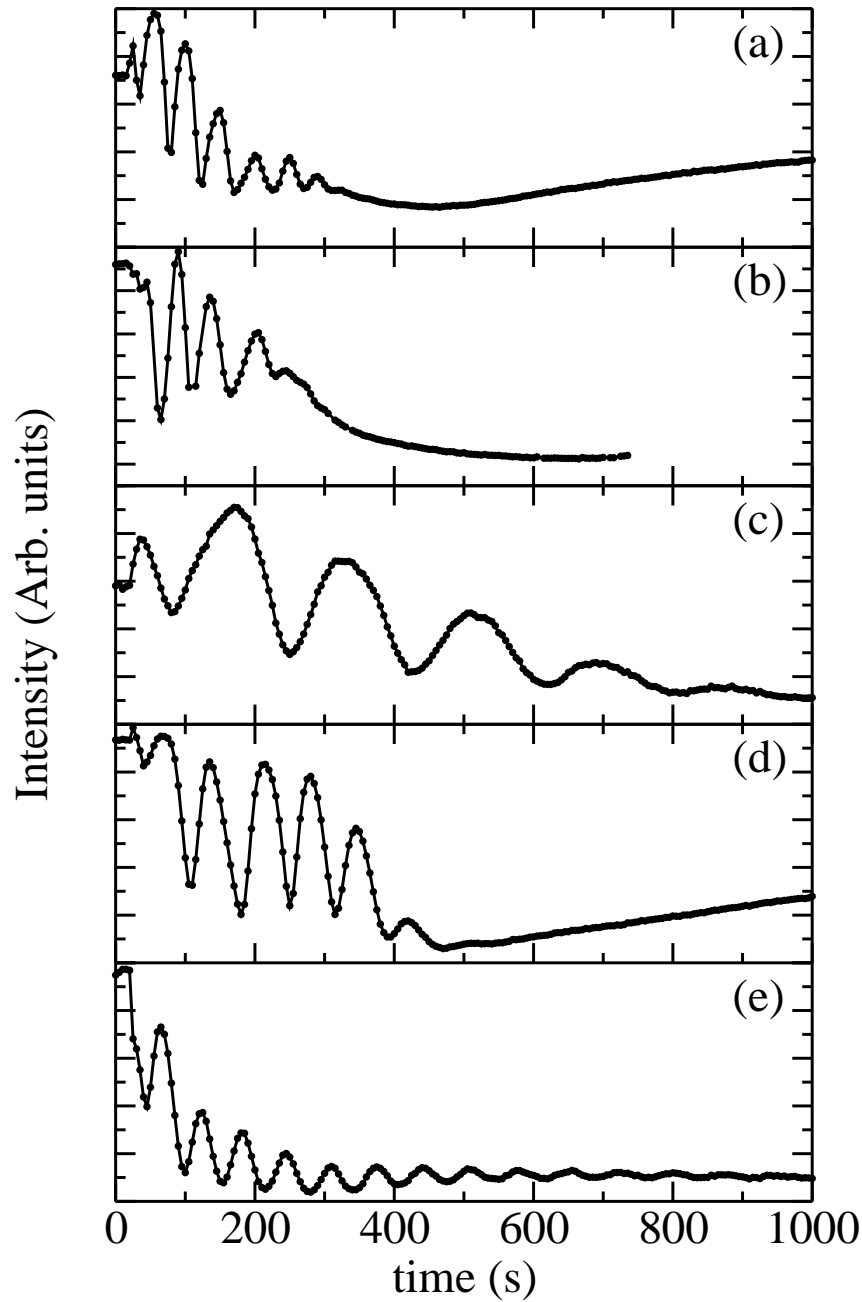}}
\caption{\label{euoepi6} Intensity of the specular reflected
RHEED beam versus time during growth of EuO on YSZ. The beam is
parallel to the [110] direction. (a) Sample \#3,
T$_{osc}=47\pm$5~s (b) Sample \#4, (c) Sample \#5,
T$_{osc}=183\pm$9~s (d) Sample \#6, T$_{osc}=70\pm$4~s (e) Sample
\#7, T$_{osc}=67\pm$8~s.}
\end{figure}
\begin{figure}[!htb]
\centerline{\includegraphics[width=11.25cm]{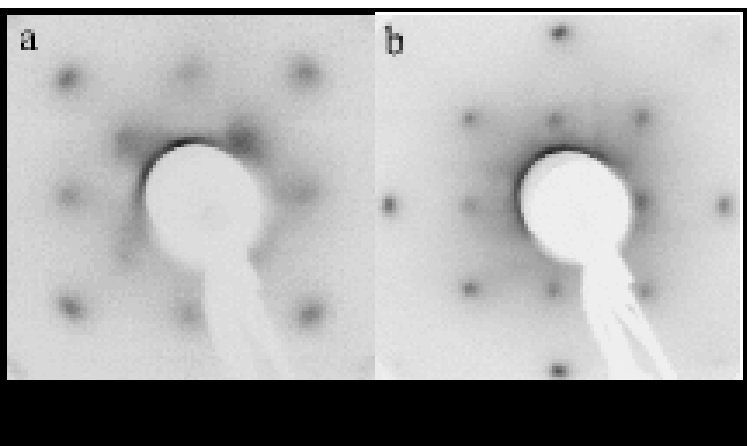}}
\caption{\label{euoepi7} LEED of the surface of EuO films grown
on (a) the (001) surface of MgO at 400$^\circ$C (sample \#1,
incoming electron energy E$_p$ = 221 eV) and on (b) the (001)
surface of YSZ at 300$^\circ$C (sample \#4, E$_p$ = 224 eV). The
arrows indicate the orientation of the substrates.}
\end{figure}
During growth, the intensity of the specularly reflected RHEED
beam was recorded. For the films grown on YSZ, clear oscillations
in the intensity were observed during the initial stage of growth
as is shown in figure \ref{euoepi6}. Every oscillation
corresponds to the formation of an atomic monolayer of EuO and
therefore the oscillations are a measure of the growth rate
\cite{Neave83}. The observed oscillation period was not
completely constant and had the tendency to increase during
growth, this is indicated by the error bars in the caption of the
figure. Films were grown at 300$^\circ$C, 350$^\circ$C,
450$^\circ$C and 550$^\circ$C. Figures \ref{euoepi6}(a) and
\ref{euoepi6}(b) show the intensity oscillations of films \#3 and
\#4. During the intensity oscillations a streaky RHEED pattern
was observed, like in figure \ref{euoepi5}(b). This pattern
gradually changed to that of figure \ref{euoepi5}(c) or
\ref{euoepi5}(f). The change was accompanied by a disappearance
of the specularly reflected beam due to surface roughness and
thus quenched the oscillations. In general, films grown above
400$^\circ$C showed about 6-8 oscillations before the specularly
reflected beam disappeared, while films grown below 400$^\circ$C
only showed 3-4 oscillations. To prevent the formation of
Eu$^{3+}$, the film growth was always started by first opening
the Eu effusion cell, followed 15 s later by the admission of
oxygen gas to the chamber, this might explain the slightly
irregular behavior during the first seconds of growth. We used
the RHEED oscillation period (T$_{osc}$) to investigate the
effect of the partial oxygen pressure on the growth rate. In
figures \ref{euoepi6}(c-e) we show the oscillations for oxygen
background pressures of respectively 1$\times 10^{-8}$ mbar,
2$\times 10^{-8}$ mbar and 4$\times 10^{-8}$ mbar, with T$_{osc}$
in the caption. The growth rate is more than doubled by
increasing the oxygen flux from 1 to 2$\times 10^{-8}$ mbar
indicative of an oxygen limited growth. When doubling the oxygen
pressure again, there is only a small increase in the growth rate
(figure \ref{euoepi6}(e)), but as many as 14 oscillations were
observed, with streaky RHEED patterns, suggesting a less rough
surface. However when investigating this film with XPS, a clear
signature of Eu$^{3+}$ was observed in addition to the Eu$^{2+}$
feature like in figure \ref{euoepi2}(c), indicating that an
excess flux of oxygen had led to the formation of Eu$_2$O$_3$. In
this case the growth rate was probably limited by the Eu flux,
thus explaining why the growth rate did not increase much upon
increasing the oxygen flux. These observations indicate that the
presence of RHEED oscillations and streaky patterns are not a
reliable indication for a good stoichiometry. For the films grown
on MgO no RHEED oscillations could be detected, which is probably
a result of large defect densities and island growth during the
initial stages of growth which are necessarily present to relax
from the MgO to the EuO lattice constant.

The surface structure of the films was also studied by LEED as is
shown in figure~\ref{euoepi7}. Both the films grown on MgO(001)
and on YSZ(001) show clear and stable 1$\times$1 LEED patterns.
The surface lattice constant of the EuO films as determined by
comparing substrate and film LEED patterns, is within error bars
equal to that of bulk EuO. Similar LEED patterns have been
recorded on cleaved bulk EuO crystals \cite{Bas74,Grazhulis88}.
Interestingly, despite the surface roughness and the unclear
hexagonal-like pattern observed in the RHEED of sample \#4, a
sharp 1$\times$1 LEED pattern was still observed, indicative of
epitaxial growth of the film on the substrate. In fact, samples
grown at lower temperatures generally showed sharper LEED, and
were less susceptible to charging effects. This may be related to
the fact that charging problems are less likely to occur for
these low temperature films, as they tend to be much less
insulating.

\subsection{Transport and magnetic properties}

\begin{figure}[!htb]
\centerline{\includegraphics[width=11.25cm]{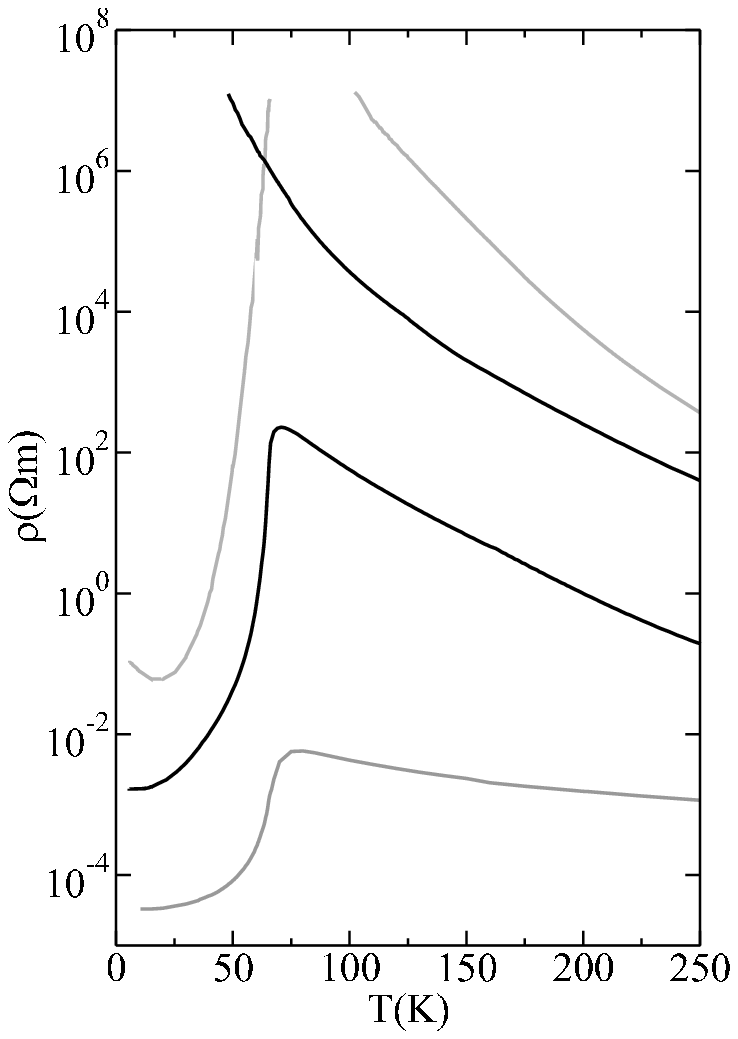}}
\caption{\label{euoepi8} Resistivity of EuO films on (100) YSZ
substrates (black lines) and ($1\bar{1}02$) Al$_2$O$_3$
substrates (grey lines). Films \#8 (upper black curve) and \#10
(upper gray curve) were grown at 350$^\circ$C, films \#9 (lower
black curve) and \#11 (lower gray curve) at 300$^\circ$C. The
thickness of all films was 20-30 nm.}
\end{figure}

To study the effect of the growth conditions on the transport
properties, we have measured the resistivity of films grown at
different temperatures as is shown in figure \ref{euoepi8}.
A huge effect of the growth temperature on the transport
properties is observed, as the films grown at 350$^\circ$C have a
much higher resistivity than those grown at 300$^\circ$C. Films
\#8 and \#9 were grown on YSZ(100) under conditions which were
known to result in epitaxial growth, whereas films \#10 and \#11
were grown on Al$_2$O$_3$($1\bar{1}02$), which was known to
result in polycrystalline films (circular patterns were observed
in RHEED). The epitaxial films had a much higher resistivity than
the polycrystalline films grown at the same temperature. Film \#8
did not show a metal-insulator transition (this was also the case
for films grown at 400$^\circ$C on Al$_2$O$_3$) indicative of
(nearly) stoichiometric EuO \cite{Oliver72}. The low temperature
resistivity and the magnitude of the metal-insulator transition
depend strongly on the substrate temperature during growth and on
the fact if the film is epitaxial or not. The resistivity of the
films was also strongly modified by heating them {\it in situ} to
temperatures $\geq 350^\circ$C, at which evaporation of excess Eu
could occur.
\begin{figure}[!htb]
\centerline{\includegraphics[width=11.25cm]{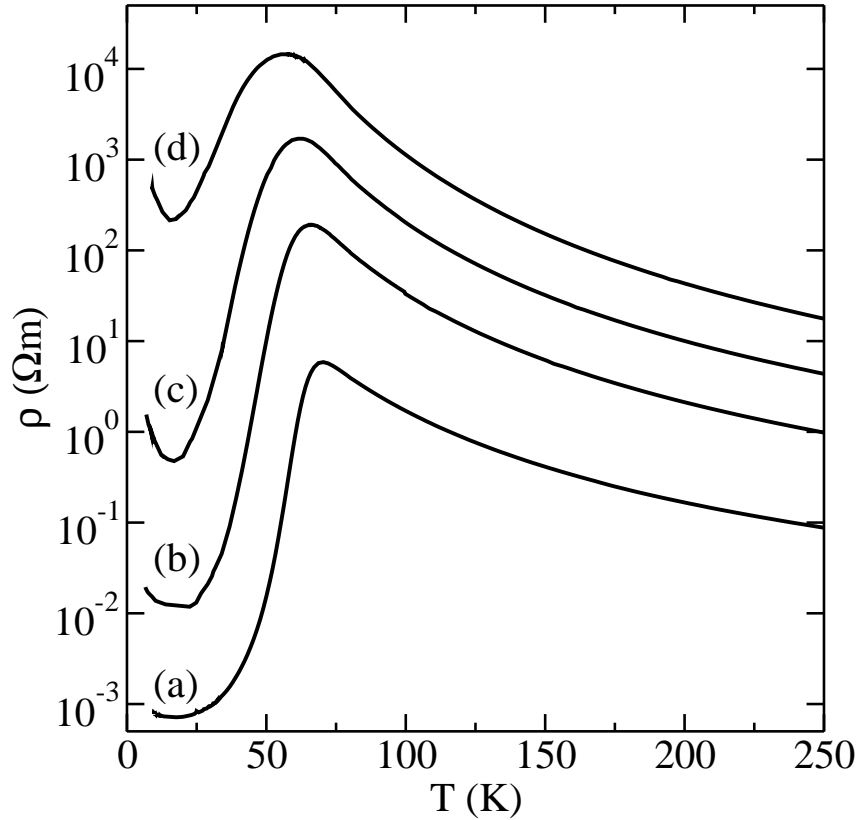}}
\caption{\label{euoepi9} (a) Resistivity of film \#12 grown at
340$^\circ$C on Al$_2$O$_3$. (b) Resistivity after heating the
film to 350$^\circ$C for 30 minutes. (c) Resistivity after
heating the film to 360$^\circ$C for 30 minutes. (d) Resistivity
after heating the film to 370$^\circ$C for 30 minutes.}
\end{figure}
In figure \ref{euoepi9} we show how the resistivity of a film
increases by heating it successively for 30 minutes at high
temperatures from 350-370$^\circ$C under UHV conditions. A
possible explanation for the increased resistivity could be the
evaporation of excess Eu atoms from the film, thus resulting in a
lowering of the free electron concentration. However, the removal
of Eu atoms could also increase the disorder and the number of
defects in the film. Therefore, in addition to a decreased
carrier concentration a larger defect scattering would be
expected in the heated films, which might explain why such
annealed films have a higher low-temperature resistivity than
films that were grown with an intrinsically small carrier
concentration. This effect also explains why we noticed that the
transport properties of the films could depend on the cooling
rate (faster cooling generally led to lower resistance), as the
reevaporation of Eu atoms still continues after growth is stopped
(i.e. when Eu and O fluxes are set to zero). Another remarkable
and not well-understood effect is the shift of the peak
resistivity in figure \ref{euoepi9} to lower temperatures upon
heating at higher temperatures. Maybe this lowering is a result
of a reduced Curie temperature, although we did not check this.
\begin{figure}[!htb]
\centerline{\includegraphics[width=10.25cm]{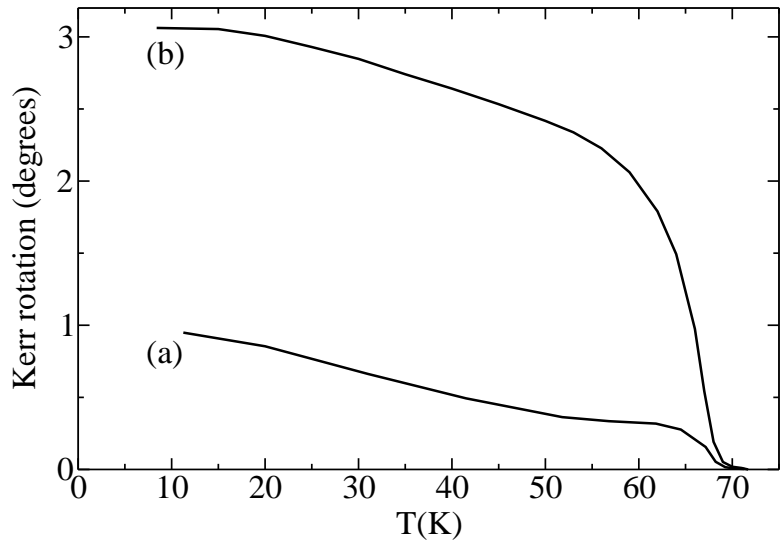}}
\caption{\label{euoepi10} Remanent "Kerr rotation" of EuO films:
(a) EuO film \#8, grown at 350$^\circ$C on YSZ(001), thickness
$\sim$ 25 nm. (b) Polycrystalline EuO film of $\sim$ 30 nm, grown
at 290$^\circ$C on a 100 nm Cr metal film on ($1\bar{1}02$)
Al$_2$O$_3$. See footnote in section \ref{experimental} for
details on the determination of the Kerr rotation.}
\end{figure}
\begin{figure}[!htb]
\centerline{\includegraphics[width=10.25cm]{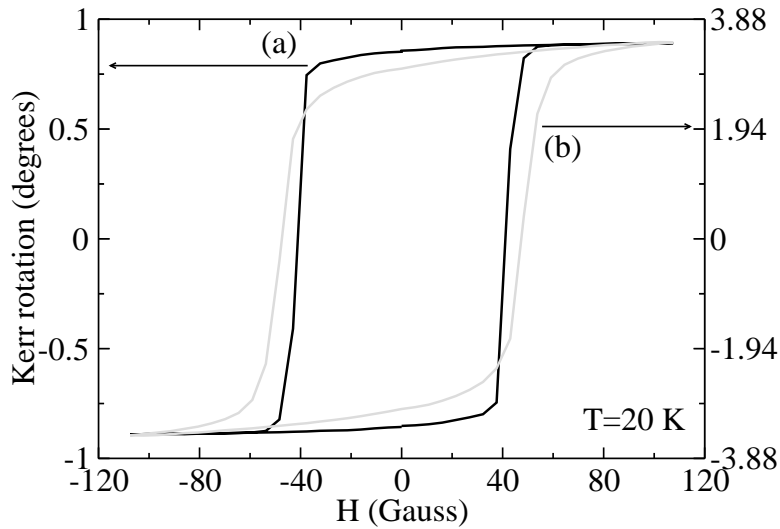}}
\caption{\label{euoepi11} Hysteresis "Kerr rotation" curves at 20 K of
(a) epitaxial EuO film on YSZ and (b) polycrystalline EuO film on Cr
metal.}
\end{figure}
In figure \ref{euoepi10} we show \emph{in situ} measurements of
the remanent Kerr rotation of films grown on YSZ and Cr metal.
The films show a Curie temperature of 69 K, similar to that
observed in bulk samples. Figure \ref{euoepi11} shows a
hysteresis curve for both an epitaxial film on YSZ and a
polycrystalline film on Cr. The epitaxial film shows a more
rectangular curve. The less rectangular hysteresis of the
polycrystalline films might be understood from a random
distribution of magnetic anisotropies of the polycrystalline
domains, like in the Stoner-Wohlfarth model \cite{Stoner48}.

\section{Discussion}

From photoemission measurements on EuO films grown under
different conditions we verified that there is a large range of
growth parameters which result in EuO films without large
fractions of Eu$^{3+}$ or Eu metal. To obtain such films, the
main conditions are:
\begin{enumerate}
\item J$_{Eu} \geq J_{O}$ to avoid the formation of Eu$^{3+}$.
\item T$_s \geq 280^\circ$C to ensure reevaporation of excess Eu atoms.
\end{enumerate}
Based on these observations we propose the following growth
mechanism for EuO. Due to their high reactivity, the oxygen atoms
have a sticking ratio of close to 1 when they encounter a
europium atom which is not yet fully oxidized. Therefore when
J$_{O} > J_{Eu}$ it is very likely that Eu$_2$O$_3$ will be
formed. If on the other hand J$_{Eu} > J_{O}$, excess Eu atoms
will be present and a mixture of EuO and Eu metal will form. When
the substrate temperature is high enough, the vapor pressure of
Eu metal will be large and Eu metal atoms will be reevaporated,
unless they react with oxygen (as is strongly supported by figure
\ref{euoepi3}).  Therefore these two conditions guarantee close
to stoichiometric EuO films without requiring accurate tuning of
the fluxes. The behavior of Eu on an EuO surface is similar to
that of Ga atoms on a GaAs surface \cite{Arthur68}, although a
difference exists in the fact that Ga can at maximum bind 1 As
atom, whereas Eu can also bind 1.5 O atoms.

The proposed mechanism seems to operate by oxygen limited growth
at $J_{Eu}/J_{O} >$~1, thus the number of Eu atoms that stick to
the film is equal to the number of oxygen atoms and the growth
rate is determined by the oxygen flux. This is supported by the
RHEED intensity oscillations of samples \#5 and \#6 in figure
\ref{euoepi6}(c) and (d). The oscillation period strongly
increases upon doubling the oxygen pressure. It is interesting to
compare the growth rates as deduced from the RHEED oscillations
with those expected when assuming a sticking fraction of oxygen
of 1. The period to grow a monolayer of EuO is then given by
$T_{osc}=7.6 \times 10^{14}/J_{O}$~[cm$^{-2}$s$^{-1}$]~s. Thus we
find T$_{osc, \#5}=141$ s and T$_{osc, \#6}=69$ s in quite good
agreement with the measured periods of respectively 183$\pm$9 s
and 70$\pm$4 s. The results from Iwata \emph{et al.}
\cite{Iwata00} can be understood, if we assume an oxygen limited
growth mechanism. From their data we deduce that when they use a
flux J$_{O}=1.1 \times 10^{13}$ (cm$^{-2}$s$^{-1}$) they measure
a growth rate of the EuO film of $9.0 \times 10^{-2}$ \AA/s.
However to have enough oxygen atoms for the reported growth rate
of EuO, an oxygen flux of at least J$_{O}$ = 2.7$\times 10^{13}$
(cm$^{-2}$s$^{-1}$) is required. Therefore it is not unlikely
that their actual oxygen flux was 2.4 times higher than the
measured flux. The sticking probability of Eu atoms in their
figure 6 would then approach 1 close to J$_{O}$/J$_{Eu}$ = 1, and
the rate of sticking Eu atoms at low oxygen fluxes would be
approximately equal to J$_{O}$, consistent with a mechanism of
oxygen limited growth. Moreover they also found that increasing
the oxygen flux further resulted in the formation of Eu$_2$O$_3$
and a europium limited growth, similar to what we observed for
film \#7.

We found that films grown below 350$^\circ$C with $J_{Eu}/J_{O} >
1$ generally had rougher surfaces (right panels of figures
\ref{euoepi4} and \ref{euoepi5}) and lower resistances (figure
\ref{euoepi8}) than films grown at higher temperatures. Moreover
films grown at these low temperatures show metal-insulator
transitions around their Curie temperatures, which are known only
to be present in Eu-rich EuO \cite{Oliver72}. A clue as to the
origin of these effects can be found in figure \ref{euoepi3} from
which it can be concluded that the surface accumulation of Eu
metal on the EuO surface starts below 350$^\circ$C. These
superfluous Eu atoms will be built into the lattice of EuO, and
as they are probably too large to be incorporated as interstitial
atoms, this will likely result in the formation of oxygen
vacancies. A reduced growth temperature would lead to a larger
accumulation of excess Eu atoms, increasing the likeliness of the
formation of oxygen vacancies and thus forming a film with a
higher doping concentration. It might be that the accumulated Eu
metal atoms at the surface play a role in the increased surface
roughness, and contribute to the electron doping of the EuO film
being formed.


For the growth of epitaxial films, YSZ seems to be the best
substrate due to its excellent lattice matching with EuO. Despite
a large lattice mismatch, EuO films on MgO also grow crystalline
and align their crystal axes to those of the substrate, however
defects in the first monolayers due to relaxation of the film are
unavoidable and monitoring the growth rate by RHEED oscillations
seems impossible. To obtain films with a large metal-insulator
transition epitaxial growth is not necessary, in fact
polycrystalline films grown on Al$_2$O$_3$ keep their
metal-insulator up to higher growth temperatures than epitaxial
films. We did not manage to obtain metal-insulator transitions
which spanned more than 6 orders of magnitude in the epitaxial
films, while for the polycrystalline films we could obtain
conductivity changes of 10 orders or magnitude or more. These
differences between epitaxial and polycrystalline films are
probably due to the fact that oxygen vacancies, which provide
charge carriers, are more easily incorporated at grain boundaries
or defects in the polycrystalline film, than in the crystalline
epitaxial films. In fact for the epitaxial films it might even be
the case that no oxygen vacancies are incorporated in the film,
but all doping is provided by a thin Eu metal layer that floats
on the EuO surface. We suggest this possibility, as we observed
that the conductivity of Eu-rich films during growth
(polycrystalline and epitaxial), usually did not increase linear
with time, but was constant in time. This time independent
resistance showed a strong dependence on oxygen pressure. It
might be that a surface population of Eu atoms with density $n$
exists on the EuO surface during growth (a similar effect is
observed in GaAs \cite{Arthur68}) and the resistivity of the film
only depends on the number of electrons donated by this surface
population. The density of free Eu atoms on the surface should
follow:
\begin{equation}
\frac{{\rm d}n}{{\rm d}t} = \frac{-n}{\tau(T)} - J_O + J_{Eu}
\end{equation}
where $\tau(T)$ is the temperature dependent surface
lifetime\footnote{The average time a free Eu atom stays on the
surface before it is reevaporated.} of a Eu atom and where we
assume that all oxygen atoms that reach the surface react with a
Eu atom (i.e. oxygen limited growth). At steady state:
\begin{equation}
\label{surfpop} n=\tau(T)(J_{Eu} - J_O)
\end{equation}
We propose two ways in which the effect of temperature on the
doping concentration of the resulting films might be understood.
Possibly, the probability to incorporate an oxygen vacancy in the
film increases with $n$. Decreasing the temperature will increase
$\tau$ and thus increase the number of vacancies. Alternatively
it might be that no vacancies are formed, but if $\tau$ is large
and the film is cooled quickly after growth, a thin layer of Eu
metal may persist on the surface and provide free carriers.

For all the measured films, the Curie temperature is similar to
that of stoichiometric bulk samples. We have not observed
significant increases of the Curie temperature above 69 K as was
reported for Gd doped crystals \cite{Mauger80}, oxidized Eu metal
films \cite{Massenet74} and EuO films \cite{Iwata00} on
SrTiO$_3$. Possibly the high doping concentrations that are
required to increase T$_c$ cannot be reached in crystalline
films, just like in Eu-rich bulk crystals \cite{Mauger86}.
\begin{figure}[!htb]
\centerline{\includegraphics[width=11.25cm]{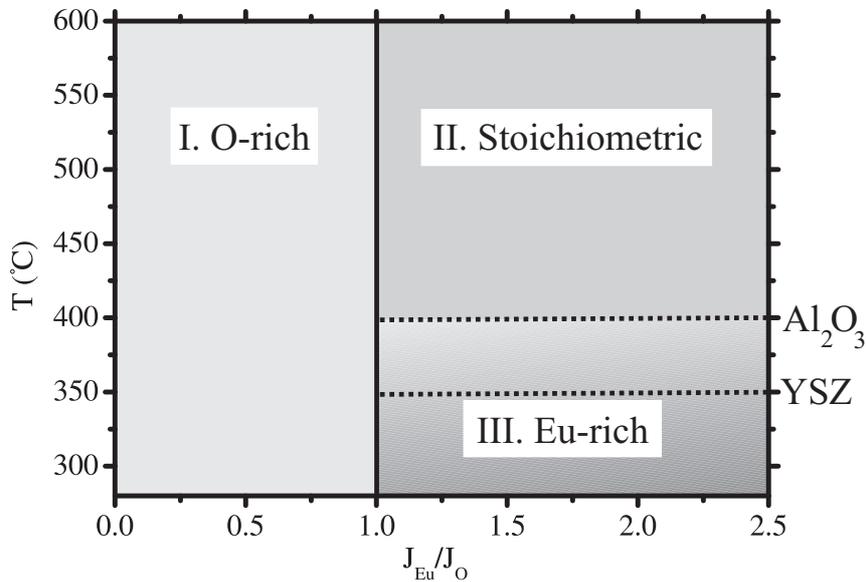}}
\caption{\label{Phasediagfilm} Proposed diagram of resulting
EuO$_{1\pm x}$ films versus growth conditions, based on
photoemission, conductivity and RHEED. The substrate temperature
T is plotted versus the ratio of Eu and O fluxes $J_{Eu}/J_O$.
The region colors darken in the direction of increasing
Eu-content. The dotted lines sketch the borders of region II and
III for polycrystalline films on Al$_2$O$_3$ and for epitaxial
films on YSZ.}
\end{figure}
Based on our observations from photoemission, transport and
RHEED, the types of resulting EuO$_{1\pm x}$ films at different
growth conditions can be divided in three groups, as is sketched
in figure \ref{Phasediagfilm}. Let us summarize their properties.
Films in region I, which are grown with $J_{Eu}/J_{O} < 1$ show
traces of Eu$^{3+}$ in XPS (fig. \ref{euoepi2}(c)) in the studied
temperature range. RHEED oscillations suggest that the growth of
such films is limited by the Eu flux (fig. \ref{euoepi6}(e)),
although the RHEED patterns are similar to those of films in
region II. The films show a semiconductor-like resistivity. These
signs are indicative of oxygen rich films. Region II represents
films grown at temperatures greater than or equal to
approximately 400$^\circ$C and with $J_{Eu}/J_{O}
> 1$. No signs of Eu$^{3+}$ are observed in these films by XPS (fig.
\ref{euoepi2}(b)), showing that they are not oxygen rich. RHEED
oscillations indicate that their growth was oxygen limited (fig.
\ref{euoepi6}(c,d)) and sharp RHEED patterns are observed (fig.
\ref{euoepi5}(a-d)). As the films still show a semiconducting
temperature dependent resistivity without a metal-insulator
transition and no spectral weight is detected near their Fermi
level, they are not Eu-rich and are thus very close to
stoichiometric. For films in region III, which are grown at lower
substrate temperatures, (below $400^\circ$C for polycrystalline
films and below $350^\circ$C for epitaxial films (dotted lines)),
a metal-insulator transition is observed near the Curie
temperature. Such films show less RHEED oscillations as the
streaks in the RHEED pattern turn into spots after $\sim 3$
monolayers. A hexagonal RHEED pattern emerges which is not well
understood. Spectral weight near the Fermi level is not well
visible in XPS, but is can be observed by He-I UPS, which is more
sensitive for the delocalized $5d$ and $6s$ electrons than XPS.
These signs are indicative of Eu-rich films. The magnitude of the
metal-insulator transition decreases upon decreasing the growth
temperature, indicative of increased doping. This is shown by the
color darkening in region III. As is indicated by equation
(\ref{surfpop}) the doping concentration might also depend on the
difference $J_{Eu}-J_O$. We have not systematically studied this
effect. In fact, the diagram in figure \ref{Phasediagfilm} has
mainly been investigated for Eu fluxes of $\sim 1.2 \times
10^{13}~$cm$^{-2}$/s and might look different at higher or lower
fluxes.

\section{Conclusions}

We have grown EuO films under varying conditions. We found a
large range of growth parameters which resulted in stoichiometric
films. The main conditions to obtain such films are a higher
europium than oxygen flux and a high enough substrate temperature
during growth. The films grow epitaxially on both YSZ and MgO
substrates, with sharp RHEED and LEED patterns, structurally
consistent with those of bulk EuO. Oscillations of the specularly
reflected RHEED electron beam during the initial stages of growth
on YSZ, indicate a layer by layer homogeneous growth, with a
growth rate which is controlled by the oxygen pressure. On the
basis of these observations we propose a mechanism of oxygen
limited growth for the growth of stoichiometric EuO, where
reevaporation of Eu avoids the formation of Eu-rich material.
This mechanism facilitates the growth of stoichiometric EuO
enormously, as it does not require accurate tuning of the
europium and oxygen fluxes. It also simplifies the growth of
slightly Eu-rich films, which are especially interesting for the
colossal metal-insulator transition and high spin-polarization of
the charge carriers in the ferromagnetic state
\cite{Steeneken02}. We found that such films can be made by
reducing the growth temperatures to a range at which slight
amounts of Eu metal can accumulate at the surface, which might
increase the probability of the incorporation of oxygen vacancies
in the films. The transport properties of such films can thus
easily be tuned by varying the growth temperatures or
alternatively by heating the films after growth. Such accurate
tuning of transport properties might prove essential for the
application of ferromagnetic semiconductors as spin-injectors or
spin-filters in semiconducting spintronics devices.

\markboth{}{}
\chapter{Exchange splitting and charge carrier spin-polarization in
EuO\footnote{P. G. Steeneken, L. H. Tjeng, I. Elfimov, G. A.
Sawatzky, G. Ghiringhelli, N. B. Brookes and D.-J. Huang, Phys.
Rev. Lett. {\bf 88}, 047201 (2002).}} \label{Spinxas}


{\sl High quality thin films of the ferromagnetic semiconductor
EuO have been prepared and were studied using a new form of
spin-resolved spectroscopy. We observed large changes in the
electronic structure across the Curie and metal-insulator
transition temperature. We found that these are caused by the
exchange splitting of the conduction band in the ferromagnetic
state, which is as large as 0.6~eV. We also present strong
evidence that the bottom of the conduction band consists mainly
of majority spins. This implies that doped charge carriers in EuO
are practically fully spin polarized.}

\section{Introduction}

EuO is a semiconductor with a band gap of about 1.2~eV and is one
of the very rare ferromagnetic oxides \cite{Tsuda91,Mauger86}.
Its Curie temperature ($T_c$) is around 69 K and the crystal
structure is rocksalt (fcc) with a lattice constant of 5.144 \AA.
Eu-rich EuO becomes metallic below $T_c$ and the metal-insulator
transition (MIT) is spectacular: the resistivity drops by as much
as 8 orders of magnitude \cite{Oliver72,Shapira,Shapira73b}.
Moreover, an applied magnetic field shifts the MIT temperature
considerably, resulting in a colossal magnetoresistance (CMR)
with changes in resistivity of up to 6 orders of magnitude
\cite{Shapira,Shapira73b}. This CMR behavior in EuO is in fact
more extreme than in the now much investigated
La$_{1-x}$Sr$_x$MnO$_3$ materials \cite{Ramirez97,imada}. Much
what is known about the basic electronic structure of EuO dates
back to about 30 years ago and is based mainly on optical
measurements \cite{Busch,Freiser,Schoenes} and band structure
calculations~\cite{Cho70}. With the properties being so
spectacular, it is surprising that very little has been done so
far to determine the electronic structure of EuO using more
modern and direct methods like electron spectroscopies.

Here we introduce spin-resolved x-ray absorption spectroscopy, a
new type of spin-resolved electron spectroscopy technique to
study directly the conduction band of EuO where most of the
effects related to the MIT and CMR are expected to show up.
Spin-resolved measurements of the conduction band could
previously only be obtained by spin-polarized inverse
photoemission spectroscopy. Spin-resolved x-ray absorption
spectroscopy is an alternative technique which is especially well
suited to study ferromagnetic oxides, a currently interesting
broad class of materials. Using this technique we observed large
changes in the conduction band across $T_c$ and we were able to
show experimentally that these are caused by an exchange
splitting of the conduction band below $T_c$. Moreover, we found
that this splitting is as large as 0.6 eV and show that the
states close to the bottom of the conduction band are almost
fully spin-polarized, which is very interesting for basic
research in the field of spintronics.

\section{Experimental}

The experiments were performed using the helical undulator
\cite{Elleaume94} based beamline ID12B \cite{Goulon95} at the
European Synchrotron Radiation Facility (ESRF) in Grenoble.
Photoemission and Auger spectra were recorded using a 140~mm mean
radius hemispherical analyzer coupled to a mini-Mott 25~kV spin
polarimeter \cite{Ghiringhelli99}. The spin detector had an
efficiency (Sherman function) of 17\%, and the energy resolution
of the electron analyzer was 0.7~eV. The photon energy resolution
was set at 0.2~eV. The measurements were carried out at normal
emission with respect to the sample surface and at an angle of
incidence of the x-rays of 60$^\circ$. The sample was magnetized
remanently in-plane using a pulsed magnetic coil, the
magnetization direction was alternated to eliminate the effect of
instrumental asymmetries \cite{Getzlaff98}. The pressure of the
spectrometer chamber was better than 1x10$^{-10}$~mbar.

\begin{figure}[!htb]  
\centerline{\includegraphics[width=11.25cm]{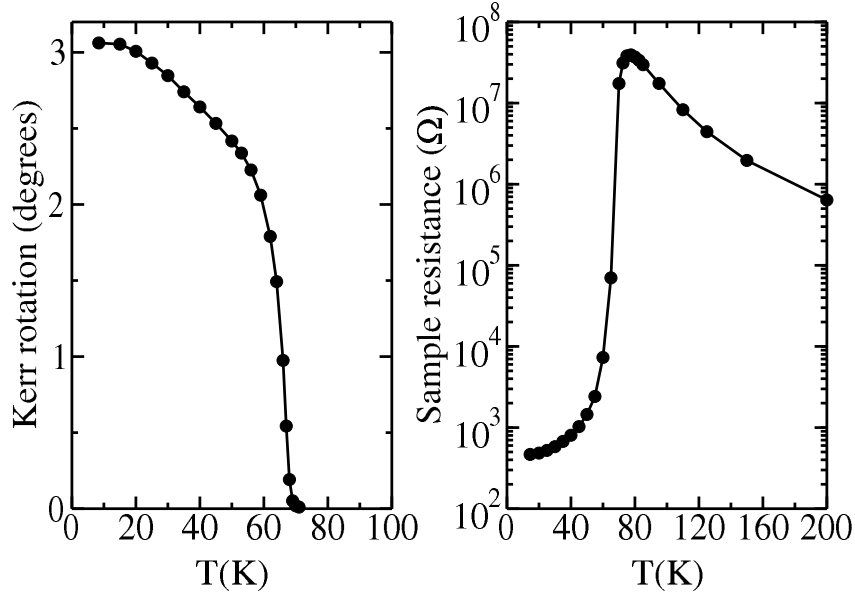}}
\caption{\label{fig1} Left panel: Remanent longitudinal
Kerr-rotation of a 50 nm EuO thin film as a function of
temperature using p-polarized light at $h\nu$ = 1.96~eV and
$\theta_{in}$ = 45$^\circ$. Right panel: Metal-insulator
transition in the temperature dependent resistivity of a EuO
film.}
\end{figure}

EuO samples with a film thickness of $\approx$ 200 \AA\ were
grown {\it in situ} by evaporating Eu metal from a Knudsen cell
at a rate of $\approx$ 3 \AA\ per minute in an oxygen atmosphere
of 1x10$^{-8}$~mbar on top of a Cr covered, chemically polished
Al$_2$O$_3$ substrate kept at 280$^{\circ}$C. In a separate
experiment in the Groningen laboratory, we have verified that
this recipe provides us with high quality polycrystalline EuO
films. Valence band and core level x-ray photoemission
spectroscopy (XPS) show no detectable presence of Eu$^{3+}$ ions,
and ultra-violet photoemission experiments reveal that there is
no detectable density of states at the Fermi level, demonstrating
that we have indeed obtained nearly stoichiometric semiconducting
EuO without detectable traces of Eu$_2$O$_3$, Eu$_3$O$_4$, or Eu
metal. The left panel of figure \ref{fig1} displays the
magneto-optical Kerr-rotation as a function of temperature on our
films using a He-Ne laser ($h\nu$ = 1.96~eV) and shows that these
films have indeed the correct T$_c$ of 69 K. The right panel of
figure \ref{fig1} depicts the resistivity of the films as a
function of temperature, and it demonstrates clearly the presence
of a metal-insulator transition at T$_c$. The very large change
in resistivity, namely 5 orders in magnitude, indicates that the
carrier concentration due to oxygen defects, i.e. the
off-stoichiometry, is of the order of 0.3\% or less
\cite{Shapira}.

\section{Results and Discussion}

\begin{figure}[!htb] 
\centerline{\includegraphics[width=11.25cm]{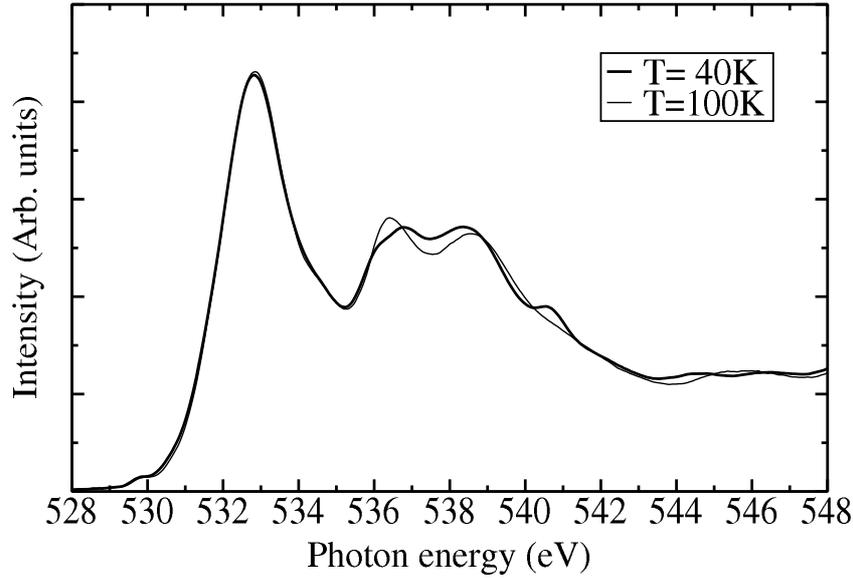}}
\caption{\label{fig2} O $K$ x-ray absorption spectrum of EuO,
above (thin solid line) and below (thick solid line) the Curie
temperature ($T_c$ = 69 K).}
\end{figure}

To study the conduction band of EuO, we use O $K$ edge x-ray
absorption spectroscopy (XAS). This technique probes the O $2p$
character of the conduction band, which is present because of the
covalent mixing between the O $2p$ and Eu $5d$-$6s$ orbitals.
Figure \ref{fig2} displays the O $K$ XAS spectra, recorded by
collecting the total electron yield (sample current) as a
function of photon energy. Large changes over a wide energy range
can be clearly seen between the spectra taken above and below
T$_c$. The low temperature spectrum contains more structures and
is generally also broader. We note that the spectra also show a
very small feature at 529.7~eV photon energy, with a spectral
weight of not more than 0.1\% relative to the entire spectrum.
Since the intensity of this feature is extremely sensitive to
additional treatments of the sample surface it is probably
related to surface states \cite{Schiller01d} or imperfections at
the surface.

It is evident that the spectral changes across T$_c$ cannot be
explained by a phonon mechanism, since the changes involve more
than a simple broadening, and above all, since the spectrum
becomes broader upon temperature lowering. Because there are also
no changes in the crystal structure across T$_c$, we attribute
the spectral changes to the appearance of a spin-splitting in the
Eu $5d$ like conduction band below T$_c$. To prove this, we have
to determine the spin-polarized unoccupied density of states. To
this end, we measured the O $K$ XAS spectrum no longer in the
total electron yield mode, but in a partial electron yield mode
in which we monitor the O $KL_{23}L_{23}$ Auger peaks that emerge
at a constant kinetic energy from the XAS process. By measuring
the spin-polarization of this Auger signal while scanning the
photon energy across the O $K$ edge, we can obtain the {\it
spin-resolved} O $K$ XAS spectrum.

\begin{figure}[!htb] 
\centerline{\includegraphics[width=11.25cm]{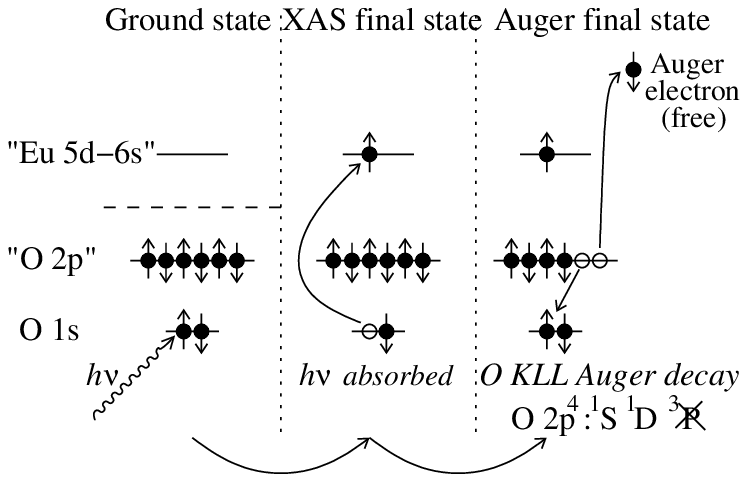}}
\caption{\label{fig3} Mechanism of spin-resolved x-ray absorption
spectroscopy. The spin of the outgoing O $KL_{23}L_{23}$ Auger
electron is opposite to the spin of the electron that is excited
in the x-ray absorption process. This scheme illustrates the
observation of a spin-up conduction band state.}
\end{figure}

The underlying concept of this new type of experiment is
illustrated in figure \ref{fig3}. This figure shows the O $1s$
core level, the occupied O $2p$ valence band and the unoccupied
Eu $5d$-$6s$ conduction band. Quotation marks indicate that due
to covalent mixing the conduction band also has some O $2p$
character and the valence band some Eu $5d$-$6s$ character. This
mixing allows an x-ray to excite an O $1s$ electron to the
conduction band, leaving a spin-polarized core hole if the
conduction band is spin-polarized (middle panel of figure
\ref{fig3}). The subsequent $KL_{23}L_{23}$ Auger decay of the
XAS state leads to O($2p^4$) like final states \cite{Tjeng}, and
the outgoing Auger electron will now also be spin-polarized
(right panel of figure \ref{fig3}). Unique to a $KL_{23}L_{23}$
Auger decay is that the entire two-hole final states are of pure
singlet ($^1S$ and $^1D$) symmetry since the triplet $^3P$
transitions are forbidden by Auger selection rules
\cite{SawatzkyAug,Fug}. This implies that the O $KL_{23}L_{23}$
Auger electrons will have a degree of spin-polarization which is
equal to that of the conduction band, but which has an opposite
sign due to the singlet character of the Auger transition. Thus,
the measurement of the spin of the O $KL_{23}L_{23}$ Auger
electrons across the O $K$ edge will reflect the
spin-polarization of the unoccupied conduction band states. We
note that spin-resolved XAS is different from a magnetic circular
dichroism (MCD) experiment. In the latter the helicity of the
circularly-polarized light is varied and the dichroic signal
contains a more convoluted information about the spin and the
orbital moments of the unoccupied states \cite{Tanaka}. We also
note that the availability of EuO in thin film form is crucial
for the measurement of spin-polarized electron spectroscopies
because the remanent magnetic field created by the very small
amount of material involved is negligibly small, and thus a
perturbation of the trajectories of the emitted electrons can be
avoided.

\begin{figure}[!htb] 
\centerline{\includegraphics[width=11.25cm]{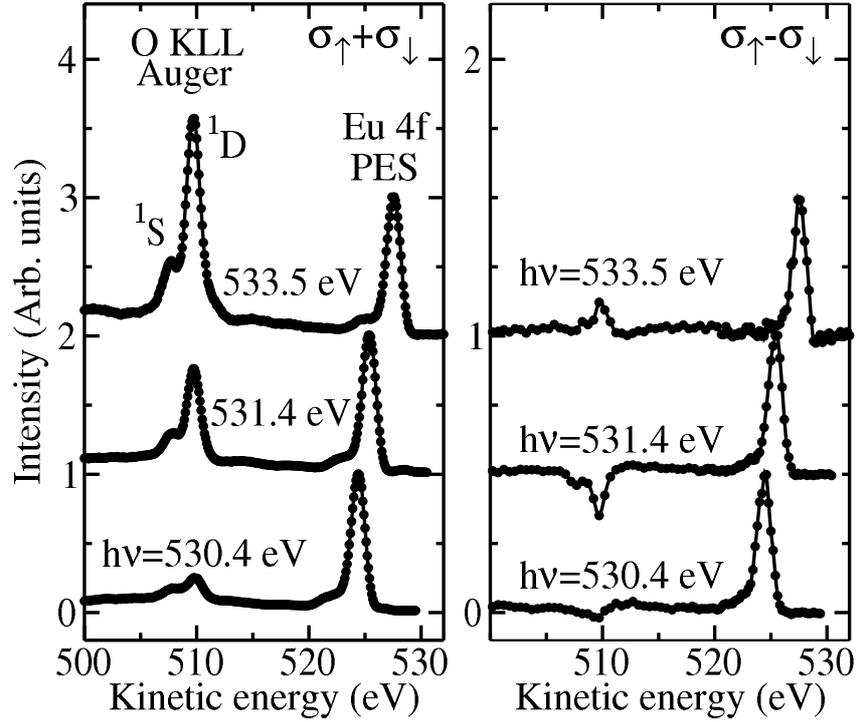}}
\caption{\label{fig4} Left panel: Spin-integrated valence band
photoemission and O $KL_{23}L_{23}$ Auger spectra of EuO. Right
panel: Difference spectra between the spin-up and spin-down
channels for the valence band and O $KL_{23}L_{23}$ Auger.}
\end{figure}

As an illustration, we show in figure \ref{fig4} a small
selection of the photoelectron spectra which we have taken from
the EuO valence band and the O $KL_{23}L_{23}$ Auger as a
function of photon energy at T=20 K. The left panel displays the
unpolarized spectra while the right panel gives the difference
spectra between the spin-up and spin-down channels (the spin-up
direction is parallel to the magnetization direction). We can
clearly distinguish the narrow Eu $4f^7$$\rightarrow$$4f^6$
photoemission (PES) peak in the valence band spectrum
\cite{Eastman69,Sattler72,Sattler75}, and observe that its
spin-polarization is about 50\%, indicating that the remanent
magnetization of the EuO films at this temperature is only half
of the saturation magnetization. This remanence is confirmed by
analyzing the magnetic circular dichroism measurements at the Eu
$M_{45}$ ($3d$$\rightarrow$$4f$) photoabsorption edges and is
comparable to magnetization measurements on EuO films
\cite{Iwata00}.

\begin{figure}[!htb]  
\centerline{\includegraphics[width=11.25cm]{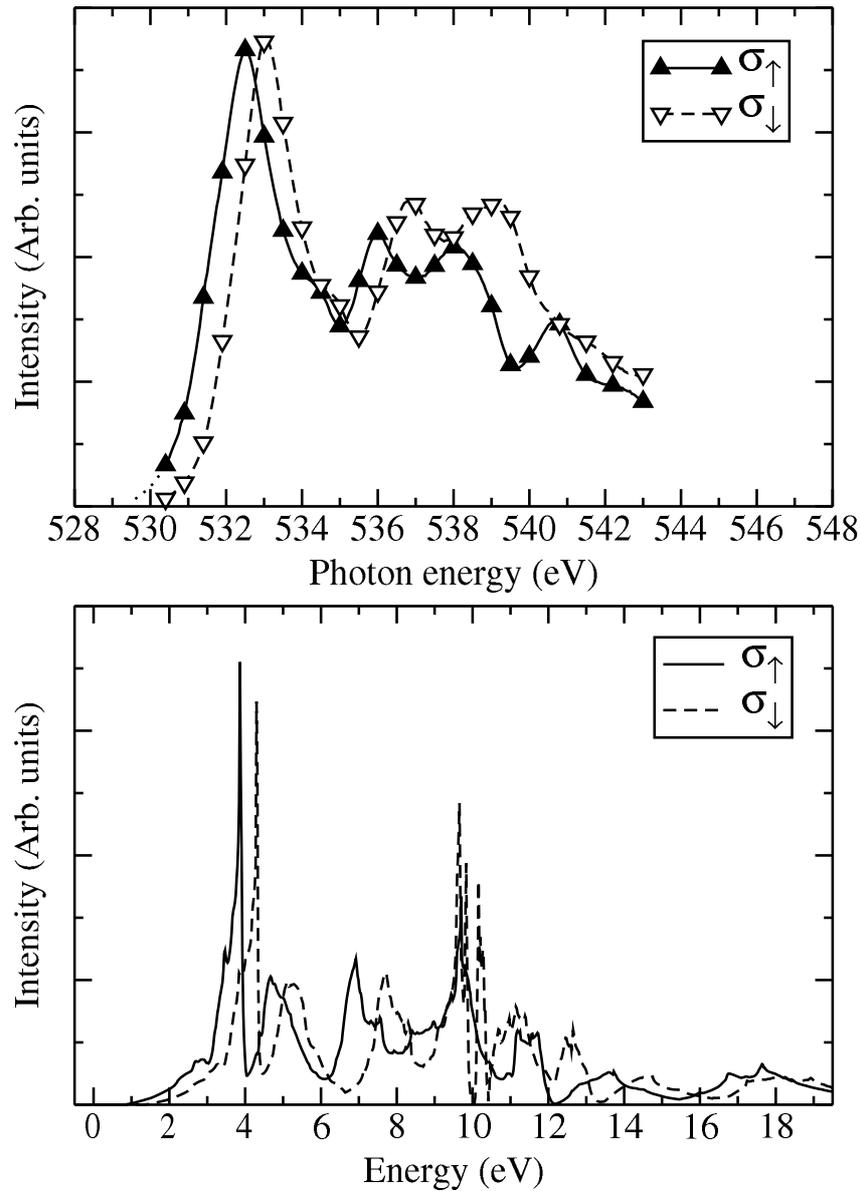}}
\caption{\label{fig5} Top panel: spin-resolved O $K$ x-ray
absorption spectrum of EuO taken at 20~K. Bottom panel:
spin-resolved unoccupied O 2p partial density of states from
LSDA+U calculations (U = 7.0 eV). The zero of energy corresponds
to the top of the valence band. There is a difference of (528.5
$\pm$ 0.5) eV between the XPS low binding energy O $1s$ onset and
valence band onset. The O $1s$ onset energy is taken as being 1
eV below the peak value. }
\end{figure}

It is interesting to see the strong photon energy dependence of
the magnitude and spin-difference of the O $KL_{23}L_{23}$ Auger.
By measuring these spin-resolved O $KL_{23}L_{23}$ Auger spectra
across the entire O $K$ edge region with closely spaced photon
energy intervals, we can construct an accurate spin-resolved O
$K$ XAS spectrum of EuO. The results are shown in the top panel
of figure \ref{fig5}. Here a normalization for full sample
magnetization has been made using the measured spin-polarization
of the Eu $4f$. We observe an almost rigid splitting between the
spin-up and spin-down peaks near the bottom of the conduction
band, which is as large as 0.6 eV. Extrapolation of the data
strongly suggests a very high spin-polarization at the bottom of
the band. It is also interesting to note that the spin polarized
features at 536.0 eV and 536.8 eV correspond with features in the
low temperature XAS scan of figure \ref{fig2} and that above
T$_c$ these peaks seem to merge into one feature at 536.4 eV,
indicating that the changes in the density of states below T$_c$
can indeed largely be attributed to a spin-splitting, which is a
shift of the spin-up band to lower energy and the spin-down band
to higher energy. However at higher energies the spin-behavior
seems to be less simple: the spin-up feature at 540.7 eV, for
example, does not seem to have a spin-down counterpart, possibly
due to the near presence of the $4f^7$$\rightarrow$$4f^8$
electron addition peak in the conduction band. To strengthen our
understanding of the experimental findings, we have also carried
out band structure calculations in the local spin density
approximation including the on site Hubbard U (LSDA+U)
\cite{Anisimov} for EuO in the ferromagnetic state. The bottom
panel of figure \ref{fig5} displays the results for the
spin-resolved unoccupied O $2p$ partial density of states. The
agreement between this mean-field theory and the experiment is
remarkable: the approximate position and spin-splitting of most
peaks is well reproduced, including the more intricate features
that arise at about 10 eV above the Fermi level which are
associated with the unoccupied 4$f$ states. Recently we have
become aware of calculations that obtain results similar to ours
\cite{Schiller01c,Schiller3}. When comparing XAS with band
structure calculations it should be noted that the interaction
between the core hole and the conduction band electron can in
principle lead to exitonic effects in the XAS spectra. However
these effects generally lead to sharp white lines which we do not
observe here. Moreover because the core hole is on the oxygen
atom while the conduction band consists of Eu orbitals, the
interaction with an O 1$s$ core hole is so small as compared to
the 5$d$ band width that its effect will be negligible. This
conclusion is also based on a vast amount of data on the O $K$
edge XAS of transition metal oxides.

These experimental results clearly demonstrate that the large
changes observed in the conduction band structure below T$_c$ in
figure \ref{fig2} are caused by a spin-splitting. We attribute
this splitting to the direct exchange interaction between the
localized $4f$ moments and the delocalized $5d$-$6s$ conduction
band states. We note that these measurements suggest that the red
shift of the optical absorption edge is also due to this
spin-splitting rather than to a broadening of the conduction
band. These results support the following picture for the
metal-insulator transition in Eu-rich EuO \cite{Oliver72}. Above
$T_c$, defect or impurity states have their energy levels located
slightly below the bottom of the conduction band, and the
material behaves like a semiconductor: the resistivity decreases
with increasing temperatures as a result of a thermal activation
of the electrons from the defect states into the conduction band.
Below $T_c$, the conduction band splits due to the exchange
interactions, and the defect states now fall into the conduction
band. The electrons of these defects can then propagate in the
spin-polarized bottom of the conduction band without needing any
activation energy, and the system behaves like a metal. As we
estimate the depolarization of the conduction band due to
spin-orbit coupling ($\xi_{5d}$ = 0.067~eV) to be small (less
than 5\%), we expect the doped charge carriers in ferromagnetic
EuO to be almost fully spin polarized, an observation that is
very interesting for fundamental research projects in the field
of spintronics.

\section{Conclusions}

To conclude, our experiments have revealed large changes in the
conduction band states of EuO if the temperature is varied across
$T_c$. Using new spin-resolved measurements we have shown that
these changes are caused by a splitting between the spin-up and
spin-down unoccupied density of states. This exchange splitting
is appreciable, about 0.6~eV. From this we conclude that electron
doped EuO in the ferromagnetic state will have charge carriers
with an almost 100\% spin-polarization.




\chapter{Temperature-dependent exchange splitting in EuO}
\label{Tempxas}


{\sl The temperature dependence of the x-ray absorption spectrum
near the O $K$ edge in single crystals of the ferromagnetic
semiconductor EuO is studied. Clear signatures of an exchange
splitting proportional to the long range magnetic order are
observed in the XAS spectra. This temperature dependence is very
different from that of the optical absorption edge, which was
generally believed to be a measure of the exchange splitting. The
present results contradict this and lead to the conclusion that
the temperature dependent exchange splitting of conduction band
electrons in EuO is similar to that in Gd metal despite the
different origin of the magnetic interactions in these
materials.}

\section{Introduction}

Since their discovery \cite{Tsubokawa, Matthias} around 1960, the
properties of ferromagnetic semiconductors have continuously
fascinated scientists. The magnetic order in this class of
compounds is induced by indirect and superexchange interactions
\cite{Kasuya70} and is thus of a different origin than the
itinerant electron ferromagnetism of the elemental ferromagnets.
The transition to the magnetically ordered state is accompanied
by large changes in transport and magneto-optical properties,
which include colossal mageneto-resistance and a giant
Kerr-effect. Moreover, a giant red-shift in the optical
absorption edge is observed upon entering the ferromagnetic state
of EuO~\cite{Busch}. This shift appeared to be a universal
feature of ferromagnetic semiconductors
\cite{Freiser,Schoenes,Arai,Demin} and has been attributed to a
splitting of the conduction band as a result of the exchange
interactions between localized magnetic moments and conduction
electrons. Remarkably this temperature dependent red-shift does
not follow the long range order and part of the shift is even
realized above the Curie temperature (see figure \ref{Optics:a}).
This part of the shift in the paramagnetic state has generally
been ascribed to a splitting or lowering of the conduction band
due to interactions between conduction band electrons and
spin-fluctuations in the paramagnetic state
\cite{Rys67,Alexander76,Haas68,Mauger86,Nolting79,Nagaev01}.

More recent developments in photo-electron spectroscopy
techniques have allowed direct observation of the exchange
splitting, a characteristic parameter in most models for
ferromagnetism. During the last decades this has resulted in
intensive efforts to measure the temperature dependence of this
splitting in the elemental ferromagnets. These studies have shown
that the temperature dependence of the Gd $5d$ bands is well
described by an exchange splitting which follows the long range
magnetization \cite{Kim92,Li95,Weschke96,Donath96}. However more
locally sensitive spectroscopies show that short range order
induces a local exchange splitting which can persist even above
T$_c$ \cite{Bode99,Tober98}. For the elemental transition metal
ferromagnets with their narrower $3d$ bands the situation seems
to be even more complicated, with an exchange splitting which
persists in the paramagnetic state for some bands and collapses
at T$_c$ for others \cite{Kirschner84}.

Although the ferromagnetic exchange interactions in
stoichiometric EuO are not mediated by itinerant electrons like
in Gd metal, but by virtual excitations of valence electrons
\cite{Kasuya70,Mauger86}, similarities are still expected as in
both materials the magnetic interactions which split the
conduction band are mainly generated by the ferromagnetic
alignment of the $4f$ spins. However the strong differences
between the temperature dependence of the red-shift in EuO and
that of the exchange splitting in Gd metal seem to suggest
otherwise.

To clarify this issue we present temperature dependent O $K$
x-ray absorption spectroscopy (XAS) data on single crystals of
EuO. We show strong evidence that the exchange splitting of the
Eu $5d$ conduction band \emph{does} follow the long range
magnetic order. In contrast to optical absorption measurements we
do not find evidence for a splitting above the Curie temperature.
We discuss these results and point out important differences
between the interpretation of data obtained with optical and
x-ray photon energies.

\section{Experimental}

Stoichiometric single crystals of EuO were grown by a solution
sintering process~\cite{Fischer}. Measurements were performed
using the Dragon beamline at the Synchrotron Radiation Research
Center (SRRC) in Taiwan. The crystals were cleaved \emph{in situ}
after which x-ray absorption spectra were recorded in
fluorescence yield mode. The resolution of the x-rays
was $\sim$ 0.2 eV. 
\begin{figure}[!htb]
\centering \vspace{0.2cm}
\includegraphics[width=11.25cm]{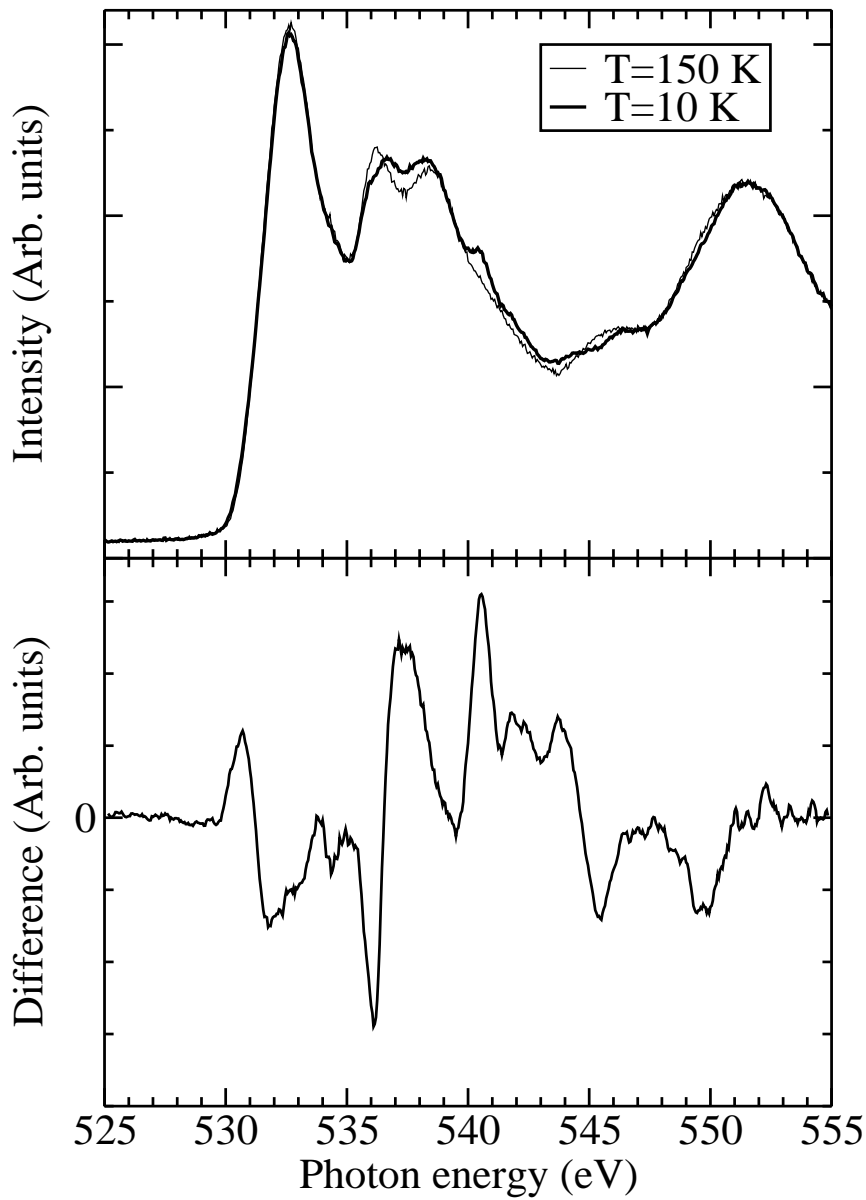}
\caption{\label{figure1} Upper panel: O $K$ XAS spectrum at 10 K
and 150 K of a EuO single crystal. Lower panel: Difference data
(smoothed) between the intensity at 10 K and 150 K.}
\end{figure}
In the upper panel of figure \ref{figure1} we show O~$K$ XAS
spectra taken above and below the Curie temperature (T$_c=69$ K).
This technique probes the O $2p$ character of the conduction
band, which is present as a result of covalent mixing between the
O $2p$ and Eu $5d-6s$ orbitals. The spectra are very similar to
those observed on thin EuO films in total electron yield mode
(compare with figure \ref{fig2}). However in fluorescence yield
on single crystals the feature at 529.5~eV related to surface
defects is absent, due to the lower surface sensitivity of this
technique and possibly due to a higher stoichiometry of the
single crystals. Clear changes are observed upon lowering the
temperature. To highlight these changes, the difference spectrum
is plotted in the lower panel of figure \ref{figure1}.
Interestingly the maxima in the high temperature spectrum
generally seem to coincide with the minima of the difference
spectrum, indicative of a splitting or broadening of spectral
weight. The feature at 536.2 eV which transforms into two peaks
at low temperature strongly suggests that it concerns a
splitting. The splitting may also be present in the main peak at
532.7 eV, but it is less obvious here due to the larger intrinsic
width of this peak. Spin-resolved XAS measurements
\cite{Steeneken02} indeed confirmed that these changes in the XAS
spectrum are strongly related to a spin splitting of the spin-up
and spin-down density of states due to the $4f-5d$ exchange
interaction. On the other hand the peaks which appear in the
ferromagnetic state between 540.0 and 543.0 eV do not follow this
behavior and are probably related to changes in the unoccupied Eu
$4f$ density of states. Qualitative agreement was found between
the x-ray absorption spectrum and the density of states as
obtained by LSDA+U calculations (see figure \ref{fig5}). These
calculations also showed that the exchange splitting depends on
momentum and energy. To extract quantitative data on this
splitting the bottom of the band is therefore preferable, as its
density of states is not a sum of contributions from different
energies or parts of the Brillouin zone.

\section{Discussion}

In the simplest mean field picture the exchange splitting can be
described as follows. If $D(E)$ is the unoccupied density of
states of the $5d$ conduction band in the absence of $4f-5d$
exchange interactions, the exchange interaction $\Delta_{\rm ex}$
will lower the spin-up levels by an energy $\Delta_{\rm ex}$,
while the spin-down states will be raised by the same amount. The
total unoccupied density of states $D'(E,\Delta_{\rm ex})$ will
thus become:
\begin{equation}
\label{DOS} D'(E,\Delta_{\rm ex}) = \frac{1}{2}(D(E+\Delta_{\rm
ex}) + D(E-\Delta_{\rm ex}))
\end{equation}
Therefore, in a one electron picture one expects the absorption
spectrum to have a shape similar to $D'(E)$ if the electron is
excited to the conduction band from a sharp energy level. As the
valence band becomes spin-polarized in ferromagnetic state,
optical spin-flip transitions are much less probable than
spin-conserving transitions and the transition probabilities to
the spin-up and spin-down bands are changing when the sample is
magnetized\footnote{The reduction in the optical absorption peak
energy in the ferromagnetic state (see figure \ref{freiser}),
which is not expected from equation (\ref{DOS}), might be related
to this effect.}. Also, excitonic effects can play an important
role, as the excited electron can increase the exchange coupling
between Eu atoms in the vicinity of the $4f$ hole. This might
lead to an increased local magnetization and exchange splitting
inside the exciton and thus modify the position of the absorption
edge. These considerations show that although the shift of the
absorption edge strongly suggests the presence of exchange
splitting it is not a very reliable measure of the size or
temperature dependence of this splitting.

\begin{figure}[!htb]
\vspace{0.5cm} \centering
\includegraphics[width=11.25cm]{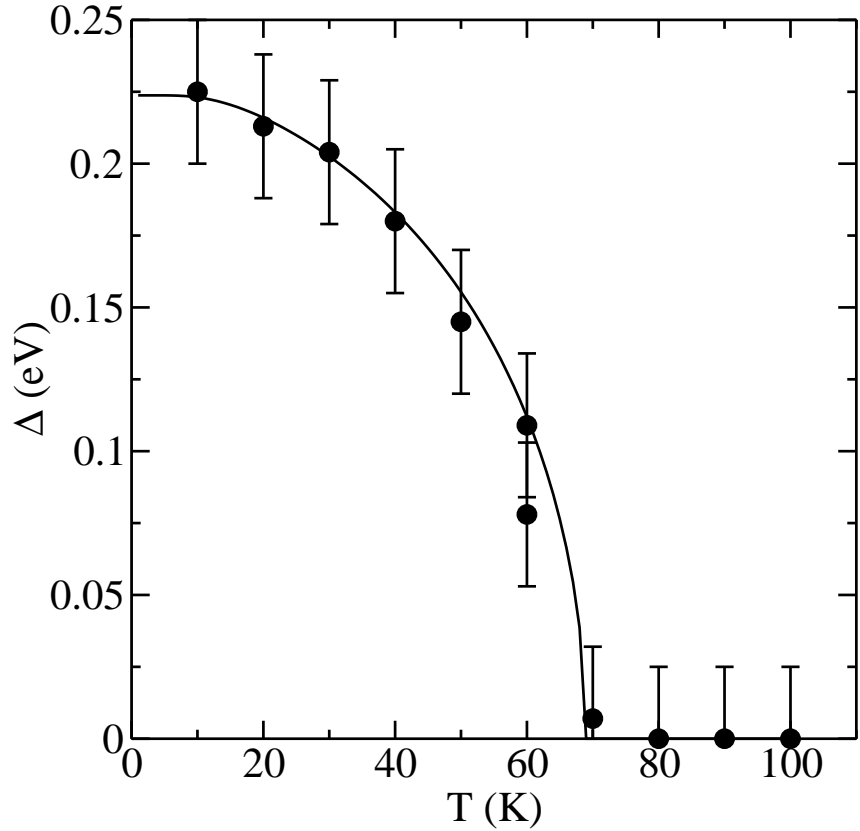}
\caption{\label{figure3} $\Delta_{\rm ex}$ as obtained from least
square fits of the low temperature data with the splitted 150 K
XAS spectrum following equation (\ref{DOS}). Also shown is a
solid curve which has the shape of the magnetization in a mean
field model for a $J=7/2$ ferromagnet with T$_c$ = 69 K.}
\end{figure}
\begin{figure}[!htb]
\vspace{0.5cm} \centering
\includegraphics[width=11.25cm]{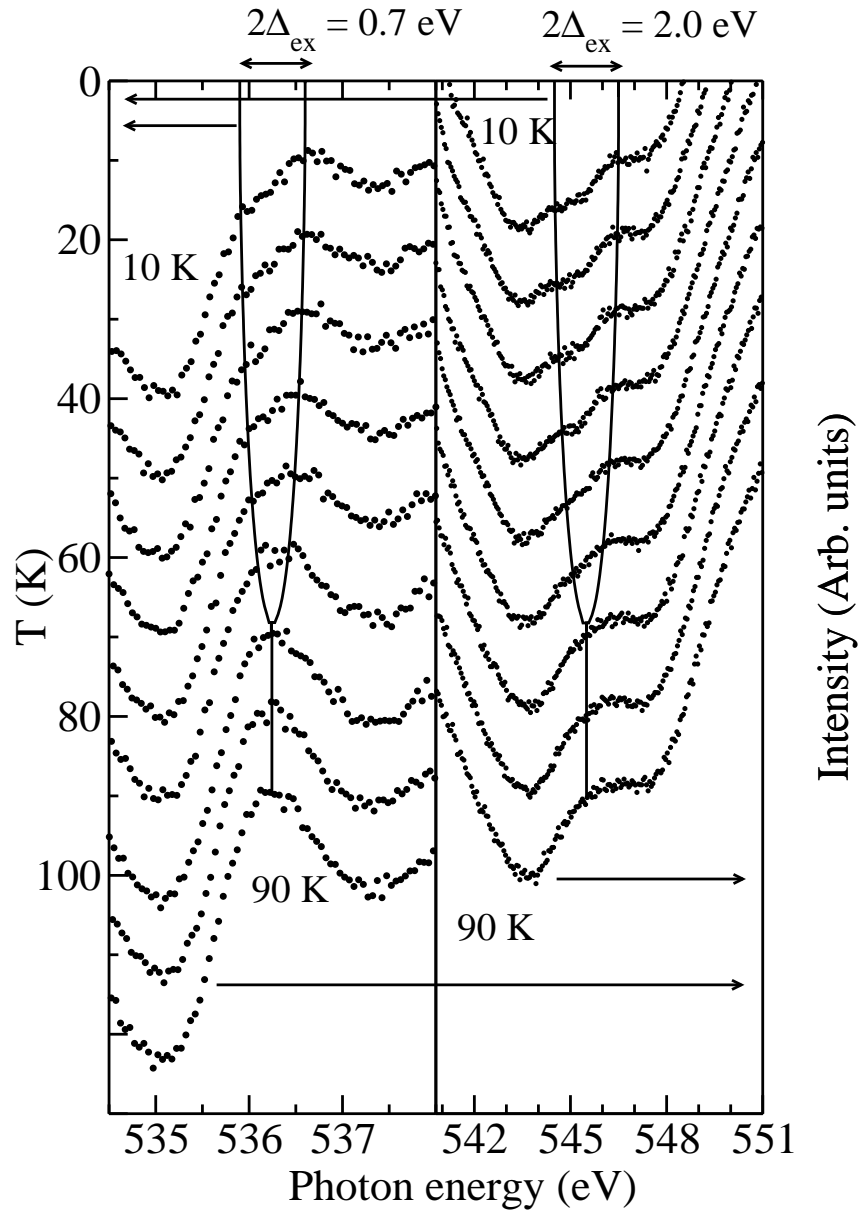}
\caption{\label{figure4} Two energy regions of the O $K$ XAS
spectra of EuO at temperatures from 10 to 90 K, taken at
intervals of 10 K. Spectra are offset to align the split peak
with its temperature on the left axis. Solid lines are split
proportional to the magnetization in a mean field model as in
figure \ref{figure3}.}
\end{figure}
The interpretation of the O $K$ x-ray absorption spectrum is
easier because it involves excitations from a dispersionless and
spinless core-level. Therefore we expect that the XAS data should
approximately follow equation (\ref{DOS}). To obtain a
quantitative estimate of the exchange splitting from the XAS data
it is thus not sufficient to measure the edge shift as the
measured shift will depend on the shape $D(E)$. A least squares
fit of the various low temperature spectra was made in the photon
energy region 529.0-534.7 eV using $D'(E,\Delta_{\rm ex})$
according to equation (\ref{DOS}), with $\Delta_{\rm ex}$ as the
only fitting parameter and $D(E)$ assumed equal to the XAS
spectrum\footnote{A smoothed curve of the 150 K spectrum, which
was obtained by fitting it with multiple Gaussians, was used for
$D(E)$ to eliminate smoothing effects when applying equation
(\ref{DOS}) to a discrete data set which contains statistical
noise.} at 150 K. Excellent fits were obtained in this energy
region. To verify if a temperature independent center of gravity
indeed resulted in the best fit we also fitted the spectra (in
the same energy region) allowing an arbitrary energy shift in
addition to the arbitrary splitting $\Delta_{\rm ex}$. For all
spectra, best fits were obtained for shifts $\leq 0.02$~eV, very
near to the shift of the center of gravity as expected. In figure
\ref{figure3} we show the fitted $\Delta_{\rm ex}$ versus
temperature. A maximum splitting of $2\Delta_{\rm ex}=0.45$~eV is
found, which is somewhat smaller than the exchange splitting of
$2\Delta_{\rm ex}=0.6$~eV found in the spin-polarized XAS
measurements \cite{Steeneken02}. This smaller value might be due
to statistical errors as a result of the fact that it is very
difficult to resolve a splitting of two peaks which is smaller
than their linewidth. However, the smaller value of the
$\Delta_{\rm ex}$ might also very well be a result of differences
between the mean field assumption of $S=0$ above T$_c$ and the
actual random spin-orientation of local $S=7/2$ moments in the
paramagnetic state, which can broaden the high-temperature peak
with respect to the spin-polarized peak width in the
ferromagnetic state (compare figures \ref{figure1} and
\ref{fig5}). In the same figure a solid curve is plotted with the
shape of the spontaneous magnetization of a $J=7/2$ ferromagnet
with T$_c=$69 K as was obtained from solving the mean/molecular
field model. The magnetization in EuO is known to follow such a
mean field model to a very high accuracy (see figure \ref{EuOmag}
\cite{Mauger78}). The exchange splitting as determined from the
XAS data very nearly follows the calculated magnetization curve.
Additional information on the exchange splitting in EuO was found
by examining the temperature dependence of the split features at
536.2~eV and 545.5 eV, which are shown in figure \ref{figure4}.
In the same figure solid lines split proportional to the mean
field magnetization of a $J=7/2$ ferromagnet are shown. There is
clear qualitative agreement between the observed peak position
and that expected from a mean field model with an exchange
splitting proportional to the long range order. A maximum
splitting of $2\Delta_{\rm ex}=0.7\pm 0.1$ eV is observed in the
peak at 536.2 eV, corresponding well to an energy separation of
the spin-polarized peaks of about 0.8 eV. The peak at 545.5 eV
seems to show a maximum splitting of $2.0\pm 0.1$ eV, which is
remarkable as it is considerably larger than the splitting
predicted by LSDA+U calculations. As we have not studied this
structure with spin-polarized XAS it is unclear if these large
changes are due to the separation of two spin-polarized bands, or
if they are related to more intricate changes in the spectral
weight as a result of the ferromagnetic ordering. The XAS data
above suggest that the exchange splitting in the EuO conduction
band follows the long range magnetic order, with a collapse into
a single band at T$_c$. This is in remarkable contrast to the
behavior of the optical absorption edge, which does not follow
the long range magnetic order \cite{Freiser,Schoenes}, and
already starts shifting in the paramagnetic state. This shift has
been subject of many theoretical studies
\cite{Mauger86,Nagaev01,Alexander76,Rys67,Haas68,Nolting79}. Most
effort focused on calculations of the changes in the
quasiparticle energy due to dressing of conduction electrons by
spin-fluctuations in the paramagnetic state of the ferromagnet.
The shift of the optical absorption edge in the paramagnetic
state could however have a different origin as the Coulomb
attraction of the $4f$ hole might localize the excited electron
in an $s$-like orbital with radius $R$. Kasuya \cite{Kasuya72}
suggested that the bottom of the absorption edge is formed by
such a magnetic exciton. This suggestion was supported by the
exponential shape of the absorption edge \cite{Freiser}. The
radius $R$ of this exciton might very well be of the order of the
lattice constant. In this case the exchange splitting of the
electron will not depend on the long range order, but on the
short range order within its radius. In fact it was shown that
the shift of the absorption edge follows the nearest neighbor
spin-correlation function $<{\mathbf S_i\cdot S_j}>/S^2$, for $i$
and $j$ at nearest neighbor sites, quite closely
\cite{Freiser,Schoenes}, thus suggesting that the excited
electron is more sensitive to the short-range order than to the
long-range order. Indications for a similar magnetic short-range
($\sim$ 20~\AA) order above T$_c$ were found in Gd metal
\cite{Bode99,Tober98}.

Unlike the local optically excited $4f-5d$ exciton, excitonic
effects are expected to be much smaller in O $K$ XAS with the
final state hole on an oxygen atom. Although on-site excitations
from a core-hole can lead to large excitonic effects, the
conduction band of EuO consists mainly of Eu $5d$ orbitals and
the excited electron will therefore be effectively separated from
the oxygen atom and thus have a much smaller Coulomb attraction
to the core hole. The absence of a sharp exciton white line in
the XAS spectra supports this conclusion.

\section{Conclusions}

In summary, we have investigated the conduction band of EuO by
XAS. We did not find evidence for an onset of the splitting above
T$_c$ due to spin-fluctuations
\cite{Mauger86,Nagaev01,Alexander76,Rys67,Haas68,Nolting79}. This
led us to suggest that the red-shift in the optical absorption is
related to excitonic effects, as a result of which the absorption
edge will follow the local order instead of the splitting
experienced by conduction band electrons. In contrast to the
temperature dependence of the absorption edge we found strong
evidence that the temperature dependent exchange splitting in EuO
follows the long-range order, similar to the exchange splitting
in Gd metal.




%

\chapter{Strong O $K$ edge MCD in EuO}
\label{oxmcd}


{\it We have studied the magnetic circular dichroism (MCD) in the
O $K$ edge x-ray absorption spectrum of EuO. We observed a
remarkably large MCD signal on this "non-magnetic" atom. As there
is no spin-orbit splitting in the $K$-shell, the MCD is a result
of the spin-orbit coupling of the spin-polarized conduction band.
In this chapter we consider several mechanisms that can account
for this large O $K$ MCD signal and investigate its relation to
the spin-polarization of the conduction band.}

\section{Introduction}
Magneto-optical effects have been known to exist for a long time
\cite{Hulme32}. However, in the last decades a surge in the
interest for materials with large magneto-optical properties has
occurred. This surge was driven by technological interest for
magneto-optical recording media, and by large developments in
synchrotron technology, allowing the study of magnetic materials
with x-ray magnetic circular dichroism (XMCD) measurements
\cite{Schutz87,Goedkoop} following predictions by Erskine and
Stern \cite{Erskine75} and Thole {\it et al.} \cite{Thole85}.
This technique is extremely useful as it yields element-specific
information on the magnetization state of atoms in a compound. It
especially gives large signals when transitions are made between
orbitals on the same atom for which at least one orbital involved
has a large spin-orbit coupling. As the XMCD signals are
generally largest for transitions from spin-orbit split core
levels to partially filled $d$ or $f$ orbitals, most research has
focussed on transition metal and rare earth metal compounds.
Measurements on the Fe $K$ edge have shown that transitions from
an $s$-orbital result in a much smaller MCD signal that is
related to the spin-polarization of the conduction band
\cite{Schutz87,Stahler93}. Moreover, it was found that such
measurements are also sensitive to the magnetic moment of
neighboring atoms \cite{Chaboy96}. XMCD measurements at the
oxygen $K$ edge of La$_{1-x}$Sr$_x$MnO$_3$ and at the sulfur $K$
edge of EuS have shown that strong MCD signals can also be
detected at $K$ edges of "non-magnetic" atoms
\cite{Pellegrin97,Rogalev99}.

Here we present MCD measurements near the O $K$ edge of EuO. A
relatively large MCD signal $\sim 1\%$ is found on this
"non-magnetic" atom, of similar magnitude as the MCD at the iron
$K$ edge. We propose three possible effects that could account
for this MCD: the combination of spin-orbit and exchange
interactions on the atomic $2p$ final states, a Fano-like effect,
and the splitting of orbital moments in the O $2p$ orbital as a
result of the hybridization with the exchange and spin-orbit
split Eu $5d$ orbitals. We find that the first two effects will
result in an MCD which is approximately proportional to the
spin-polarization of the conduction band, whereas the third
effect results in a more convoluted relation between momentum,
spin and orbital moments. The measured MCD spectrum is compared
to calculations.

Polycrystalline EuO films with a thickness of $\sim$ 200 \AA
~were grown \emph{in situ} as described in chapter \ref{Spinxas}.
X-ray absorption spectra where recorded in total electron yield
mode with a photon resolution of 0.2 eV. A photon beam with a
circular polarization of 92\% reached the sample at an angle of
incidence 60$^\circ$ with respect to the surface normal. The
sample was remanently magnetized by a pulsed magnetic coil. The
magnetization direction was in the film plane and in the plane of
the light's Poynting vector. Both the magnetization direction and
the circular polarization of the light were alternated to measure
the absorption coefficient for photon helicities parallel
($\mu_+$) and antiparallel ($\mu_-$) to the magnetization.

In general spin-orbit interactions are very weak for the outer
electrons of light atoms as their strength scales approximately
as $Z^4/n^3$, where $Z$ is the atomic number and $n$ the
principal quantum number. Although the conduction band of EuO has
a strong spin-polarization which is almost 100\% near the bottom
of the band \cite{Steeneken02}, the spin-orbit coupling effects
in the O $2p$ band are small. As the O $2p$ orbitals hybridize
strongly with the Eu $5d$ orbitals, interaction with the heavy Eu
nuclei might provide an increased spin-orbit coupling on oxygen,
clarifying the size of the MCD. To our knowledge the spin-orbit
splitting of states on a light atom due to relativistic
interactions with neighboring atoms has not been theoretically
investigated. Although spin-orbit coupling effects can be
incorporated in relativistic LDA calculations
\cite{Anderson75,Kunes01}, such calculations have not yet been
performed for the O $K$ edge of EuO or similar materials. Let us
therefore continue to discuss several mechanisms that could be
responsible for the relatively large MCD at the $K$ edge of EuO.

\section{Atomic O $2p$ final state with $\mathcal{H}_{so}$ and $\mathcal{H}_{ex}$}
\label{atomoxmcd}

\begin{figure}[!htb]
\vspace{0.8cm}
\centerline{\includegraphics[width=11.25cm]{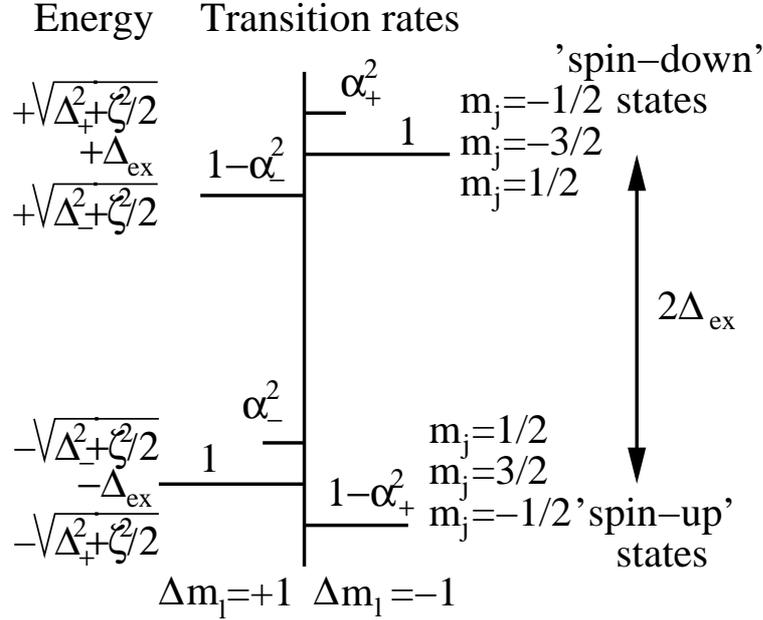}}
\caption{\label{oxmcdfig0} $1s^2\rightarrow1s^12p^1$~~transition
rates and $2p$ energy levels in a simple atomic model with
exchange splitting and spin-orbit coupling. Quotation marks
indicate that the final states are not pure spin states.}
\end{figure}

Let us first consider the simple case of an atomic $p$-electron
which experiences an exchange interaction with neighboring spins
$\mathcal{H}_{ex}$ and an effective spin-orbit interaction
$\mathcal{H}_{so}$. If we treat the exchange interactions in the
mean field approximation, the Hamiltonian is given by:
\begin{equation}
\mathcal{H} = \mathcal{H}_{ex} + \mathcal{H}_{so} =
-2\Delta_{ex}\mathbf{S_z} + \zeta\mathbf{S}\cdot\mathbf{L}
\end{equation}
As $\mathcal{H}$ only commutes with $\mathbf{J_z}$, only the
total angular momentum $j_z$ is conserved. The spin-orbit
coupling splits both exchange split levels into 3 levels. The
normalized wavefunctions for the 'spin-up' levels in terms of
$|~m_l,m_s>$ $p$-orbitals are given by:
\begin{eqnarray}
\label{psijz}
\psi_{m_j=-\frac{1}{2}} &=& -\sqrt{1-\alpha_+^2}|-1,\frac{1}{2}>+\alpha_+|~0,-\frac{1}{2}> \nonumber \\
\psi_{m_j=+\frac{3}{2}} &=& |~1,\frac{1}{2}> \nonumber\\
\psi_{m_j=+\frac{1}{2}} &=& -\sqrt{1-\alpha_-^2}|~0,\frac{1}{2}>+\alpha_-|~1,-\frac{1}{2}> \\
\alpha_\pm=\frac{\beta_\pm}{\sqrt{1+\beta_\pm^2}}&\mbox{,}&
\beta_\pm =
\frac{\zeta/\sqrt{2}}{\Delta_\pm+\sqrt{\Delta_\pm^2+\zeta^2/2}}
\nonumber
\end{eqnarray}
Where $\Delta_\pm = \Delta_{ex}\pm\zeta/4$. The energy of
$\psi_{m_j=\frac{3}{2}}$ is $-\Delta_{ex}$ and that of
$\psi_{m_j=\mp\frac{1}{2}}$ is $-\sqrt{\Delta_\pm^2+\zeta^2/2}$.
For $\zeta^2\ll\Delta_{ex}^2$ this gives a level splitting of
$\zeta/4$ which is probably too small to be resolved in our XMCD
measurements. From Fermi's Golden Rule we obtain that the total
$1s\rightarrow 2p$-'spin-up' transition rates for circularly
polarized light with $\Delta m_l=\pm 1$ are
$\mu_\pm\propto1\pm\alpha_\mp^2$. For small $\zeta$ we thus
expect that the normalized MCD signal
$P_\mu=(\mu_+-\mu_-)/(\mu_++\mu_-)=\frac{\alpha^2_++\alpha^2_-}{2-\alpha^2_++\alpha^2_-}\approx
\zeta^2/8\Delta_{ex}^2$. For the 'spin-down' levels the MCD
signal is of equal magnitude but of reversed sign, resulting in a
zero integrated MCD signal. The transition rates and energy
levels for this simple model are schematically drawn in figure
\ref{oxmcdfig0}.
\begin{figure}[!htb]
\centerline{\includegraphics[width=11.25cm]{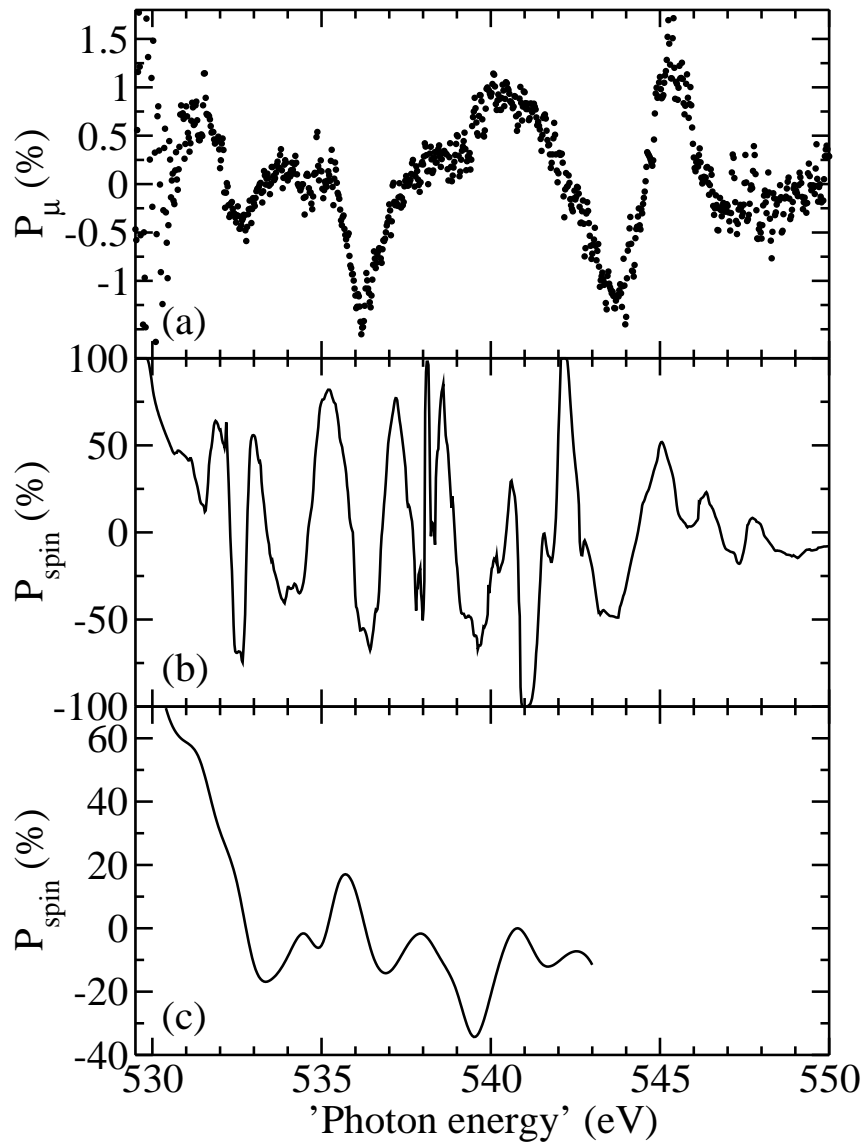}}
\caption{\label{oxmcdfig1} (a) Normalized MCD at the O $K$ edge
of EuO. (b) LDA+U calculation of the spin-polarization of the O
$2p$ density of states (top of valence band at 528.5~eV). (c)
Spin-polarization of the empty O $2p$ states as measured by
spin-polarized XAS \cite{Steeneken02}.}
\end{figure}
In EuO the empty oxygen $2p$ orbitals form a band which has an
energy dependent spin-polarization \cite{Steeneken02} as a result
of the exchange interactions with the spin-polarized $4f$
electrons via hybridization with the $5d$ orbitals. Let us assume
that the effect of the spin-orbit coupling on the levels is well
described by the atomic situation above. As the MCD signal is
positive for 'spin-up' levels and negative for 'spin-down'
levels, the intensity of the normalized atomic MCD signal $P_\mu$
obtained above should be multiplied by the spin-polarization of
the density of states $P_{\rm spin}$. Thus we expect $P_\mu
\approx \zeta^2P_{\rm spin}/8\Delta_{ex}^2 $.

To compare this simple model to our measurements we show in
figure \ref{oxmcdfig1}(a) the normalized MCD spectrum on the O
$K$ edge of EuO. The spectrum is corrected for the circular
polarization of the light and for the angle between the
magnetization and the incident light beam. A remarkably large and
structured MCD signal is observed which is even larger than that
at the $K$ edge in iron metal \cite{Schutz87}. For comparison we
show the spin-polarization of the O $2p$ partial density of
states as calculated by LDA+U and as measured by spin-polarized
XAS \cite{Steeneken02} in figures \ref{oxmcdfig1}(b) and (c). The
top of the valence band in the calculated spectra has been
aligned at a photon energy of 528.5 eV. The peak energies and
signs of the MCD agree quite nicely with the calculated and
measured spin-polarizations. Only in the region around 540 eV the
MCD sign is clearly opposite to that of the spin-polarization.
This is possibly related to the fact that the XAS effects in this
energy region are complicated by the empty Eu $4f$ orbitals. The
ratio $P_\mu/P_{\rm spin,LDA+U}$ at the peak positions varies
between approximately 0.008 and 0.025. Using
$\Delta_{ex}\approx0.3$ eV and $P_\mu \approx
\zeta^2/8\Delta_{ex}^2 P_{\rm spin}$ we find that the effective
spin-orbit coupling of the empty $2p$ states $\zeta\sim80-140$
meV which seems too large to be accounted for by the atomic
$\zeta_{2p} \approx 43$ meV \cite{Cowan81}. Moreover we note that
although the peak positions and signs qualitatively follow $P_\mu
\propto P_{\rm spin}$, the quantitative agreement is rather poor,
showing that the model is too simple to fully capture the
relativistic effects. Therefore let us explore alternative
mechanisms.

\section{Description in terms of Fano effect}
\label{Fano} An approach adopted by Sch{\"u}tz {\it et al.}
\cite{Schutz87,Stahler93}, to explain the $K$ edge MCD on iron is
to use a two step model in which one first obtains the effective
spin-polarization of the excited electron by considering the
transition from the $s$-orbital of an atom to spherically
symmetric continuum states with $j=1/2$ and 3/2. As was shown by
Fano \cite{Fano69}, the spin-orbit coupling affects the radial
part of the continuum wave functions and thus yields different
transition rates to the $j=1/2$ and 3/2 continuum states.
Additionally, an electron which has its spin parallel to the
photon's angular momentum, will end up in a $j=3/2$ final state,
whereas an electron of opposite spin can both have a $j=1/2$ or
$j=3/2$ excited state. The combination of these effects will lead
to a spin-polarization of the continuum electrons parallel to the
angular momentum of the light\footnote{The Fano effect from
orbitals with $l\neq 0$ or on magnetic atoms is more involved
\cite{Thole91,Ebert97} and has recently been used to explain our
circularly polarized spin-resolved photoemission measurements on
Cu and Ag metal \cite{Minar01}.}. In the second step the
transition rate of these polarized continuum electrons to a
spin-polarized conduction band state will depend on their spin
and thus on the circular polarization of the light. Therefore the
resulting MCD spectrum will approximately follow the
spin-polarization of the conduction band. It is difficult to
estimate if the magnitude of the MCD caused by the Fano effect is
sufficient to explain the observed MCD. However, it cannot be
excluded that the mixing of the orbital and spin moments $m_l$
and $m_s$ as a result of spin-orbit coupling and exchange
splitting, which was described in the previous section, gives an
at least as important contribution to the MCD signal.

\section{Hybridization via Eu $5d$}
\label{so5d} From Cowan \cite{Cowan81} (pp. 238 and 588) we find
that the atomic spin orbit coupling parameter for the oxygen $2p$
orbitals $\zeta_{2p} \approx$ 43 meV whereas $\zeta_{5d} \approx$
67 meV for the Eu $5d$ orbitals \cite{deGroot00}. However, as the
oxygen holes only exist as a result of hybridization with the Eu
$5d$ orbitals, they have a strong $5d$ character. Instead of
using an atomic spin-orbit parameter $\zeta_{2p}$, like in
section \ref{atomoxmcd}, we consider the possibility that the MCD
is dominated by $\zeta_{5d}$ in this section. To estimate these
effects, we have performed a Linear Combination of Atomic
Orbitals (LCAO) calculation \cite{Slater54}, including exchange
splitting and spin-orbit coupling. Eu $5d$, Eu $6s$ and O $2p$
orbitals were taken into account. The $4f$ orbitals are very
localized, and it is thus reasonable to neglect their effect.
Only nearest neighbor Eu-Eu and Eu-O integrals were considered.
\begin{figure}[!htb]
\centerline{\includegraphics[angle=-90,width=15cm]{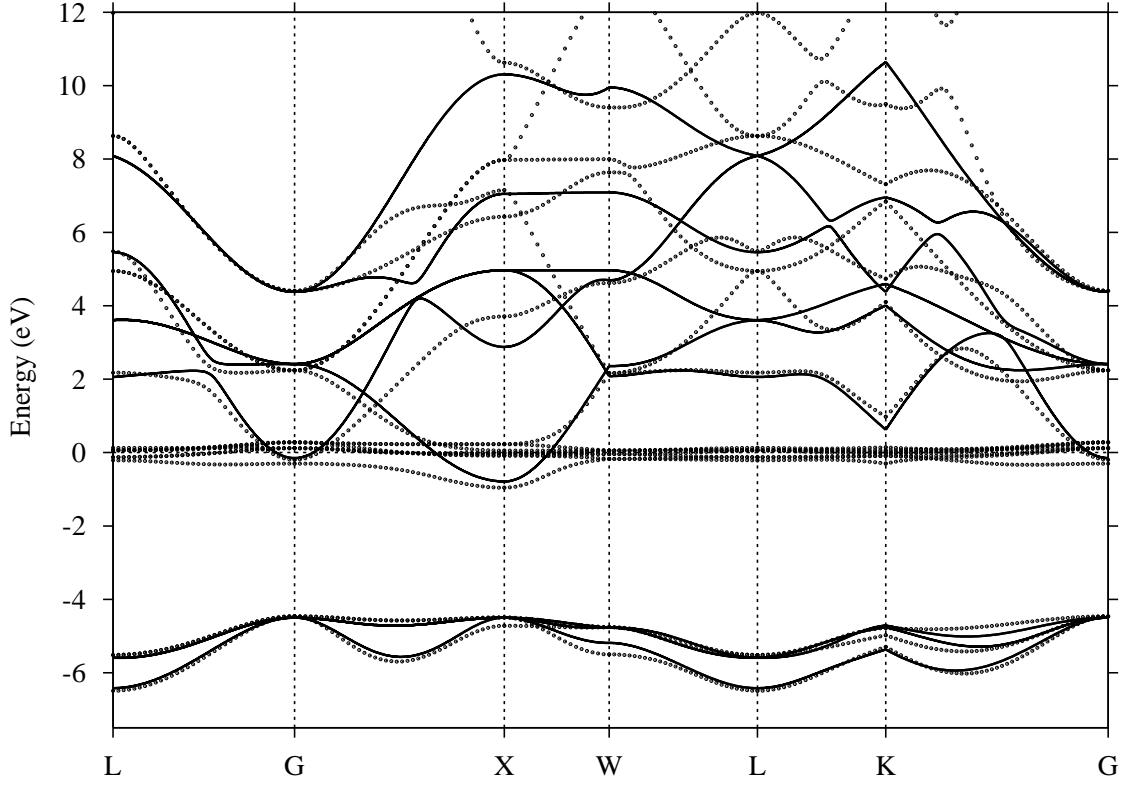}}
\caption{\label{bandstruct} Band structure calculation by LDA
(circles) and fitted band structure from LCAO (solid lines).}
\end{figure}
\begin{table}[!htb]
\centerline{
\begin{tabular}{|c|c|c|c|c|c|c|c|c|c|c|c|}
\hline $s_0$ & $p_0$ & $d_0$ & ss$\sigma$ & pp$\sigma$ &
dd$\sigma$ & dd$\sigma e_g$ & dd$\pi$ & sp$\sigma$ & pd$\sigma$ &
pd$\pi$ &
D$_q$ \\
\hline 5.25 & $-4.5$ & 4.31 & $-0.45$ & 0.2 & $-0.74$ & $-2.3$ &
0.4 & 0.2 & $-0.45$ & 0.2
& 0.39 \\
\hline
\end{tabular}}
\caption{\label{lcaopar} Fitted energies (in eV) of LCAO
two-center integrals for EuO.}
\end{table}
First we obtained the two-center integrals by fitting the LCAO
band structure without exchange and spin-orbit coupling to an LDA
($U=0$) calculation by Elfimov, which included also the Eu $4f$
and $6p$ and the O $2s$ and $3d$ orbitals. The resulting fit is
shown in figure \ref{bandstruct}. In the fit the integrals
sd$\sigma$, pp$\pi$ and dd$\delta$ were kept zero and the ratios
pd$\sigma$/pd$\pi$ and dd$\sigma$/dd$\pi$ were kept at
respectively $-2.2$ and $-1.85$ as given by Harrison
\cite{Harrison80}. The crystal field splitting 10D$_q$ was fitted
at 3.9 eV, which is somewhat larger than the value of 3.1 eV
found from optical measurements of EuO by G\"untherodt
\cite{Guntherodt74}. The values of the fit parameters are
tabulated in table \ref{lcaopar}. To get a reasonable dispersion
of the $e_g$ orbitals near the $\Gamma$-point (G), we had to
increase the value of dd$\sigma$ integral between the $e_g$
orbitals. Possibly this is due to the deformation of the $e_g$
orbitals by the oxygen charge which pushes the $e_g$ orbitals
away from the Eu-O axis resulting in a larger Eu-Eu overlap. Pen
\cite{Pen97} similarly found that the $e_g$ bands could not be
fitted well by a LCAO model in LiVO$_2$.

\noindent The spin-orbit coupling Hamiltonian which is required
to include the relativistic effects can also be written as:
\begin{equation}
\mathcal{H}_{so} =  \frac{\zeta}{2}(\mathbf{J}^2 - \mathbf{L}^2 -
\mathbf{S}^2)
\end{equation}
By transforming the d-orbitals to eigenstates of $\mathbf{J}^2$
and $J_z$ we find the following spin orbit and exchange
Hamiltonian $\mathcal{H}$ in terms of the basis $\psi_{5d}=\sum_i
a_i \phi_i$ used for the LCAO calculation:
\begin{equation}
\nonumber \scriptsize
H\psi_{5d}=\left(\begin{tabular*}{0.75\textwidth}
     {@{\extracolsep{\fill}}c@{}c@{}c@{}c@{}c@{}c@{}c@{}c@{}c@{}c@{}c@{}c}
 $-\Delta_{ex}$ &  0 &  0 &  $ -i\zeta$ &  0 &  0 & $i\zeta/2$ & $\zeta/2$ & 0 & 0 \\
 0 &  $-\Delta_{ex}$& $i\zeta/2$& 0 & 0 & $-i\zeta/2$ & 0 & 0 & $-\zeta/2$ & $\zeta\sqrt{3}/2$ \\
 0 & $-i\zeta/2$ & $-\Delta_{ex}$& 0 & 0 & $ -\zeta/2$ & 0 & 0 &$ i\zeta/2$ & $i\zeta\sqrt{3}/2$ \\
 $i\zeta$ & 0 & 0& $-\Delta_{ex}$& 0 & 0 & $\zeta/2$ &  $-i\zeta/2$ & 0 & 0 \\
 0& 0 & 0 & 0 & $-\Delta_{ex}$ & 0 & $-\zeta \sqrt{3}/2$ & $-i\zeta \sqrt{3}/2$ & 0 & 0 \\
 0 & $i\zeta/2$ & $-\zeta/2$ & 0 & 0 & $\Delta_{ex}$& 0 & 0 & $i\zeta$ & 0 \\
 $-i\zeta/2$ & 0 & 0 & $\zeta/2 $&$ -\zeta\sqrt{3}/2$ & 0 & $\Delta_{ex}$ & $-i\zeta/2$ & 0 & 0 \\
 $\zeta/2$ & 0 & 0 & $i\zeta/2$ & $i\zeta\sqrt{3}/2$ & 0 & $i\zeta/2$& $\Delta_{ex}$ & 0 & 0 \\
 0 &$ -\zeta/2 $& $-i\zeta/2$ & 0 & 0 & $-i\zeta$ & 0 & 0 & $\Delta_{ex}$ & 0 \\
 0 & $\zeta \sqrt{3}/2$ & $-i\zeta \sqrt{3}/2$ & 0 & 0 & 0 & 0 & 0 & 0 & $\Delta_{ex}$ \normalsize \\
\end{tabular*} \right) \left(
\begin{tabular}{@{}c@{}}
$a_{xy\uparrow}$ \\
$a_{xz\uparrow}$ \\
$a_{yz\uparrow}$ \\
$a_{x^2-y^2\uparrow}$ \\
$a_{3z^2-r^2\uparrow}$ \\
$a_{xy\downarrow}$ \\
$a_{xz\downarrow}$ \\
$a_{yz\downarrow}$ \\
$a_{x^2-y^2\downarrow}$ \\
$a_{3z^2-r^2\downarrow}$ \\
\end{tabular}\right)
\end{equation}
\begin{figure}[!htb]
\centerline{\includegraphics[width=11.25cm]{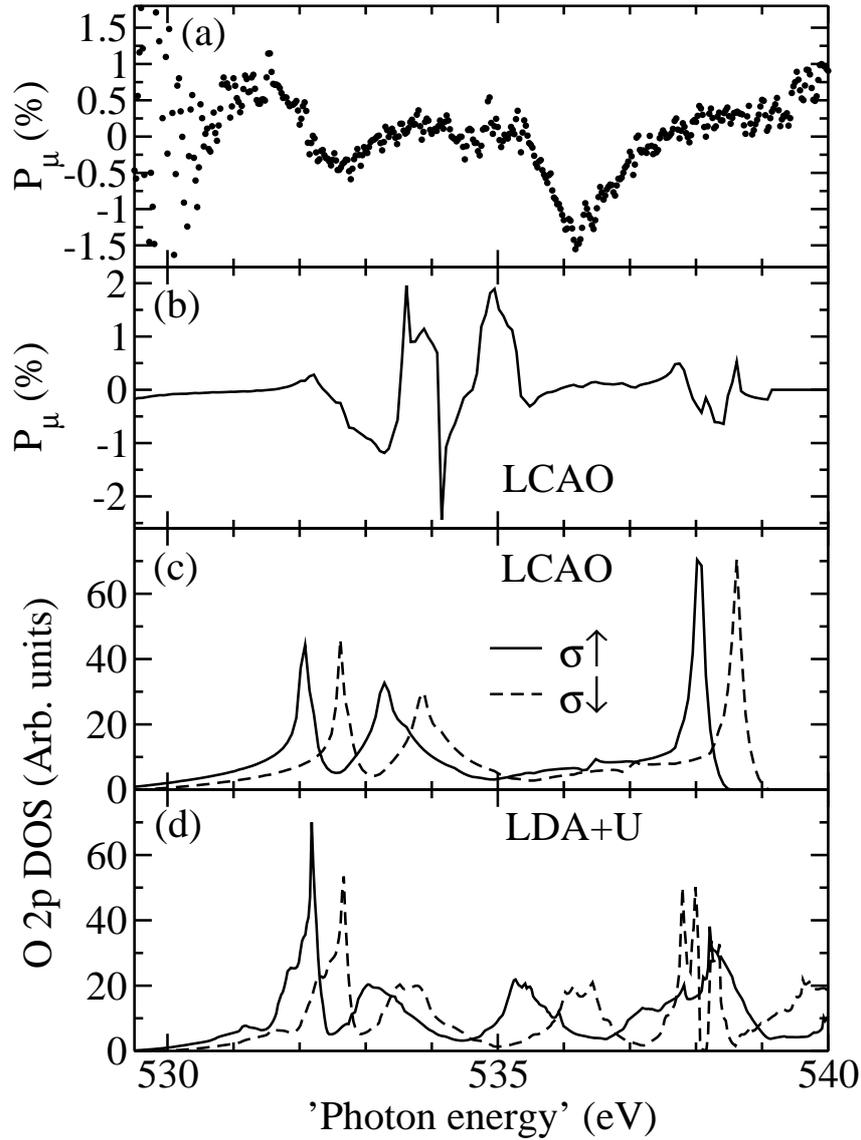}}
\caption{\label{mcdcalc} a) Normalized MCD at the O $K$ edge of
EuO. b) LCAO calculation of MCD at $K$ edge using $\zeta_{5d}$=15
meV and $\Delta_{ex}$ = 0.3 eV. c) and d) LCAO and LDA+U
calculation of the spin-polarized O $2p$ density of states. The
LCAO spectra were shifted by 530.25 eV with respect to figure
\ref{bandstruct} to align the low energy peaks in the DOS with
those in the LDA+U spectrum (which was aligned to the XAS spectra
using a comparison with XPS \cite{Steeneken02}).}
\end{figure}
When this Hamiltonian is added to the LCAO program, the spin-up
and spin-down oxygen states are mixed via the O-Eu hybridization
and the spin-orbit coupling of the Eu $5d$. This will induce an
effective spin-orbit coupling on the oxygen atoms and will
therefore lead to a $K$ edge oxygen MCD even though the atomic
spin-orbit coupling on oxygen is still assumed to be zero. By
projecting the LCAO eigenvectors $\psi_n(\mathbf{k,r})$ on the
localized atomic wavefunctions $\phi_{m_l,m_s}(\mathbf{r})$ we
obtain the partial oxygen density of states for $\Delta m_l = \pm
1$. The energy dependent oxygen $K$ MCD can thus be obtained by
summing the following expression over the Brillouin zone:
\begin{equation}
\mu_\pm (E) \propto \sum_{n,E=E_k} (<\psi_n(\mathbf{k})|\phi_{\pm
1,\uparrow}>^2+<\psi_n(\mathbf{k})|\phi_{\pm 1,\downarrow}>^2)
\end{equation}

Thus we calculated the normalized MCD $P_\mu$ using
$\zeta_{5d}$=15 meV and $\Delta_{ex}$ = 0.3~eV. The results are
shown in figure \ref{mcdcalc}. Although there is some qualitative
correspondence between the measured and calculated energy
dependent MCD in figures \ref{mcdcalc}(a) and (b), it is not very
convincing. The calculated O $2p$ density of states from LCAO in
figure \ref{mcdcalc}(c) follows the LDA+U calculation in figure
\ref{mcdcalc}(d) quite well up to 535 eV, however the
correspondence above this energy is worse and the calculated MCD
above 535 eV should therefore not be relied on. Although the
agreement between calculated and measured MCD is weak, we can
draw the following conclusions from this calculation:
\begin{itemize}
\item If the spin-orbit coupling is induced via hybridization with neighbors
the energy dependence of the $K$ edge MCD does not necessarily
follow the spin-polarization of the conduction band as one might
conclude from sections \ref{atomoxmcd} and \ref{Fano}.
\item A small spin-orbit coupling constant of only 15 meV in combination with
strong hybridization and exchange splitting can induce a large
effective spin-orbit coupling and $K$ edge MCD on neighboring
atoms.
\end{itemize}
The LCAO calculations in this section do not take into account
the momentum dependence of the exchange splitting and spin-orbit
coupling. This might explain why the agreement between the
measured and calculated spectra is not optimal. More
sophisticated LMTO methods including relativistic effects
\cite{Anderson75,Kunes01} could probably improve this.

\section{Conclusions}

In addition to the Fano effect, which is the result of a
modification of the radial wavefunction by the spin-orbit
coupling and was previously held responsible for the $K$ edge MCD
in iron \cite{Schutz87,Stahler93}, we have shown two alternative
mechanisms which might lead to $K$ edge MCD. In the first
mechanism the combination of exchange interactions and spin-orbit
coupling leads to a mixing of spin-states, which can result in a
$1s \rightarrow 2p$ MCD of $\sim \zeta^2 P_{spin}/8
\Delta_{ex}^2$. A more intricate effect is the induction of MCD
by spin-orbit coupling on a neighbor atom, which can lead to an
appreciable MCD effect with an energy dependence which does not
follow the spin-polarization of the conduction band. These
different mechanisms were compared to the remarkably high MCD
signal observed at the O $K$ edge of EuO. The MCD spectrum seemed
to follow the spin-polarization to a reasonable extent. However,
the oxygen $\zeta_{2p}$ is probably too small to fully account
for the observed MCD intensity. Therefore a significant
contribution as a result of hybridization with Eu might also be
present. For a complete understanding of MCD at $K$ edges of
magnetic compounds all three effects should generally be
considered.



%
\markboth{}{}
\chapter{Transport properties of EuO thin films}
\label{Magres}


{\it We present magnetoresistance and photoconductivity
measurements on thin films of Eu-rich EuO. The data are compared
to a model in which changes in the free carrier concentration are
due to overlap and thermal excitations between vacancy levels and
the Eu $5d$ conduction band. We show that the exchange splitting
of the conduction band can account for the large increase in the
free carrier concentration and corresponding insulator-metal
transition that is observed upon entering the ferromagnetic
state. Interestingly we find that the magnetoresistance shows a
universal scaling with magnetization $\rho(B,T) \propto
e^{-bM(B,T)/M_{sat}}$ in excellent agreement with the model.
Similar behavior in La$_{0.7}$Ca$_{0.3}$MnO$_3$ and EuB$_6$ seems
to indicate that such scaling is an intrinsic property of
colossal magnetoresistance (CMR) materials. Moreover a
metal-insulator transition is observed in the photoconductivity
of EuO. It is argued that this transition in the
photoconductivity can be understood by electron-hole separation
of excitonic electron-hole pairs in the ferromagnetic state.}

\section{Introduction}
Soon after the discovery of ferromagnetism in the europium
chalcenogides \cite{Matthias}, it was found that these compounds
exhibited spectacular transport properties with huge
metal-insulator transitions (MIT) and colossal magnetoresistance
(CMR) effects around the Curie temperature
\cite{Suits63,Heikes64,vonMolnar67}. The explanation of these
effects is of much interest as there appears to be a large class
of ferromagnetic materials with similar transport properties
among which are the manganites \cite{vanSanten50},
(Hg,Cd)Cr$_2$Se$_4$ \cite{Lehmann67}, EuB$_6$ \cite{Guy80} and
diluted ferromagnetic semiconductors like Ga$_{1-x}$Mn$_x$As
\cite{Ohno98}. The europium chalcenogides are model compounds for
the study of colossal magnetoresistance behavior, as their
structure is very simple and because there is a considerable
doping range in which their magnetic properties do not depend
strongly on doping. Besides this, they show the largest CMR
effects of all these compounds. Despite these favorable
conditions the transport properties of the doped Eu chalcenogides
have been very difficult to understand and a large number of
mechanisms has been proposed to explain these effects.

Early studies \cite{Molnar70,Methfessel68} have attributed the
resistivity anomaly near T$_c$ to scattering due to spin-fluctuations.
Magnetic scattering was predicted to result in a maximum resistivity
near the Curie temperature \cite{deGennes58,Fisher68,Haas68} which was
indeed observed in several rare-earth metals at their magnetic
ordering temperature \cite{Legvold53,Curry60,Colvin60}. Optical
absorption measurements on EuO indicated that the scattering rate
indeed changes by about 1 order of magnitude across the Curie
temperature \cite{Oliver72}. Hall effect measurements \cite{Shapira}
show a similar change in the mobility, although interpretations are
complicated by the anomalous Hall effect. These results indicate that
a reduction of the electron mobility due to magnetic scattering indeed
provides a considerable contribution to the metal-insulator transition
in EuO. However the scattering rate is not expected to depend strongly
on doping concentration and is thus not expected to be much larger in
EuO than in Gd metal. Therefore, it seems unlikely that magnetic
scattering is responsible for the size of the metal-insulator
transition which can span as many as 13 orders of magnitude
\cite{Penney72} (see figure \ref{trans:d}).

The metal-insulator transition is not observed in stoichiometric
EuO but only in Eu-rich EuO, in which oxygen vacancies provide
electron doping \cite{Oliver72,Shafer72}. This strong doping
dependence of the MIT soon resulted in a general acceptance of
the notion that the metal-insulator transition is mainly due to
the delocalization of trapped electrons at oxygen vacancies as a
result of changes in the magnetic structure. This notion was
supported by the strong increase in the free carrier density in
the ferromagnetic state found by optical \cite{Oliver72}, Hall
effect \cite{Shapira} and Seebeck effect \cite{Samokhvalov88}
measurements. Several mechanisms have been proposed to explain
this delocalization. Some authors have attributed it to the
exchange splitting of the conduction band edge (which was
observed in optics \cite{Busch,Freiser} and x-ray absorption
\cite{Steeneken02}) with respect to a fixed vacancy level in the
gap \cite{Oliver70,Oliver72,Laks76}. They argued that when the
conduction band crosses the vacancy level, the vacancy electrons
will delocalize in the conduction band. As the oxygen vacancies
can bind two electrons, this model is usually called the
He-model.

Most authors have related the metal-insulator transition to the
properties of bound magnetic polarons (BMP)
\cite{Kasuya68,Molnar70b,Petrich71,Torrance72,Leroux72,Kubler75,Mott79,Mauger83,Mauger86,Nagaev99}.
A BMP consist of an oxygen vacancy or doped impurity ion which
binds electrons by its Coulomb potential. These electrons will
increase the ferromagnetic coupling of the surrounding Eu moments
via indirect exchange. The resulting increased local
magnetization will enhance the binding energy of the electron to
the vacancy site. The metal-insulator transition was attributed
to the occurrence of a Mott \cite{Leroux72,Kubler75,Nagaev99} or
Anderson \cite{Kasuya68,Mott79,Mauger86} transition of the
randomly distributed bound magnetic polarons upon magnetic
ordering.

%
%
As the attempts to quantitatively compare these models to
transport experiments have been rather limited
\cite{Oliver72,Laks76,Mauger83}, it is still unresolved which of
the proposed mechanisms describes the properties of doped Eu
chalcogenides most accurately. In this paper we will consider the
BMP and He-model models, and investigate the effects of doping on
these models. We will try to make a quantitative estimate of the
resistivity and compare it to resistance and magnetoresistance
measurements on Eu-rich EuO films. We discuss similarities with
other CMR compounds. Finally the photoconductivity of the films
is discussed.

\section{Theory}
\label{model}
\subsection{The vacancy levels: BMP and He-model}
\label{vaclev} As Eu atoms are too large to be incorporated in
the lattice as interstitials, it is generally believed that
Eu-rich EuO contains oxygen vacancies. Because the
characteristics of the metal-insulator transition depend strongly
on the concentration of oxygen vacancies, it is very likely that
it originates from the delocalization of electrons captured by
the positive effective charge at the vacancy sites. Infrared,
Hall and Seebeck measurements
\cite{Oliver72,Shapira,Samokhvalov88} indeed indicated a large
increase in the carrier density in the metallic state. Before
considering the delocalization of electrons from the vacancy
sites in the ferromagnetic phase, it is important to know in what
electronic state they are in the paramagnetic phase. Oliver {\it
et al.} \cite{Oliver72} proposed that the two vacancy electrons
form a singlet which derives its binding energy mainly from the
Coulomb attraction ($\sim 0.3$ eV) of the vacancy. Mauger
\cite{Mauger83,Mauger86b} proposed a completely different
approach, in which the vacancy electrons are mainly bound by the
local magnetization which is induced as a result of their
exchange interactions ($\Delta_{ex} \sim$ 0.3 eV) with the
surrounding Eu moments. The electrons are thus bound in a BMP up
to temperatures which can exceed room temperature. The Coulomb
attraction in this model is assumed to be very small ($\sim 17$
meV). Moreover the spin-polarization inside the polaron is
assumed to remain 100\%, a point that is certainly questionable
\cite{Dietl86}. Although not much experimental evidence exists
for polarons which are stable up to room temperature, the polaron
could theoretically be stable as long as the contribution of the
exchange interaction ($\sim 0.3$ eV) to the free energy
 exceeds that of the magnetic entropy of the Eu local moments in the
 polaron. For a polaron consisting of an octahedron of 6 Eu moments neighboring
 an oxygen vacancy the magnetic
contribution to the free energy $F_M$ can be estimated as
\begin{equation}
F_M = -6 k_B T \ln \left( \sum_{J_z=-J}^J e^{-g \mu_B H J_z/k_B
T}\right)
\end{equation}
Therefore in the paramagnetic state, in the absence of a magnetic
field, the effective stabilization energy of a polaron is
approximately given by:
\begin{equation}
\label{bmpbind} E_d \approx (\Delta_{\rm ex,pol}-\Delta_{\rm
ex,LR}(T))-6 k_B T \ln 8
\end{equation}
Where $\Delta_{\rm ex,pol}$ and $\Delta_{\rm ex,LR}(T)$ are the
exchange splitting in the polaron and the long-range exchange
splitting. Thus in this simple approximation the polaron will
remain magnetized up to a temperature of T$_p\approx \Delta_{\rm
ex}/(6 k_B \ln 8) \approx 280$ K.

Several experimental efforts have been made to distinguish
between the BMP and He-model. As was argued by Torrance {\it et
al.} \cite{Torrance72}, the fact that the Curie temperature of
doped EuO can be higher than that of undoped EuO seems to favor a
BMP model. Because if the vacancy electrons are in a triplet
configuration, their energy is lowered by the polarization of the
cluster of surrounding Eu$^{2+}$ spins. The formation of such
ferromagnetic clusters would result in an increased Curie
temperature. On the contrary, if the electrons at the vacancy
site are in a singlet state, like in the He-model, the
ferromagnetic arrangement of the surrounding $4f$ moments is not
favored by their presence.


Instead one can however also argue that the increased Curie
temperature in EuO is solely caused by increased exchange via
\emph{delocalized} electrons. In fact, the Curie temperature is
only significantly higher in samples which have a relatively high
conductivity and thus a large concentration of free electrons.
Thus the He-model can not be excluded by Torrance's argument.

In essence, electron spin or paramagnetic resonance (ESR or EPR)
measurements are sensitive for the difference between a
diamagnetic or paramagnetic configuration of the vacancy
electrons. However the high density of paramagnetic Eu $4f^7$
local moments and the large exchange interaction between the
doped electrons and these moments, precludes the observation of
resonances related to the vacancy or conduction electrons
\cite{Auslender86,Massenet74}. Instead only an EPR of the
Eu$^{2+}$ ions is observed and the effect of doping is to change
the $g$-factor (only 1 or 2\% \cite{Shamokhvalov74}) and line
width (almost 50\% \cite{Massenet74}) of the resonance as a
result of an increased exchange coupling. Such exchange narrowing
effects \cite{Vanvleck48,Anderson53} were indeed observed
\cite{Massenet74,Samokhvalov88,Auslender86}, and do not allow to
distinguish between BMP or He-model.

Recently, Raman measurements \cite{Nyhus97,Snow01,Rho02} were
interpreted as evidence for the existence of magnetic polarons.
However, as these polarons also seem to exist in undoped EuO
\cite{Snow01}, they apparently do not contain doped electrons and
might thus be unrelated to the metal-insulator transition.

\subsection{Effect of magnetization}

Oliver {\it et al.} first realized \cite{Oliver70} the influence
of the exchange splitting of the conduction band edge on the
transport properties of EuO. First a model was proposed in which
the vacancy level was fixed, later \cite{Oliver72} it was
correctly realized that the electrons in the vacancy level could
also be exchange split by a certain energy. Recently, LDA+U
calculations \cite{Elfimov03} of EuO containing oxygen vacancies
have confirmed that the impurity level associated with an oxygen
vacancy in EuO is indeed exchange split. The total exchange
splitting of the vacancy level was found to be comparable to that
of the conduction band edge in these calculations. Temperature
dependent XAS measurements have suggested (see chapter
\ref{Tempxas}) that the conduction band exchange splitting
follows the long-range order and starts at the Curie temperature
T$_c$. The onset of the vacancy level exchange splitting might
occur at a higher temperature T$_p$, as the ferromagnetic
coupling around the vacancy is enhanced by the indirect exchange
via the vacancy electrons, thus forming a bound magnetic polaron.
Mauger suggested that this effect is so large that T$_{p}$ is
larger than room temperature and the reduction of the electron
binding energy in the ferromagnetic state is mainly due to the
reduced difference between the polaron and long-range exchange
splitting $\Delta_{\rm ex,pol}$ and $\Delta_{\rm ex,LR}(T)$, like
in equation (\ref{bmpbind}).

\begin{figure}[!htb]
   \vspace{1cm}
   \centering
   \begin{minipage}[b]{0.5\textwidth}
      \centering
      \subfigure[One-electron energy levels in the conduction band and vacancy/impurity level.]{
         \label{IMTmech:a}
         \includegraphics[width=7.5cm]{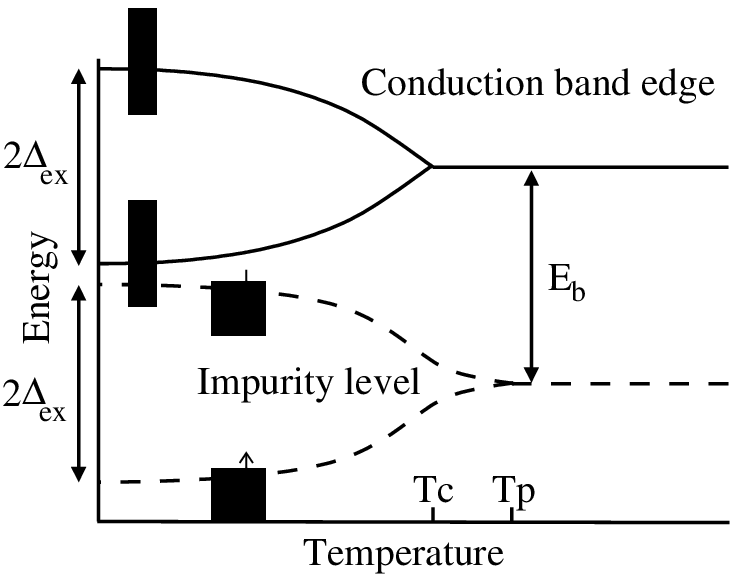}}
   \end{minipage}%
   \begin{minipage}[b]{0.5\textwidth}
      \centering
      \subfigure[Two-electron total energy levels, circled electrons are in the vacancy state.]{
         \label{IMTmech:b}
         \includegraphics[width=7.5cm]{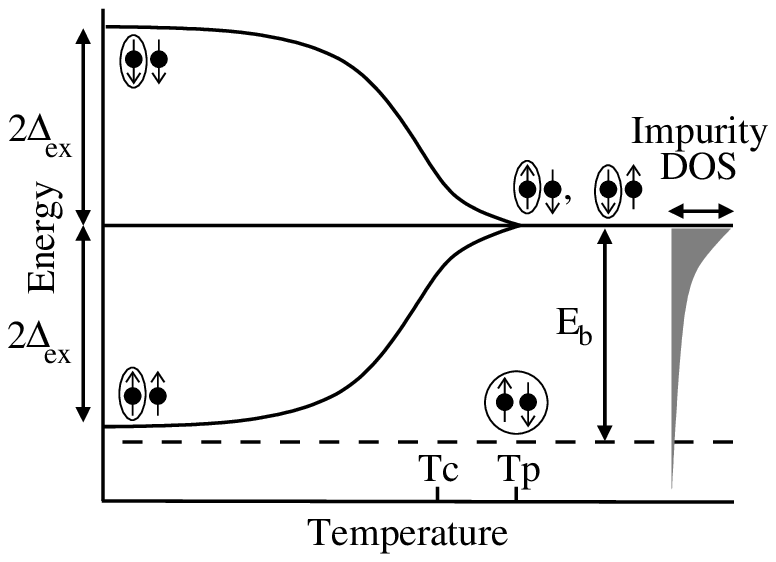}}
   \end{minipage}%
   \\
   \begin{minipage}[t]{\textwidth}
      \centering
      \subfigure[Calculated variation of the ionization energy $E_d$ of the BMP in EuO.
      From \cite{Mauger83}.]{
         \label{IMTmech:c}
      \includegraphics[width=9cm]{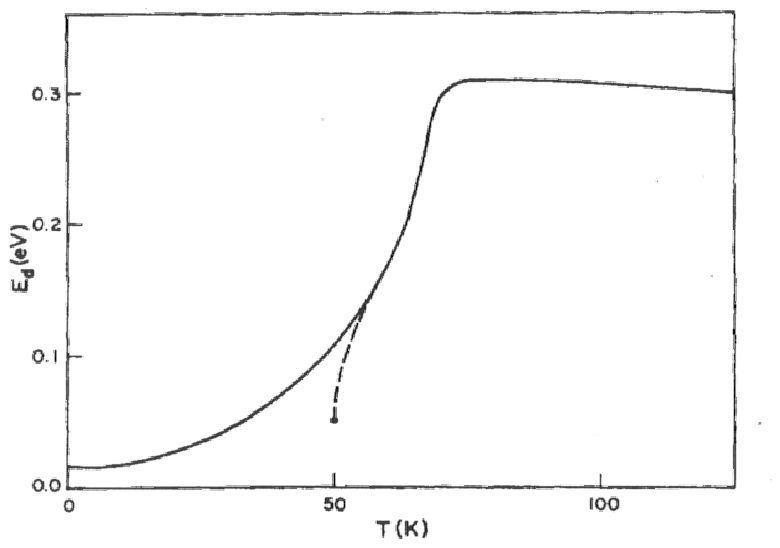}
      }
   \end{minipage}
   \label{IMTmech}
   \caption{Temperature dependence of the total energy in the He-model for one-electron (a) and
      two-electron (b) configurations, compared to (c) calculations of Mauger for the ionization energy
      of the BMP in EuO.}
\end{figure}
In figure \ref{IMTmech:a} the temperature dependence of the one
electron energy levels in the He-model is schematically depicted.
Shown are the conduction band edge and the impurity level, both
of which are exchange split $\Delta_{\rm ex}$. When the
magnetization increases, the spin-down vacancy level approaches
the spin-up conduction band edge and the binding energy $E_b$ of
the spin-down vacancy electron decreases\footnote{The
magnetization direction is parallel to the spin-up spins.}. This
exchange splitting of the vacancy results in an approach rate
which is twice as high as in the model for which the vacancy
level is fixed \cite{Oliver70}. In figure \ref{IMTmech:b} we show
the variation of the total energy levels for configurations with
two electrons in the vacancy level (dashed line) and with one
electron ionized in the conduction band (solid lines). Coulomb
interactions between the two electrons in the vacancy are
neglected. Clearly, the ionization energy of the vacancy singlet
strongly reduces upon magnetization. In figure \ref{IMTmech:c}
the polaron binding energy as calculated by Mauger is shown which
follows approximately equation (\ref{bmpbind}) with $\Delta_{\rm
ex,LR} (T)$ proportional to the magnetization.

Despite the different assumptions of the models of Oliver and
Mauger, reasonable agreement with experimental transport
measurements was obtained by both
\cite{Oliver72,Laks76,Mauger83}. This is largely due to the fact
that the resulting temperature dependent binding energy of the
electrons is similar as can be seen by comparing figures
\ref{IMTmech:b} and \ref{IMTmech:c}. It should be noted that the
He-model can only lead to a metal-insulator transition if both
the spin-up and the spin-down band contain an electron. This is
only the case if the impurity potential is large enough to bind
two electrons\footnote{The He-model can thus not account for the
metal-insulator transition in Gd$^{3+}$ or photo-doped materials
if the exchange splitting at the impurity site is equal to the
long range splitting. If the exchange splitting at the impurity
site is smaller than the long range splitting, the He-model might
however still be valid as the binding energy of the vacancy
electrons would still be reduced upon magnetization (see
\cite{Oliver72}).}, moreover these electrons have to be in a
singlet state, which will only be the case if the ground-state is
non-degenerate. In fact, if the vacancy ground state is indeed
non-degenerate, the BMP model of Mauger, with a low Coulomb
potential at the vacancy site, cannot account for the
metal-insulator transition, as the formation of the BMP will then
result in one exchange-bound electron, but will repel the
electron with opposite spin, thus leading to a delocalization of
half of the donor electrons when the BMP is formed, which can be
$\sim 280$ K according to equation (\ref{bmpbind}). If the ground
state, on the other hand, is degenerate, both electrons will
occupy a spin-up state, as this will reduce their exchange energy
with the $4f$ electrons and with each other.

\subsection{Lowest energy levels of an oxygen vacancy}
Thus to check wether the He-model or the BMP model is most
applicable to the metal-insulator transition in Eu-rich EuO, one
should know what are the lowest energy levels in an oxygen
vacancy as this will determine whether the two electrons in the
vacancy level are in a singlet or triplet state. In this section
we will try to estimate the lowest energy levels by a cluster
calculation of the oxygen vacancy states. For this purpose, we
assumed that the wave function of the vacancy electron is well
described by a superposition of $d$-orbitals on the octahedron of
six neighboring Eu atoms at sites $\pm x$,$y$,$z$. The LCAO
Hamiltonian \cite{Slater54} of this cluster was then constructed
from the nearest neighbor two-center integrals between the $d$
orbitals\footnote{See Harrison \cite{Harrison80} figure 19-4 for
a diagram of the dd$\sigma$, dd$\pi$ and dd$\delta$ two-center
integrals, and \cite{Harrison80} table 20-1 or \cite{Slater54}
table I for their relation to the interatomic energy matrix
elements.}, including the Hamiltonian $\mathcal{H}_{cf}$ of the
D$_{4h}$ crystal field\footnote{The O$_h$ symmetry is broken by
the missing oxygen atom.}. Next to the oxygen vacancy, the
crystal field for the $d$-orbitals on the Eu atom at $-z$ is
given by:
\begin{eqnarray}
\mathcal{H}_{cf,xy}&=&-4D_q - D_t + 2D_s \nonumber \\
\mathcal{H}_{cf,xz}=\mathcal{H}_{cf,yz}&=&-4D_q + 4D_t - D_s \nonumber\\
\label{crystf} \mathcal{H}_{cf,x^2-y^2}&=&6D_q - D_t + 2D_s\\
\mathcal{H}_{cf,3z^2-r^2}&=&6D_q - 6D_t - 2D_s \nonumber
\end{eqnarray}
It can be shown by a derivation similar to that given in
\cite{Sugano70} and using equations for $<r^2>/<r^4>$ from
\cite{Cowan81}, that in the case of a missing point charge:
$D_t=\frac{2}{7}D_q$ and $D_s=\frac{6}{7} a^2
\frac{<r^2>_{5d}}{<r^4>_{5d}} D_q \approx \frac{4 a^2
Z_{eff}^2}{8575 a_0^2} D_q \approx 0.1 D_q$, where $a$ is the
Eu-O distance, $a_0$ the Bohr radius and $Z_{eff}$ is the
effective charge of the Eu$^{2+}$. Using the ionization energy of
the $5d$ electron of lutetium we estimated $Z_{eff}\approx 3$ and
assumed hydrogenic $5d$ wavefunctions.

The 30$\times 30$ Hamiltonian was diagonalized, with $dd\delta =
0$\footnote{As suggested by Harrison \cite{Harrison80}.}. Two
different ground states were found for $dd\sigma \neq 0$ and
$dd\pi \neq 0$:
\begin{eqnarray}
dd\sigma \neq 0 &:& \psi_{GS,dd\sigma}=
\frac{1}{2}(\phi_{xy,+x}+\phi_{xy,-x}+\phi_{xy,+y}+\phi_{xy,-y})
\nonumber \\
 && E_{GS,dd\sigma}=\frac{3}{2}dd\sigma \mbox{    (threefold degenerate)} \\
dd\pi \neq 0 &:& \psi_{GS,dd\pi} = \frac{1}{\sqrt{18}}
\sum_{i=x,y,z}
\left(\phi_{xy,+i}-\phi_{xy,-i}+\phi_{xz,+i}-\phi_{xz,-i}+\phi_{yz,+i}-\phi_{yz,-i}\right)
\nonumber \nonumber\\
 & & E_{GS,dd\pi}=-4 dd\pi \mbox{ (non-degenerate)} \nonumber
\end{eqnarray}

Here, the basis functions $\phi$ are designated by the type of
$d$-wavefunction and the site at which they are located, e.g.
$\phi_{xy,+x}$ is an $xy$ orbital at the Eu atom next to the
vacancy in the $+x$ direction. When both the dd$\sigma$ and
dd$\pi$ two-center integrals are non-zero the wavefunctions mix,
however the lowest energy states still have mainly
$\psi_{GS,dd\sigma}$ and $\psi_{GS,dd\pi}$ character with the
same degeneracy as before. Therefore we will keep designating
them by $\psi_{GS,dd\sigma}$ and $\psi_{GS,dd\pi}$. The condition
that determines which is the lowest state is still very well
approximated by $E_{GS,dd\sigma}-E_{GS,dd\pi}=\frac{3}{2}
dd\sigma + 4 dd\pi = 0$. As Harrison \cite{Harrison80} gives
$dd\sigma/dd\pi=-1.85$ this strongly indicates that the lowest
energy level is non-degenerate as the calculation shows that a
degenerate ground state only exists for $dd\sigma/dd\pi <
-\frac{8}{3}$. If we calculate the ground state energies with
respect to the center of the band $d_0$ including crystal field
and using the parameters of table \ref{lcaopar} and equation
(\ref{crystf}), we find a splitting between degenerate and
non-degenerate ground states of 0.37 eV with
$E_{GS,dd\sigma}=-2.59 $ eV and $E_{GS,dd\pi}=-2.96$ eV.
Increasing $D_s$ reduces the splitting, e.g. setting
$D_s=D_q=0.39$ eV, we find $E_{GS,dd\sigma}=-2.78$ eV and
$E_{GS,dd\pi}=-3.0$ eV, if we instead set $D_t=D_q$ the splitting
increases, $E_{GS,dd\sigma}=-2.18$ eV and $E_{GS,dd\pi}=-2.88$
eV. In this treatment, we have not taken into account that the
orbitals will be deformed by the attractive potential of the
vacancy. This might significantly increase the two-center
integrals and therefore the splitting between the lowest energy
states. If we take the normal crystal field and for example
double the two center-integrals to $dd\sigma=1.48$~eV and
$dd\pi=0.80$ eV, we find a splitting of 0.68 eV with
$E_{GS,dd\sigma}=-4.60$ eV and $E_{GS,dd\pi}=-3.92$ eV.

This analysis indicates that the ground state of the oxygen
vacancy is probably a singlet and that the energy splitting with
the first excited state seems to be larger than $\Delta_{ex}$,
and is possibly even larger than $2\Delta_{ex}$, thus supporting
the He-model. Moreover, the lowest energy conduction band states
at the $\Gamma$ point in the EuO band structure have strong Eu
$6s$ character (see figure \ref{bandstruct}), which will could
also stabilize a singlet state. Recent LDA+U calculations
\cite{Elfimov02} of CaO with oxygen vacancies have found that the
lowest energy vacancy orbital had a non-degenerate $s$-like
character which was 0.7 eV below the first degenerate state,
presenting additional support for the singlet picture. However,
up to now we have neglected the Coulomb and Hund's rule exchange
interactions between the two electrons in the vacancy, which
would favor the triplet state. These electron-electron
interactions might thus also result in a triplet ground state,
although an exact solution of this problem is difficult.
Therefore, as long as no more accurate experimental or
theoretical estimates for the singlet-triplet splitting are
found, the issue of the lowest energy ground state in the oxygen
vacancy will remain unsolved.


In the rest of this section we choose to discuss the
metal-insulator transition in EuO along the lines of the
He-model. We want to emphasize that this choice is not based on
our conviction that this is the correct model, as in our opinion
there is still too much uncertainty as to the origin of the
metal-insulator transition to make such a decision. However,
Mauger \cite{Mauger83} has shown that reasonable agreement with
experiment can be obtained from the BMP model. We will take a
different starting point and also come to close agreement with
experiment. We note that, our derivation would be similarly
applicable to the case of the BMP, as the temperature dependent
binding energy of the vacancy electrons is almost the same for
both models (compare figures \ref{IMTmech:b} and
\ref{IMTmech:c}). The analysis could therefore be more generally
applicable to any model in which the binding energy of vacancies
varies (approximately) proportional to the magnetization,
although the numerical estimates will be different.

\subsection{The binding energy of vacancy levels}
\label{bindvac} The high temperature conductivity of Eu-rich EuO
shows an activated behavior $\sigma \propto e^{-E_{ac}/k_BT}$
with $E_{ac}\sim 0.3$ eV \cite{Torrance72}. If the vacancy
electrons have a binding energy of $E_b$ with respect to the
bottom ($\varepsilon=0$) of a parabolic conduction band with a
density of states \cite{Ashcroft76}
$g(\varepsilon)=(m^{\frac{3}{2}}/\hbar^3\pi^2)\sqrt{2\varepsilon}$,
the density of conduction band electrons $n_c$ is obtained from
the equilibrium rate equation:
\begin{equation}
\label{rate} \frac{{\rm d}n_c}{{\rm d}t}=-\beta n_c(N-n_i) +
\beta n_i \int_0^\infty g(\varepsilon)
e^{-(\varepsilon+E_b)/k_bT} {\rm d}\varepsilon = 0
\end{equation}
Here $\beta$ is the transition probability, $m$ the effective
electron mass, $N$ the density of vacancy states (number of
vacancies is $N/2$ as each vacancy can contain two electrons) and
$n_i$ the density of electrons occupying the vacancy states. In a
EuO crystal which only contains oxygen vacancies and no other
impurities or imperfections, charge neutrality imposes that the
total density of electrons $n = n_c + n_i = N$. If only a small
fraction of vacancy electrons is thermally excited to the
conduction band ($n_c \ll n$) we find, similar to an intrinsic
semiconductor \cite{Ashcroft76}, that $n_c = \sqrt{n N_c(T)}
e^{-E_b/2 k_BT}$, with $N_c(T) = \frac{1}{4}(\frac{2mk_BT}{\pi
\hbar^2})^{3/2}$, i.e. this corresponds to an effective
activation energy of E$_b$/2. The situation changes when a
compensating density $N_A$ of acceptor levels is present in the
crystal for example as Eu$^{3+}$ ions. In that case $N_A$
electrons will be bound to these sites and $n = n_c + n_i = N -
N_A$. If $N_A \gg n_c$ we find from equation (\ref{rate}) $n_c =
(n N_c(T)/N_A) e^{-E_b/k_BT}$, i.e. an activation energy of
$E_b$.

As Eu-rich EuO samples show an activated behavior of ~0.3 eV many
authors have assumed that the binding energy of the electrons at
the vacancy sites is about 0.3 eV in the paramagnetic phase
\cite{Petrich71,Laks76,Oliver72,Torrance72,Penney72,Godart80}.
The derivation above shows that an activated behavior of 0.3 eV
might instead very well correspond to $E_b$ = 0.6~eV. In the
presence of increasing concentrations of acceptor sites a
transition should occur towards an activation energy of 0.6 eV.
Such behavior was indeed observed \cite{Torrance72} in oxygen
rich EuO samples which possibly still contain oxygen vacancies.
In fact a range of activation energies from ~0.6 to ~0.3 eV was
found in doping ranges between oxygen-rich and europium-rich
samples \cite{Shafer72,Godart80}.

Experimental signs related to the electronic state of the vacancy
electrons also come from the infrared absorption spectrum of EuO.
Shafer {\it et al.} first observed two peaks at ~0.58 eV and ~0.66 eV
in the optical absorption of Eu-rich EuO \cite{Shafer72}. It was found
that the intensity and energy of these peaks is almost temperature
independent \cite{Helten77}. The peaks have been attributed to
spin-flip transitions of electrons in magnetic polarons between
exchange split states
\cite{Torrance72b,Lascaray76,Escorne79,Coutinho84}. As the peaks are
still present at room temperature this would implicate that the
vacancy exchange splitting at room temperature is still similar to
that that in the ferromagnetic state. Moreover such spin-flip
transitions are only very weakly allowed as spin-orbit interactions
are probably quite small. Helten {\it et al.} on the other hand held
transitions from the $4f$ valence band to the vacancy states
responsible for the peaks \cite{Helten77}. However in the slightly
Eu-rich EuO samples which show the peaks, the vacancy states are
almost fully occupied except for a small fraction of thermally excited
electrons. The number of empty states to excite valence band electrons
to is thus very small. In our view the possibility (which seems not to
have been considered by other authors) that the peaks correspond to
transitions from the vacancy states to the conduction band cannot be
excluded. This would suggest that the vacancy electron binding energy
is $\sim 0.6$ eV. It should be noted that, to our knowledge, the
infrared absorption peaks have not been observed in Gd doped EuO. Gd
doping of EuO seems to result in a Drude-like low frequency optical
spectrum (see figure \ref{epsilon:c}).

If the peaks in the IR-absorption spectrum of Eu-rich EuO are
spin-flip excitations in the BMP, the fact that they remain visible up
to room temperature would support the model of Mauger, as it shows
that the exchange splitting inside the polaron persists at high
temperatures. However if the excitations are spin-conserving
transitions from the vacancy/polaron level to the conduction band like
we propose, they would show that the vacancy electrons are bound at
$\sim 0.6$ eV. Moreover this would show that the exchange splitting in
the conduction band and in the BMP have approximately the same
temperature dependence.

Although the determination of the vacancy electron binding energy
remains ambiguous, both the activated behavior and the peaks in
the infrared spectrum seem to indicate a vacancy binding energy
of around 0.6 eV, which we will thus take as a numerical estimate
for the determination of the vacancy electron density of states
in the next section.



\subsection{Effect of doping}
\label{hybridizationeffect}

To our knowledge, the only serious attempt to quantitatively
model the effect of doping on the temperature dependent
resistivity of EuO was by Mauger \cite{Mauger83}. In that BMP
model the doping dependence of the transport properties is mainly
introduced by the configurational entropy associated with the
various repartitions of the bound magnetic polarons over the
vacancy sites. Therefore, in that model, the shape of the
resistivity curve is mainly determined by the ratio of acceptors
to donors $N_A/N$. However, it might very well be that the
activation energy to go from one configuration to the other is so
high (i.e. low probability of polaron hopping) that the system is
essentially in one configuration, instead of being in
thermodynamic equilibrium for the relevant time scales. The
reduction of the free energy as a result of configurational
entropy might thus be overestimated.

\begin{figure}[!htb]
\centerline{\includegraphics[width=11.25cm]{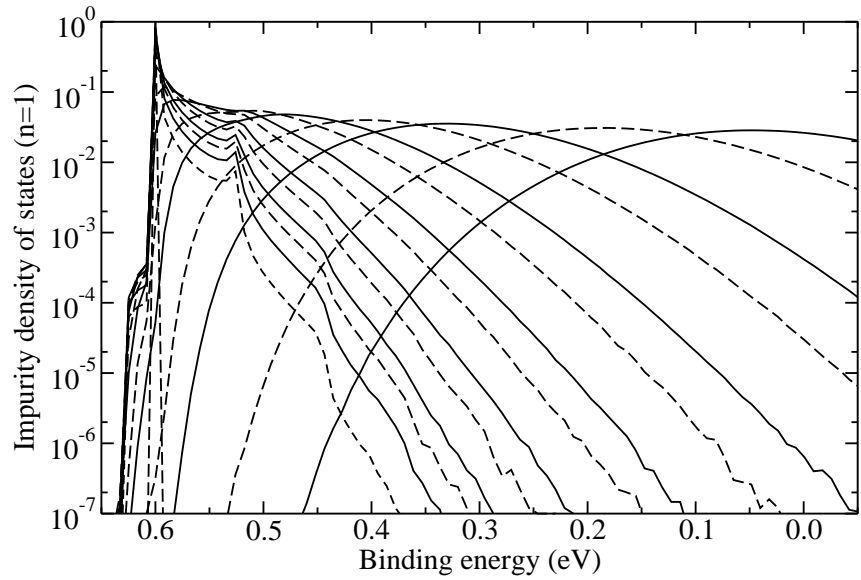}}
\caption{\label{impdos} Density of vacancies $N_j$ versus their
electron binding energy $E_{b,j}$, normalized to the total number
of vacancy states $N$. Calculations were performed at vacancy
concentrations, from left to right respectively: 0.025\%, 0.05\%,
0.075\%, 0.1\%, 0.15\%, 0.2\%, 0.3\%, 0.4\%, 0.6\%, 0.8\%, 1.2\%,
1.6\%, 2.4\% and 3.2\%.}
\end{figure}

In the present chapter we will therefore not consider such
thermodynamic effects. Instead we will treat the doping dependent
reduction in ionization energy as a result of interactions
between vacancies that are in each others vicinity along the
lines of Mott \cite{Mott79}. It is well known that the electron
binding energy in color centers can strongly be modified by
neighbors (formation of M-centers or R-centers)
\cite{Ashcroft76}. Although the spatial distribution of oxygen
vacancies in Eu-rich EuO has not been studied experimentally, we
assume that they are randomly distributed. If two of such
vacancies are in each others vicinity, the binding energy of
their electrons will be changed. To model this effect we treat
each of the oxygen vacancies as a Helium atom with two electrons
in a $1s$-like orbital. As two He atoms do not form a bond their
energy $E_{{\rm He}_2}(R)$ does barely depend on their separation
distance except for very small separations \cite{Beck68}. The
energy of an ionized He$_2^+$ molecule $E_{{\rm He}_2^+}(R)$ will
depend much more strongly on the separation distance as it will
form a bond with two electrons in a bonding orbital and only 1 in
an anti-bonding orbital. Therefore the binding energy of an
electron in the He molecule varies as a function of separation
distance $R$: $E_b(R) = E_{{\rm He}_2^+}(R) - E_{{\rm He}_2}(R)
\approx E_{{\rm He}_2^+}(R) + 2 E_1$. Here $E_1$ is the first
ionization energy of a vacancy electron in a single vacancy as a
result of the effective screened core charge $Z_{\rm eff} e$. The
energy decrease of a bonding electron will approximately cancel
the energy increase of an anti-bonding electron when the atoms
approach each other. Therefore $E_{{\rm He}_2^+}(R) \approx
E_{H_2^+}(R)$, the well known \cite{Gasiorowicz74} energy of the
ionized hydrogen molecule with one electron in a bonding orbital.
Thus we find:
\begin{eqnarray}
\label{Eb} E_b(R)&=&E_1+\frac{Z_{\rm eff}e^2\left[\frac{S(R)}{R}+
 \left(1+\frac{R}{a_e}\right)\left(\frac{e^{-2R/a_e}}{R}-\frac{e^{-R/a_e}}{a_e}\right)\right]}
 {1+S(R)} \nonumber \\
S(R)&=&\left(1+\frac{R}{a_e}+\frac{R^2}{3a_e^2}\right)e^{-R/a_e}
\end{eqnarray}
Approximately, $E_1=Z_{\rm eff}^2 \times 13.6$ eV = 0.6 eV and
thus we find for the orbital radius $a_e = a_0/Z_{\rm eff}$ with
$Z_{\rm eff}=0.21$ and the Bohr radius $a_0$.
Using equation (\ref{Eb}) we can approximate the binding energy
distribution for vacancies randomly situated in the crystal. We
did this by generating a large number of random ensembles with
periodic boundary conditions and calculating the binding energy
for each vacancy, assuming that changes in the binding energy due
to neighbors were additive. The random distributions were
generated with the constraints that distances between vacancies
should be at least the nearest neighbor separation and not more
than two neighbor vacancies should be contained within the second
nearest neighbor spacing. The calculation was also performed for
random vacancies generated on an fcc lattice with the lattice
constant of EuO, resulting in essentially the same results. The
results of the calculation are displayed in figure \ref{impdos}
for a range of concentrations. At small concentrations most
vacancies do not have a neighbor at a distance for which equation
(\ref{Eb}) differs significantly from $E_1$. Therefore the
density of states is strongly peaked around 0.6 eV, the binding
energy for a single vacancy. However upon increasing the vacancy
concentration, the average nearest neighbor distance decreases
and as a result, the average binding energy also decreases. For
concentrations above $\sim$ 0.3\% most vacancies have a binding
energy significantly smaller than $E_1$. Increasing the
concentration above 0.6\% leads to a significant fraction of
vacancies which have a binding energy less than 0 eV. One of the
electrons from such a vacancy will delocalize into the conduction
band\footnote{We neglected the effect of this delocalization on
the binding energy distribution of the remaining vacancy
electrons.}. These electrons will contribute to the conduction
and if their density is much larger than that of thermally
excited electrons, the temperature dependent resistivity will be
non-activated as is indeed observed at comparable doping
concentrations in Eu-rich EuO.

\subsection{Shift of conduction band edge}
\label{condshift} To properly model the effect of the vacancy
binding energy distribution on the transport properties we need
to extend equation (\ref{rate}) to a distribution of impurity
states $j$ with densities of state $N_j$ and electron occupation
density $n_{i,j}$. From equation (\ref{rate}) we now obtain:

\begin{equation}
\label{nij} n_{i,j,\sigma=\pm} = \frac{n_c
N_{j,\sigma=\pm}}{n_c+\frac{N_c(T)}{2}e^{-E_{b,j}/k_BT}(1+e^{\mp
E_b(M)/k_BT})}
\end{equation}

For impurity states of positive and negative spin $\sigma$
($\sigma=+$ corresponds to spin-up in figure \ref{IMTmech:a}) .
Thermal excitations to both spin conduction bands are taken into
account. As indicated by figure \ref{IMTmech:b} the ionization
energy of the vacancy levels of spin $\sigma$ with respect to the
opposite spin conduction band varies with magnetization as
$E_{b,j,\sigma=\pm}(M)=E_{b,j}\pm E_b(M)$ with
$E_b(M)=\frac{M}{M_{sat}}\times 0.52$ eV (using $\Delta_{\rm
ex,max}=0.26$~eV). The equation of charge neutrality now becomes:

\begin{equation}
\label{chargeneutrality} -N_A + \sum_{j,\sigma} N_{j,\sigma} = n
= n_c + \sum_{j,\sigma} n_{i,j,\sigma}
\end{equation}

Where the upper limit of the last sum is $E_{b,j,-}(M)=0$ for
$\sigma=-$ and $E_{b,j,+}=0$ for $\sigma=+$. Using
(\ref{nij}),(\ref{chargeneutrality}) and the calculated impurity
density of states $N_j(E_{b,j})$ from figure \ref{impdos}, we can
now numerically solve for the density $n_c$ of conduction
electrons. The dependence of the magnetization $M(B,T)$ on
temperature and field is obtained from a mean field model for a
ferromagnet with $J=7/2$ and $T_c=69$~K. The magnetization of EuO
is known to follow such a mean field model to a very high
accuracy \cite{Mauger78}. The resistivity is then found to be
$\rho = 1/(n_c e \mu)$ with $\mu = e\tau/m$. The scattering time
$\tau$ in EuO was found \cite{Oliver72} to be of the order of
$3\times 10^{-14}$ s, using the free electron mass we find a
mobility $\mu=53$ cm$^2$V$^{-1}$s$^{-1}$ that was used in the
calculation (similar values were found from Hall effect
measurements \cite{Shapira,Shafer72}).

\subsection{Resistivity curves}
\begin{figure}[!htb]
\centerline{\includegraphics[width=10cm]{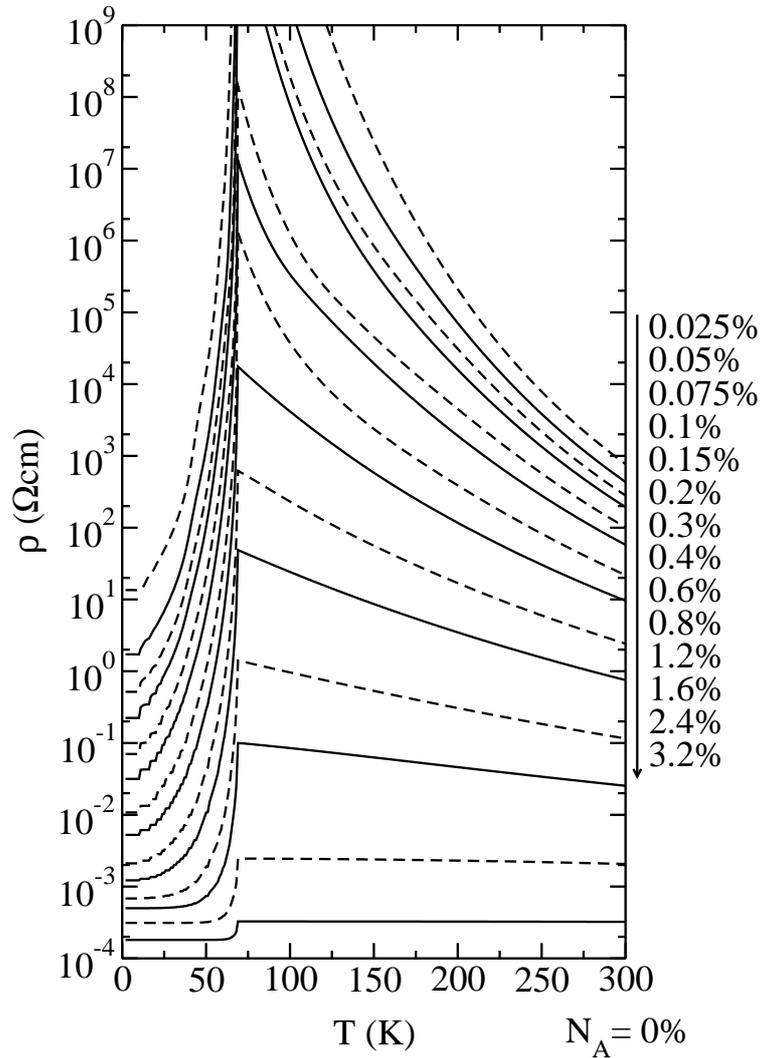}}
\caption{\label{noacc} Calculated resistivity for doping
concentrations $N=0.025-3.2$\% in order of descending resistance
as indicated next to the graph.}
\end{figure}

Thus we have calculated the temperature dependence of the
resistivity for a range of doping concentrations at magnetic
field $B=0$ T. The results with $N_A = 0$\% are displayed in
figure \ref{noacc}. At low doping concentrations the high
temperature resistivity shows an activated behavior $\rho \propto
e^{-E_{ac}/k_BT}$ with $E_{ac} \sim 0.3$ eV as discussed in
section \ref{vaclev}. Upon increasing the doping, the importance
of thermal excitations from energy levels with a reduced binding
energy due to hybridization increases. Therefore the effective
activation energy decreases with doping. Above a doping level of
about 0.4\%, the number of electrons which have zero binding
energy due to hybridization becomes substantial. As $N_j
e^{-E_{b,j}/k_BT}$ decays exponentially with $E_{b,j}$ the main
contribution to the thermally excited conduction electrons will
come from a region $\sim k_BT$ below the zero binding energy
level and $n_c \approx n_0 + \int_0^{k_BT} N_{j}(E_{b,j}) {\rm
d}E_{b,j}$. As $\rho(T) \propto n_c^{-1}$ the exponential decay
of the high temperature resistivity with temperature in figure
\ref{noacc} reflects the exponential increase of the tails of the
impurity density of states with binding energy in figure
\ref{impdos}.

Below the Curie temperature the calculated resistivity is
dominated by the shift of the conduction band and vacancy levels
due to exchange. At low concentrations the energy overlap between
the conduction band and the impurity DOS is essentially zero in
the paramagnetic phase. A decrease $\Delta$ of the energy
splitting between vacancy and conduction band will therefore
initially result in an increase of the number of thermally
excited carriers by a factor $e^{\Delta/k_BT}$. If $\Delta$
becomes larger, substantial overlap between spin-down impurity
and spin-up conduction band will occur and an even larger
contribution to the conduction will be made by vacancy electrons
which have their binding energy reduced below zero. For higher
doping concentrations this second contribution will be dominant
over the whole magnetization range. As $\Delta = \alpha M$, ${\rm
d}\sigma/{\rm d}M \propto N_{j}(\alpha M)$. Interestingly, the
conductivity $\sigma$ is therefore fully determined by the
magnetization and does not depend explicitly on temperature
anymore. The application of a magnetic field that increases the
magnetization thus has the same effect on the resistivity as
increasing the magnetization by a reduction of the temperature as
we have indeed observed experimentally (see section
\ref{magres}).
\begin{figure}[!htb]
\centerline{\includegraphics[width=10cm]{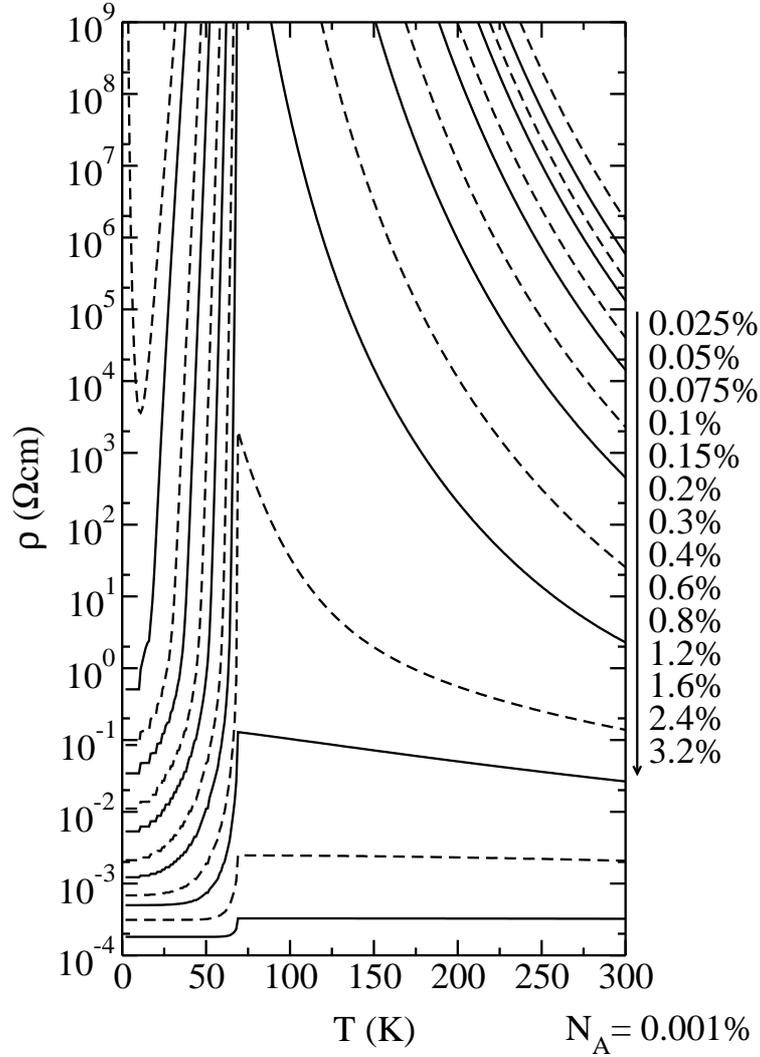}}
\caption{\label{acc} Calculated resistivity for doping
concentrations $N=0.025-3.2$\% in order of descending resistance
as indicated next to the graph. All curves were calculated for
acceptor density $N_A=0.001$\%.}
\end{figure}
The introduction of a small concentration of acceptor sites to
the model has a large effect on the obtained conductivity as is
shown in figure \ref{acc} for $N_A=0.001\%$. The high temperature
resistivity at low electron densities now shows an activated
temperature dependence with activation energy $E_{ac}\sim 0.6$ eV
for reasons discussed in section \ref{vaclev}. For small
densities all vacancy electrons which have a reduced binding
energy due to hybridization will be captured by the acceptor
sites. However for higher doping concentrations the number of
such electrons will exceed the number of acceptor sites and
gradually fill the impurity DOS up to the zero binding energy
level for increasing concentrations. This results in an
activation energy which decreases from 0.6 eV and reaches 0 eV at
about $N=1.6$\%. Above this concentration the effect of the
acceptor impurities is negligible. Now, it takes a higher
magnetization to reduce the binding energy of the highest
occupied impurity levels to zero as the lowest binding energy
states have been depleted by the acceptor sites . Therefore,
especially for low vacancy concentrations, the temperature range
in which thermal excitations dominate the resistivity is larger
than for $N_A=0$. For concentrations of less than 0.075\% the
density of vacancies with a binding energy which is reduced by
more than 0.08 eV~($=E_1-2\Delta_{ex,max}$) due to hybridization
is smaller than the number of acceptor sites. Therefore an
effective gap will persist even at saturation magnetization like
in figure \ref{IMTmech:b}. As the low temperature magnetization
does not vary much, the low temperature resistance shows an
activated behavior across this gap. For the 0.025\% and 0.05\%
vacancy concentration the number of electrons excited across this
gap is so small that the low temperature resistivity is outside
the range of the graph. For the 0.075\% sample an activated
behavior is observed with a resistivity minimum around 12 K.
Similar low temperature activated behavior has often been
observed in slightly europium rich EuO samples. It was studied by
Oliver {\it et al.} \cite{Oliver72} who found an activation
energy $E_{ac} = $18.3 meV. In our model the low temperature
activation energy can have any value between 0 and 80 meV
depending on the density of vacancy states $N$ and acceptors
$N_A$. The upper limit being determined by
$E_1-2\Delta_{ex,max}$.

\section{Experiment}

Thin films of EuO with thicknesses of 20-30 nm were grown by
evaporating Eu metal in a low background pressure of oxygen. Both
polycrystalline films grown on (1$\overline{1}02$) Al$_2$O$_3$
and epitaxial films grown on the (001) plane of 16 mol\%
Y$_2$O$_3$ stabilized cubic ZrO$_2$ (YSZ) were studied. A
detailed account of the growth has been given in chapter
\ref{euoepi}. The film resistivity was determined by 2 and 4
point DC measurements or by a low-frequency lock-in technique,
using Cr contacts. For the photoconductivity measurements the EuO
film area ($\sim 0.5$ mm$^2$) between the contacts was
illuminated by a 1.5 mW He-Ne laser. The intensity of the laser
was varied by inserting neutral density filters in the light
path. All resistivity and photoconductivity measurements were
performed without removing the films from the UHV chamber with a
base pressure below 5$\times 10^{-11}$ mbar. Conductivity
measurements were performed in the dark, to exclude
photoconductivity effects. For the magnetoresistance measurements
the films were protected by a MgO cap layer of $\sim 15$ nm,
which slightly increased their resistivity. The films were
transported in air to a Quantum Design PPMS in which
magnetoresistance measurements were performed with the magnetic
field parallel to the film plane and perpendicular to the current
direction. The in-plane field has the advantage of eliminating
effects of demagnetizing fields as the demagnetization factor in
the plane of the film is zero \cite{Osborn45}.



\subsection{Resistivity}
Many studies of the resistivity of Eu-rich EuO have shown a
metal-insulator transition (see figure \ref{trans}), with a
resistivity peak that is usually located several degrees above
the Curie temperature. Upon increasing the doping concentration a
decrease of the magnitude of this transition is observed,
accompanied by a change of the high temperature resistivity from
an activated behavior with $E_{ac}$ between 0.3 and 0.6~eV to a
non-activated behavior. At low doping concentrations the low
temperature resistivity also often shows an activated behavior.
Besides these similarities between the different studies there
are however a number of differences. Extensive resistivity
measurements on single crystals
\cite{Shafer72,Oliver72,Godart80,Shapira,Llinares74,Shamokhvalov74}
show a large spread in both the detailed shape as in the absolute
value of the resistivity. Besides the concentration of vacancies,
the temperature dependence of the resistivity also seems to
depend on the detailed growth procedure \cite{Shafer72} including
the way the crystal is cooled to room temperature. This might be
explained by the fact that the resistivity curve does not just
depend on the density of vacancies, but also on their spatial
distribution. This spatial distribution could very well be
affected by the way the crystal is grown. Moreover the growth
procedure might affect the density of acceptor sites.

\begin{figure}[!htb]
\centerline{\includegraphics[width=10cm]{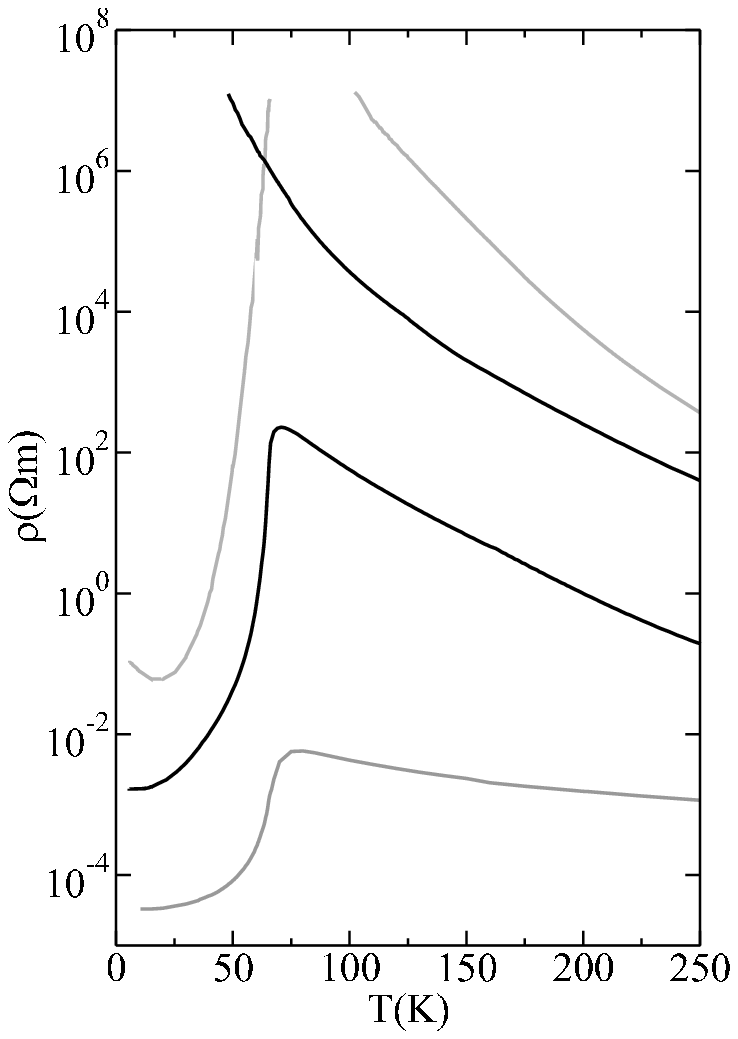}}
\caption{\label{euoepi8b} Resistivity of EuO films on (100) YSZ
substrates (black lines) and ($1\bar{1}02$) Al$_2$O$_3$
substrates (grey lines) grown at 350$^\circ$C (high resistivity)
and 300$^\circ$C (lower resistivity). The thickness of all films
was $\sim$ 20-30 nm.}
\end{figure}

The number of transport studies of EuO thin films
\cite{Paparoditis71,Massenet74,Konno98,Iwata00b} is much smaller
than that on bulk samples. In figure \ref{euoepi8b} we show the
resistivity curves of both epitaxial and polycrystalline films.
The resistivity curves most resemble those of Massenet {\it et
al.} \cite{Massenet74} (see figure \ref{trans:c}). Under the same
growth conditions the epitaxial films generally show a higher
resistivity than the polycrystalline films. The reason for this
might be that it is less probable to incorporate vacancies in a
crystalline lattice whereas they are naturally created at the
grain boundaries in polycrystalline films. However, the
resistance curve of the epitaxial films is qualitatively not very
different of those of polycrystalline films. Raising the growth
temperature seems to reduce the number of oxygen vacancies in the
films, eventually leading to a disappearance of the
metal-insulator transition. The resistivity curves of films show
several important differences from those of bulk crystals. Bulk
crystals usually show a peak in the resistivity around the Curie
temperature even in the rather strongly electron doped crystals
which show a non-activated high temperature behavior. We have
never observed such a peak in our films, and as far as we know it
has never been observed in other resistivity measurements on thin
films. The second difference concerns the decrease of the
resistivity below T$_c$. For bulk crystals this slope often seems
to consist of 2 transitions at 70 K and 50 K, between which ${\rm
d}\rho/{\rm d}T$ shows a minimum (see figure \ref{trans:a}).
Oliver {\it et al.} \cite{Oliver72} attributed the high
temperature transition partly to a change in mobility related to
scattering by magnetic fluctuations, whereas the second
transition is fully due to an increase in the free carrier
concentration. For thin films such a minimum has never been
observed, with ${\rm d}\rho/{\rm d}T$ decreasing monotonically
from close to T$_c$ to the resistivity minimum. These differences
between bulk and film samples might be caused by the fact that
the growth procedure of thin films is very different from that of
single crystals as thin films can be grown at low temperatures
($\sim 350^\circ$C) whereas bulk crystals are grown at
temperatures around 1800$^\circ$C. Therefore the oxygen vacancies
might be much more mobile during the growth of bulk samples. This
could result in differences in the spatial distribution of the
vacancies. Possibly the vacancies are ordered or even phase
separated with small clusters of Eu metal in the EuO lattice of
the bulk samples. On the other hand the vacancies in the thin
films are probably more or less randomly frozen in. This could
result in a different impurity DOS and thus a different
temperature dependent resistivity. Therefore it would be very
interesting to study the vacancy distribution using microscopic
techniques. Thin films might also have a higher intrinsic
scattering rate than bulk crystals due to surface scattering.
Therefore the effect of scattering by magnetic fluctuations might
be relatively less important.

There is a good agreement between the calculated resistivity
curves in figure \ref{noacc} and the measured curves in figure
\ref{euoepi8b}. The high temperature resistivity of the films
does not follow an activated behavior but decays exponentially
with temperature, as is expected when substantial overlap exists
between the tail of the impurity density of states with the
conduction band. Also the shape of the resistivity curves below
the Curie temperature is similar to the calculated curves. The
cut-off around T$_c$ of the measured resistivity curves is less
sharp than that in the calculated curves. Moreover the maximum in
the measured curves is often located several degrees above the
Curie temperature. These differences are possibly due to the
formation of bound magnetic polarons below T$_p$.

\subsection{Magnetoresistance}
\label{magres} 
\begin{figure}[!htb]
\centerline{\includegraphics[width=14cm]{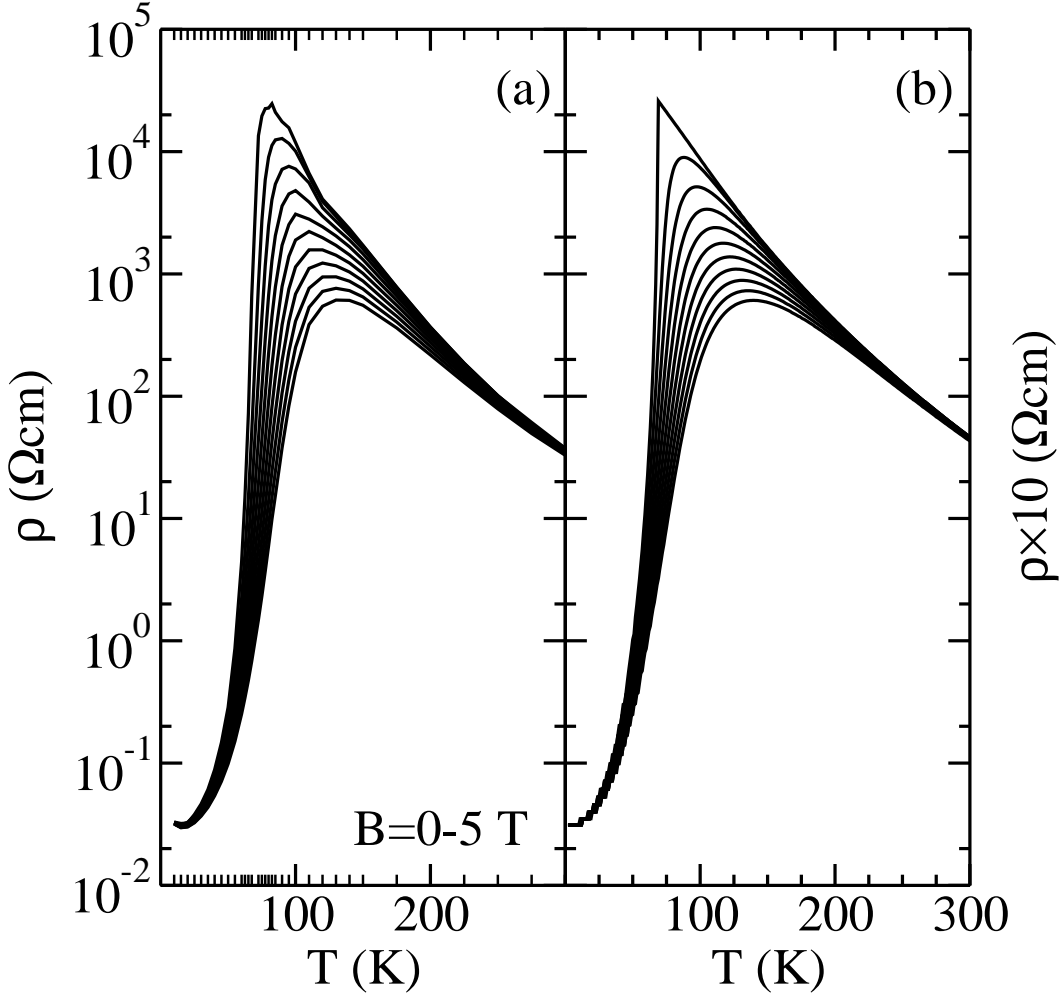}}
\caption{\label{Tmagres} (a) Magnetoresistance at in-plane fields
from 0 T (top curve) to 5 T (bottom curve) with 0.5 T intervals
of an epitaxial EuO film on YSZ protected by a MgO cap layer. The
tick marks on the horizontal axis indicate the temperatures at
which the resistance was measured. (b) Calculated
magnetoresistance with $N=0.5\%$ for $B=0-5$ T and $N_A=0$\%.}
\end{figure}

In figure \ref{Tmagres}(a) we show the
magnetoresistance of an epitaxial EuO film. Around the Curie
temperature a field of 5 T strongly reduces the resistivity by a
factor of more than 10$^4$. In figure \ref{Tmagres}(b) we show
the calculated magnetoresistance for a vacancy concentration of
0.5\%. In the calculation the magnetic field increases the
magnetization and thus reduces the vacancy binding energy $E_b$.
Except for a factor of 10 in the absolute value, which might be
related to an intrinsically lower mobility in films due to
surface effects, the calculated resistivity corresponds very well
to the measured magnetoresistance at all fields.

\begin{figure}[!htb]
\centerline{\includegraphics[width=14cm]{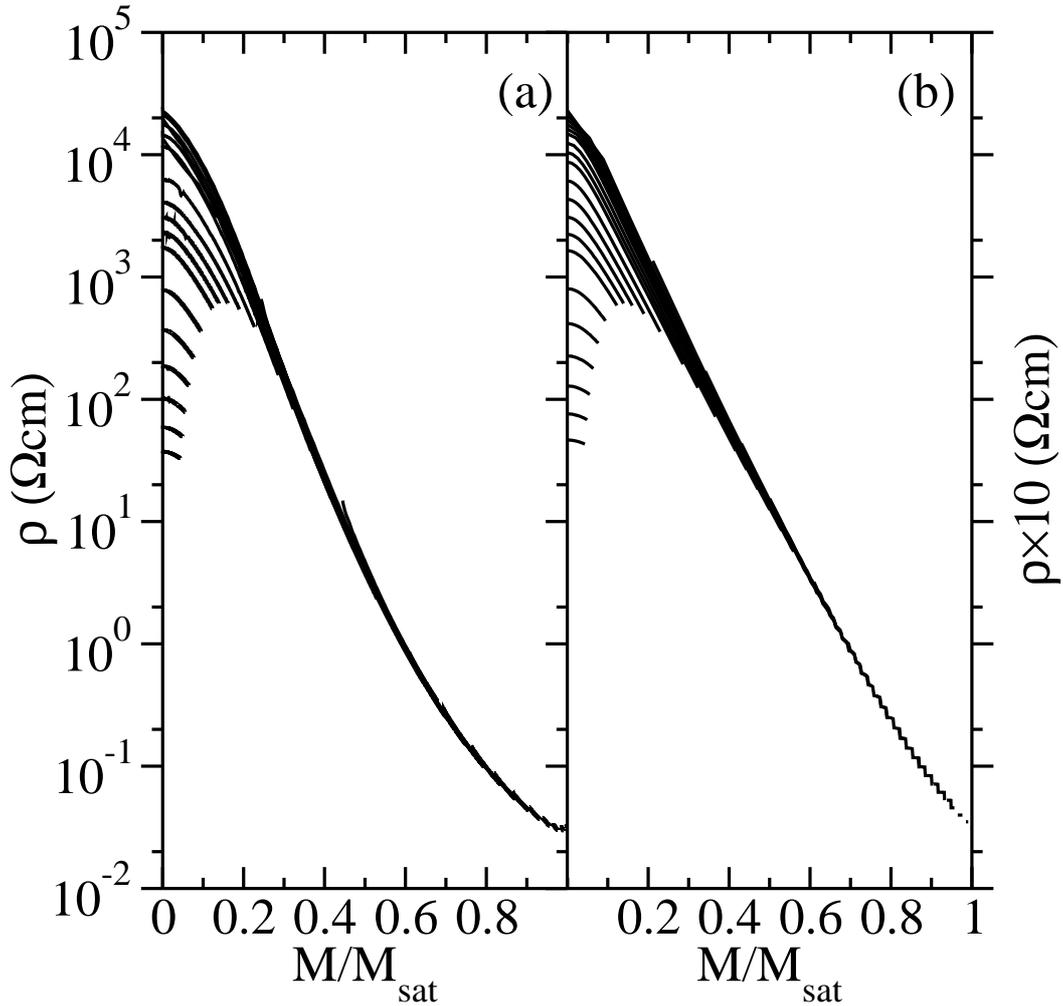}}
\caption{\label{Mmagres} (a) Curves $\rho(B)$ for fields from 0-5
T at 34 fixed temperatures. All data points are plotted versus
$M(B,T)/M_{sat}$ as calculated from a mean field model with
T$_c=69$ K and $J=7/2$. (b) $\rho(B)$ with $N=0.5\%$ for $B=0-5$
T calculated at the same 34 temperatures also plotted versus
$M/M_{sat}$.}
\end{figure}
In figure \ref{Mmagres}(a) we show the same data as in figure
\ref{Tmagres}(a), for $\rho(B)$ scans for $B=0-5$ T at 34
different temperatures from 10-300 K. However, this time we
plotted the resistivity versus the normalized magnetization
$M(B,T)/M_{sat}$ as was calculated from the experimental $B$ and
$T$ using a mean field model with T$_c=69$~K. Remarkably, the
figure shows that all the resistivity data below ~100 K seem to
superimpose on a universal curve. This strongly suggests that the
resistance of Eu-rich EuO does not explicitly depend on
temperature and magnetic field, but is fully determined by the
magnetization. Interestingly this observation imposes strict
limits on models for the conductivity and magnetoresistance in
EuO. Above 100~K the $\rho(B)$ data leave the universal curve,
thus indicating that the contribution of thermally activated
electrons to the conduction has become non-negligible. In figure
\ref{Mmagres}(b) we show the calculated resistivity with
$N=0.5\%$, which shows that these results are excellently
reproduced by a decrease of the impurity electron binding energy
proportional to the magnetization as discussed in section
\ref{condshift}. As argued in this section d$\sigma/$d$M \propto
N_{j}(\alpha M)$ and the universal curve that is found therefore
directly reflects the integrated impurity density of states. In
fact as the measured conductivity $\sigma(M)$ is approximately
proportional to $e^{bM/M_{sat}}$ this indicates that
$N_{j}(E_{b,j}) \propto e^{\frac{b}{\alpha} E_{b,j}}$, with
$\alpha \sim 0.52$ eV and $b \approx 13.3$. Therefore the
experimental data are consistent with the calculated
exponentially decaying tails of the vacancy density of states.
Hundley {\it et al.} \cite{Hundley95} have found a similar
behavior $\rho(B,T) \propto e^{-bM/M_0}$ in
La$_{0.7}$Ca$_{0.3}$MnO$_3$ films. They explain this behavior by
a polaron hopping conductivity $\sigma \propto e^{-M/k_B T}$ with
a hopping transfer integral proportional to the magnetization $t
\propto M$. A weak point in this assignment is that the
conductivity should still depend explicitly on temperature
through the factor $k_B T$ in the exponent, an effect that would
strongly deviate the data from $\rho(B,T) \propto e^{-bM/M_0}$,
especially at low temperatures\footnote{For similar reasons it is
difficult to reconcile a polaron hopping model with the observed
$\rho(B,T) \propto e^{-M/M_0}$ behavior in EuO.}. The success of
the model of an exchange split conduction band (a red shift of
the absorption was also found in the manganites \cite{Demin})
that moves through an occupied (impurity) density of states
suggests that it should also be considered as an origin for the
metal-insulator transition in the manganites, as was also argued
by Nagaev \cite{Nagaev01}. The same remark might be made about
the metal-insulator transition in EuB$_6$, as it was recently
found from optical measurements that the number of effective
charge carriers in this compound also shows a universal scaling
with magnetization. Broderick {\it et al.} \cite{Broderick02}
found that $M^2 \propto (\omega_p^2)^2$, i.e. $M \propto
n_{eff}/m^{*}$, a behavior that seems difficult to reconcile with
polaron models. This behavior is however consistent with a model
in which the shift of the impurity binding energy is proportional
to the magnetization, if the (impurity) DOS is approximately
constant. In that case our model gives d$n_c/$d$M \propto
N_{j}(E) \approx const$. Recently it was argued that EuB$_6$
intrinsically is a semiconductor and that structural defects seem
to play an important role \cite{Werheit00}. This suggests that
the transport properties of EuB$_6$ might be dominated by doping
from B$_6$ vacancies in a way similar to Eu-rich EuO. The large
spread in resistivity and carrier concentration for different
crystals \cite{Aronson99} might be the result of different
vacancy densities. Anyhow, even disregarding the origin for the
shape of the occupied density of states in Eu-rich EuO, EuB$_6$
and La$_{0.7}$Ca$_{0.3}$MnO$_3$, scaling laws for
magnetoresistance $\sigma \propto f(M(B,T))$ or $n_c \propto
f(M(B,T))$, seem to be obeyed by all these compounds. This
universal behavior seems to be naturally explained as a crossing
of occupied states and conduction bands as a result of an
exchange splitting proportional to $M$.

\subsection{Photoconductivity}
\label{Photocond} 
\begin{figure}[!htb]
\centerline{\includegraphics[width=11.25cm]{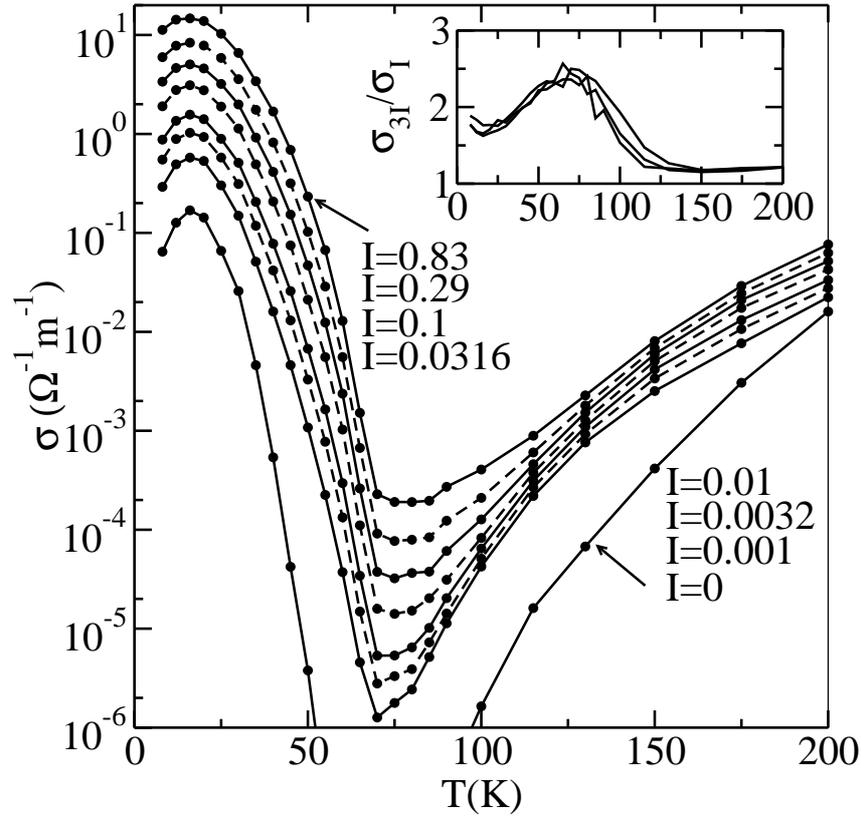}}
\caption{\label{trphoto} Photoconductivity of a 20 nm EuO film
for different laser intensities $I$ as indicated in the graph.
Inset: ratio of neighboring photoconductivity curves
$\sigma_{I=0.83}/\sigma_{I=0.29}$,
$\sigma_{I=0.29}/\sigma_{I=0.1}$ and
$\sigma_{I=0.1}/\sigma_{I=0.0316}$.}
\end{figure}

\begin{figure}[!htb]
\centerline{\includegraphics[width=11.25cm]{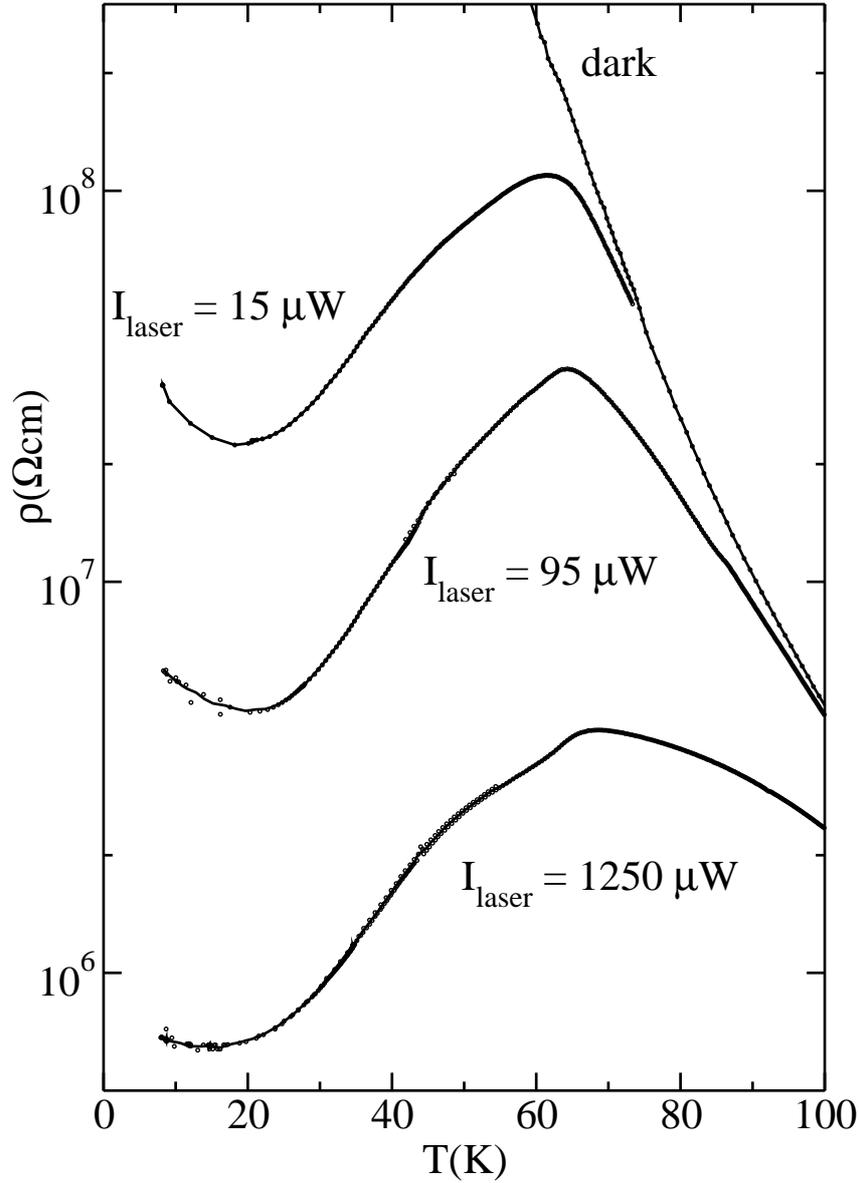}}
\caption{\label{photoMIT} Resistivity of a thin epitaxial EuO
film grown on stabilized ZrO$_2$(001). Although no
metal-insulator transition is observed in the dark conductivity,
the film ($\sim 0.3 \times 1$ mm$^2 \times 35$ nm) shows metallic
behavior below $\sim $65 K upon illumination with a He-Ne laser.}
\end{figure}

The photoconductivity of EuO is very
remarkable. In figure \ref{trphoto} we show the
photoconductivity\footnote{$\sigma_I$ is just the conductivity of
the film when it is illuminated by a laser of intensity $I$.} of
a thin EuO film on an Al$_2$O$_3$ substrate for different laser
intensities $I$. A striking effect is that the magnitude of the
metal-insulator transition is only slightly reduced by the
illumination with light. A related interesting effect that we
observed in films without a normal metal-insulator transition was
a light induced metal-insulator transition (shown in figure
\ref{photoMIT}). A similar effect has been observed in EuO single
crystals \cite{Llinares75,Desfours76}, although these authors did
not show that the metal-insulator transition was absent in the
dark. Our data shows for the first time that d$\rho$/d$T$ in
stoichiometric EuO films stays negative in the dark whereas it
becomes positive (i.e. metallic) below T$_c$ when it is
illuminated by a laser.

Previous studies
\cite{Bachmann68b,Bachmann68,Penney71,Llinares75} have described
the photoconductivity using the formula $\Delta \sigma_{I} =
\sigma_{I}-\sigma_{I=0} \propto I \mu(T) \tau_{ph}$, where
$\tau_{ph}$ is the lifetime of photoexcited carriers. This
formula is based on the assumptions that the density of
photoexcited carriers is proportional to the light intensity:
$n_{ph}\propto I$ and the conductivity is proportional to the
number of photoexcited carriers: $\sigma_I \propto n_{ph}$. The
authors assumed that $\tau_{ph}$ is temperature independent.
Therefore they have interpreted the photoconductivity curves as
evidence that the metal-insulator transition in EuO is completely
due to changes in the mobility $\mu(T)$, in contrast with Hall,
Seebeck and optical measurements that assign it to changes in the
carrier density \cite{Oliver72,Shapira,Shamokhvalov74}. It is
however important to note that the mobility of light induced
carriers can be very different from that of free carriers as the
excited carriers probably form excitons. Such excitons could show
similar behavior as chemically doped electrons, with electron
hole separation only occurring in the ferromagnetic phase. The
large increase in photoconductivity in the ferromagnetic state
could thus also be understood as a change in the number of
\emph{free} photoelectrons. This would also explain the
occurrence of the photoinduced metal-insulator transition (figure
\ref{photoMIT}). The photoconductivity measurements therefore do
not prove that the metal-insulator transition is caused by a
mobility change.
\begin{figure}[!htb]
\centerline{\includegraphics[width=11.25cm]{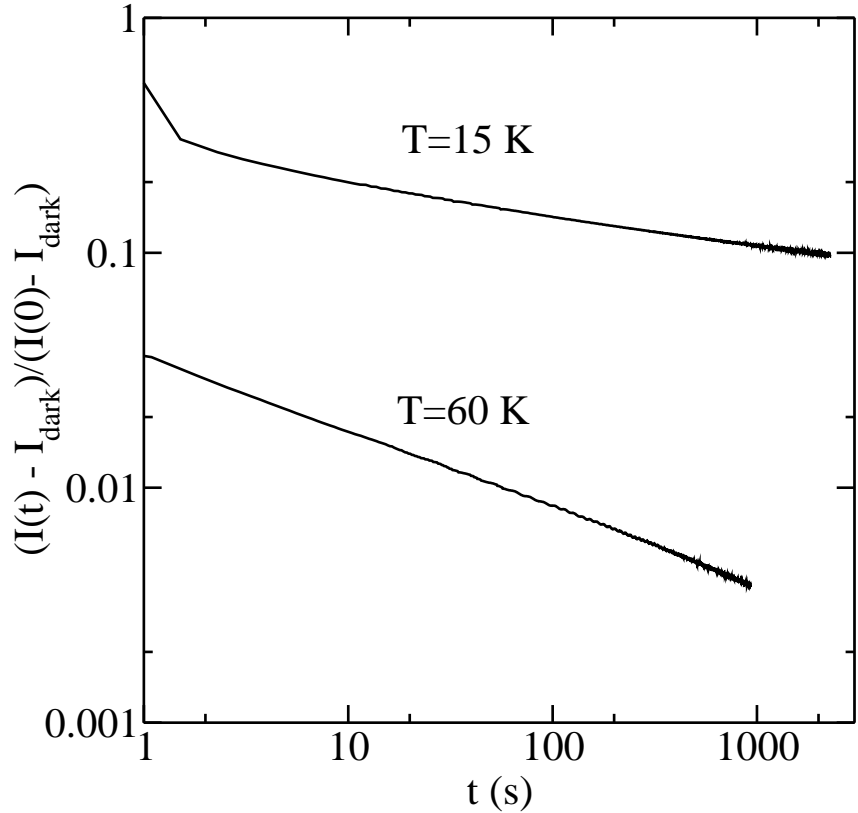}}
\caption{\label{phototrans} Normalized film current at constant
voltage after removing full illumination of a He-Ne (1250 $\mu$W)
at $t=0$. After a rapid decay in conductivity in less than one
second, a very slow decay of the conductivity sets in.}
\end{figure}
We have studied the slow transient decay of the photoconductivity
after the illumination of the sample was stopped as is shown in
figure \ref{phototrans}. Near the Curie temperature the
conductivity quickly ($\tau_{ph} < 1$ s) returns to within 1\% of
its dark value after removing the light, after which a slow
conductivity decay sets in. However at 15 K we found that the
photocurrent in our films decays even more slowly. Even $\sim
3000$ s after blocking the laser the (photo)conductivity $\Delta
\sigma$ was still at $\sim$ 10\% of the photoconductivity $\Delta
\sigma_I$ at full illumination. Possibly these extremely long
decay times are caused by electrons that are excited to the
minority-spin conduction band. The lifetime of such
photoelectrons will be much increased in the ferromagnetic state
as they can not recombine with Eu $4f$ holes which all have
opposite spin at saturation magnetization. Another cause for the
slow decay rates might be the thermal excitation of electrons
which are trapped in deep impurity levels. These long relaxation
times had not been observed in previous studies of the transient
photoconductivity in which only short decay times
($\tau_{ph}<0.1$ s) were examined \cite{Llinares75,Penney71}.

The relaxation rate $\tau_{ph}$ seems to depend on the
illumination $I$. To show this we have plotted
$\sigma_{I=0.83}/\sigma_{I=0.29}$,
$\sigma_{I=0.29}/\sigma_{I=0.1}$ and
$\sigma_{I=0.1}/\sigma_{I=0.0316}$ in the inset of figure
\ref{trphoto}. For all these curves $\sigma_I \gg \sigma_0$, and
therefore $\sigma_{3I}/\sigma_{I}\approx \Delta
\sigma_{3I}/\Delta \sigma_{I}$. The three curves in the inset
show very similar behavior indicating that
$\sigma_{3I}(T)/\sigma_{I}(T)\approx (3I/I)^{\eta(T)}$, i.e.
$\sigma_I \propto I^\eta(T)$. $\eta$ is related to the mechanism
that is responsible for the recombination of the photocarriers.
At low temperatures $\eta \approx \frac{1}{2}$ indicating
recombination between photocarriers and valence band holes,
higher values of $\eta$ often indicate that levels in the gap
(traps) play an important role in the recombination dynamics
\cite{Bube92}. The dependence of the relaxation rate $\tau_{ph}$
on $I$ and the large values of $\tau_{ph}$ at low temperatures
complicate the interpretation of the temperature dependent
photoconductivity.

\section{Conclusions}
We have compared the He-model \cite{Oliver72} and the BMP model
\cite{Mauger83} for the metal-insulator transition in EuO. We
argued that the difference between these models is the
diamagnetic or paramagnetic configuration of the two electrons in
the oxygen vacancy. This configuration strongly depends on the
degeneracy of the lowest energy levels in the vacancy. We
performed a simple cluster calculation which seems to indicate
that the diamagnetic configuration is more stable. The similarity
between the temperature dependent binding energy in the polaron
and He-model as indicated by figures \ref{IMTmech:b} and
\ref{IMTmech:c} shows why both models are capable of obtaining a
good agreement with experimental resistance curves. Instead of
modelling the effect of doping as a change in configurational
entropy at the vacancy sites \cite{Mauger83}, we calculated the
reduction in binding energy as a result of the formation of
bonding and anti-bonding orbitals between vacancy sites at small
distances from each other. Thus we calculated the transport
properties as a function of temperature, magnetic field and
doping concentration.

Despite the fact that the model neglects several potentially
important effects, like the effects of scattering by magnetic
fluctuations and the possibility of polaron hopping/band
formation, we found good agreement between the temperature
dependent resistivity of epitaxial and polycrystalline EuO thin
films and the calculated resistivity. In particular the model
accounts well for the observed exponential decay of the high
temperature resistivity $\rho \propto e^{-\alpha T}$. We have
measured the magnetoresistance of an epitaxial EuO film and have
shown that it can be scaled on a universal curve which gives
approximately $\rho \propto e^{-b M(B,T)/M_{\rm sat}}$ below
$\sim$ 100 K. The proposed model excellently reproduces this
behavior, whereas polaron hopping models do not seem to be
consistent with it as they should show a temperature activated
dependence. Interestingly similar scaling relations with
magnetization were observed in La$_{0.7}$Ca$_{0.3}$MnO$_3$ and
EuB$_6$, suggesting that similar mechanisms could be responsible
for CMR effects in these compounds. We measured the
photoconductivity of EuO films and also found a metal-insulator
transition in the photoconductivity. We argued that this
metal-insulator transition is not necessarily due to a mobility
change, but might also be due to a transition in the number of
delocalized photocarriers, similar to that of chemically doped
carriers.

\chapter[Crossing the gap from $p$- to $n$-type doping in the cuprates]{Crossing the gap from $p$- to $n$-type doping: nature of the
states near the chemical potential in \lsco\ and \ncco}
\label{NCCO}
\def\ncco{Nd$_{\rm 2-x}$Ce$_{\rm x}$CuO$_{\rm 4-\delta}$}
\def\nca{Nd$_{\rm 1.85}$Ce$_{\rm 0.15}$CuO$_{\rm 4-\delta}$}
\def\ncb{Nd$_{\rm 1.92}$Ce$_{\rm 0.08}$CuO$_{\rm 4}$}
\def\nco{Nd$_{\rm 2}$CuO$_{\rm 4}$}
\def\lsco{La$_{\rm 2-x}$Sr$_{\rm x}$CuO$_{\rm 4}$}
\def\lscoc{La$_{\rm 1.85}$Sr$_{\rm 0.15}$CuO$_{\rm 4}$}
\def\lco{La$_{\rm 2}$CuO$_{\rm 4}$}
\def\bisco{Bi$_{\rm 2}$Sr$_{\rm 2}$CaCu$_{\rm 2}$O$_{\rm 8}$}

{\sl We report on an x-ray absorption and resonant photoemission
study on single crystal \lsco\ and \ncco\ high $T_c$ cuprates.
Using an internal energy reference, we find from the
photoemission data that the chemical potential of \lsco\ lies
near the top of the \lco\ valence band and of \ncco\ near the
bottom of the \nco\ conduction band. The x-ray absorption data
establish clearly that the introduction of Ce in \nco\ results in
electrons being doped into the CuO$_2$ planes. We infer that the
states closest to the chemical potential have a Cu $3d^{10}$
singlet origin in \ncco\, and $3d^{9}\underline{L}$ singlet in
\lsco.}

\section{Introduction}
One of the long standing puzzles in the field of high $T_c$
superconductivity concerns the nature of the charge carriers in
doped high $T_c$ cuprates \cite{imada}. Several photoemission
studies \cite{allen,namatame,suzuki} have revealed very little
difference in the valence band spectra of the \ncco\ system as
compared to the \lsco. In particular, the position of the
chemical potential in the two different systems seem to be quite
similar and is located in the middle of the gap of the parent
compounds, while one would expect that the chemical potential
should shift from the top of the valence band to the bottom of
the conduction band by changing from $p$-type to $n$-type doping.
From this unexpected behavior it was concluded that both hole and
electron doping result in new states that fill the band gap.
Several explanations have been proposed for this mid gap pinning
of the chemical potential \cite{allen,allen2,harima,damascelli},  
with perhaps the scenario involving the occurence of phase        
separation to be currently the most discussed                     
\cite{allen2,emery,damascelli}. In addition, for \ncco\, it       
becomes even an issue whether it can be regarded as a really
electron doped system \cite{jansen,blackstead,hirsch}, since
transport measurements on optimally doped \ncco\ have revealed a
positive Hall coefficient \cite{wang,jiang}. This has led to
propositions that the charge carriers relevant for the
superconductivity in \ncco\ are hole like.

In this paper we present a comparative x-ray absorption (XAS) and
resonant photoemission (RESPES) study on \ncco\ and \lsco\ single
crystals. To ensure that the spectra collected are representative
for the bulk material, we rely on the bulk sensitivity of the
x-ray absorption technique as well as of valence band
photoemission measurements using high photon energies
\cite{suga}. In addition, we use an internal energy reference
within the CuO$_2$ planes as proposed earlier \cite{allen2}, in
order to ensure a reliable measurement of the position of the
chemical potential relative to the valence band. We find
unambiguously that the chemical potential $\mu$ of \lsco\ lies
near the top of the \lco\ valence band and $\mu$ of \ncco\ lies
near the bottom of the \nco\ conduction band, and that the
introduction of Ce results in electrons being doped into the
CuO$_2$ planes. Resonance data suggest that the charge carriers
in \ncco\ are different in nature than in \lsco, in the sense
that they are singlets of Cu $3d^{10}$ character rather than of
$3d^{9}\underline{L}$, where $\underline{L}$ denotes an oxygen
ligand hole. In addition we find that the presence of Nd $4f$
states and a different O $2p$ to Cu $3d$ charge transfer energy
cause \ncco\ to have an \textit{apparently} similar chemical
potential position as \lsco\ if the leading edge of the main
valence band structure is used as reference.

\section{Experimental}
Single crystals of \nco, \nca, \lco, and \lscoc\ were grown by
the travelling solvent floating zone method
\cite{hien,nugroho,li}. The onset of the superconducting
transition in the \lscoc\ sample was found to be {35~\nolinebreak
K} from ac-susceptibility measurements. The \nca\ crystals show a
critical temperature of {21~\nolinebreak K} after reducing them
in flowing N$_2$ gas at 900$^{\circ}$C for 30 hours. XAS and
RESPES experiments were performed at the ID12B beamline of the
European Synchrotron Research Facility (ESRF). The overall energy
resolution was set to 0.3 eV for the XAS and 0.5 eV for the
RESPES. The samples were both cleaved and measured at 20 K in a
chamber with a base pressure below 5$\times$10$^{\rm -11}$ mbar.
Core level x-ray photoemission (XPS) measurements were carried
out in Groningen using a Surface Science Instruments system
equipped with a monochromatized Al-$K\alpha$ source. The XPS
energy resolution is 0.5 eV, and the samples were cleaved and
measured at room temperature in a chamber with a base pressure
below 1$\times$10$^{\rm -10}$~mbar.

\section{Is \nca\ electron doped?}
\begin{figure}[!htb]  
\centering
\includegraphics[width=14cm]{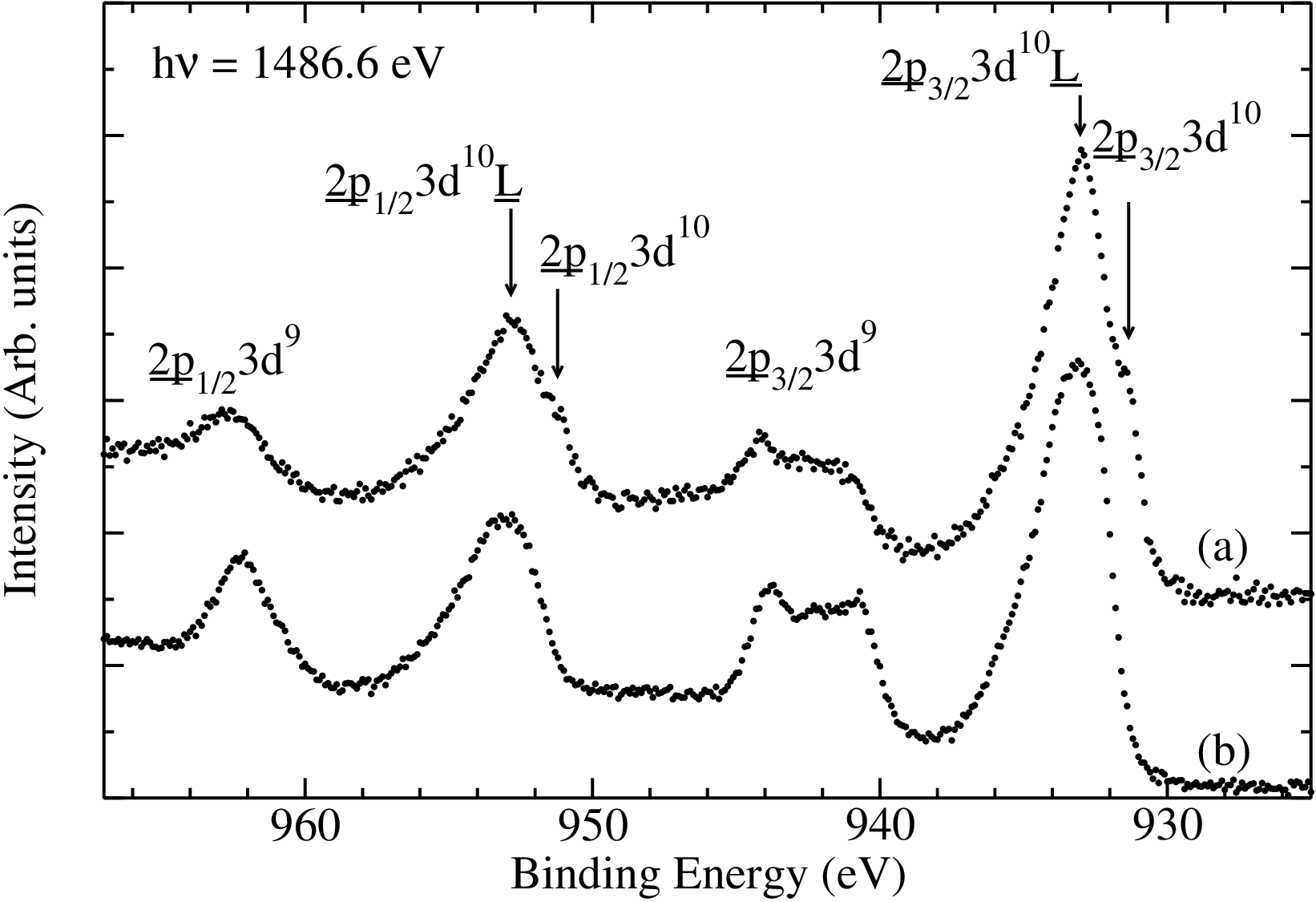}
\caption{Cu $2p$ XPS spectra of (a) \nca\ and (b) \nco.}
\label{cu2pxps}
\end{figure}

Figure \ref{cu2pxps} shows the Cu $2p$ core level XPS of \nco\
and \nca\, where we have used the standard labelling for the
structures \cite{ishii,suzuki,cummins,yamamoto}. We clearly
observe that the $\underline{2p}3d^{9}$ 'satellites' decrease in
intensity with Ce doping, and that also new structures start to
appear on the low binding energy side of the
$\underline{2p}3d^{10}\underline{L}$ 'main lines', where the
underline denotes a hole. These new structures have been assigned
as $\underline{2p}3d^{10}$. These spectra imply that doping \nco\
with Ce results in a decrease of the Cu$^{2+}$ content with the
$3d^{9}$ and $3d^{10}\underline{L}$ configurations, and
simultaneously in an increase of the Cu$^{1+}$ with the $3d^{10}$
configuration. In other words, Ce doping indeed introduces
electrons into the CuO$_2$ planes in \ncco.
\begin{figure}[!htb]  
\centering
\includegraphics[width=11.25cm]{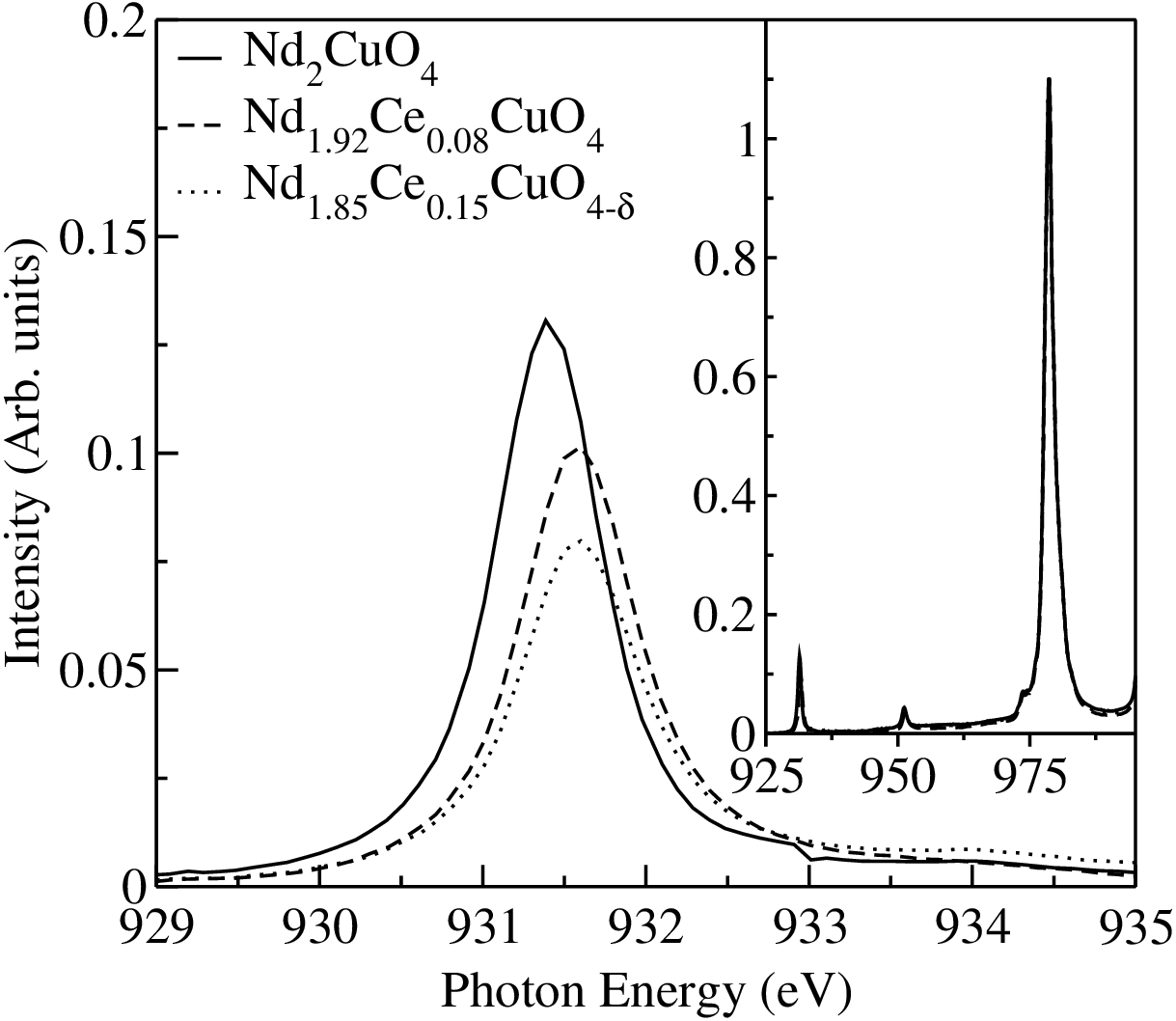}
\caption{Cu $L_3$ XAS spectra of \ncco\ normalized on the Nd
$M_5$ absorption intensity as shown in the inset.}
\label{ndcucom}
\end{figure}

Figure \ref{ndcucom} depicts the Cu $L_{3}$ XAS spectrum of \nco\
and \nca\, normalized to the Nd $M_{5}$ absorption line as shown
in the inset. The Cu white line is given by the transition
$3d^{9} + h\nu \rightarrow \underline{2p}3d^{10}$. One can
clearly see that the Cu white line intensity has decreased
appreciably when Ce is introduced into the system, consistent
with the results from earlier electron energy loss studies on
polycrystalline samples \cite{alexander,nucker}. This indicates
again directly that the Cu$^{2+}$ content in the CuO$_2$ plane is
reduced, i.e. Ce doping indeed means electron doping.
\begin{figure}[!htb]  
\vspace{0.3cm} \centering
\includegraphics[width=11.25cm]{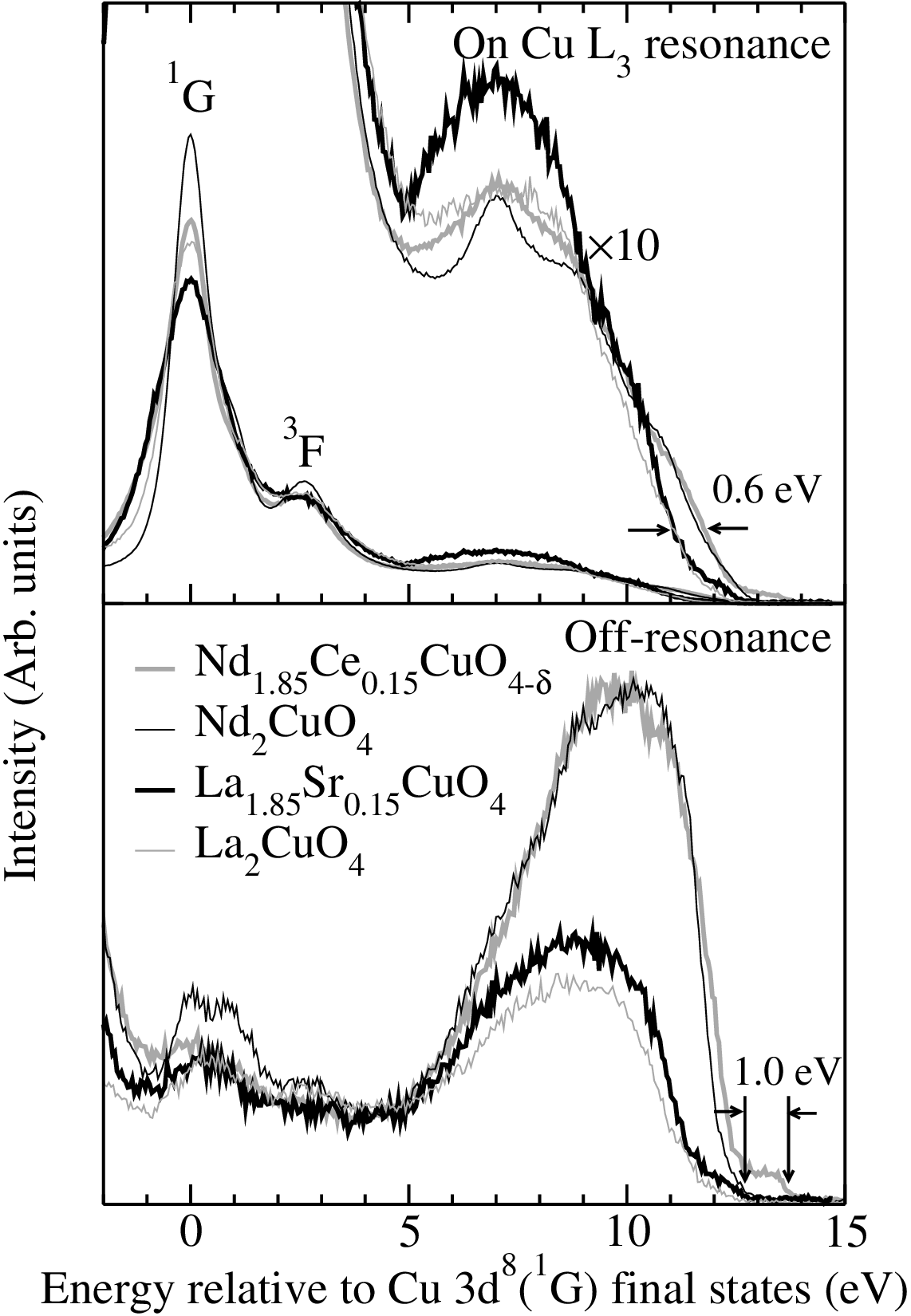}
\caption{Upper panel: resonant photoemission of \nca, \nco, \lco\
and \lscoc\ at the Cu $L_3$ edge. The energy of the spectra has
been aligned on the Cu $3d^8$ $^1G$ final states. Lower panel:
off-resonant spectra of the same samples at a photon energy 5 eV
below the resonance. The arrows indicate the valence band onset
of \nco\ and the Fermi level of \nca.} \label{valcom}
\end{figure}

\section{Position of the chemical potential}
Having established that \ncco\ is an electron doped material, we
now will shift our focus to the issue of the position of the
chemical potential in \ncco\ relative to that in the \lsco\ as
well as their parent materials. Figure \ref{valcom} shows the on-
and off-resonant photoemission spectra of \nca, \nco\, \lco\ and
\lscoc. The on-resonant spectra are taken with the photon energy
tuned at the Cu $L_{3}$ XAS white line, and the off-resonant
spectra with a photon energy which is $5.0 \pm 0.1$ eV lower. It
has been demonstrated earlier \cite{tjeng91,tjeng92,brookes01}
that the Cu $L_3$ resonant photoemission spectrum of Cu$^{2+}$
oxides is dominated by the auto-ionization process of the type
$3d^{9} + h\nu \rightarrow \underline{2p}3d^{10} \rightarrow
3d^{8}$, thereby enhancing the spectral weight of the $3d^{8}$
final states considerably. Indeed, the spectra depicted in the
top panel contain the characteristic $^{1}G$ and $^{3}F$
structures of a $3d^{8}$ \cite{antonides77,haak78}.

We now will use the $3d^{8}$ $^{1}G$ peak to set the energy scale
for all the photoemission spectra \cite{allen2}, since this will
give us an energy zero that refers directly to the electronic
structure of the CuO$_2$ plane, which is at the heart of the
current discussion. Such an internal energy reference is much
more reliable than using the leading edge of the main peak of the
off-resonant valence band spectrum \cite{allen,namatame}, since
in comparing the \ncco\ with the \lsco\ system one can suffer
from differences due to the presence of different chemical
species and charge transfer energies, as we will show later. It
is also better to use the internal energy reference rather than
the Fermi level of the spectrometer \cite{harima}, since the
chemical potential of the undoped \nco\ and \lco\ is not well
defined and in fact will even be pinned by impurities or defects
of unknown nature and quantities.

With the $^{1}G$ peak position set to zero, we can clearly
observe in the lower panel of figure \ref{valcom}, that the first
ionization state of \nca\ is about 1.0 eV further away than that
of the undoped \nco. Recalling that the onset of the optical gap
in \nco\ is approximately 1.0 eV (with the first peak in the
optical spectrum at 1.5 eV) \cite{cooper,uchida,arima}, one can
immediately conclude that the chemical potential of the Ce doped
system lies near the bottom of the conduction band of the undoped
\nco. Contrary to earlier claims \cite{allen,namatame}, we find
that the chemical potential for the doped material is not pinned
in the middle of the gap of the undoped material. A similar
conclusion can also be reached from recent doping dependent
angle-resolved photoemission experiments \cite{armitage},
although there one needs to assume that one can align the
spectrum of the undoped \nco\ with respect to those of the \ncco\
by using the leading edge of the valence band for small doping
levels.
\begin{figure}[!htb]  
\vspace{0.3cm} \centering
\includegraphics[width=11.25cm]{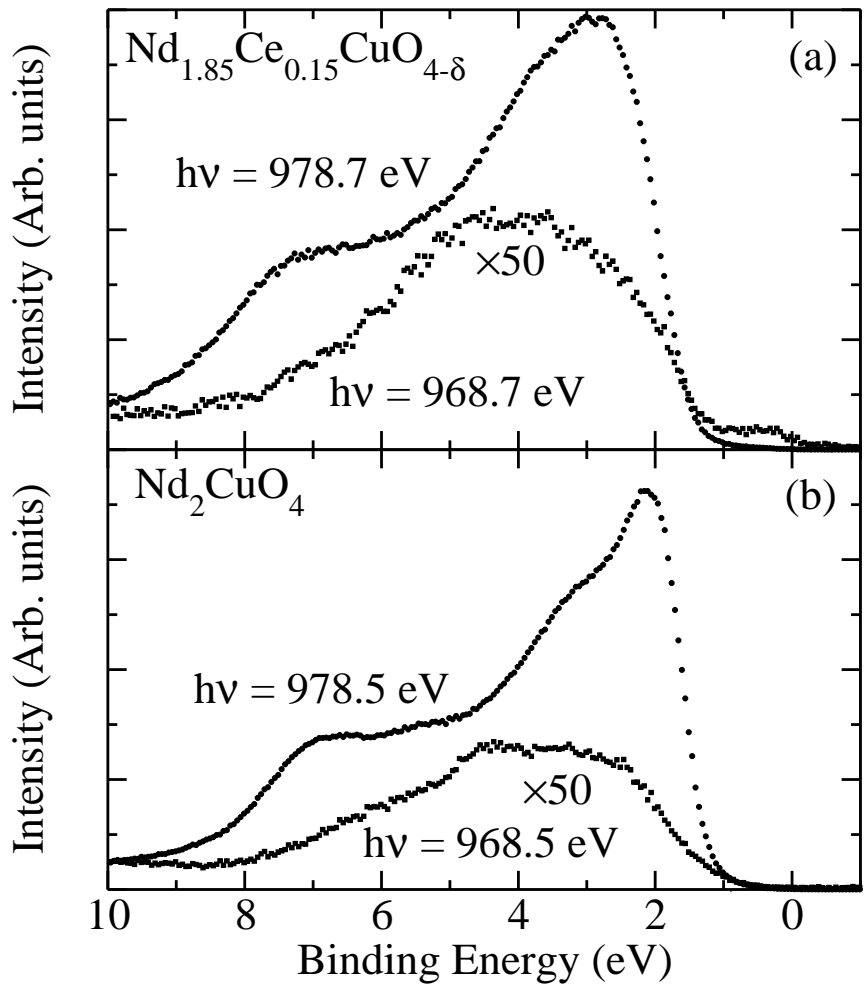}
\caption{On and off resonant photoemission spectra at the Nd
$M_5$ absorption edge.} \label{ndres}
\end{figure}

In comparing \lsco\ with \lco\ using the $^{1}G$ internal
reference, we can also clearly see from the lower panel of figure
\ref{valcom}, that the chemical potential in \lscoc\ resides in
the vicinity of the top of the valence band of \lco, contrary to
earlier reports in which the chemical potential was concluded to
be pinned in the middle of the insulator gap
\cite{namatame,harima}. We suspect that this discrepancy is
caused by the different method for the energy referencing, which
can lead to hidden energy shifts, as we have explained above.

The lower panel of figure \ref{valcom} also shows that the
valence band spectrum of \nco\ is very different in intensity and
also in energy position as compared to the spectrum of \lco. We
can identify two reasons for this. The first is that in the
\ncco\ system the presence of Nd contributes significantly to the
valence band spectrum, while this is not the case for La in
\lsco. From photo-ionization cross-section tables \cite{yeh}, one
can estimate that the Nd $4f$ is responsible for more than 50\%
of the signal in \ncco\ at the photon energies used, while in
\lsco\ the spectrum is dominated by the Cu spectral weight. It is
also important to realize that the Nd $4f$ signal is present over
the entire valence band energy range. This is demonstrated by
figure \ref{ndres}, where we plot the resonant photoemission
spectrum of the \nco\ and \nca\ valence bands, with the photon
energy tuned on the Nd $3d$ absorption edge. A double peak
structure is clearly visible, of which the higher-binding energy
feature can be assigned to screening by charge transfer from the
O $2p$ orbitals (4f$^3$\underline{L} final states) and the
lower-binding energy peak which resonates most strongly is due to
pure Nd $4f$ hole states (4f$^2$ final states)
\cite{namatame,fujimori88}. Similar resonances were found at the
Nd $M_{5}$ \cite{klauda93,klauda94} and Nd $N_{45}$ absorption
edge \cite{grassmann,allen,namatame,weaver}. From these data we
can clearly see that the Nd $4f$ component continues all the way
up to the top of the valence band in \nco. It is not
inconceivable that the top of the valence band in the undoped
\nco\ consists of Nd $4f$ states and not of Cu
$3d^{9}\underline{L}$. It is interesting to note that this could
provide a possible explanation why it is difficult to dope holes
into the CuO$_2$ planes in cuprates with the $T'$-structure, as
these holes would be doped into the Nd-O planes.

Another cause for differences in the valence band spectrum of
\nco\ as compared to \lco\ is laid out in the upper panel of
figure \ref{valcom}. A blow-up of the Cu $L_3$ on-resonant
photoemission spectra shows that the leading edge of the valence
band in \nco\ and also \nca\ is higher in energy by about 0.6 eV
as compared to that of \lco. This number agrees very well with
the fact that the optical gap in \nco\ is about 0.5 eV smaller
than in \lco (which has its first peak in the optical spectrum at
2.0 eV)
\cite{uchida,uchida2}. One could infer that this 0.6 eV difference 
reflects the difference in the effective O $2p$ - Cu $3d$ charge
transfer energy, since this parameter determines to a large
extent the magnitude of the band gap of the insulating correlated
insulator. On hindsight, it is perhaps not surprising to find
such differences in view of the fact that \nco\ and \lco\ have
different crystal structures and O-Cu bond lengths, resulting in
differences in Madelung potentials and O $2p$ - Cu $3d$ hopping
integrals. In fact, there is no compelling reason to believe that
a comparative study of the development of the chemical potential
in \ncco\ can be carried out using \lsco\ as a reference.
\begin{figure}[!htb]  
\vspace{0.3cm} \centering
\includegraphics[width=10.5cm]{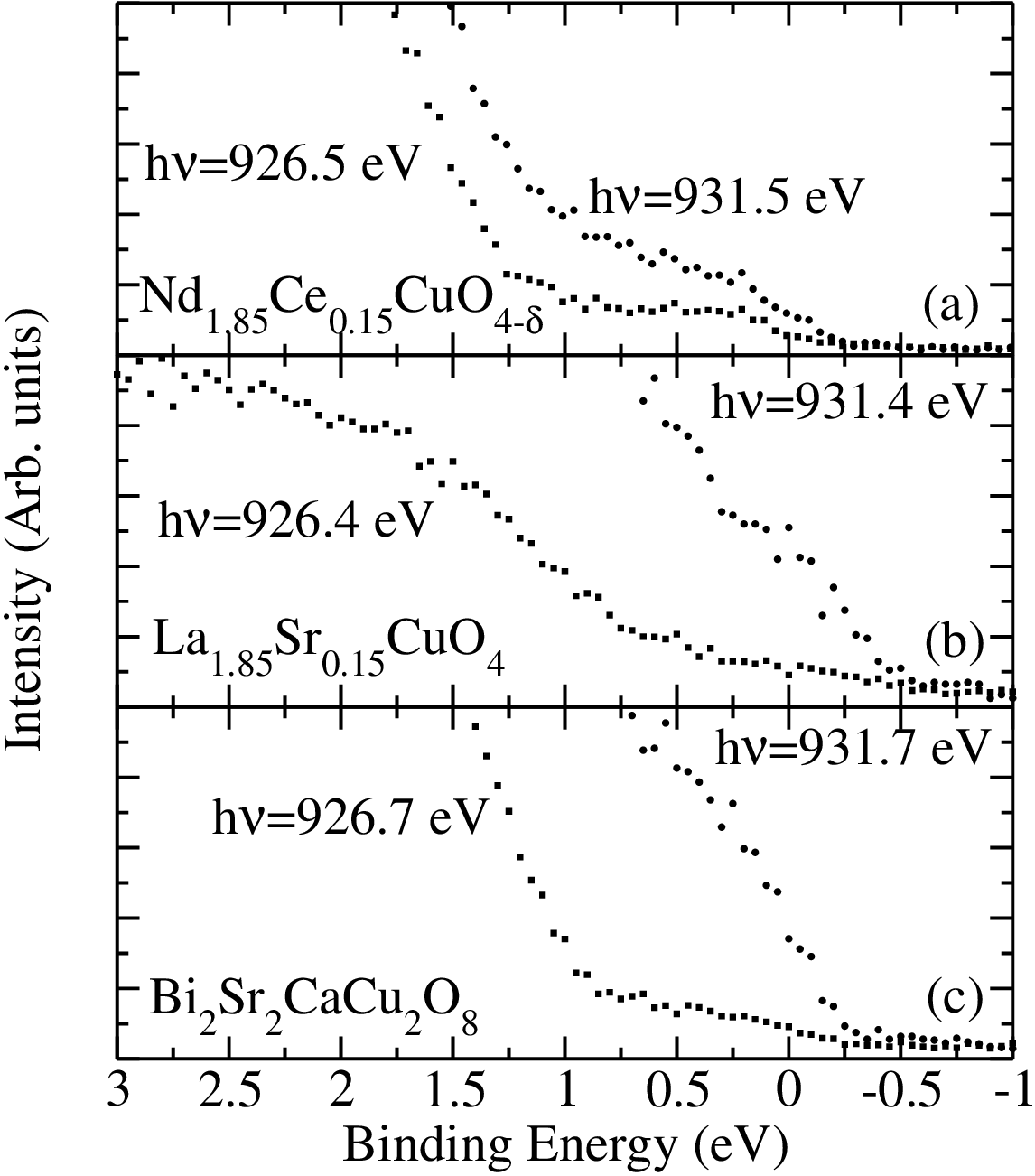}
\caption{On and off resonant photoemission spectra at the Cu
$L_3$ absorption edge.} \label{curesFermi}
\end{figure}

\section{Character of the states near the Fermi level}
Finally, we will study the character of the states closest to the
Fermi level (E$_F$) in the electron doped \ncco\ system. Figure
\ref{curesFermi} shows the on- and off-resonant photoemission
spectrum of \nca\ in the vicinity of the top of the valence band
and the near E$_F$ region. As expected, the on-resonant intensity
is enhanced as compared to the off-resonant one. However, the
enhancement is substantially smaller than that for the hole doped
cuprates, as one can clearly see from figure \ref{curesFermi},
which also includes the on- and off-resonant data for \lscoc\ and
\bisco. Since at the $L_3$ resonance the Cu $3d^{8}$ spectral
weight is being measured, the enhancements for the near E$_F$
states basically indicate the amount of $3d^{8}$ character that
is mixed into these states. For the undoped cuprates, cluster
calculations have shown that the top of the valence band consists
of Zhang-Rice singlets having mainly a $3d^{9}\underline{L}$
character, and also about 8\% $3d^{8}$ admixture due to the
direct hybridization between these two local configurations
\cite{zhang88,eskes91,eskes90}. Subsequent RESPES studies
\cite{tjeng91,tjeng97} have confirmed this picture. Also RESPES
experiments on the hole doped cuprates have shown that the top of
the valence band as well as the near E$_F$ intensity are of
$3d^{9}\underline{L}$ Zhang-Rice singlet nature \cite{brookes01}.
The much smaller resonant enhancement factor for the near E$_F$
intensity in the \nca\ implies a much smaller admixture of the
$3d^{8}$ states, indicating that these near E$_F$ states do not
have a direct hybridization with the $3d^{8}$. We therefore
conclude that these states are not of local $3d^{9}\underline{L}$
character, and we infer that they are in fact doublet $3d^{9}$
final states that can be reached by the photoemission process
starting from a $3d^{10}$ singlet initial state. In this
framework, the doublet $3d^{9}$ has to hybridize first with the
$3d^{9}\underline{L}$ on a neighboring cluster, before it can
acquire the $3d^{8}$ character of the neighboring cluster on
which the resonance process takes place. These data therefore
support the scenario that the charge carriers in \ncco\ are
electrons of singlet $3d^{10}$ like nature.

\section{Conclusions}
To conclude, we have measured that the CuO$_2$ planes in the
\ncco\ high T$_c$ cuprates are electron doped. Using a reliable
internal energy reference we have established that the chemical
potential in \lsco\ and \ncco\ is not pinned in the middle of the
gap of the insulating parent compounds. Instead, it is located in
the vicinity of the top of the valence band for \lsco, and near
the bottom of the conduction band for \ncco. One might say that
it crosses the gap upon going from $p$- to $n$-type doping.
Important is to note that the existence of phase separation
\cite{allen2,harima,damascelli,emery} in the doped materials
could involve the pinning of the chemical potential inside the
gap, but that our results indicate that this pinning could not
bring the chemical potential very far away from the band edges,
with perhaps a value of 0.3 eV as an upper limit judging from our
experimental resolution and uncertainties in determining the
onset of the optical gap. Resonance data suggest that the charge
carriers in \ncco\ are electrons of Cu $3d^{10}$ character, while
in \lsco\ and \bisco\ they are $3d^{9}\underline{L}$ like holes.
As far as symmetry is concerned, the charge carriers in both the
electron and hole doped cuprates are singlet in nature.


\markboth{}{}
\chapter*{Samenvatting} \label{samenvatting}
\markboth{Samenvatting}{} \addtocounter{chapter}{1}
\setcounter{figure}{0}
\addcontentsline{toc}{chapter}{Samenvatting}
\begin{otherlanguage}{dutch}

Een vaste stof bestaat uit atomen, die elk uit een kern met
elektronen zijn opgebouwd. De vaste stof kan ontstaan doordat de
elektronen als een soort lijm werken tussen de kernen. De kernen
zitten daardoor vast in een kristalrooster. De elektronen zitten
echter minder vast en kunnen relatief makkelijk door licht of
door een elektrisch of magnetisch veld bewogen
worden\footnote{Met bewegen bedoel ik zowel veranderingen in
positie als veranderingen in de toestand van het elektron. Het
moet hier worden opgemerkt dat het ook heel weinig energie kost
om de magnetische en trillingstoestanden van de kernen te
veranderen, dit is echter alleen meetbaar met zeer gevoelige
technieken.}, waardoor ze in belangrijke mate de optische,
elektrische en magnetische eigenschappen van een materiaal
bepalen. Zo wordt de kleur van een materiaal grotendeels
veroorzaakt door de interactie van licht met elektronen, is de
elektrische geleiding gekoppeld aan de mobiliteit en concentratie
van elektronen en worden de magnetische eigenschappen bepaald
door de verdeling van spin-op en spin-neer elektronen in het
materiaal\footnote{Spin is een eigenschap van een elektron die
twee waarden kan aannemen: op en neer. Een spin-op elektron
genereert een klein magneetveld omhoog en een spin-neer elektron
eenzelfde veld omlaag.} en hun draaisnelheid om de kernen. Daarom
kunnen de eigenschappen van een materiaal vaak pas begrepen
worden door de gezamenlijke toestand van de elektronen te
onderzoeken.

In dit proefschrift zijn de eigenschappen van europium monoxide
(EuO) bestudeerd. EuO is opgebouwd uit evenveel zuurstof als
europium\footnote{Europium is het element met atoomnummer 63.}
atomen die om en om (NaCl structuur) een kubisch kristalrooster
vormen. Voordat we de eigenschappen van EuO konden bestuderen was
het echter essentieel om kwalitatief hoogwaardige EuO kristallen
te kunnen groeien. Bovendien moesten we in staat zijn om deze
kristallen in dunne lagen lagen (vari\"erend van $\sim$ 1 nm tot
$\sim$ 1 $\mu$m dikte) te kunnen maken om er spin-afhankelijke
metingen aan te kunnen doen. Dit was echter niet de enige reden
waarom dunne lagen onze voorkeur hadden. Dunne lagen van EuO
hebben als voordeel dat ze bij veel lagere temperaturen
(300-400$^\circ$C) gegroeid kunnen worden dan dikke kristallen
(1800$^\circ$C). Bovendien zijn zulke lagen zeer geschikt voor
technologisch onderzoek. Op dunne magnetische lagen kunnen
magnetische gegevens namelijk met hoge dichtheid worden
opgeslagen. Ook zijn er idee\"en om materialen als EuO te
gebruiken in het veld van de spintronica. In dit nieuwe
onderzoeksgebied worden elektrische componenten ontwikkeld (zoals
transistors) die niet alleen gebruik maken van de lading, maar
ook van de spin van de elektronen. Om deze componenten klein
genoeg te maken om op chips te passen is het van groot belang dat
dunne lagen gegroeid kunnen worden. Bovendien zijn de lage
groeitemperaturen van EuO lagen goed te combineren met bestaande
fabricage processen.

Om de dunne lagen te groeien, verdampten we europium metaal op
een substraat, waarbij we zuurstof toelieten\footnote{De groei
van EuO lagen wordt gedetailleerd beschreven in hoofdstuk
\ref{euoepi}}. Hierdoor konden de europium en zuurstof atomen
reageren op het oppervlak van het substraat. We verwachten dat
voor de groei van EuO een heel nauwkeurige dosering van Eu en O
atomen nodig was om precies evenveel europium atomen als zuurstof
atomen met elkaar te laten reageren. Bij het onderzoek naar de
optimale groei-omstandigheden deden we echter een nuttige
ontdekking. Het bleek dat boven 350$^\circ$C de europium atomen
alleen op het oppervlak bleven plakken als ze reageerden met
zuurstof. Wanneer ze niet reageerden verdampten ze namelijk weer.
Dit effect maakte het een stuk makkelijker om de goede verhouding
van zuurstof en europium te krijgen aangezien er nooit een
overschot aan europium atomen kon ontstaan als de temperatuur van
het substraat boven de 350$^\circ$C bleef, omdat het overschot
dan verdampte. Door nu met opzet een overschot van europium
atomen aan te bieden, konden we ook voorkomen dat er een
overschot aan zuurstof in de laag ontstond. Zo konden we
kwalitatief hoogwaardige kristallagen groeien met een nauwkeurige
1:1 Eu:O verhouding. Een ander resultaat was dat lagen met net
iets meer europium dan zuurstof atomen (Eu:O = $\sim$ 1.005:1)
ook gemakkelijk verkregen konden worden door de temperatuur van
het substraat onder de 350$^\circ$ te brengen. Zulke lagen met
iets teveel europium atomen zijn belangwekkend, omdat hun
elektrische geleiding met een factor van meer dan een miljard kan
veranderen wanneer ze magnetisch worden.

Ofschoon EuO een relatief onbekend materiaal is, zijn de
eigenschappen van EuO in veel opzichten zeer interessant. Ten
eerste wordt het ferromagnetisch\footnote{Magnetisch op dezelfde
manier als ijzer: ferrum.} onder een temperatuur van 69 K
($=-204^\circ$C). Dit ferromagnetisme wordt veroorzaakt doordat
de spins van de elektronen steeds meer in dezelfde
richting\footnote{Spins die in deze richting wijzen zullen we
verder spin-op noemen, elektronen met een tegengestelde spin
noemen we spin-neer.} gaan wijzen, waardoor EuO beneden deze
temperatuur meer spin-op dan spin-neer elektronen bevat.
Aangezien elk elektron een magneetveld veroorzaakt, zal dit
verschil tussen de spin-op en spin-neer elektronen dus een
magneetveld tot gevolg hebben.

Op het eerste gezicht lijkt dit misschien niet bijzonder, toch is
ferromagnetisme een zeldzame eigenschap. Dit heeft te maken met
de toestanden waarin elektronen zich bevinden. Een elektron kan
zich in veel verschillende toestanden in een vaste stof bevinden.
Het heeft echter de voorkeur voor toestanden met lage energie.
Alle elektronen zouden dus bij voorkeur in de laagste
energietoestand blijven, ware het niet dat er een fundamentele
restrictie bestaat die verbiedt dat \'e\'en toestand meer dan
\'e\'en elektron van dezelfde spin-soort bevat. In totaal biedt
elke toestand dus maximaal plaats aan twee elektronen. Deze
restrictie wordt het Pauli principe genoemd. Een vaste stof bevat
enorm veel ($\sim 10^{23}$/gram) elektronen. Aangezien de
toestanden met lage energie gevuld raken, zullen de elektronen
gedwongen worden om de hogere toestanden op te vullen tot een
bepaalde energie. Dit is vergelijkbaar met een zee die tot op een
zeker (energie)niveau gevuld is met water. Elk water-molecuul
heeft de laagste energie op de bodem van de zee, maar kan deze
niet bereiken doordat zich daar al andere water-moleculen
bevinden. De 'zee' van elektronen wordt ook wel de Fermi-zee
genoemd en het energieniveau op het oppervlak van de zee de
Fermi-energie. Een belangrijk verschil tussen een echte zee en de
Fermi-zee is dat de Fermi-zee eigenlijk uit twee zee\"en bestaat:
\'e\'en voor spin-op en \'e\'en voor spin-neer elektronen. Omdat
een elektron uit de spin-op zee in de spin-neer zee kan stromen,
door zijn spin om te keren, zullen de oppervlakken van beide
zee\"en bij dezelfde energie liggen.

Nu heeft het elektron behalve zijn spin ook een negatieve lading.
Deze lading zorgt ervoor dat elektronen elkaar afstoten en dus de
laagste energie hebben als ze zich zo ver mogelijk uit elkaar
bevinden. Omdat elektronen met \emph{ongelijke} spin zich volgens
het Pauli principe in dezelfde ruimtelijke toestand kunnen
bevinden, zullen ze zich over het algemeen dichter bij elkaar
bevinden. Dit resulteert erin dat elektronen zoveel mogelijk
\emph{gelijke} spin aannemen, aangezien ze zich hierdoor verder
uit elkaar bevinden en dus een lagere energie hebben, omdat ze
elkaar minder afstoten. Dit mechanisme wordt \emph{exchange}
genoemd. Een materiaal krijgt echter alleen meer spin-op dan
spin-neer elektronen als de energiewinst door exchange groter is
dan het energie-verlies door het verhogen van het Fermi-oppervlak
van de spin-op Fermi-zee ten opzichte van de spin-neer
Fermi-zee\footnote{In dit geval resulteert exchange erin dat de
spin-op Fermi-zee effectief dieper is. Een waarschuwing is hier
op zijn plaats. Eigenlijk is de energie van \'e\'en elektron niet
meer goed bepaald als het zich in een systeem met veel elektronen
bevindt, aangezien zijn energie kan afhangen van de toestand van
de andere elektronen (zoals in het geval van exchange). Daarom is
de Fermi-zee waarin de energie van elk elektron bepaald wordt
door zijn diepte een conceptuele benadering. Deze benadering
wordt in dit geval ter simplificatie toegepast, maar dit is niet
voor alle vaste stoffen een goede benadering.}.

Vaak is het effect van exchange klein, doordat elektronen met een
energie in de buurt van het Fermi-oppervlak zich meestal al ver
uit elkaar bevinden. Daarom bevatten de meeste materialen
evenveel spin-op als spin-neer elektronen. In EuO is echter iets
bijzonders aan de hand: de elektronen met een energie dichtbij
het Fermi-oppervlak bevinden zich in een zogenaamde $4f$ schil
die in de buurt van de sterk positief geladen atoomkern van
europium ligt. Deze atoomkern trekt de negatief geladen
elektronen sterk naar zich toe. Dit heeft als bij-effect dat deze
$4f$-elektronen ook sterk naar elkaar toegetrokken worden en
daardoor een veel groter effect van exchange ondervinden. Deze
exchange is zo groot dat alle $4f$ elektronen die zich bij
dezelfde atoomkern bevinden dezelfde spin krijgen. Dit is echter
nog niet genoeg voor ferromagnetisme, aangezien de spins van de
elektronen bij verschillende atoomkernen nog verschillend kunnen
zijn. Doordat de $4f$-elektronen op verschillende europium atomen
echter ook een (in dit geval indirecte) exchange wisselwerking
met elkaar hebben, krijgen alle $4f$ elektronen dezelfde spin en
is EuO ferromagnetisch\footnote{Het feit dat deze indirecte
wisselwerking voor ferromagnetisme zorgt, is eigenlijk een van de
meest bijzondere eigenschappen van EuO en wordt nog steeds niet
precies begrepen.}.

Deze indirecte exchange wisselwerking tussen de spins van
$4f$-elektronen van verschillende europium atomen is echter een
factor $\sim 1000$ kleiner dan de exchange tussen $4f$-elektronen
op hetzelfde atoom. Bij temperaturen boven de Curie temperatuur
van 69 K, is de indirecte wisselwerking dan ook niet meer sterk
genoeg om te zorgen dat de spins van verschillende atomen in
dezelfde richting wijzen, omdat de energie lager is als de spins
van verschillende atomen willekeurig in alle richtingen kunnen
bewegen. Dit is te vergelijken met het smelten van een vaste
stof, waarbij de ordening van het kristalrooster verdwijnt om de
bewegingsvrijheid van de atomen te vergroten. Boven de Curie
temperatuur 'smelt' het spin-rooster waardoor het ferromagnetisme
verdwijnt.

Er is nog een belangrijk verschil tussen een normale zee en een
Fermi-zee. Ofschoon er in een normale zee over het algemeen op
elke diepte water te vinden is, zijn de diepten van de Fermi-zee
veel grilliger. Op sommige diepten zijn veel elektronen te
vinden, terwijl er op veel andere diepten geen toestanden voor de
elektronen zijn. De Fermi-zee bestaat dus eigenlijk uit vele, met
elkaar verbonden, grotten op verschillende diepten, die ieder
toestanden bevatten. Deze grillige structuur geeft de
energieverdeling weer van de vele mogelijke banen die een
elektron door het rooster\footnote{Net als voor elektronen geldt
ook voor licht dat het zich niet met alle energie\"en (=kleuren)
door een rooster kan voortplanten. Een voorbeeld hiervan zijn de
kleuren die door het rooster van lijnen op een compact disc
worden gereflecteerd.} van atoomkernen en andere elektronen kan
volgen. De energie van een elektronenbaan wordt bepaald door de
lading van het elektron, die ervoor zorgt dat het wordt
aangetrokken door de kernen, maar ook door zijn bewegingsenergie,
die juist verhoogd wordt door in de buurt van \'e\'en kern te
blijven.

Als dit grottenstelsel waaruit de Fermi-zee bestaat gevuld is met
alle elektronen in de vaste stof kan het Fermi-oppervlak ergens
halverwege een grot liggen. Het kan echter ook zo zijn dat de
toestanden in \'e\'en grot volledig gevuld zijn, maar dat de grot
die daar direct boven ligt, geen enkel elektron bevat. Deze twee
situaties zijn essentieel verschillend, omdat ze heel andere
elektrische geleidingseigenschappen tot resultaat hebben.
Elektrische geleiding kan namelijk alleen optreden als er golven
over het oppervlak van de Fermi-zee kunnen lopen. Als het
Fermi-oppervlak ergens halverwege een grot ligt kunnen er golven
lopen en gedraagt het materiaal zich als een metaal. Als de grot
volledig vol is zijn golven echter onmogelijk en is het materiaal
elektrisch isolerend. Toch kan er een (kleine) elektrische
geleiding optreden als de geheel gevulde grot slechts weinig
onder de daarboven gelegen lege grot ligt. Bij hoge temperaturen
kunnen elektronen uit de lager gelegen grot namelijk opspatten
naar de hogere grot en door over de bodem van deze grot te golven
geleiden ze toch elektrische stroom. Materialen met deze
eigenschap worden halfgeleiders genoemd omdat ze alleen bij hoge
temperaturen elektriciteit geleiden.

Dit brengt ons bij de tweede interessante eigenschap van EuO: EuO
is een halfgeleider. Op zich is dit niets bijzonders, er bestaan
vele halfgeleiders, maar een ferromagnetische halfgeleider is
zeer zeldzaam. Het is zelfs zo dat het bestaan van materialen die
zowel ferromagnetisch als halfgeleidend zijn sterk betwijfeld
werd voor de ontdekking van EuO in het begin van de zestiger
jaren. De excentriciteit van zulke materialen kan begrepen worden
als we de voorwaarden beter bekijken. Ten eerste moet het
materiaal meer spin-op dan spin-neer elektronen bevatten en ten
tweede moeten alle grotten vol of leeg zijn omdat het anders een
geleider is. Deze combinatie van factoren kan alleen voorkomen
als er meer spin-op grotten volledig gevuld zijn dan spin-neer
grotten. Daarom moet de exchange zo groot zijn dat het verschil
in diepte tussen een spin-op en neer grot groter is dan de
afstand tussen de bodem en het plafond van deze grot.
Natuurkundigen noemen het \ grottenstelsel van een vaste stof de
\emph{toestandsdichtheid} en een grot wordt een \emph{band}
genoemd. We zullen daarom verder deze termen gebruiken.

\begin{figure}[!htb]
\centerline{\includegraphics[width=12cm]{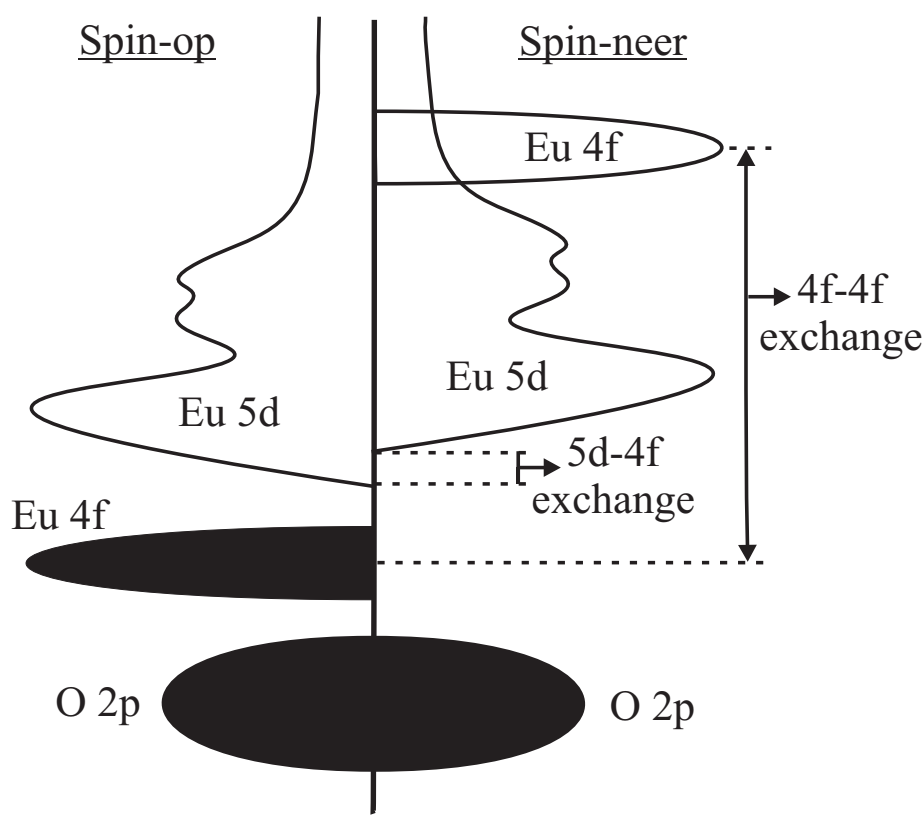}}
\caption{\label{DOSsimpel} Schematische weergave van de
toestandsdichtheid van EuO bij een temperatuur van 0 K, zoals bepaald
met spin-afhankelijke fotoemissie en r\" ontgen absorptie
spectroscopie. De gevulde banden zijn zwart gekleurd. De verschillen
tussen de spin-op en spin-neer toestandsdichtheid als gevolg van
exchange zijn duidelijk zichtbaar.}
\end{figure}

Een belangrijk resultaat van het onderzoek dat in hoofdstuk
\ref{Spinxas} van dit proefschrift beschreven wordt, is dat het
ons gelukt is om de toestandsdichtheid van EuO te meten voor
zowel spin-op als spin-neer elektronen. Om de gevulde banden van
EuO te meten hebben we gebruik gemaakt van de foto\"emissie
techniek. Bij deze techniek wordt het materiaal beschenen met
fotonen (lichtdeeltjes) van een precies bekende energie. De
energie van de fotonen kan gebruikt worden om een elektron uit
het materiaal te schieten. Door nu de energie van het elektron
buiten het materiaal te meten, kunnen we bepalen hoeveel energie
verloren is gegaan bij het verwijderen van het elektron uit zijn
band, waardoor we weten hoe diep de band ligt waarin het elektron
zich bevond. Door bovendien de spin van het elektron te meten
kunnen we de bezette toestandsdichtheid bepalen van zowel de
spin-op als spin-neer elektronen.

Het is veel lastiger om de onbezette toestandsdichtheid voor
verschillende spins te meten, omdat een onbezette toestand alleen
gemeten kan worden door hem eerst te bezetten met een elektron.
Bovendien moeten de energie en spin van zo'n sonde-elektron
nauwkeurig bepaald zijn. Om dit te doen hebben we een nieuwe
spectroscopische techniek ontwikkeld: spin afhankelijke r\"ontgen
absorptie spectroscopie. Deze techniek werkt als volgt: we
gebruiken weer een r\"ontgen-foton met een bekende energie, die
een sonde-elektron vanuit een diepe band\footnote{Er bestaat ook
een andere techniek om de onbezette toestanden te meten: inverse
foto\"emissie. Deze techniek maakt gebruik van een sonde-elektron
dat van buiten op het materiaal wordt geschoten. Beide technieken
hebben hun voor- en nadelen.} omhoog schiet naar \'e\'en van de
onbezette banden (welke onbezette band bereikt wordt, kan
geregeld worden met de energie van het foton). Er blijft nu een
gat over dat zich in het spin-op of spin-neer kanaal van de diepe
band bevindt afhankelijk van de spin van het sonde-elektron. Om
te bepalen wat de spin van de onbezette band is, moeten we de
spin van het sonde-elektron meten. Hiervoor maken we gebruik van
een proces waarbij een elektron naar beneden valt om het
achtergelaten gat in de diepe band te vullen. Dit gat kan
namelijk alleen opgevuld worden als het elektron dezelfde spin
heeft als het gat. De energie die dit elektron hierbij wint,
gebruikt het om een ander elektron met de \emph{omgekeerde} spin
uit het materiaal te lanceren,\footnote{Dit kan vergeleken worden
met een acrobaat die op een wip-wap springt en zo een andere
acrobaat die op zijn kop staat lanceert.} waarna we de spin van
dit gelanceerde elektron detecteren. De spin van het elektron dat
uit het materiaal komt, is dus omgekeerd aan dat van het gat dat
achtergelaten was en daarom ook omgekeerd aan de spin van de
onbezette band waarnaar het sonde-elektron was geschoten (zie
figuur \ref{fig3} voor een schema van deze techniek). Door nu bij
veel verschillende foton energie\"en de spin van deze elektronen
die uit het materiaal komen te meten, kunnen we de
spin-afhankelijke toestandsdichtheid van de onbezette banden
bepalen.

Een schematische weergave van de aldus bepaalde spin-afhankelijke
toestandsdichtheid in EuO van zowel de bezette als onbezette
banden is gegeven in figuur \ref{DOSsimpel}. De banden zijn
gemerkt met Eu of O om aan te geven of de elektronen in de band
zich vooral in de buurt van de europium of de zuurstof
atoomkernen bevinden. Bovendien is er een index ($2p$, $4f$ of
$5d$), die aangeeft in wat voor soort banen de elektronen in de
band om de atoomkern bewegen. Onze metingen laten zien dat het
ferromagnetisme in EuO inderdaad samengaat met een groter aantal
spin-op dan spin-neer elektronen in de $4f$ band. Bovendien is
het duidelijk dat EuO een halfgeleider is, aangezien de
scheidslijn tussen bezette en onbezette toestanden (het
Fermi-oppervlak) zich niet \emph{in} een band bevindt, maar
\emph{tussen} de geheel gevulde $4f$ spin-op en de lege Eu $5d$
banden ligt. Het grote verschil in energie tussen de spin-op en
spin-neer $4f$ band is grotendeels een gevolg van de grote
$4f-4f$ exchange wisselwerking tussen deze elektronen. Een
opmerkelijk resultaat is echter dat we ook een
energieverschuiving tussen de onbezette spin-op en spin-neer $5d$
banden waarnemen. We schrijven deze verschuiving toe aan de
exchange wisselwerking tussen de $5d$ band en de elektronen in de
$4f$ band. Ofschoon de $5d$ elektronen andere banen beschrijven
die een stuk verder van de Eu kernen liggen dan de banen van de
$4f$ elektronen, komen ze toch nog behoorlijk dicht bij de $4f$
elektronen. Daarom is er ook een exchange effect tussen de $4f$
en $5d$ elektronen, alhoewel dit ongeveer een faktor 10 kleiner
is dan dat tussen de $4f$ elektronen onderling. De exchange
tussen $5d$ en $4f$ elektronen is grotendeels verantwoordelijk
voor de interessante optische en elektrische eigenschappen van
EuO en de studie van de kenmerken, oorzaken en gevolgen van deze
exchange loopt dan ook als een rode draad door dit proefschrift.

De optische eigenschappen van EuO zijn interessant omdat de kleur van
EuO aanzienlijk verandert als het ferromagnetisch wordt. Deze
kleurverandering is ge\"\i nter-preteerd als een gevolg van de $5d-4f$
exchange. Om een beter beeld van de veranderingen in de $5d-4f$
exchange te krijgen hebben we de toestandsdichtheid bij veel
temperaturen gemeten met r\"ontgen absorptie spectroscopie. We vonden
aanwijzingen dat de energieverschuiving tussen de spin-op en spin-neer
$5d$ band niet zozeer een resultaat is van de exchange tussen een
$5d$-elektron met $4f$ elektronen op hetzelfde atoom (zoals bij de
$4f-4f$ exchange), maar dat de exchange voor zo'n 75\% het gevolg is
van de wisselwerking tussen een $5d$-elektron en de $4f$ elektronen op
een groot aantal verschillende Eu atomen. Dit komt doordat een $5d$
elektron niet sterk gebonden wordt aan \'e\'en atoom, maar vrij is om
zich door het EuO rooster te verplaatsen. Daardoor komt het in de
buurt van veel verschillende Eu atomen en heeft dus ook een $4f-5d$
exchange bijdrage van de $4f$ elektronen van al deze atomen. De
consequentie hiervan is, dat als de $4f$ spins van de afzonderlijke
atomen zich in willekeurige richtingen gaan bewegen, boven de Curie
temperatuur, de gemiddelde bijdrage aan de $5d-4f$ exchange van het
$5d$ elektron naar nul gaat. Daarom is de energieverschuiving als
gevolg van de $5d-4f$ exchange, evenredig met de magnetisatie van
EuO\footnote{De magnetisatie geeft aan hoe groot het verschil is
tussen het aantal spin-op en spin-neer $4f$-elektronen. Bij een
ferromagnetisch materiaal is de magnetisatie 100\% (dat wil zeggen:
alle $4f$-elektronen zijn spin-op) bij temperaturen ver onder de Curie
temperatuur en is de magnetisatie nul boven de Curie temperatuur. De
magnetisatie kan echter vergroot worden door het materiaal in een
magneetveld te plaatsen.}.

De kleurverandering van EuO bij de Curie temperatuur is een
gevolg van de temperatuurafhankelijkheid van de $5d-4f$ exchange,
omdat de kleur sterk bepaald wordt door de absorptie van een
foton door een $4f$ elektron, dat de energie van het foton
gebruikt om in een hogere band te komen. De laagste energie
waarbij dit proces kan plaatsvinden is gelijk aan het
energieverschil tussen het plafond van de $4f$ band en de bodem
van de $5d$ band. De verschuiving van de $5d$ band als gevolg van
$5d-4f$ exchange verkleint dit energieverschil en resulteert dus
in een kleurverandering\footnote{De grote effecten van een extern
magneetveld en de licht-polarisatie op de kleur van EuO worden in
hoofdstuk \ref{optprop} gemodelleerd en uitgerekend.}.

Misschien wel het meest spectaculaire effect treedt op in de
geleidingseigenschappen van EuO. Als EuO iets meer europium dan
zuurstof atomen bevat (Eu-rijk EuO), ontstaat er een klein
aantal, volledig gevulde toestanden, die zich vlak onder de $5d$
band bevindt. Deze toestanden ontstaan doordat elk zuurstof atoom
normaliter twee elektronen van een europium atoom naar zich toe
trekt. Als er echter meer europium dan zuurstof atomen zijn,
blijven er een aantal elektronen op europium over, die niet naar
het zuurstof gaan. De toestanden die door deze elektronen gevormd
worden zijn volledig gevuld, waardoor Eu-rijk EuO boven de Curie
temperatuur nog steeds een halfgeleider is. Als de Eu $5d$ band
echter gaat verschuiven als gevolg van $5d-4f$ exchange, kruist
hij deze extra toestanden, waardoor de elektronen uit deze
toestanden vrijkomen. De toestanden smelten namelijk samen met de
$5d$ band en er ontstaat een veel grotere band, waarin zich
alleen het kleine aantal elektronen bevindt dat eerst in de extra
toestanden zat. De grotere band is nu dus slechts voor een klein
deel gevuld waardoor er golven kunnen stromen. Daarom gaat
Eu-rijk EuO zich als een geleider gedragen beneden de Curie
temperatuur. Zo'n overgang waarbij een materiaal verandert van
halfgeleider naar geleider, wordt wel een halfgeleider-metaal of
isolator-metaal overgang genoemd. Deze overgang is bijzonder
spectaculair bij EuO, omdat de weerstand met een faktor van meer
dan een biljoen kan veranderen. In hoofdstuk \ref{Magres}
presenteren we metingen van deze metaal-isolator overgang in EuO
lagen, die een goede overeenkomst vertonen met een model op basis
van de bovenstaande beschrijving. Bovendien laten we in dit
hoofdstuk metingen zien die aantonen dat een gelijksoortige
isolator-metaal overgang ge\"\i nduceerd kan worden in EuO, dat
evenveel Eu als O atomen bevat, als dit beschenen wordt door een
lichtbundel. Ook hebben we de geleidingseigenschappen van Eu-rijk
EuO bij verschillende aangelegde magneetvelden bestudeerd. Zeer
opmerkelijk was de ontdekking dat de elektrische geleiding alleen
afhangt van de magnetisatie, onafhankelijk van het feit of deze
ge\"\i nduceerd is door het materiaal af te koelen of door het
aanleggen van een magneetveld. Deze universele afhankelijkheid
van de elektrische geleiding van alleen de magnetisatie is een
sterke aanwijzing dat de geleidingseigenschappen van EuO
voornamelijk veroorzaakt worden door de ge\"\i nduceerde $5d-4f$
exchange interactie, zoals in het berekende model.

Het onderzoek in dit proefschrift heeft geleid tot een beter
begrip van de spin- en temperatuurafhankelijke toestandsdichtheid
van de ferromagnetische halfgeleider EuO en de relatie hiervan
met de optische, elektrische, opto-elektrische en
magneto-elektrische eigenschappen. Daarbij hebben we het
groeimechanisme voor EuO dunne lagen ontwikkeld en verklaard en
is een nieuwe spectroscopische techniek ontwikkeld. Het onderzoek
roept echter ook nieuwe vragen over EuO op (zie met name
hoofdstuk \ref{Introduction} en \ref{Magres}). Bovendien zou het
zeer interessant zijn om te onderzoeken of de mechanismen die de
eigenschappen van EuO bepalen, ook verantwoordelijk zijn voor
vergelijkbare fenomenen in andere materialen. Ik hoop daarom dat
dit proefschrift niet alleen een bijdrage aan het begrip van EuO
heeft geleverd, maar ook een steun en impuls zal zijn voor
toekomstig fundamenteel en toegepast onderzoek met EuO en
gelijksoortige materialen.
\end{otherlanguage}
\markboth{References}{}
\bibliographystyle{apsrev}

\markboth{}{}
\chapter*{Publications}
\label{publist} \markboth{Publications}{}
\addcontentsline{toc}{chapter}{Publications}
\begin{itemize}
\item T. Mizokawa, L.~H. Tjeng, P.~G. Steeneken, N.~B. Brookes, I. Tsukada, T. Yamamoto and K.
Uchinokura. \\
{\it Photoemission and x-ray-absorption study of misfit-layered
(Bi,Pb)-Sr-Co-O compounds: Electronic structure of a hole-doped Co-O
triangular
lattice} \\
Phys. Rev. B {\bf 64}, 115104 (2001)
\item K. Schulte, M.~A. James, P.~G. Steeneken, G.~A. Sawatzky, R. Suryanarayanan, G. Dhalenne and A.
Revcolevschi. \\
{\it Weight of zero-loss electrons and sum rules in extrinsic processes that can influence photoemission spectra} \\
Phys. Rev. B {\bf 63}, 165429 (2001)
\item K. Schulte, M.~A. James, L.~H. Tjeng, P.~G. Steeneken, G.~A. Sawatzky, R. Suryanarayanan, G. Dhalenne and A.
Revcolevschi. \\ {\it Work function changes in the double layered
manganite La$_{1.2}$Sr$_{1.8}$Mn$_2$O$_7$} \\ Phys. Rev. B {\bf 64},
134428 (2001)
\item J. Min\'ar, H. Ebert, L.~H. Tjeng, P. Steeneken, G. Ghiringhelli, O. Tjernberg and N. B.
Brookes. \\ {\it Theoretical description of  the Fano-effect in the
angle-integrated valence-band photoemission of paramagnetic solids} \\
Appl. Phys. A {\bf 73}, 663 (2001)
\item P.~G. Steeneken, L.~H. Tjeng, I. Elfimov, G.~A. Sawatzky, G. Ghiringhelli, N.~B. Brookes and D.~-J.
Huang. \\ {\it Exchange splitting and charge carrier spin-polarization in EuO} \\ Phys. Rev. Lett. {\bf 88}, 047201 (2002) \\
\url{http://xxx.lanl.gov/abs/cond-mat/0105527}
\item P.~G. Steeneken. \\ Oral presentation on Talksplanet website \\
\url{http://www.talksplanet.com/topics/phys0108001}
\newpage
\item G. Ghiringhelli, N.~B. Brookes, L.~H. Tjeng, T. Mizokawa, O. Tjernberg, P. G. Steeneken and A. A.
Menovsky. \\
{\it Probing the singlet character of the two-hole states in cuprate superconductors} \\
Physica B {\bf 312-313}, 34 (2002)
\item O.~Tjernberg, L.~H. Tjeng, P.~G. Steeneken, G. Ghiringhelli, A.~A. Nugroho, A.~A. Menovsky and N.~B.
Brookes. \\
{\it Existence and stability of the Zhang-Rice singlet in
Nd$_{2-x}$Ce$_x$CuO$_{4-\delta}$} \\ Submitted to Phys. Rev. Lett.
\item P.~G. Steeneken, L.~H. Tjeng, M.~V. Tiba, R. Bhatia, W. Eerenstein, T. Hibma and G.~A.
Sawatzky. \\ {\it Growth of EuO films with controlled properties}
\\ Chapter \ref{euoepi} of this thesis, in preparation.
\item P.~G. Steeneken, L.~H. Tjeng, G.~A. Sawatzky, A. Tanaka, O.
Tjernberg, G. Ghiringhelli, N.~B. Brookes, A.~A. Nugroho and A.~A.
Menovsky. \\ {\it Nature of the states near the chemical potential in
Nd$_{2-x}$Ce$_x$CuO$_{4-\delta}$} \\
Chapter \ref{NCCO} of this thesis, in preparation.
\item P.~G. Steeneken, L. H. Tjeng, G. A. Sawatzky {\it et al.} \\ Parts of chapters \ref{Tempxas},
\ref{oxmcd} and \ref{Magres} of this thesis. \\ To be published.
\end{itemize}

\markboth{}{}
\chapter*{Acknowledgements}
\markboth{Acknowledgements}{}
\addcontentsline{toc}{chapter}{Acknowledgements}
\begin{otherlanguage}{dutch}
Ofschoon artikel 1 van het promotiereglement het proefschrift
betitelt als een proeve van bekwaamheid tot het
\emph{zelfstandig} beoefenen van de wetenschap, lijkt mij
bekwaamheid in het beoefenen van de wetenschap in
samenwerkingsverband minstens zo belangrijk. Bij het schrijven
van dit proefschrift ben ik dan ook niet onafhankelijk geweest
van de samenwerking, steun en hulp van vele anderen, die ik in
dit hoofdstuk hiervoor hartelijk wil bedanken.

Allereerst wil ik mijn beide promotores George Sawatzky en Hao
Tjeng bedanken. Beste George, toen ik begon in jouw groep met als
opdracht om overgangsmetaaloxiden met behulp van
elektronenspectroscopie te bestuderen, had ik nooit kunnen
vermoeden waar dit toe zou leiden. Jouw idee om te proberen een
spin polarizator van EuO te maken was de aanzet tot de studie van
dit interessante materiaal. Een polarizator heb ik er niet van
kunnen maken, toch heeft dit idee een geweldige 'spin-off' gehad
in de vorm van dit proefschrift. Jouw intu\"\i tie voor
fascinerende onderzoeksonderwerpen, gecombineerd met je enorme
inzicht, kennis en ervaring hebben mij heel erg geholpen en veel
geleerd. Bedankt voor alle aanmoediging, discussies en idee\"en.

Hao, jouw begeleiding heeft niet alleen mijn onderzoek, maar ook
mijn manier van onderzoeken enorm verbeterd. Jouw idee van de
experimenteel fysicus als all-rounder die een experiment van de
bouw van de opstelling en de sample-groei tot de theoretische
analyse van de metingen beheerst, heeft grote indruk op mij
gemaakt. Ik heb ook veel geleerd van jouw zorgvuldige werkwijze,
waarbij het experiment niet zozeer bepaald wordt door de
beschikbare apparatuur, maar waarbij de apparatuur aangepast
wordt om de gewenste meting te kunnen doen. Hierdoor ben ik
namelijk met een groot aantal experimentele technieken in
aanraking gekomen. Hoogtepunt in onze samenwerking vond ik de
verscheidene metingen bij het synchrotron in Grenoble. Niet
alleen zijn de belangrijkste metingen voor dit proefschrift daar
gedaan, ook veel van de concepten voor de interpretatie van deze
metingen zijn tijdens onze nachtelijke discussies bij de beamline
ontstaan. Bovendien heeft ook jouw brede kennis en inzicht van
spectroscopie en vaste stof fysica een zeer belangrijke bijdrage
aan het tot stand komen van dit proefschrift geleverd. Bedankt
voor dit alles en voor je onuitputtelijke enthousiasme voor de
fysica, het was en is een grote stimulans voor me.

Een groot deel van de transport en fotoemissie studies aan EuO
zijn gedaan in de mooie opstelling die Ronald Hesper tijdens zijn
promotieonderzoek geconstrueerd heeft. Ronald, bedankt voor al je
hulp bij mijn onderzoek. Jouw kunde wat betreft elektronica,
chemie en computers, gecombineerd met je zorgvuldigheid, precisie
en indrukwekkende hulpvaardigheid, maken je een grote aanwinst
voor elk laboratorium. Verder heb ik ook je gezelschap en onze
vele interessante discussies in de mensa zeer gewaardeerd.

Arend, ik heb veel geprofiteerd van jouw technische ervaring en
de vele opstellingen die je ontwikkeld hebt, bedankt! Karina, ik
heb onze gesprekken en de samenwerking tijdens mijn eerste jaar
bij de EELS metingen aan Bi 2212 op prijs gesteld.
\end{otherlanguage} Alex, thanks for coping with an experimentalist
as a roommate and for our fascinating conversations about science
and life. Ilya, your calculations and our discussions on EuO were
a great support and stimulant for this work, I hope you will
continue your EuO research. Dear Oana, thanks for the happiness
you brought to the lab. Salvatore, your pioneering work with the
mini-MBE setup has been a great support for the growth of the EuO
films. Takashi, it was a privilege to benefit from your large
scientific experience during the EELS measurements and the work
on the cobaltates. Veronica and Rakesh, thanks for your
contributions to the EuO project. Christian, I appreciate your
help with the ellipsometry measurements very much, thanks.
\begin{otherlanguage}{dutch} Wilma, bedankt voor je hulp bij de
groei en magnetoweerstandsmetingen van de EuO lagen. Bedankt
Harry, dat je vervangende groepsleider bent geweest het laatste
jaar.
\end{otherlanguage}

Thanks to the people at beamline ID12B (now ID08) of the ESRF,
and especially to Nick, Giacomo, Oscar and Kenneth. Without your
nice beamline with spin-detector, many of the measurements in
this thesis could not have been done. Also your comments on the
EuO and \ncco\ work were very useful.

The high quality samples of the cuprates which were studied in chapter
\ref{NCCO}, were provided by Agung Nugroho and A. A. Menovsky. Thanks
Agung, for the many discussions of the work and for your company at
the conference in Trieste. K.~J. Fischer kindly provided us with the
nice EuO crystals that were studied in chapter \ref{Tempxas}.

\begin{otherlanguage}{dutch}
De hulp bij verschillende experimentele facetten heb ik zeer op
prijs gesteld. Bedankt Ferry, Frans, Cor, Arjen, Luc, Minte,
Siemon, Bernard, Henk (2$\times$) en Jacob. Ook de kunststukjes
die door de mechanische en electronische werkplaats werden
afgeleverd waren voor dit werk van belang. In het bijzonder wil
ik Jan Kappenburg en Leo Huisman hiervoor bedanken.
\end{otherlanguage}

Interpretations and results in this work have benefitted much
from discussions with and comments from many people. In
particular I want to thank the reading committee Meigan Aronson,
Ronald Griessen and Dick van der Marel for their careful reading
and comments on the manuscript. Useful discussions with Daniil
Khomskii, Tjip Hibma, Frank de Groot, Arata Tanaka, D.-J. Huang,
Paul van Loosdrecht, Thom Palstra, Bart van Wees, Maurits
Haverkort, Peter Armitage, Zhi-Xun Shen, Atsushi Fujimori and Jim
Allen are also gratefully acknowledged.

I would also like to thank all the other friendly and helpful
people who make the Groningen physics department such a good and
interesting place to be. My (former) group members Ruth, Sjoerd,
Andri, Mark, Peter, Yutaka, Sarker, Michel, Mirwais, Hans,
Anna-Maria, Anita, Renate en Brigitta. The people from the
optical solid state physics group, made me feel like I was still
a member of their group. Thanks Hajo, Patricio, Alexey,
Katarzyna, Artem, Johan, Andrea, Markus, Jeroen, Herpertap,
Christine and Sagar. Thanks to the people from the physics of
nanodevices and physical chemistry groups: Friso, Jochem, Diana
D., Andrei, Michele, Hubert, Mayke, Jorden, Michail, Sjoerd,
Szilly, Diana R., Frans, Bas, Christine and Gaby.
\begin{otherlanguage}{dutch} Verder wil ik de mensen van didactiek en
opleiding bedanken voor de prettige samenwerking en Whee Ky voor
zijn aanstekelijke enthousiasme.

Mijn dank gaat ook uit naar al mijn familie en vrienden, die een
belangrijke bijdrage aan dit proefschrift hebben geleverd door
hun interesse en aanmoediging. Bedankt Marleen, Arjen, Michiel,
Janneke, oma, Eugen, Ika, Caroline, Floris (ook voor je hulp als
paranimf) en alle anderen!

In het bijzonder wil ik mijn ouders bedanken voor hun
onvoorwaardelijke vertrouwen, steun, en advies. Zonder jullie
stimulatie al die jaren was dit proefschrift er niet gekomen.

Lieve Nicole, ik besef dat mijn nachtelijke metingen, weekenden
dat ik doorwerkte en de vele momenten dat ik afwezig was, omdat
ik aan mijn onderzoek dacht, niet altijd gemakkelijk voor jou
geweest zijn. Ontzettend bedankt voor je begrip de laatste 4.5
jaar en voor al je liefde, steun, vrolijkheid, hulp en optimisme!
\end{otherlanguage}
\markboth{}{}
\newpage
\begin{center}
\end{center}
\newpage
\begin{center}
\end{center}
\thispagestyle{empty}
\newpage
\begin{center}
\end{center}
\thispagestyle{empty}
\end{document}